\documentclass[aps,pre,preprint,floatfix]{revtex4}

\usepackage{latexsym}

\usepackage{amsmath}    
\usepackage{graphicx}

\begin{document}






\begin{center}
\Large{Mathematical Physics of Cellular Automata}\\
\large{Vladimir Garc\'{\i}a-Morales}\\
\small{Institute for Advanced Study -  Technische Universit\"{a}t M\"{u}nchen,\\ Lichtenbergstr. 2a, D-85748 Garching, Germany \\
vmorales@ph.tum.de}
\end{center}
\small{\noindent A universal map is derived for all deterministic 1D cellular automata (CA) containing no freely adjustable parameters. The map can be extended to an arbitrary number of dimensions and topologies and its invariances allow to classify all CA rules into equivalence classes. Complexity in 1D systems is then shown to emerge from the weak symmetry breaking of the addition modulo an integer number $p$. The latter symmetry is possessed by certain rules that produce Pascal simplices in their time evolution. These results elucidate Wolfram's classification of CA dynamics.}
\tableofcontents

\section*{Summary of the main results and conclusions}



\begin{figure}
\includegraphics[width=0.85\textwidth]{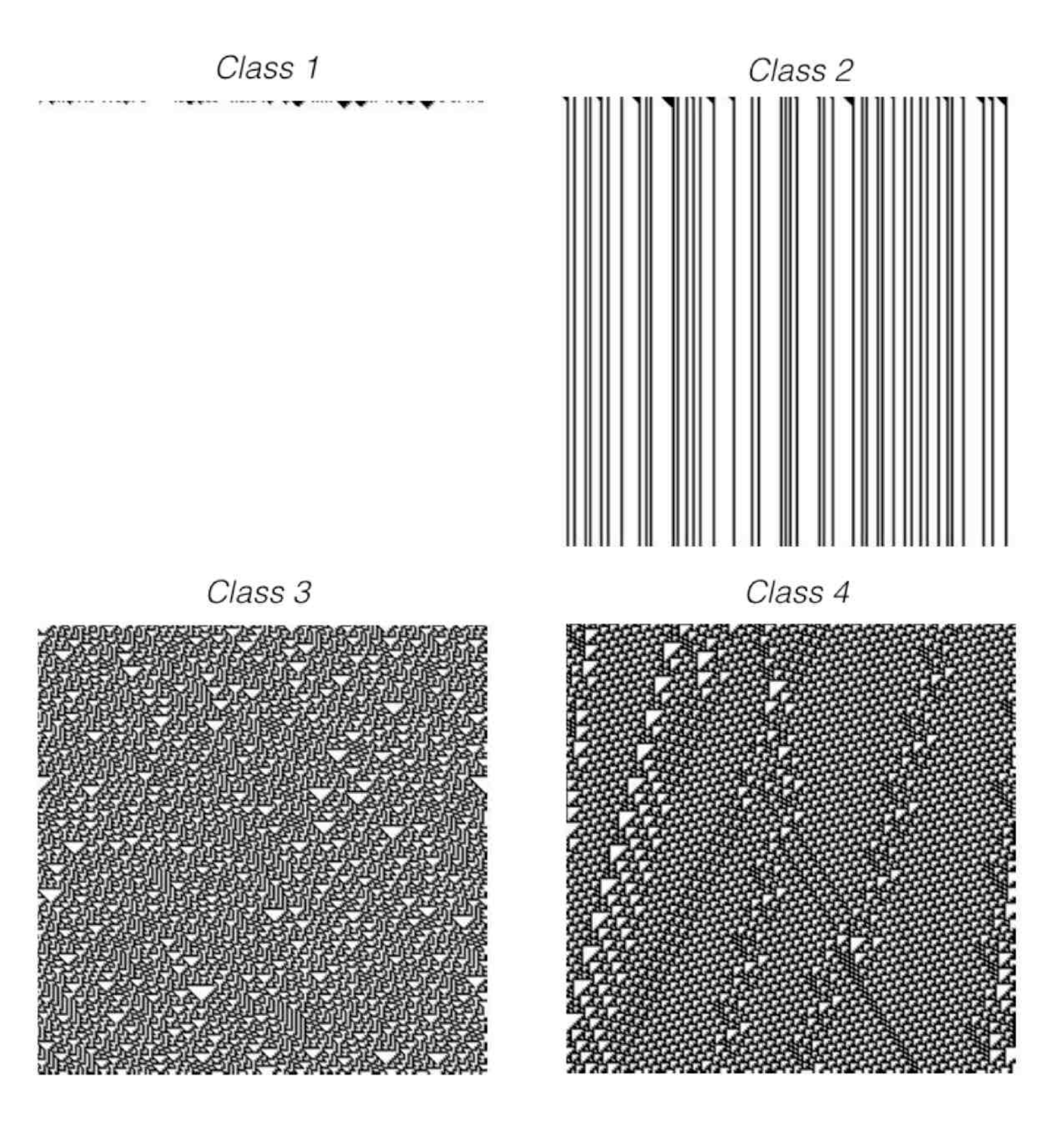}
\caption{Spatiotemporal evolution of one-dimensional CA rules representative of each of the Wolfram classes calculated from Eq. (\ref{CAr}). Rule $^{1}254^{1}_{2}$  (Class 1), rule $^{1}232^{1}_{2}$ (Class 2), rule $^{1}30^{1}_{2}$ (Class 3) and rule $^{1}110^{1}_{2}$ (Class 4). Time flows from top to bottom.} \label{classes}
\end{figure}

Cellular automata (CA) constitute paradigmatic models of complexity in nature, from snowflakes, patterns in mollusc seashells and spiral waves in the Belousov-Zhabotinsky reaction to neural networks and the fundamental physical reality \cite{Wolfram1, WolframB, PhysicaD}. CA serve as models for complex natural systems made of large numbers of  identical parts \cite{Wolfram1, WolframB, PhysicaD, Wolfram2, Wolfram3, Wolfram4}.  They consist of a discrete lattice of sites, with a finite set of possible values each, although they even allow the dynamics of continuum systems to be accurately described \cite{Gerhardt}. The site values evolve synchronously in discrete time steps according to identical rules, being determined by the previous values on the sites of their neighborhood. The concept of CA dates back over half a century to the efforts of John von Neumann to design self-replicating artificial systems capable of universal computation \cite{Neumann} (see also \cite{Codd}). It was however in the early eighties, when Wolfram published a series of groundbreaking papers on the subject \cite{Wolfram2,Wolfram3,Wolfram4,Wolfram5,Wolfram6}, that CA received wide attention from the scientific community. Wolfram classified CA into four classes of increasing complexity \cite{Wolfram2}. For a random initial condition, he found through computer experiments that a CA can evolve into a single homogeneous state (Class 1), a set of separated simple stable or periodic structures (Class 2), a chaotic, aperiodic or nested pattern (Class 3) and complex, localized structures, some times long-lived (Class 4), see Fig.\ref{classes}. Despite some monographs and further important work \cite{Wuensche, Adamatzky, Chua, Langton, Li, Toffoli, Ilachinski, Griffeath, group, Mcintosh, Fredkin} and tantalizing hints \cite{Chua}, no theory provided neither a satisfactory explanation for Wolfram's observations nor analytical expressions valid for arbitrary neighborhood sizes, number of site values and dimensions. Mathematical tools for these discrete systems with the practical value that partial differential equations have for continuum systems \cite{Toffoli} were  absent. This has been a major drawback precluding significant progress: although quite a tour de force, no general conclusions could be sharply drawn other than the computational ones \cite{Wolfram1, Wuensche, Langton}. It has indeed long been thought that no simple mathematical expressions could be given for most CA and a strong emphasis has been put on computational aspects instead of standard mathematics \cite{Wolfram1}. This preprint tackles this problem providing the missing mathematical tool for CA. A universal map encompassing all 1D deterministic first-order in time CA  is presented here. This is to be considered as the discrete counterpart of partial differential equations in continuum systems. The map can be easily extended to an arbitrary number of dimensions and topologies. The advantages of having such a mathematical expression become soon apparent: symmetry arguments applied to the map allow to classify all dynamical CA rules into equivalence classes and a most surprising result is that a theorem can be proved which establishes how a CA rule is constructed in terms of rules of lower range, relating its behavior to them. This theorem and its consequences allow the rationale behind Wolfram's classification sketched above to be established. The crucial result is that the most complex rules (those with Class 4 behavior) can be found with a simple prescription, starting from rules possessing the symmetry upon addition modulo an integer number $p$, and weakly breaking this symmetry through the additional degree of freedom involved in constructing a rule of higher range. I illustrate how this  mechanism is the origin of complexity in 1D systems.

I consider a general rule whose action is described by giving a table of configurations indexed by an integer $n$ so that to each $configuration_{n}$ corresponds an $output_{n}$. The rule is deterministic and each of these configurations is \emph{exclusive}, i.e., a given input cannot be simultaneously equal to two different $configuration_{n}$ within a certain $tolerance=(configuration_{n+1}-configuration_{n})/2$. The $output$ of the rule is then given by
\begin{eqnarray}
&&output= \\
&&\sum_{n\in table}output_{n} \cdot \mathcal{B}(configuration_{n}-input,\ tolerance)  \nonumber
\end{eqnarray}
where the function $\mathcal{B}(x,\ \epsilon)$ returns one if $-|\epsilon| \le x \le |\epsilon|$ and zero otherwise and, thus, coincides  with the boxcar function  
\begin{equation}
\mathcal{B}(x,\epsilon)=\frac{1}{2}\left(\frac{x+\epsilon}{|x+\epsilon|}-\frac{x-\epsilon}{|x-\epsilon|}\right) \label{d1}
\end{equation}
(see Fig. S1 in the Supporting Information \cite{supp}).

I show now how CA rules fit always in this structure. I focus first on a 1D ring containing a total number of $N_{s}$ sites. An input is given as initial condition in the form of a vector $\mathbf{x}_{0}=(x_{0}^{1},...,x_{0}^{N_{s}})$. Each of the $x^{i}_{0}$ is an integer in the range $0$ through $p-1$ where superindex $i$ specifies the position of the site on the 1D ring. At each $t$ the vector $\mathbf{x}_{t}=(x_{t}^{1},...,x_{t}^{N_{s}})$ specifies the state of the CA. Inputs and outputs from the rule are integers on the interval $[0,p-1]$. Periodic boundary conditions are considered so that $x_{t}^{N_{s}+1}=x_{t}^{1}$ and $x_{t}^{0}=x_{t}^{N_{s}}$. Let $x_{t+1}^{i}$ be taken to denote the value of site $i$ at time step $t+1$.  Formally, its dependence on the values at the previous time step is given through the mapping $x_{t+1}^{i}=\phi(x_{t}^{i-r},x_{t}^{i-r+1},..., x_{t}^{i},...,x_{t}^{i+l-1},x_{t}^{i+l})$ or, equivalently $x_{t+1}^{i}=\ ^{l}R_{p}^{r}(x_{t}^{i})$, where $\phi(...) \equiv \ ^{l}R_{p}^{r} $ is the function of the site values which specifies the rule. Here $r$ and $l$ denote the number of cells to the right and to the left of site $i$ respectively. The range $\rho=l+r+1$ of the rule is the total number of sites in the neighborhood. There are $p^{\rho}=p^{r+l+1}$ different configurations on the table for each possible combination of site values. Each $configuration_{n}$ in the table is then simply given by the integer number $n$ which runs between $0$ and $p^{r+l+1}-1$ (then $tolerance=1/2$). They compare to the dynamical configuration reached by site $i$ and its $r$ and $l$ first-neighbors at time $t$ and given by $\sum_{k=-r}^{l}p^{k+r}x_{t}^{i+k}$. The latter is the $input$ of the rule.  The outputs $a_{n}$ for each configuration $n$ are also integers $\in[0,p-1]$. An integer number $R$ can then be given in base 10 to fully specify the rule $^{l}R_{p}^{r} $ as $R=\sum_{n=0}^{p^{r+l+1}-1}a_{n}p^{n}$.  With all these correspondences the following expression is obtained
\begin{equation}
x_{t+1}^{i}=\sum_{n=0}^{p^{r+l+1}-1}a_{n}\mathcal{B}\left(n-\sum_{k=-r}^{l}p^{k+r}x_{t}^{i+k},\frac{1}{2}\right) \label{CAr}
\end{equation}
Eq. (\ref{CAr}) describes \emph{all first-order-in-time deterministic CA rules in 1D} with no freely adjustable parameters: the $p^{r+l+1}$ coefficients $a_{n}$ directly specify the dynamical rule. For example, for Wolfram's rule $^{1}110_{2}^{1}$  $(a_{0},a_{1},...a_{7})=(0,1,1,1,0,1,1,0)$ (see Fig. S2 in the Supporting Information \cite{supp}, where all above notations are also clarified).

Since each site on the ring satisfies Eq. (\ref{CAr}), the whole set of equations is globally invariant upon translation modulo $N_{s}$. As shown in the Supporting Information \cite{supp}, Eq. (\ref{CAr}) is also invariant under certain permutations of the integers in $[0,p-1]$ (change of colors), reflection,  change of base, change of range, and shift (Galilean invariance) (see Figs. S3-S5 in the Supporting Information \cite{supp}). These invariances allow to classify all CA rules into equivalence classes reducing enormously the number of rules to a fewer representative ones. Polynomial maps involving integers can also be directly $derived$ from Eq. (\ref{CAr}). For example, the map for all 256 Wolfram rules $^{1}R_{2}^{1}$ is 
\begin{eqnarray}
x_{t}^{i+1}&=&a_{0}(1-x_{t}^{i+1})(1-x_{t}^{i})(1-x_{t}^{i-1})+a_{1}x_{t}^{i-1}(1-x_{t}^{i+1})(1-x_{t}^{i}) + a_{2}x_{t}^{i}(1-x_{t}^{i+1})(1-x_{t}^{i-1}) \nonumber \\
&&+a_{3}x_{t}^{i}x_{t}^{i-1}(1-x_{t}^{i+1})+a_{4}x_{t}^{i+1}(1-x_{t}^{i})(1-x_{t}^{i-1})+a_{5}x_{t}^{i+1}x_{t}^{i-1}(1-x_{t}^{i})\nonumber \\
&&+a_{6}x_{t}^{i+1}x_{t}^{i}(1-x_{t}^{i-1})+a_{7}x_{t}^{i+1}x_{t}^{i}x_{t}^{i-1}
\end{eqnarray}
where all variables and parameters involved are either '0' or '1' (since $p=2$), see Theorem S11 in the Supporting Information \cite{supp}. Thus, rule $^{1}110_{2}^{1}$, has the remarkably simple expression $x_{t+1}^{i}=x_{t}^{i}+x_{t}^{i-1}(1-x_{t}^{i}-x_{t}^{i}x_{t}^{i+1})$. Many Wolfram rules can be analytically solved for the orbit so that the solution at each site value $i$ is known in closed form as a function of the initial condition and time. These orbits are the CA analogy for the trajectories in differential calculus. For example, rule $^{1}12_{2}^{1}$ has orbit $x_{t}^{i}=x_{0}^{i}-x_{0}^{i}x_{0}^{i+1}$ so that each site value $i$ is known in closed form as a function of the initial condition and time. Then, since rule $^{1}34_{2}^{1}$ is related to rule $^{1}12_{2}^{1}$ through a shift transformation, the orbit of the former is automatically known as $x_{t}^{i}=x_{0}^{i-t}-x_{0}^{i-t}x_{0}^{i-t+1}$.  All 4 trivial local rules $^{0}R_{2}^{0}$ (see Fig. S6 in the Supporting Information \cite{supp}) and all 32 rules $^{0}R_{2}^{1}$ and $^{1}R_{2}^{0}$ (which implement the 16 logical functions, see Figs. S7 to S9 in the Supporting Information \cite{supp}) can be solved for the orbits as well. Tables S1 to S4 in the Supporting Information \cite{supp} show the polynomial maps for all these simple rules and the orbits which solve the maps for all rules $^{0}R_{2}^{0}$, $^{0}R_{2}^{1}$ and $^{1}R_{2}^{0}$. Remarkably, rule $^{0}6_{2}^{1}$ (and its class equivalents $^{1}6_{2}^{0}$, $^{0}9_{2}^{1}$, $^{1}9_{2}^{0}$) is, although predictable, far more complex than the rest of the $\rho=2$ rules (see Figs. S8 and S9 in the Supporting Information \cite{supp}). While the latter are $monotonic$ (changing the input value of one site makes the output to increase (decrease) or to remain constant for every possible input value of the other site), rule $^{0}6_{2}^{1}$ is non-monotonic. It implements the addition modulo 2 of the input site values. I call this kind of rules Pascal rules: they reproduce, for given $\rho$ and $p$ the Pascal simplex modulo $p$ (see Figs. S10 and S11 in the Supporting Information \cite{supp}) and are therefore invariant upon addition modulo $p$. Since the outputs of these rules depend only on the sum of the previous site values, these rules constitute also a particular case of the so-called $totalistic$ rules \cite{Wolfram1, WolframB} (which are themselves a subset of all the rules described by the universal map, Eq. (\ref{CAr})). Pascal rules are also a subset of the so-called  $additive$ CA rules. In \cite{Martin} an algebraic theory is provided for additive CA. 

Eq. (\ref{CAr}) allows to prove a theorem (see the Supporting Information \cite{supp}, Theorem S6) which shows how any rule is constructed from rules of lower range. It implies, for example, that two rules with $p=2$ and vectors $(a_{0},a_{1},...,a_{2^{r+l+1}-1})$ and $(a_{0},a_{1},...,a_{2^{r+l+1}-1}, a_{0},a_{1},...,a_{2^{r+l+1}-1})$ describe $identical$ behavior (although the latter has a  neighborhood with one site  to the left more than the former). Thus rules $^{0}6_{2}^{1}$ and $^{1}102_{2}^{1}$ with vectors $(0,1,1,0)$  and $(0,1,1,0,0,1,1,0)$ respectively, describe the same dynamics. The additional degree of freedom of rule $^{1}102_{2}^{1}$ can be used to break the symmetry of the Pascal rule $^{0}6_{2}^{1}$ from which it is entirely constructed. \emph{If the symmetry is broken so that some few more (but not all) configurations within the Pascal simplex change their output in a monotonic manner, the most complex rules are generated.}  This is the origin of complexity in 1D CA. Class 4 rules can be $derived$ through this process. In the case of Pascal rule $(0,1,1,0)$, the symmetry can be weakly broken by changing the output of configuration '011' from zero to one. This amounts to change $a_{3} \to 1$ in rule $^{1}102_{2}^{1}$ (see Fig. 2 and Fig S12 in the Supporting Information \cite{supp}) obtaining $(0,1,1,1,0,1,1,0)$ i.e. Wolfram's celebrated Class 4 rule $^{1}110_{2}^{1}$, which is known to be capable of universal computation \cite{Wolfram1, Wolfram5, Cook}.   

\begin{figure}
\includegraphics[width=0.5\textwidth, angle=270]{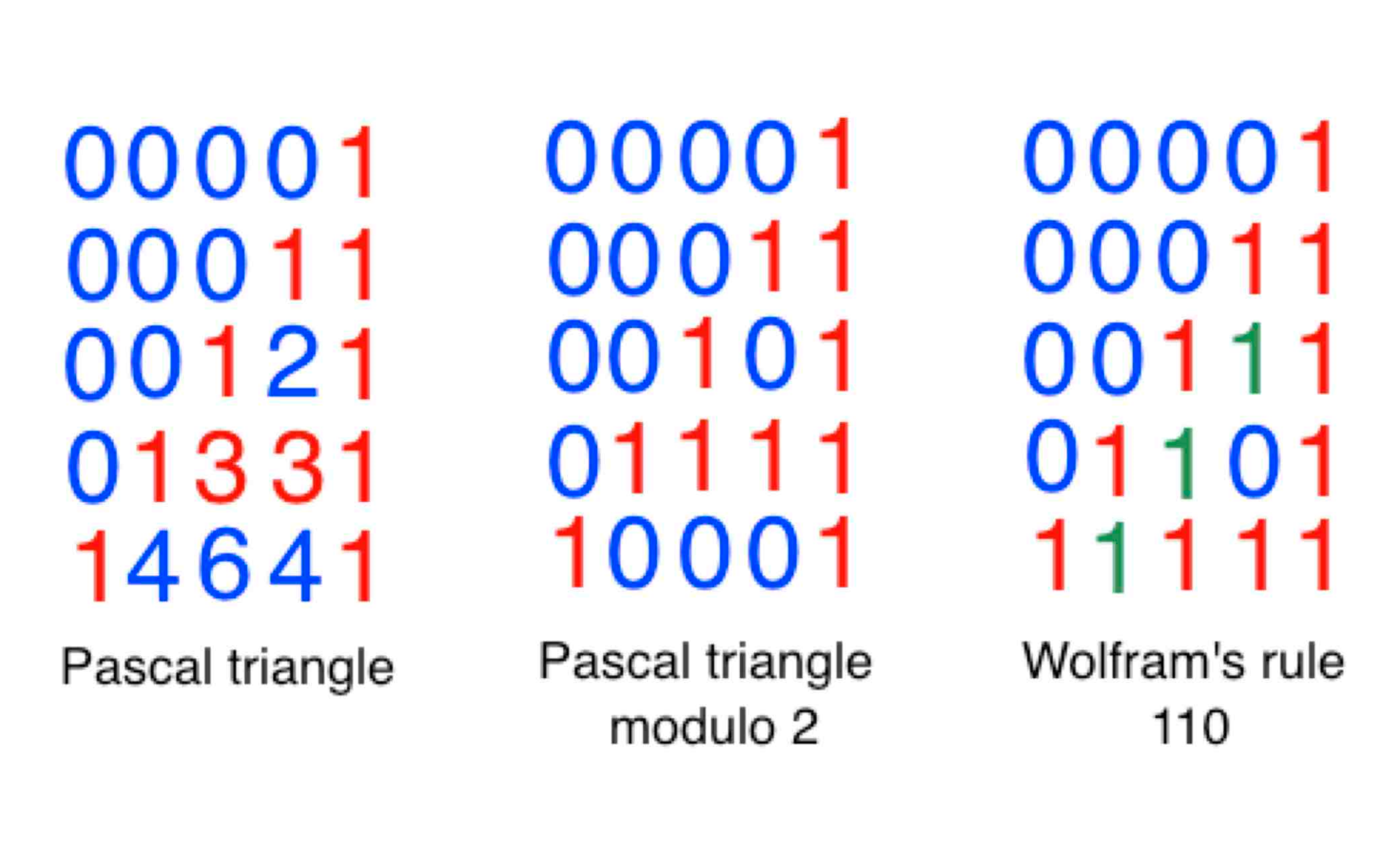}
\caption{Scheme showing the relationship of the Class 4 Wolfram's $^{1}110^{1}_{2}$ rule to the Pascal triangle modulo 2. A weak symmetry breaking of the addition modulo 2 within the triangle is introduced: All the outputs but one are still dictated by the same prescription that generates the Pascal triangle. The only exception is configuration '011' that returns '0' as output in the Pascal rule while it returns '1' in Wolfram's rule. This slight difference (indicated in green) propagates within the triangle of expansion of the rule and is enough to make the resulting pattern both complex and unpredictable.} 
\end{figure}

\begin{figure}
\includegraphics[width=0.65\textwidth, angle=270]{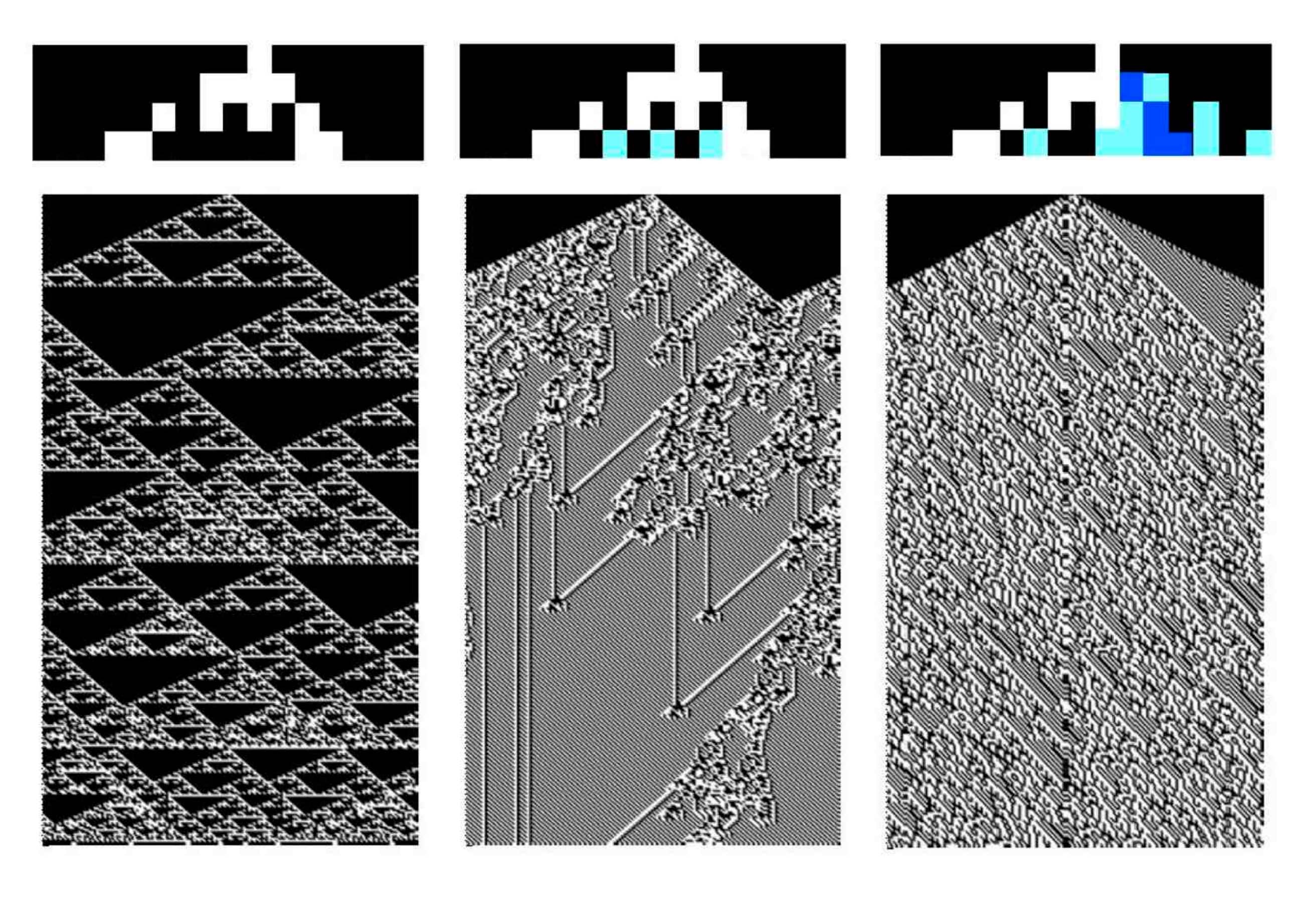}
\caption{Spatiotemporal evolution, obtained from Eq. (\ref{CAr}), of Pascal rule $^{1}27030^{2}_{2}$ (left) and of the Class 4 rule $^{2}1771466134^{2}_{2}$ (center) and the Class 3 rule $^{2}1771469206^{2}_{2}$ (right) derived from the former by a weak and a strong symmetry breaking of the addition modulo 2, respectively. A detail of the triangle of expansion for each rule is shown on the top. In the case of the Class 4 rule, the outputs in the initial time steps are the same as in the Pascal rule and only three site values (indicated with light blue, denoting '1') are different at $t=3$. In the Class 3 rule the symmetry breaking is strong, and 12 sites (indicated with light blue, denoting '1' and dark blue, denoting '0') have already different values compared to the Pascal rule at $t=3$.} 
\end{figure}

Figures 3 (center), and 4, show some other examples of Class 4 rules constructed in this way: glider guns (coherent structures that radiate other coherent structures periodically) and other elements necessary for universal computation are observed in all cases.  All these Class 4 rules are derived from the Pascal rule $^{1}27030^{2}_{2}$ with vector $(0,1,1,0,1,0,0,1,1,0,0,1,0,1,1,0)$ (see Fig. 3 left). The triangle of expansion shows that certain configurations yield their output within the triangle of expansion and can be tuned breaking non-monotonic turns. Examples are configurations '1010','0011','1111','0010','0001'. The rule is then copied to the higher range $\rho=5$ as $(0,1,1,0,1,0,0,1,1,0,0,1,0,1,1,0,0,1,1,0,1,0,0,1,1,$ $0,0,1,0,1,1,0)$  which corresponds to rule $^{2}1771465110^{2}_{2}$, a Pascal rule, as well, which coincides with the original Pascal rule $^{1}27030^{2}_{2}$ as a result of Theorem S6 in the Supporting Information \cite{supp}. Now, if any of the above configurations within the triangle of expansion is considered with the added degree of freedom one has the configurations: '01010','00011','11111','00010','00001', which correspond to numbers '10', '3', '31', '2','1' in the decimal system. Then any of the following positions (or several of them) in the vector characterizing the Pascal rule $^{2}1771465110^{2}_{2}$ can be tuned to yield Class 4 behavior: $a_{10}$, $a_{3}$, $a_{31}$, $a_{2}$, and $a_{1}$. Figure S13 shows some of these possible symmetry breaking tunings and the resulting Class 4 behavior. This result is remarkable: finding these rules by brute force simulations might imply the evaluation of thousands of millions of rules. And with the simple prescription given above, these rules can be \emph{directly} found! If addition modulo $p$ is broken in a way that the borders of the Pascal simplex are affected (or that non-monotonic turns are created within) randomness and Class 3 behavior occurs instead (see Figs. 3 (right) and Fig. S14 in the Supporting Information \cite{supp}). 

\begin{figure}
\includegraphics[width=0.85 \textwidth]{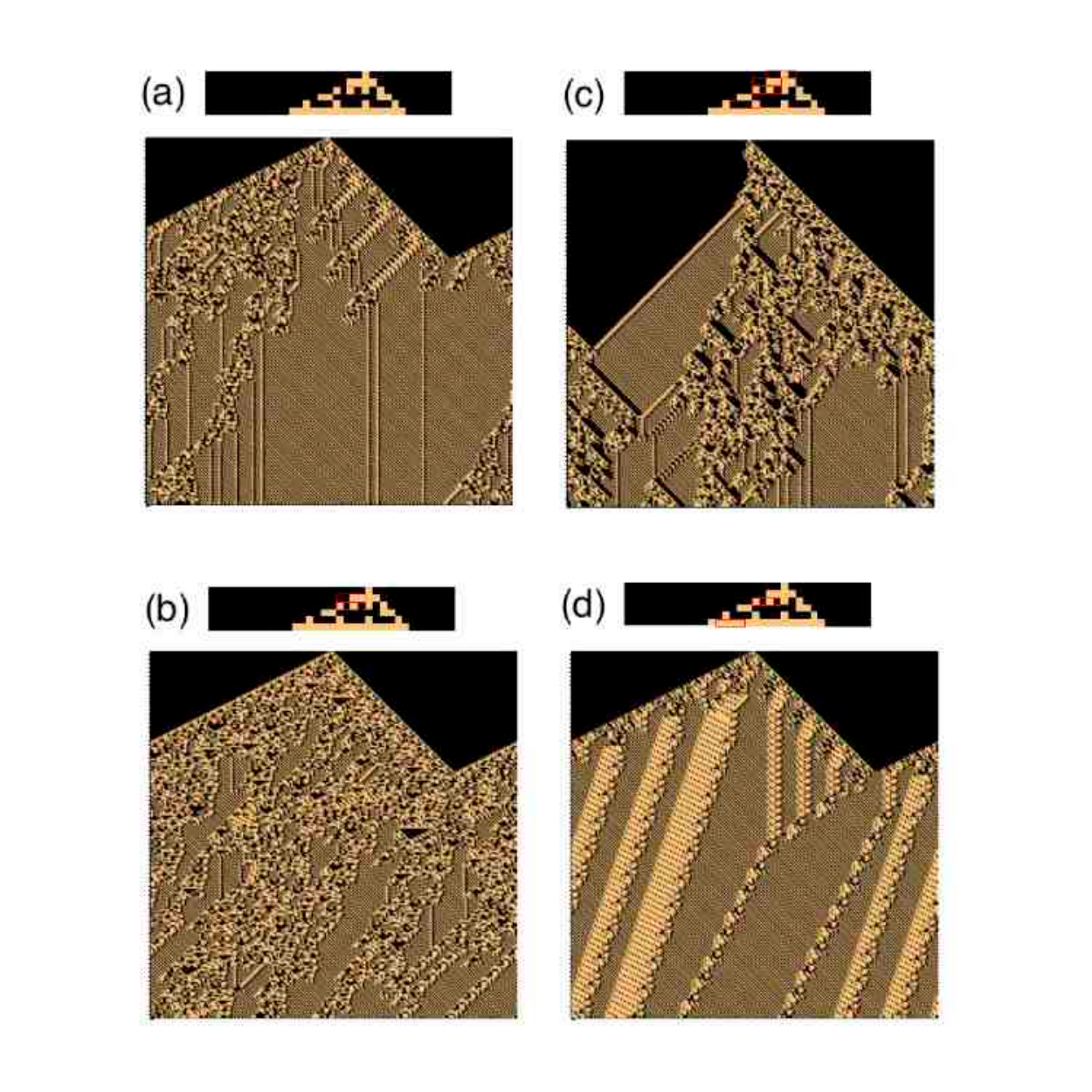}
\begin{center}
\caption{Spatiotemporal evolution of Class 4 rules obtained from a weak symmetry breaking of the Pascal rule $^{1}27030^{2}_{2}$ with vector $(0,1,1,0,1,0,0,1,1,0,0,1,0,1,1,0)$, whose triangle of expansion is shown above each rule. In red are indicated the configurations that are tuned after the rule is copied to higher range. The rules have following codes: (a) $^{2}1771466134^{2}_{2}$, (b) $^{2}1771466142^{2}_{2}$, (c) $^{2}1771466136^{2}_{2}$ and (d) $^{2}3918949782^{2}_{2}$.}  
\end{center} 
\end{figure}

Another consequence of the Theorem S6 in the Supporting Information \cite{supp} is that if two rules are monotonic with the same trend, the constructed rule is also monotonic. This allows to define a complexity index $\kappa$ for every CA, with values $\kappa=1,2,3$ depending whether the rule is monotonic, non-monotonic, or weakly non-monotonic (weakly breaking the addition modulo $p$ symmetry). $\kappa=1$ describes Class 1 and 2 behaviors, and $\kappa=2$ and $3$ Class 3 and 4 behaviors, respectively. This complexity index is tabulated for all Wolfram rules in Table S5 in \cite{supp}. This complexity index should not be confused with the one introduced by Chua in \cite{Chua}, which is related to the complexity of implementing CA rules by means of cellular neural networks. The complexity index introduced in this preprint is directly related to the complexity of the dynamical behavior observed and can be calculated for each rule and Theorem S6 in the Supporting Information \cite{supp} by decomposing the rule in its constructing sub-rules layer by layer. The complexity index applies not only to boolean rules with $p=2$, $r=1$ and $l=1$, but to rules of any range and any number of symbols in the alphabet.

Eq. (\ref{CAr}) can be generalized to higher dimensions and topologies (see Fig. S15 and the accompanying discussion in the Supporting Information \cite{supp}). For example, the popular Conway's "Game of Life" \cite{Gardner1, Gardner2} is given by the map $x_{t+1}^{i,j}=\mathcal{B}\left(3-\sum_{k,m=-1}^{1}x_{t}^{i+k,j+m},\frac{1}{2}\right)+x_{t}^{i,j}\mathcal{B}\left(4-\sum_{k,m=-1}^{1}x_{t}^{i+k,j+m},\frac{1}{2}\right)$ where $x_{t}^{i,j}$ denotes the central site of a Moore neighborhood in 2D. When $p \to \infty$  any local map involving real numbers can also be approximated by $^{0}R_{p}^{0}$ rules as briefly shown in the Supporting Information \cite{supp} for the case of the logistic equation.

The ideas and methods presented here can be fruitfully used in manifold ways to model complex systems and might be helpful in finding accurate coarse-grained models of physical systems with many degrees of freedom \cite{Goldenfeld}. The role played here by the Pascal rules is analogous to the one played by separatrices between nonlinear resonances in Hamiltonian systems. The destruction of KAM tori in Hamiltonian systems through weak perturbations, creating chaotic layers that coexist with islands of regular motion \cite{rusos}, seems to be mimicked, at the CA level, by the weak symmetry breaking of the addition modulo $p$ reported here leading to Class 4 behavior. Research on these intriguing connections might also yield valuable insight in the understanding of turbulence. 

\section*{Acknowledgments}

Thanks to Katharina Krischer and Jose Manuel Garc\'ia Sanchis for fruitful conversations. Financial support from the Technische Universit\"at M\"unchen - Institute for Advanced Study, funded by the German Excellence Initiative, is also gratefully acknowledged.

~\\

\pagebreak

\section*{SUPPLEMENTARY INFORMATION}

\section{Overview of the argument}

A mathematical function can be thought as a rule such that, given a numerical input, the output is read from a table containing a series of configurations to which the input compares. The complete table gives the rule, i.e. what to do with an appropriate input. Usually, the configurations on a table must be sharply and unambiguously distinguished from each other within a certain tolerance. Then the table provides the output corresponding to the configuration that successfully matches the input. It is then said that the output is the result of the rule acting on the input. 

The above paragraph can be seen as a computational way of thinking Mathematics. In the following, the argument is inverted, Mathematics is taken as starting point and I shall consider the most elementary mathematical structure that imitates computation as described above. Each computational process is considered here as an "experiment" and I use standard mathematics to theoretically model such an experiment. Proceeding in this way is advantageous since a general mathematical theory for cellular automata (CA) can then be rigorously established.  

I consider first the most simple rule, i.e one containing just only one configuration with one nonzero output. When $input$ coincides with $configuration$ within a certain $tolerance$, the rule returns $output$. Otherwise, the result is zero. I seek now for the simplest mathematical structure where all actors involved in the above computational process enter. It can be formally depicted in the following way:
\begin{equation}
output \times \mathcal{B}(configuration-input,\ tolerance) \label{Form1}
\end{equation}
The function $\mathcal{B}$ has two arguments and, in order for the latter expression to return $output$ or zero, it is clear that $\mathcal{B}$ should return 1 or 0, respectively. Specifically, it should return 1 if $|configuration-input| < tolerance$ and 0 otherwise. It is clear then that $\mathcal{B}$ is an even function of its first argument.  The mathematical function $\mathcal{B}(x,\epsilon)$ having all these properties is the boxcar function plotted in Fig. S1 and introduced in Definition S1 below. \\

\begin{figure}
\includegraphics[width=0.3\textwidth, angle=270]{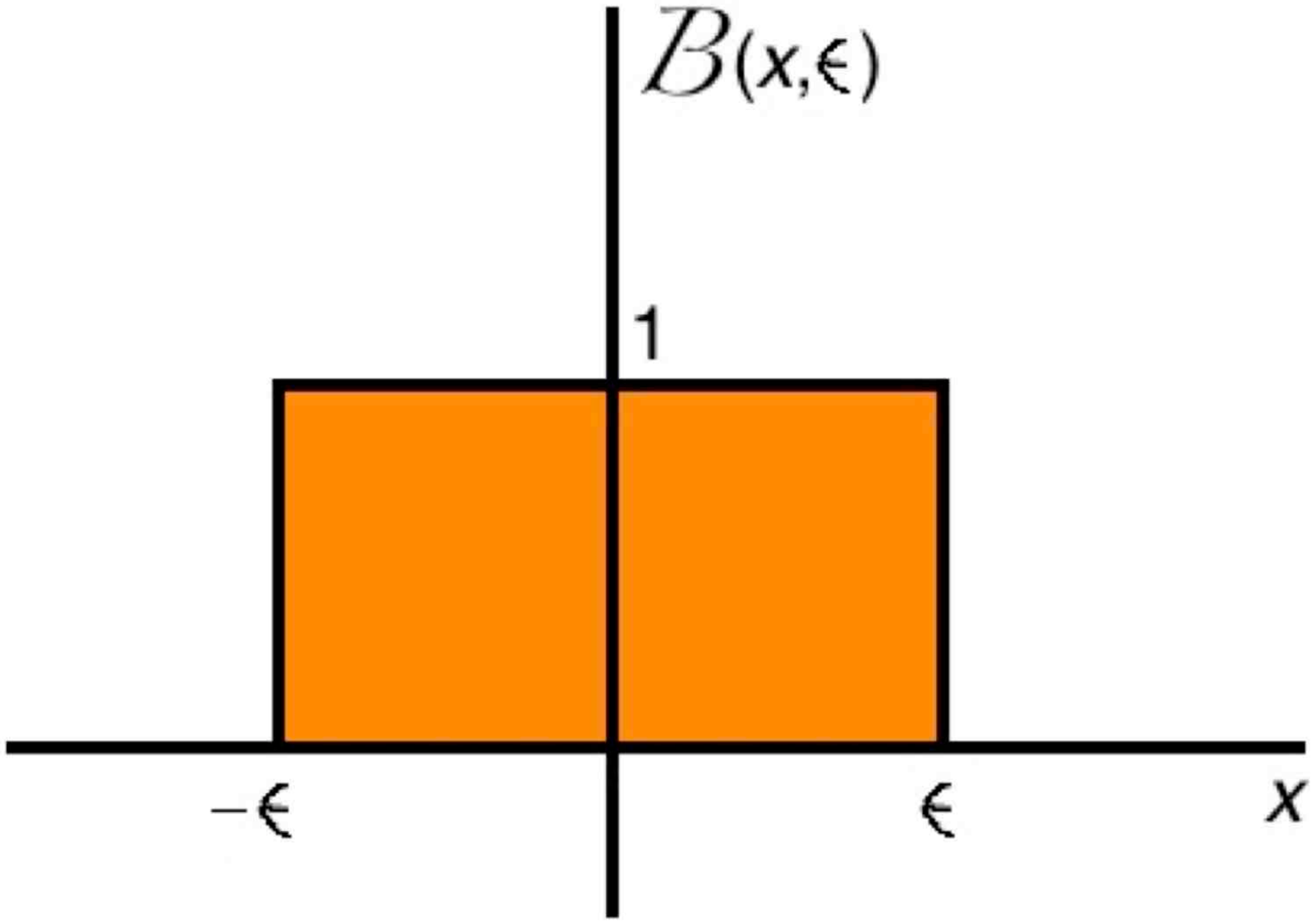}
\begin{center}
Figure S1: The boxcar function $\mathcal{B}(x,\epsilon)$ is equal to one for $-\epsilon \le x \le \epsilon$ and zero otherwise.
\end{center}
\end{figure}

\noindent \textbf{Definition S1:}
\emph{The boxcar function $\mathcal{B}(x,\epsilon)$ is defined, for $\epsilon$ and $x$ real, as\\}
\begin{equation}
\mathcal{B}(x,\epsilon)=\frac{1}{2}\left(\frac{x+\epsilon}{|x+\epsilon|}-\frac{x-\epsilon}{|x-\epsilon|}\right) \label{d1}
\end{equation} 
\emph{where $|x|=\left\{
\begin{array}{cc}
x  & \quad x \ge 0    \\
-x  & \quad x \le 0     
\end{array}
\right. $
is the absolute value function.} $\Box$ \\ 

The boxcar function is defined over the real numbers and has a discontinuity at $x = \pm |\epsilon|$. 
When $\epsilon=1/2$, the boxcar function is usually called rectangular function.  In Appendix A several alternative definitions of $\mathcal{B}$ and some useful results -some well known- are outlined.

When one has a rule with several different configurations indexed by $n$ so that to $configuration_{n}$ corresponds an $output_{n}$, the rule is completely given by summing over all elementary structures describing the action of the rule on each separate $configuration_{n}$ as given in Eq.(\ref{Form1}). Each configuration is $exclusive$, i.e., the input cannot simultaneously be equal to two different $configuration_{n}$ within $tolerance$. The $output$ of the rule is then
\begin{equation}
output=\sum_{n \in table}output_{n} \times \mathcal{B}(configuration_{n}-input,\ tolerance) \label{FormGen}
\end{equation}
The $tolerance$ is related to the distance separating two adjacent configurations on the table and is defined as
\begin{equation}
tolerance=\frac{configuration_{n+1}-configuration_{n}}{2}
\end{equation}
and so, if each $configuration_{n}$ is given by an integer number, $tolerance=1/2$ (if each configuration is instead given by a rational number separated a distance $1/d$ from the next, then $tolerance=1/(2d)$). Remarkably, and as shown in Section II, a CA can be always made to fit in the general structure given by Eq. (\ref{FormGen}).

\section{General theory for cellular automata (CA)}
\subsection{Universal CA map}

A spatially extended system consisting of a lattice of discrete sites is now considered. The topology and dimensions of this lattice can be arbitrary. I focus first on a 1D ring containing a total number of $N_{s}$ sites, and develop the complete theory for such a situation. 

An input is given as initial condition in the form of a vector $\mathbf{x}_{0}=(x_{0}^{1},...,x_{0}^{N_{s}})$. Each of the $x^{i}_{0}$ is  an integer in the range $0$ through $p-1$. The subindex labels the discrete time step in the evolution and the superindex specifies the position of the site on a 1D ring. At each time step $t$ the vector $\mathbf{x}_{t}=(x_{t}^{1},...,x_{t}^{N_{s}})$ specifies the state of the CA. Inputs and outputs from the rule are always integers on the range $0$ through $p-1$. Periodic boundary conditions so that $x_{t}^{N_{s}+1}=x_{t}^{1}$ and $x_{t}^{0}=x_{t}^{N_{s}}$. Let $x_{t+1}^{i}$ be taken to denote the value of site $i$ at time step $t+1$.  Formally, its dependence on the values at the previous time step is given through the mapping
\begin{equation}
x_{t+1}^{i}=\phi(x_{t}^{i-r},x_{t}^{i-r+1},..., x_{t}^{i},...,x_{t}^{i+l-1},x_{t}^{i+l}) \label{formal}
\end{equation}
where $\phi(...)$ is the function of the site values which specifies the rule. Here $r$ and $l$ denote the number of cells to the right and to the left of site $i$ respectively. The range $\rho$ of the rule is defined as the total number of sites involved in the rule and is therefore given by $\rho=l+r+1$. Since each site $i$ can have $p$ different values (labelled as integers between 0 and $p-1$) there is a total of $\Omega \equiv p^{\rho}$ different configurations on the table for each possible combination of symbols. For a range $\rho$ there exists a total number of $\Gamma=p^{\Omega}=p^{p^{\rho}}$ different rules that can be defined in this way. Each $configuration_{n}$ in the table is given by integer number $n$ which runs between $0$ and $\Omega-1$ (and hence $tolerance=1/2$) and which is given as a function of the site values in the table $x^{i}$ as
\begin{equation}
n=\sum_{k=-r}^{l}p^{k+r}x^{i+k} \label{confCA}
\end{equation}
The $x^{i}$ here have also values between $0$ and $p-1$ and specify all possible configurations. The above $n$ specifies each $static$ configuration given on the table. They compare to the dynamical configuration $n_{t}$ reached by site $i$ and its $r$ and $l$ first-neighbors at time $t$
\begin{equation}
n_{t}=\sum_{k=-r}^{l}p^{k+r}x_{t}^{i+k}
\end{equation}
The latter is the $input$ of the rule. 

The outputs $a_{n}$ for each configuration $n$, given by Eq.(\ref{confCA}), have integer values between $0$ and $p-1$. An integer number $R$ can then be given to specify the code of the rule in the following way
\begin{equation}
R=\sum_{n=0}^{\Omega-1}a_{n}p^{n} \label{code}
\end{equation}
i.e. the $a_{n}$ are the coefficients that accompany the powers of $p$ when writing $R$ in base $p$. Since the rule depends not only on the $a_{n}$ but also on the number of symbols $p$ involved and on the number of neighbors to the left $l$ and to the right $r$, to define an arbitrary CA rule in an unambiguous way I introduce the quantity
\begin{equation}
^{l}R_{p}^{r} \label{rule} \nonumber
\end{equation}
This latter quantity will symbolically denote in all the following the same as $\phi$ in Eq. (\ref{formal}) above. The labels $l$, $r$ and $p$ accompanying $R$ specify the rule completely and Eq. (\ref{formal}) can be written in a compact and unambiguous way as
\begin{equation}
x_{t+1}^{i}=\ ^{l}R_{p}^{r}(x_{t}^{i}) \label{rule2}
\end{equation}
with the understanding that the rule depends not only on $x_{t}^{i}$ but also on the first $r$ neighbors to the right and the first $l$ neighbors to the left. For example, Wolfram's rule 110, involves $p=2$ symbols, and one left and one right neighbors, i.e. $r=l=1$ (see Fig. S2). Such rule is then written as $^{1}110_{2}^{1}$. 

\begin{figure}
\begin{center}
\includegraphics[width=0.65\textwidth, angle=270]{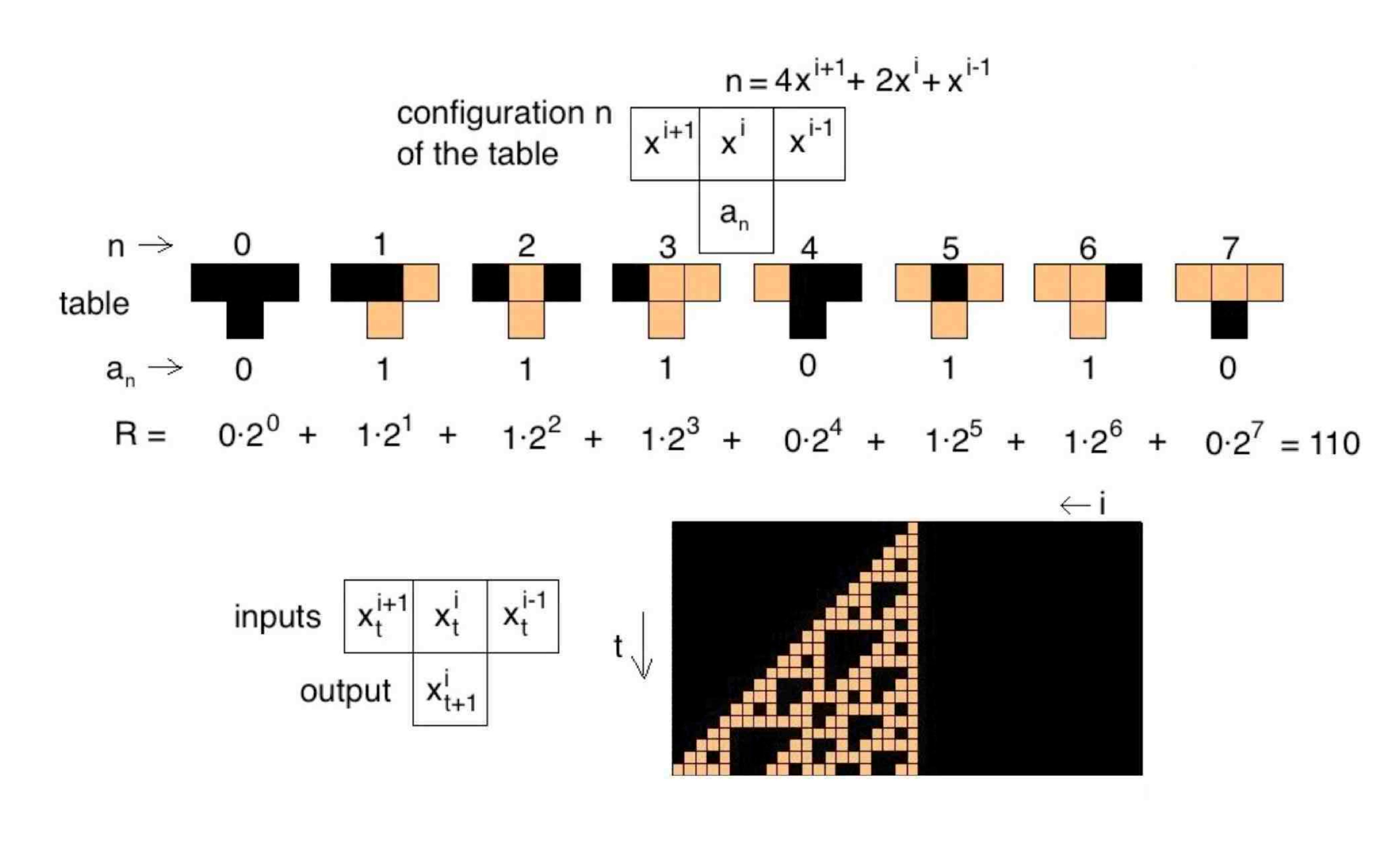}
\end{center}
\noindent Figure S2: Wolfram's rule $^{1}110_{2}^{1}$: (top) The table of the rule with each $configuration_{n}$ ($n$) and each $output_{n}$ ($a_{n}$) and code $R$; (bottom) spatiotemporal behavior of the rule running from a single initial seed on a 1D ring (time runs from top to bottom and space is plotted horizontally, under periodic boundary conditions). 
\end{figure}
~\\

Since each site $i \in [1,N_{s}]$ on the ring satisfies separately Eq. (\ref{rule2}), the whole set of $N_{s}$ equations is globally invariant upon translation modulo $N_{s}$, i.e. upon making the change $i \to i+h$ to every site $i$ (where $h$ is an integer number). 

The following set of correspondences between the quantities here defined and the ones introduced in Eq. (\ref{FormGen}) in Section I hold \\
\begin{eqnarray}
input &\Longleftrightarrow& n_{t}=\sum_{k=-r}^{l}p^{k+r}x_{t}^{i+k} \nonumber \\
configuration_{n} &\Longleftrightarrow&  n \left(\equiv\sum_{k=-r}^{l}p^{k+r}x^{i+k}\right) \nonumber \\
tolerance &\Longleftrightarrow&  1/2 \nonumber \\
output_{n} &\Longleftrightarrow& a_{n} \nonumber \\
output &\Longleftrightarrow& x_{t+1}^{i} \nonumber
\end{eqnarray}
~\\
and the universal map governing the dynamics of 1D cellular automata, by using Eq. (\ref{FormGen}), has then the form
\begin{equation}
\boxed{\boxed{x_{t+1}^{i}=\sum_{n=0}^{\Omega-1}a_{n}\mathcal{B}\left(n-\sum_{k=-r}^{l}p^{k+r}x_{t}^{i+k},\frac{1}{2}\right)}} \label{CA}
\end{equation}\\
or, equivalently, by using Definition S1

\begin{equation}
x_{t+1}^{i}=\sum_{n=0}^{\Omega-1}\frac{a_{n}}{2}\left(\frac{\frac{1}{2}+n-\sum_{k=-r}^{l}p^{k+r}x_{t}^{i+k}}{|\frac{1}{2}+n-\sum_{k=-r}^{l}p^{k+r}x_{t}^{i+k}|}+\frac{\frac{1}{2}-n+\sum_{k=-r}^{l}p^{k+r}x_{t}^{i+k}}{|\frac{1}{2}-n+\sum_{k=-r}^{l}p^{k+r}x_{t}^{i+k}|}\right) \\  \label{CAb}
\end{equation} 

Although Eq. (\ref{CA})  seems rather complex, it is to be noted that it describes \emph{all  first-order-in-time deterministic CA maps in 1D}. The map contains no freely adjustable parameters: the $a_{n}$ directly specify the dynamical rule. When most $a_{n}$ are zero, the sum reduces to just a few terms. Eq. (\ref{CA}) is the normal form for all $\Gamma=p^{\Omega}=p^{p^{\rho}}$ deterministic 1D CA. 

Sometimes, the configuration is specified not in terms of its binary code but over a sum carried over all sites in the neighborhood. The output is made in such cases dependent only on the sum of the previous values. These CA are called $totalistic$ and they are a subset of the total possibilities described by Eq. (\ref{CA}). Each $configuration_{n}$ in the table is labelled in such a case with an integer number $s$ which runs between $0$ and $\Theta \equiv \rho(p-1)$ (the latter value is the maximum value that the sum over the site values can attain)
\begin{equation}
s=\sum_{k=-r}^{l}x^{i+k} \label{confCAtot}
\end{equation}
The $x^{i}$ have as before values between $0$ and $p-1$ and specify all possible configurations of the symbols. The above $s$ specifies each $static$ configuration given on the table. They compare to the dynamical configuration $s_{t}$ reached by site $i$ and its $r$ and $l$ first-neighbors at time $t$
\begin{equation}
s_{t}=\sum_{k=-r}^{l}x_{t}^{i+k}
\end{equation}
which is the $input$ of the totalistic rule. 

The outputs $\sigma_{s}$ for each configuration $s$, given by Eq.(\ref{confCAtot}), have integer values between $0$ and $p-1$ like the inputs and the output of the rule. An integer number $R$ can then be given to specify the code of the totalistic rule in the following way
\begin{equation}
R=\sum_{s=0}^{\Theta}\sigma_{s}p^{s} \label{codetot}
\end{equation}
A totalistic CA rule is therefore labelled as
\begin{equation}
^{l}RT_{p}^{r} \label{rule} \nonumber
\end{equation}
where the label $T$ is added to avoid confusion with the coding in the normal form. The following set of correspondences between these quantities and the ones introduced in Eq. (\ref{FormGen}) in Section I holds now
\begin{eqnarray}
input &\Longleftrightarrow& s_{t}=\sum_{k=-r}^{l}x_{t}^{i+k} \nonumber \\
configuration_{n} &\Longleftrightarrow&  s \left(\equiv\sum_{k=-r}^{l}x^{i+k}\right) \nonumber \\
tolerance &\Longleftrightarrow&  1/2 \nonumber \\
output_{n} &\Longleftrightarrow& \sigma_{s} \nonumber \\
output &\Longleftrightarrow& x_{t+1}^{i} \nonumber
\end{eqnarray}
And therefore, the universal map governing the dynamics of 1D totalistic cellular automata, is obtained by using Eq. (\ref{FormGen}) as \\
\begin{equation}
\boxed{x_{t+1}^{i}=\sum_{s=0}^{\Theta}\sigma_{s}\mathcal{B}\left(s-\sum_{k=-r}^{l}x_{t}^{i+k},\frac{1}{2}\right)} \label{CAtot}
\end{equation}

Eq. (\ref{CAtot}) is a particular case of Eq.(\ref{CA}). Starting from a totalistic rule with vector $(\sigma_{0}, \sigma_{1}...\sigma_{\Theta})$ described by Eq. (\ref{CAtot}) the vector specifying the normal rule as described by Eq. (\ref{CA})  $(a_{0}, a_{1}...a_{\Omega-1})$ can be calculated from the following expression
\begin{equation}
a_{\sum_{k=-r}^{l}p^{k+r}x^{i+k}}=\sigma_{\sum_{k=-r}^{l}x^{i+k}} \label{totto}
\end{equation}

The following important theorem establishes that the universal CA maps, Eqs. (\ref{CA}) and (\ref{CAtot}) are invariant upon a bijective application that sends the set of $p$ integers $\{0,1,...,p-1\}$ to any other permutation of the same $p$ integers, provided that inputs, output and specifications in the table are subjected to the application. An analogy is provided by computer simulations, where to visualize the spatiotemporal pattern of the CA a color to each integer number is associated. The invariance under change of colors warrants that any palette can be chosen if the distinctness of the symbols is preserved, the latter having no effect on the dynamics (the same pattern is obtained albeit with different colors). This, of course, must be the case since the dynamics must be the same regardless of the labels that used for the symbols. \\

\noindent  \textbf{Theorem S1 (Invariance under change of colors):}
\emph{The universal CA map, Eq. (\ref{CA}) remains invariant after the following set of transformations
\begin{eqnarray}
x_{t}^{i+k} & \to & x_{t}^{i+k}+\sum_{m=0}^{p-1}\lambda_{m} \mathcal{B}\left(x_{t}^{i+k}-m,\frac{1}{2}\right) \label{t1} \\
x_{t+1}^{i} & \to & x_{t+1}^{i}+\sum_{m=0}^{p-1}\lambda_{m} \mathcal{B}\left(x_{t+1}^{i}-m,\frac{1}{2}\right) \label{t2} \\
n & \to & n' \equiv n+ \sum_{k=-r}^{l}\sum_{m=0}^{p-1}p^{k+r}\lambda_{m} \mathcal{B}\left(x^{i+k}-m,\frac{1}{2}\right) \label{t3} \\
a_{n} & \to & b_{n'} \equiv a_{n'}+\sum_{m=0}^{p-1}\lambda_{m} \mathcal{B}\left(a_{n'}-m,\frac{1}{2}\right) \label{t4} 
\end{eqnarray}
where the $\lambda_{m}$'s are integer numbers (not necessarily positive) so that the transformations between the sets of $p$ integers $\{0,1...,p-1\}$ implied by Eqs. (\ref{t1}), (\ref{t2}) and (\ref{t4}) are bijective so that
\begin{equation}
\sum_{k=-r}^{l}\sum_{m=0}^{p-1}p^{k+r}\lambda_{m} \left[\mathcal{B}\left(x^{i+k}-m,\frac{1}{2}\right)-\mathcal{B}\left(x_{t}^{i+k}-m,\frac{1}{2}\right)\right] \ne \sum_{k=-r}^{l}p^{k+r}\left(x_{t}^{i+k}-x^{i+k}\right) \label{condi}
\end{equation}
when both sides are different to zero.}

\noindent \emph{Proof:} By inserting Eqs.(\ref{t1}) to (\ref{t4}) in Eq. (\ref{CA}) 
\begin{eqnarray}
&&x_{t+1}^{i}+\sum_{m=0}^{p-1}\lambda_{m} \mathcal{B}\left(x_{t+1}^{i}-m,\frac{1}{2}\right)=\sum_{n' \in [0,\Omega-1]}\left[a_{n'}+\sum_{m=0}^{p-1}\lambda_{m} \mathcal{B}\left(a_{n'}-m,\frac{1}{2}\right)\right] \times \nonumber \\
&& \qquad \qquad \qquad \qquad  \times \mathcal{B}\left(n'-\sum_{k=-r}^{l}p^{k+r}\left[x_{t}^{i+k}+\sum_{m=0}^{p-1}\lambda_{m} \mathcal{B}\left(x_{t}^{i+k}-m,\frac{1}{2}\right)\right],\frac{1}{2}\right) \nonumber \\
&&\qquad \qquad \qquad \qquad =\sum_{n=0}^{\Omega-1}\left[a_{n}+\sum_{m=0}^{p-1}\lambda_{m} \mathcal{B}\left(a_{n}-m,\frac{1}{2}\right)\right] \mathcal{B}\left(n-\sum_{k=-r}^{l}p^{k+r}x_{t}^{i+k},\frac{1}{2}\right) \nonumber 
\end{eqnarray}
where to get to the penultimate equality I have used that the application between the two sets of integers $\{0,1...,p-1\}$ is bijective satisfying Eq. (\ref{condi}) and the sum has been reordered. By using result $(vii)$ from  Appendix A and the fact that $x_{t+1}^{i}=a_{n}$ for the $n \equiv \sum_{k=-r}^{l}p^{k+r}x^{i+k}$ that satisfies $n=\sum_{k=-r}^{l}p^{k+r}x_{t}^{i+k}$, 
\begin{eqnarray}
\sum_{m=0}^{p-1}\lambda_{m} \mathcal{B}\left(x_{t+1}^{i}-m,\frac{1}{2}\right)&=&
\sum_{n=0}^{\Omega-1}\sum_{m=0}^{p-1}\lambda_{m} \mathcal{B}\left(x_{t+1}^{i}-m,\frac{1}{2}\right) \mathcal{B}\left(n-\sum_{k=-r}^{l}p^{k+r}x_{t}^{i+k},\frac{1}{2}\right)
 \nonumber \\
 &=&
\sum_{n=0}^{\Omega-1}\sum_{m=0}^{p-1}\lambda_{m} \mathcal{B}\left(a_{n}-m,\frac{1}{2}\right) \mathcal{B}\left(n-\sum_{k=-r}^{l}p^{k+r}x_{t}^{i+k},\frac{1}{2}\right)
 \end{eqnarray}
and, therefore, 
\begin{equation}
x_{t+1}^{i}=\sum_{n=0}^{\Omega-1}a_{n}\mathcal{B}\left(n-\sum_{k=-r}^{l}p^{k+r}x_{t}^{i+k},\frac{1}{2}\right) 
\end{equation}
proving thus the invariance of Eq. (\ref{CA}) upon change of colors. The proof for Eq. (\ref{CAtot}) is similar. $\Box$ \\

When one considers specific rules, described by the universal maps above, they usually $break$ this symmetry as shown in the following example. \\

\begin{figure}
\begin{center}
\includegraphics[width=0.65\textwidth]{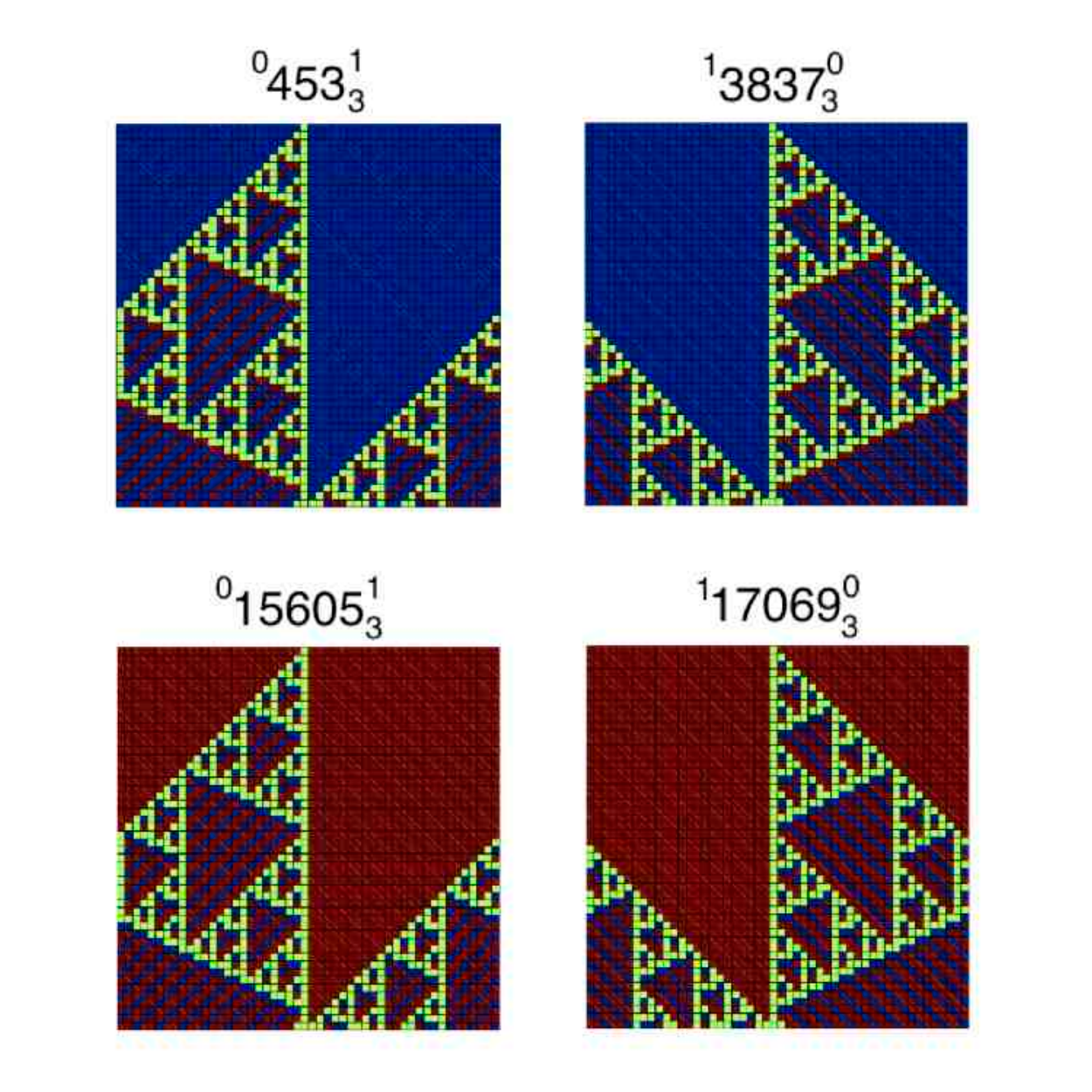}
\end{center}
\noindent Figure S3: Spatiotemporal evolution of rules $^{0}453^{1}_{3}$, $^{0}15605^{1}_{3}$, $^{1}3837^{0}_{3}$ and $^{1}17069^{0}_{3}$. Rules on the top are related to the ones at the bottom through a change of colors. Rules on the left are related to the ones on the right through reflection.
\end{figure}

\noindent \textbf{Example:} The asymmetric rule $^{0}453^{1}_{3}$ (see Fig. S3 - top left) with three symbols '0','1' and '2' has range $\rho=r+l+1=2$ (and hence $\Omega=p^{\rho}=3^{2}=9$) and coefficients $(a_{0},a_{1},...a_{8})=(0,1,2,1,2,1,0,0,0)$ (i.e. the number 453 in base 3 in inverse order) in Eq. (\ref{CA}). The neighborhood contains only the cell that is updated in the next time step and the first neighboring site to the right. I derive now the rule that operates with the symbols $2,1,0$ in the same way as rule $^{0}453^{1}_{3}$ does with symbols $0,1,2$ respectively. Clearly, $\lambda_{0}=2, \lambda_{1}=0, \lambda_{2}=-2$, and from Eqs. (\ref{t3}) and (\ref{t4}), the transformed rule $(b_{0},b_{1},...b_{8})=(2,2,2,1,0,1,0,1,2)$ is obtained, which corresponds to rule $^{0}15605^{1}_{3}$ (see Fig. S3 - bottom left). The colors corresponding to symbols with values 0 (blue) and 2 (red) are exchanged in both figures, while sites with value 1 (green) remain unchanged.

Rules $^{0}453^{1}_{3}$ and $^{0}15605^{1}_{3}$ belong then to the same equivalence class under change of colors. If $p$ is the number of symbols in a rule, there are at most $p!$ rules in its same equivalence class (note that there are rules that may be invariant upon change of some colors: in general, the cardinal $\chi$ of an equivalence class is, therefore, $1 \le \chi \le p!$).\\

\noindent  \textbf{Theorem S2 (Invariance under reflection):}
\emph{The universal CA map, Eq. (\ref{CA}) remains invariant after the following set of transformations
\begin{eqnarray}
p^{k+r}x_{t}^{i+k} & \to & p^{l-k}x_{t}^{i+k} \qquad \qquad  k \in [-r,l] \label{t1b} \\
n & \to & n' \equiv n- \sum_{k=-r}^{l}(p^{k+r}-p^{l-k})x^{i+k} \label{t2b} \\
a_{n} & \to & b_{n'}=a_{n}
\end{eqnarray}
} 

\noindent \emph{Proof:} This theorem can be checked directly by making the corresponding transformations in Eq. (\ref{CA}). $\Box$ \\

Invariance under reflection and under change of colors can be followed after the other in either direction since both commute. The dynamical behavior of all these rules is closely related and all of them belong to the same equivalence class, which, when added all rules obtained under change of colors and its reflections, contains no more as $2p!$ elements. In Fig. S3 it is shown how rules $^{0}453^{1}_{3}$ and $^{0}15605^{1}_{3}$ are related through reflection to rules $^{1}3837^{0}_{3}$ and $^{1}17069^{0}_{3}$ respectively. All these rules belong, therefore, to the same equivalence class.\\

\noindent  \textbf{Theorem S3 (Invariance under change of base):}
\emph{The universal CA map, Eq. (\ref{CA}) remains invariant after the following transformations}
\begin{eqnarray}
&&p  \to  p' \qquad \qquad (p' \ge p) \label{t1bas} \\
&& n  \to  n' \equiv \sum_{k=-r}^{l}p'^{k+r}x^{i+k} \qquad a_{n}  \to  b_{n'}=a_{n} \label{t3bas}
\end{eqnarray}

\noindent \emph{Proof:} It can be checked directly by making the transformations implied by Eqs.(\ref{t1bas}) to (\ref{t3bas}) in Eq. (\ref{CA}). $\Box$ \\

\noindent \textbf{Example:} Rules  $^{1}110_{2}^{1}$ and $^{1}590601_{3}^{1}$ are equivalent after a change of base and exhibit the same dynamical behavior.\\

The validity of the following theorem can be again checked directly employing Eq. (\ref{CA}).\\

\noindent  \textbf{Theorem S4 (Invariance under change of range):}
\emph{The universal CA map, Eq. (\ref{CA}) remains invariant after the following transformations
\begin{eqnarray}
l & \to & l' \qquad \qquad (l' \ge l) \label{t1ran} \\
r & \to & r' \qquad \qquad (r' \ge r) \label{t2ran} \\ 
n & \to& n' \equiv n+\sigma \label{t3ran} \\
a_{n} & \to & b_{n'}=a_{n} \qquad \qquad \forall \sigma \label{t4ran}
\end{eqnarray}
where 
\begin{equation}
\sigma=\sum_{k=-r'}^{-r-1}p^{k+r'}x^{i+k}+\sum_{k=l+1}^{l'}p^{k+r'}x^{i+k} \label{si}
\end{equation}
}

\noindent \textbf{Example:} Rule  $^{0}6_{2}^{1}$ has vector $(a_{0},a_{1},a_{2},a_{3})=(0,1,1,0)$. If now the range of the rule is increased adding one site from the left, i.e. $l \to l'=l+1$, then $\Omega'=p^{l'+r+1}=\Omega p=8$ and $\sigma$, from Eq. (\ref{si}), can be either $0$ or $4$. Therefore, the class equivalent rule after change of range is $(a_{0},a_{1},a_{2},a_{3},a_{0},a_{1},a_{2},a_{3})=(0,1,1,0,0,1,1,0)$ i.e. rule $^{1}102_{2}^{1}$. \\

\noindent \textbf{Example:} Rule  $^{1}6_{2}^{0}$ has also vector $(a_{0},a_{1},a_{2},a_{3})=(0,1,1,0)$. If now the range of the rule is increased adding one site from the right, i.e. $r \to r'=r+1$, then $\Omega'=p^{l+r'+1}=\Omega p=8$ and $\sigma$, from Eq. (\ref{si}), can be either $0$ or $1$. The class equivalent rule after change of range is then $(a_{0},a_{0},a_{1},a_{1},a_{2},a_{2},a_{3},a_{3})=(0,0,1,1,1,1,0,0)$ i.e. rule $^{1}60_{2}^{1}$. \\

\noindent  \textbf{Theorem S5 (Invariance under shift):}
\emph{The universal CA map, Eq. (\ref{CA}) remains invariant after the following sets of transformations
\begin{eqnarray}
l & \to & l-1 \label{t1c} \\
r & \to & r+1 \label{t2c} \\
x_{t+1}^{i} & \to & x_{t+1}^{i-1}  \label{t3b} 
\end{eqnarray}
and 
\begin{eqnarray}
l & \to & l+1 \label{t1d} \\
r & \to & r-1 \label{t2d} \\
x_{t+1}^{i} & \to & x_{t+1}^{i+1}  \label{t3d} 
\end{eqnarray}}

\noindent \emph{Proof:} By making the transformations implied by Eqs.(\ref{t1c}) to (\ref{t3b}) in Eq. (\ref{CA})
\begin{equation}
x_{t+1}^{i-1}=\sum_{n=0}^{\Omega-1}a_{n}\mathcal{B}\left(n-\sum_{k'=-r-1}^{l-1}p^{k'+r+1}x_{t}^{i+k'},\frac{1}{2}\right)=\sum_{n=0}^{\Omega-1}a_{n}\mathcal{B}\left(n-\sum_{k=-r}^{l}p^{k+r}x_{t}^{i-1+k},\frac{1}{2}\right) 
\end{equation}
where the change of the dummy variable $k' \to k-1$ has been made. The latter expression is, of course, equivalent to
\begin{equation}
x_{t+1}^{i-1}=\ ^{l}R_{p}^{r}(x_{t}^{i-1}) \label{rule3b}
\end{equation}
Making the change $i \to i+1$ (since global translation invariance holds) the invariance under shift is proved. The proof involving Eqs. (\ref{t1d}) to (\ref{t3d}) is similar. $\Box$ \\

The invariance under shift of the universal CA map implies the existence of classes of rules related by a breaking of this symmetry which contain, at most, $\rho$ elements. The rules on these classes share the same code $R$ but the neighborhoods contain different numbers of sites to the left and to the right (although the range $\rho$ is the same). These rules satisfy the following identities
\begin{eqnarray}
^{l}R_{p}^{r}(x_{t}^{i})=\ ^{l-1}R_{p}^{r+1}(x_{t}^{i-1}) \label{sh1} \\
^{l}R_{p}^{r}(x_{t}^{i})=\ ^{l+1}R_{p}^{r-1}(x_{t}^{i+1}) \label{sh2}
\end{eqnarray}
The rules are equivalent in a sense that, when known the dynamical behavior of one of them, the behavior of the others is predictable in terms of the latter just by applying a global spatial shift of the dynamical state on the ring.\\

\begin{figure}
\begin{center}
\includegraphics[width=0.85\textwidth, angle=270]{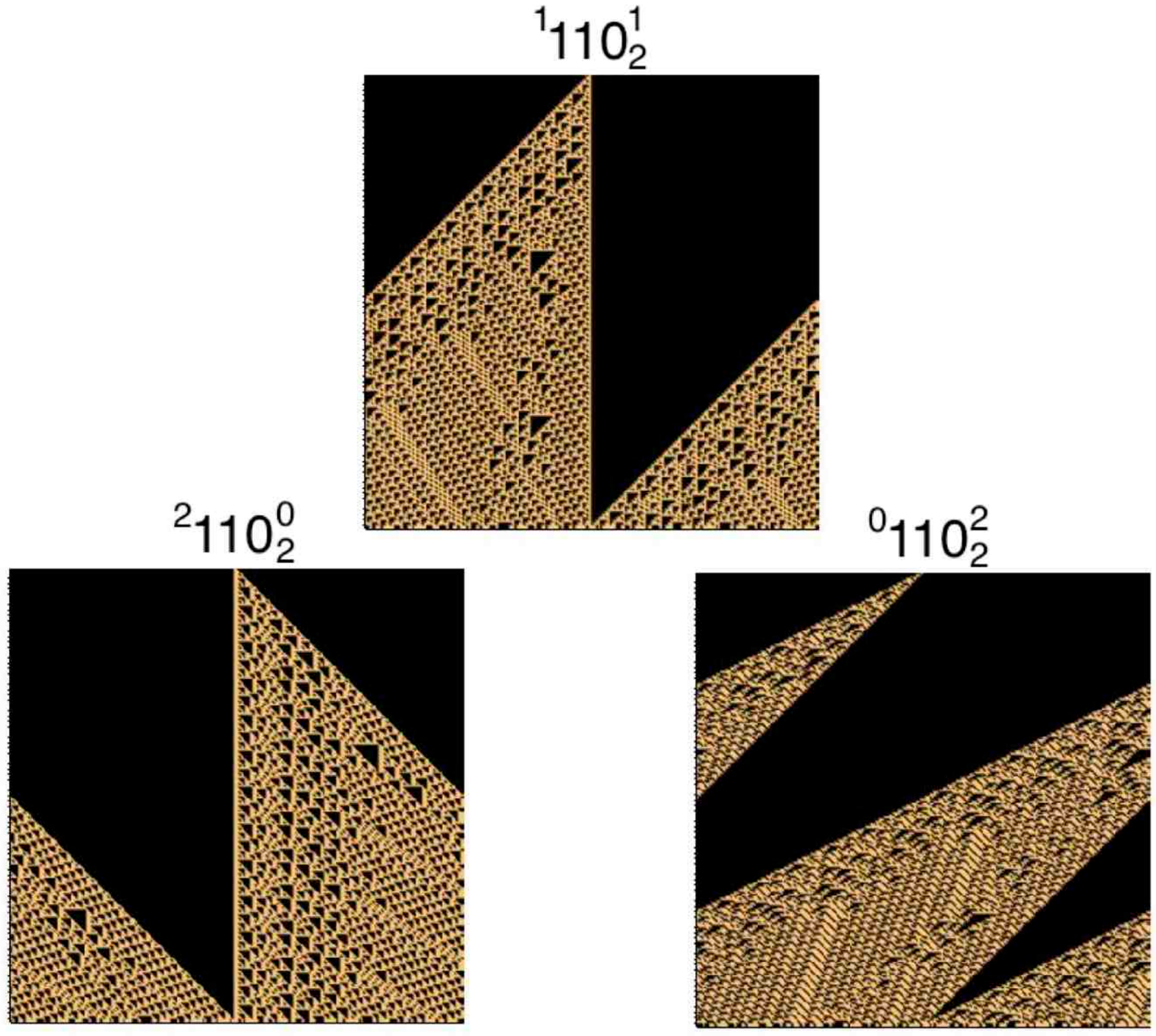}
\end{center}
\noindent Figure S4: Spatiotemporal evolution of rules $^{1}110^{1}_{2}$, $^{2}110^{0}_{2}$ and $^{0}110^{2}_{2}$. All three rules are related by a shift transformation and belong to the same equivalence class under shift. Rules related by shifting transformations correspond to the same dynamical behaviors as seen by observers moving with different (constant) velocities to either side of the ring. For a given rule $^{l}R^{r}_{p}$, the shifted rules $^{l \pm v}R^{r \mp v}_{p}$  corresponds to the same dynamics as followed by an observer moving on the ring at a constant velocity $\pm v$ (where the positive sign corresponds to motion to the left). This is the principle of Galilean invariance for cellular automata rules.
\end{figure}

\noindent \textbf{Example:} Rules $^{1}110^{1}_{2}$, $^{2}110^{0}_{2}$ and $^{0}110^{2}_{2}$ (see Fig. S4) are related by a shift transformation and belong to the same equivalence class under shift: once known the dynamical behavior of rule $^{1}110^{1}_{2}$, the ones of rules $^{2}110^{0}_{2}$ and $^{0}110^{2}_{2}$ are equivalent by globally shifting the dynamical state on the ring one site to the right or or to the left, respectively.\\

Rules related by shifting transformations correspond to the same dynamical behaviors as seen by observers moving with different (constant) velocities to either side of the ring. For a given rule $^{l}R^{r}_{p}$, the shifted rules $^{l \pm v}R^{r \mp v}_{p}$  corresponds to the same dynamics as followed by an observer moving on the ring at a constant velocity $\pm v$ (where the positive sign corresponds to motion to the left). This is the principle of Galilean invariance for cellular automata rules. \\

\begin{figure}
\begin{center}
\includegraphics[width=0.75\textwidth]{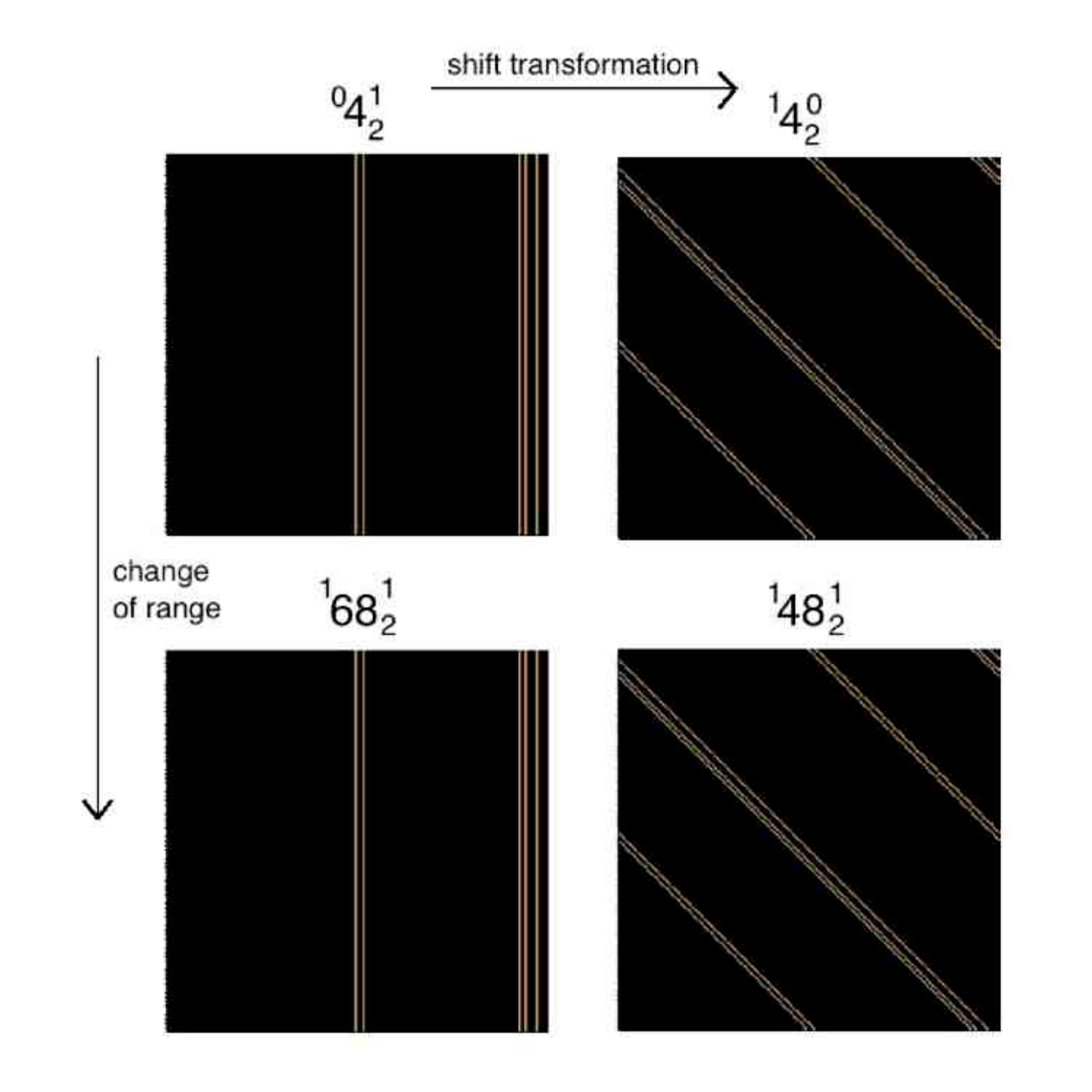}
\end{center}
\noindent Figure S5: Spatiotemporal evolution of the rules indicated in the figure and their interrelationships through shift transformation and change of range.
\end{figure}

Shift transformations and invariance under change of range are particularly powerful when combined together to relate symmetrical rules (i.e. with $l=r$) with the same range $\rho$ but a different value of $R$. For example, rules $^{0}4^{1}_{2}$ and $^{1}4^{0}_{2}$ are trivially related by a shift transformation. Since rules $^{0}4^{1}_{2}$ and $^{1}68^{1}_{2}$ are equivalent under change of range, as are the rules $^{1}4^{0}_{2}$ and $^{1}48^{1}_{2}$, it is clear that rules  $^{1}48^{1}_{2}$ and $^{1}68^{1}_{2}$ are also related by a shift transformation. This is not trivial at all, since $^{1}48^{1}_{2}$ and $^{1}68^{1}_{2}$ are both symmetrical rules, with the same range, but with a different value of $R$. Both of them represent, however, the same dynamics as followed by observers moving at different relative velocities. This is clearly evidenced in Fig. S5 where the spatiotemporal evolution of all these rules is shown.\\

\noindent  \textbf{Theorem S6 (Constructor's theorem):}
\emph{Let a set of $p$ rules each denoted by $^{l}(A_{m})^{r}_{p}$ with $m\in[0,p-1]$ with code $A_{m}=\sum_{n=0}^{\Omega-1}a_{n}^{(m)}p^{n}$, with $\Omega=p^{l+r+1}$. A rule $^{l+1}R^{r}_{p}$ can be constructed from the left in the following way
\begin{equation}
^{l+1}R_{p}^{r}= \sum_{m=0}^{p-1}\mathcal{B}\left(x_{t}^{i+l+1}-m,\frac{1}{2}\right)\ ^{l}(A_{m})_{p}^{r} \label{S3a}
\end{equation}
with $R=\sum_{n=0}^{\Omega-1}\sum_{m=0}^{p-1}a_{n}^{(m)}p^{n+m\Omega}$. and from the right as
\begin{equation}
^{l}R_{p}^{r+1}= \sum_{k=0}^{p-1}\mathcal{B}\left(x_{t}^{i-r-1}-m,\frac{1}{2}\right)\ ^{l}(A_{m})_{p}^{r}  \label{S3b}
\end{equation}
with $R=\sum_{n=0}^{\Omega-1}\sum_{m=0}^{p-1}a_{n}^{(m)}p^{np+m}.$
} \\

\noindent \emph{Proof:} By using Eqs. Eq. (\ref{CA}) and (\ref{S3a})
\begin{eqnarray}
x_{t+1}^{i}&=& \sum_{x^{i+l+1}=0}^{p-1}\sum_{n=0}^{\Omega-1}c_{n,x^{i+l+1}}\mathcal{B}\left(x_{t}^{i+l+1}-x^{i+l+1},\frac{1}{2}\right)\mathcal{B}\left(n-\sum_{k=-r}^{l}p^{k+r}x_{t}^{i+k},\frac{1}{2}\right) \nonumber \\
&=& \sum_{x^{i+l+1}=0}^{p-1}\sum_{n=0}^{\Omega-1}c_{n,x^{i+l+1}}\mathcal{B}\left(p^{l+r+1}(x_{t}^{i+l+1}-x^{i+l+1}),\frac{1}{2}\right)\mathcal{B}\left(n-\sum_{k=-r}^{l}p^{k+r}x_{t}^{i+k},\frac{1}{2}\right) \nonumber
\end{eqnarray}
where $(xiib)$ from Appendix A has been used and
\begin{equation}
c_{n,x^{i+l+1}} \equiv a_{n}^{(m)} \mathcal{B}\left(x^{i+l+1}-m,\frac{1}{2}\right) = \left\{
\begin{array}{cc}
0  & \quad m \ne x^{i+l+1}    \\
a_{n}^{(m)}  & \quad m=x^{i+l+1}     
\end{array}
\right. \nonumber
\end{equation}
has been introduced. By using now $(xii)$ from Appendix A
\begin{equation}
x_{t+1}^{i}=\sum_{n'=0}^{\Omega'-1}c_{n'}\mathcal{B}\left(n'-\sum_{k=-r}^{l+1}p^{k+r}x_{t}^{i+k},\frac{1}{2}\right) \label{constructed}
\end{equation}
where $n'=\sum_{k=-r}^{l+1}p^{k+r}x^{i+k}=n+p^{l+r+1}x^{i+l+1}$, $\Omega'=p^{l+r+2}=p\Omega$ and $c_{n'} = a_{n'-m\Omega}^{(m)}$. The new rule, Eq. (\ref{constructed}) has then the code $R=\sum_{n'=0}^{\Omega'-1}c_{n'}p^{n'}=\sum_{n=0}^{\Omega-1}\sum_{m=0}^{p-1}a_{n}^{(m)}p^{n+m\Omega}$.

The proof of the statement on the construction from the right proceeds in a similar manner. By using Eqs. Eq. (\ref{CA}) and (\ref{S3b}) 
\begin{eqnarray}
x_{t+1}^{i}&=& \sum_{x^{i-r-1}=0}^{p-1}\sum_{n=0}^{\Omega-1}c_{n,x^{i-r-1}}\mathcal{B}\left(x_{t}^{i-r-1}-x^{i-r-1},\frac{1}{2}\right)\mathcal{B}\left(n-\sum_{k=-r}^{l}p^{k+r}x_{t}^{i+k},\frac{1}{2}\right) \nonumber \\
&=& \sum_{x^{i-r-1}=0}^{p-1}\sum_{n=0}^{\Omega-1}c_{n,x^{i-r-1}}\mathcal{B}\left(x_{t}^{i-r-1}-x^{i-r-1},\frac{1}{2}\right)\mathcal{B}\left(np-\sum_{k=-r}^{l}p^{k+r+1}x_{t}^{i+k},\frac{1}{2}\right) \nonumber
\end{eqnarray}
where the result $(xiib)$ from Appendix A has been used and
\begin{equation}
c_{n,x^{i-r-1}} \equiv a_{n}^{(m)} \mathcal{B}\left(x^{i-r-1}-m,\frac{1}{2}\right) = \left\{
\begin{array}{cc}
0  & \quad m \ne x^{i-r-1}    \\
a_{n}^{(m)}  & \quad m=x^{i-r-1}     
\end{array}
\right. \nonumber
\end{equation}has been introduced. By using now $(xiii)$ from Appendix A 
\begin{equation}
x_{t+1}^{i}=\sum_{n'=0}^{\Omega'-1}c_{n'}\mathcal{B}\left(n'-\sum_{k=-r-1}^{l}p^{k+r+1}x_{t}^{i+k},\frac{1}{2}\right) \label{constructed}
\end{equation}
where $n'=\sum_{k=-r-1}^{l}p^{k+r+1}x^{i+k}=np+x^{i-r-1}$, $\Omega'=p^{l+r+2}=p\Omega$ and $c_{n'} = a_{(n'-m)/p}^{(m)}$. 
The new rule, Eq. (\ref{constructed}) has then the code $R=\sum_{n'=0}^{\Omega'-1}c_{n'}p^{n'}=\sum_{n=0}^{\Omega-1}\sum_{m=0}^{p-1}a_{n}^{(m)}p^{np+m}$ and the result is proved. $\Box$ \\

\noindent \textbf{Example:}  Theorem S6 can be straightforwardly applied to any rule. Wolfram Rule $^{1}30^{1}_{2}$ is known to be a random number generator. It has vector $(a_{0}, a_{1}, a_{2}, a_{3}, a_{4}, a_{5}, a_{6}, a_{7})=(0,1,1,1,1,0,0,0)$. To obtain the construction from the left, simply separate $(0,1,1,1,1,0,0,0)$ into $p=2$ consecutive parts with same size. Rules with vectors $(0,1,1,1)$ and $(1,0,0,0)$ are obtained, which correspond, respectively, to rules $^{0}14^{1}_{2}$ and $^{0}1^{1}_{2}$. To construct the same rule from the right, separate the odd entries and the even entries of $(0,1,1,1,1,0,0,0)$. Rules $(0,1,1,0)$ and $(1,1,0,0)$ which correspond to rules $^{1}6^{0}_{2}$ and $^{1}3^{0}_{2}$ are hence obtained.  $\Box$ \\

\noindent  \textbf{Definition S2 (Copying rules):}
\emph{The rule $^{l}A^{r}_{p}$ is said to be \textbf{copied} to a higher range if it is used as the only rule in constructing $^{l+1}R^{r}_{2}$ from the left, i.e. if $^{l}(A_{m})^{r}_{p}=\ ^{l}A^{r}_{2}$ $\forall m$ in Eq. (\ref{S3a}).} $\Box$ \\

\noindent \textbf{Example:}  To copy rule $^{0}6^{1}_{2}$ to a higher range $\rho=3$ its vector $(a_{0}, a_{1}, a_{2}, a_{3})=(0,1,1,0)$ is simply concatenated to itself, adding the same sequence of zeroes and ones after it i.e. $(0,1,1,0,0,1,1,0)$. The latter corresponds to rule $^{1}102^{1}_{2}$ which is the copy of $^{0}6^{1}_{2}$ to range $\rho=3$. This is consistent with the construction from the left of rule $^{1}102^{1}_{2}$ as established in Theorem S6. Both rules belong to the same class under change of range and display an identical dynamical behavior.  $\Box$ \\

\noindent \textbf{Example:}  To copy rule $^{0}25^{0}_{3}$ to a higher range $\rho=2$ its vector $(a_{0}, a_{1}, a_{2})=(1,2,2)$ is concatenated after itself two times (since $p=3$) i.e. $(1,2,2,1,2,2,1,2,2)$. The latter corresponds to rule $^{1}18925^{0}_{3}$ which is the copy of the original $^{0}25^{0}_{3}$ rule.   $\Box$ \\

The following theorem allows to construct a polynomial of integer variables for each CA rule contained in Eq. (\ref{CA}) and Eq. (\ref{CAtot}) and provides, indeed, the universal CA maps in polynomial form. This is an important result since it allows to pass from a description in terms of boxcar functions to polynomials involving integer variables. In order to obtain the latter a linear system of equations involving a Vandermonde matrix of $\Omega \times \Omega$ dimensions needs to be solved. Eq. (\ref{CA}),  which is computationally inexpensive and sufficient to characterize each CA rule is, therefore, more fundamental.\\

\noindent  \textbf{Theorem S7 (Polynomial maps):}
\emph{The following result holds} 
\begin{equation}
^{l}R^{r}_{p}(x_{t}^{i})\equiv x_{t+1}^{i} =\sum_{n=0}^{\Omega-1}\alpha_{n}\left(\sum_{k=-r}^{l}p^{k+r}x_{t}^{i+k}\right)^{n} \label{poly}
\end{equation}
\emph{where the coefficients $\alpha_{n}$'s (rational numbers) are the solutions of the following linear system of equations}
\begin{equation}
    \begin{bmatrix} 0 & 0 & 0 & \ldots & 0 & 1 \\ 1 & 1 & 1 & \ldots & 1 & 1 \\ 
    2^{\Omega-1} & 2^{\Omega-2} & 2^{\Omega-3} & \ldots & 2 & 1 \\    
    \vdots & \vdots & \vdots & & \vdots & \vdots \\ (\Omega-1)^{\Omega-1} & (\Omega-1)^{\Omega-2} & (\Omega-1)^{\Omega-3} & \ldots & \Omega-1 & 1 \end{bmatrix} \begin{bmatrix} \alpha_{\Omega-1} \\ \alpha_{\Omega-2} \\ \alpha_{\Omega-3} \\ \vdots \\ \alpha_0 \end{bmatrix} = \begin{bmatrix} a_0 \\ a_1 \\ a_2 \\ \vdots \\ a_{\Omega-1} \end{bmatrix}. \label{thematrix}
\end{equation}
\emph{For a totalistic rule, one has}
\begin{equation}
^{l}RT^{r}_{p}(x_{t}^{i})\equiv x_{t+1}^{i} =\sum_{s=0}^{\Theta}\alpha_{s}\left(\sum_{k=-r}^{l}x_{t}^{i+k}\right)^{s} \label{totpoly}
\end{equation}
\emph{where the coefficients $\alpha_{s}$'s (rational numbers) are the solutions of the following linear system of equations}
\begin{equation}
    \begin{bmatrix} 0 & 0 & 0 & \ldots & 0 & 1 \\ 1 & 1 & 1 & \ldots & 1 & 1 \\ 
    2^{\Theta} & 2^{\Theta-1} & 2^{\Theta-2} & \ldots & 2 & 1 \\    
    \vdots & \vdots & \vdots & & \vdots & \vdots \\ \Theta^{\Theta} & \Theta^{\Theta-1} & \Theta^{\Theta-2} & \ldots & \Theta & 1 \end{bmatrix} \begin{bmatrix} \alpha_{\Theta} \\ \alpha_{\Theta-1} \\ \alpha_{\Theta-2} \\ \vdots \\ \alpha_0 \end{bmatrix} = \begin{bmatrix} \sigma_0 \\ \sigma_1 \\ \sigma_2 \\ \vdots \\ \sigma_{\Theta} \end{bmatrix}. 
\end{equation}
~\\
\noindent \emph{Proof:} Eqs. (\ref{poly}) and (\ref{totpoly}) are, respectively, the polynomial interpolations of Eqs. (\ref{CA}) and (\ref{CAtot}). For the topic of polynomial interpolation see for example \footnotemark \footnotetext{M.J.D. Powell (1981). \emph{Approximation Theory and Methods}, Chapter 4. Cambridge University Press.} $\Box$ \\

\subsection{Results for boolean $(p=2)$ CA rules}

The simplest CA rules involve two symbols $p=2$ and only the cell $x_{t}^{i}$ that is updated by the rule ($r=l=0$). Since the input can be just only zero or one, and for each posibility the output can be also only zero or one there is a total of four such rules, labelled as $^{0}R_{2}^{0}$. \\

\noindent  \textbf{Theorem S8 (Universal CA map for the 4 most simple boolean rules):}
\emph{The universal CA map, Eq. (\ref{CA}), reduces to the following expression
\begin{equation}
x_{t+1}^{i}=a_{0}(1-x_{t}^{i})+a_{1}x_{t}^{i}
\label{CAsimpl}
\end{equation}
for the 4 most simple boolean rules (p=2, l=0, r=0), i.e. the rules $^{0}R_{2}^{0}$, where $R=\sum_{n=0}^{1}a_{n}2^{n}$.} \\

\noindent \emph{Proof:} By inserting $l=r=0$, $p=2$, $\Omega=2$ in Eq. (\ref{CA})
\begin{equation}
x_{t+1}^{i}=\sum_{n=0}^{1}a_{n}\mathcal{B}\left(x_{t}^{i}-n,\frac{1}{2}\right)=a_{0}\mathcal{B}\left(x_{t}^{i},\frac{1}{2}\right)+a_{1}\mathcal{B}\left(x_{t}^{i}-1,\frac{1}{2}\right) \label{CASi} \nonumber
\end{equation}
whence, by simply applying $(xv)$ from Appendix A, the result is proved. An alternative proof is given by using Theorem S7, Eq. (\ref{poly}), which takes in this case the simple form
\begin{equation}
x_{t+1}^{i}=\alpha_{0}+\alpha_{1}x_{t}^{i}
\end{equation}
where $\alpha_0$ and $\alpha_1$ can be obtained by solving Eq. \ref{thematrix} 
\begin{equation}
    \begin{bmatrix} 0 & 1 \\ 1 & 1 \end{bmatrix} \begin{bmatrix} \alpha_{1} \\ \alpha_{0} \end{bmatrix} = \begin{bmatrix} a_0 \\ a_1  \end{bmatrix}. \label{thematrixpoor}
\end{equation}
which yields $\alpha_0=a_0$ and $\alpha_1=a_1-a_0$ thus proving again Eq. (\ref{CAsimpl}).
$\Box$ \\

\begin{table}[htdp]
\begin{center}
\begin{tabular}{cclcccl}
Rule & $ \qquad (a_{0},a_{1})$ & \qquad Map \\
$^{0}0_{2}^{0}$ & \qquad (0,0) & $\qquad x_{t+1}^{i}=0$ \\
$^{0}1_{2}^{0}$ & \qquad (1,0) & $\qquad x_{t+1}^{i}=1-x_{t}^{i}$ \\
$^{0}2_{2}^{0}$ & \qquad (0,1) & $\qquad x_{t+1}^{i}=x_{t}^{i}$ \\
$^{0}3_{2}^{0}$ & \qquad (1,1) & $\qquad x_{t+1}^{i}=1$ \\
\end{tabular}
\end{center}
Table S1: The four cellular automata maps for the simplest boolean rules $^{0}R_{2}^{0}$
\end{table}

The maps for the $^{0}R_{2}^{0}$ rules are summarized in Table S1 and plotted for an initial condition consisting on a single seed (Fig. S6, top) and a random initial vector (Fig. S6, bottom). 

\begin{figure}
\includegraphics[width=0.7\textwidth]{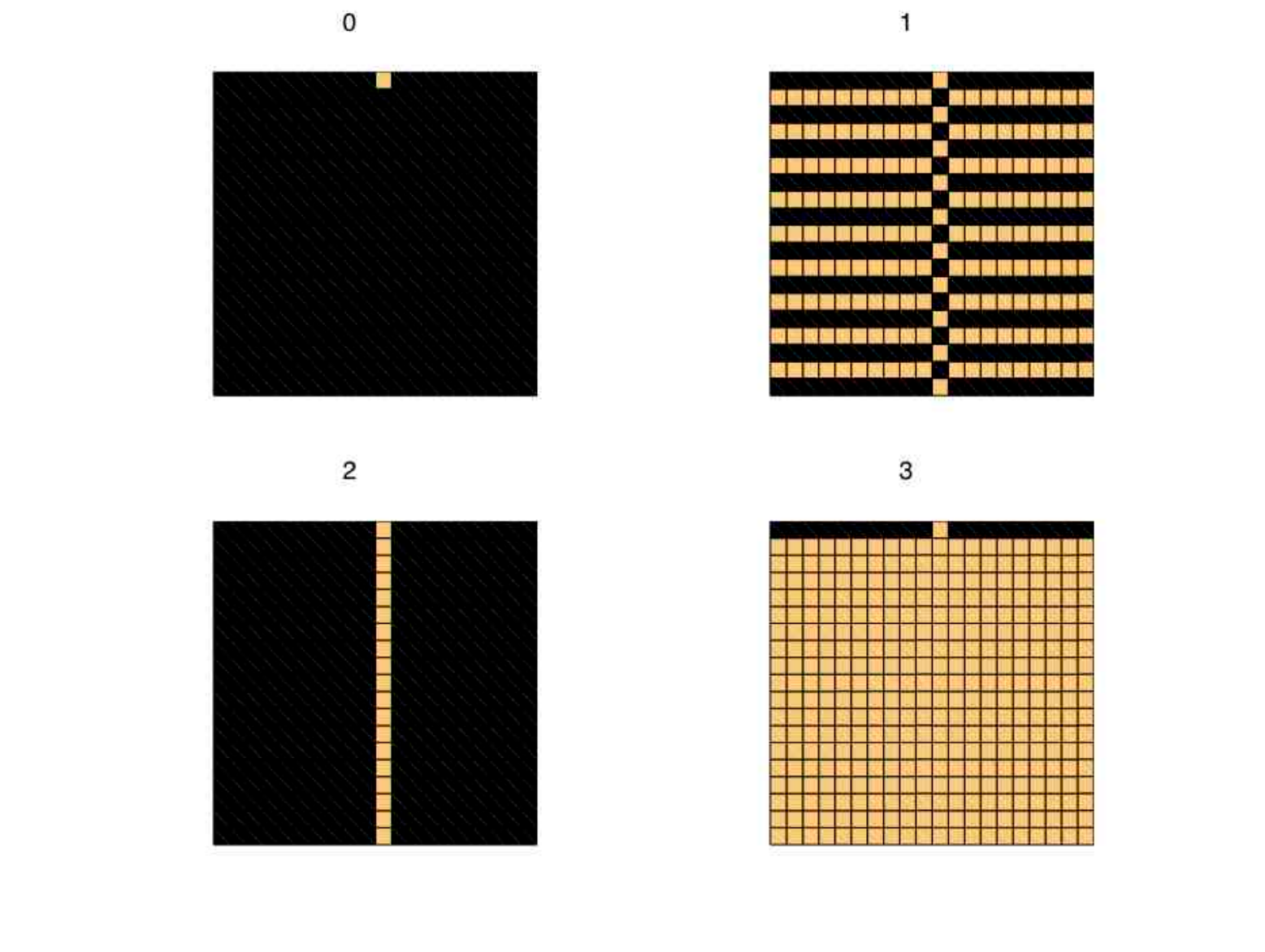}
~\\~\\
\includegraphics[width=0.7\textwidth]{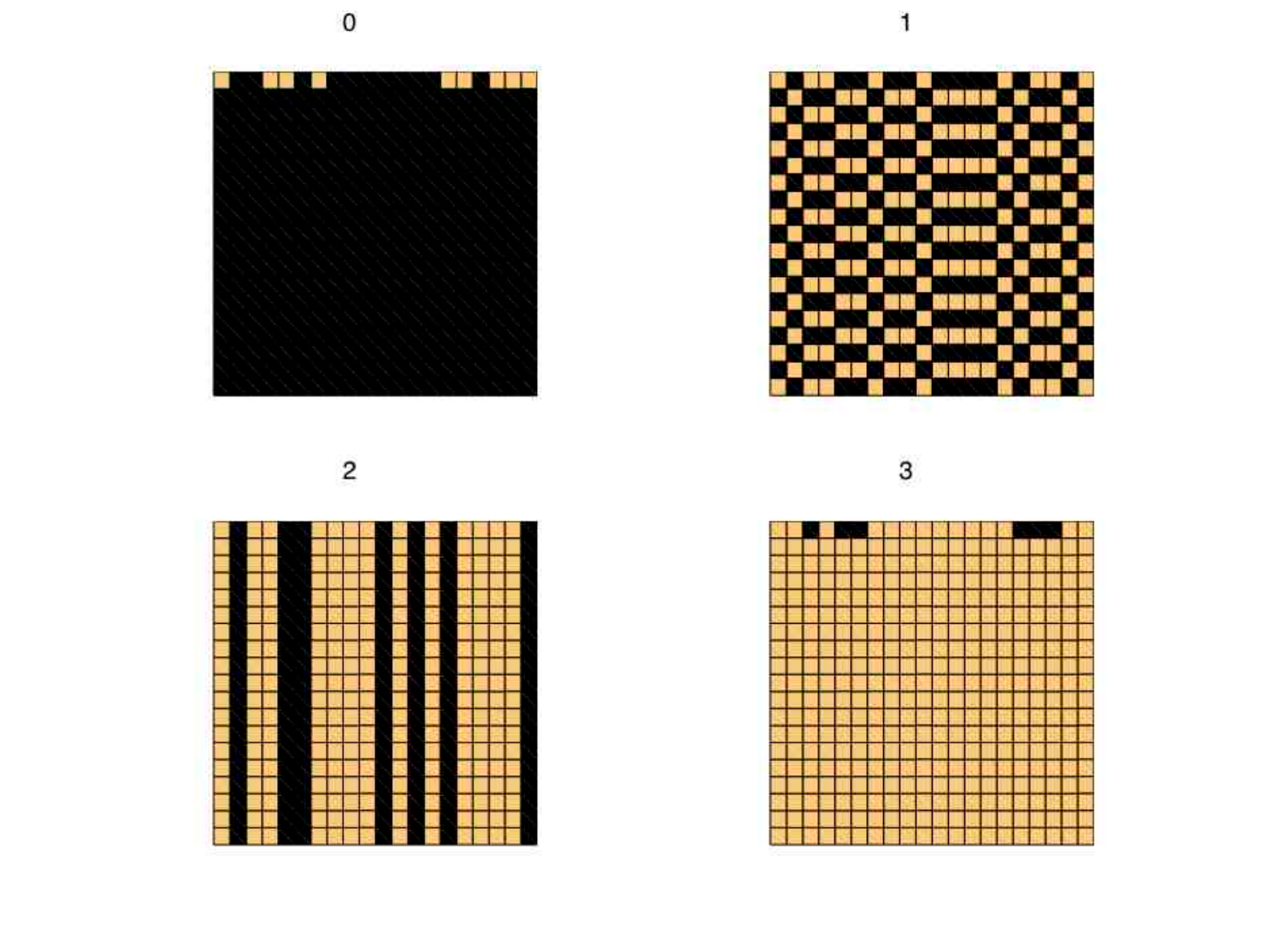}
\begin{center}
Figure S6: Spatiotemporal evolution of the four rules $^{0}R_{2}^{0}$ (values of $R$ are indicated over each panel) for an initial condition consisting on a single seed (top) and for a random vector (bottom). 
\end{center} 
\end{figure}

The local and global dynamics of the repetitive application of these rules is trivially predictable: given a vector $\mathbf{x}_{0}=(x_{0}^{1},...,x_{0}^{N_{s}})$ as initial condition, the vector $\mathbf{x}_{t}=(x_{t}^{1},...,x_{t}^{N_{s}})$ can be determined as a function of $\mathbf{x}_{0}$ and $t$ by means of a closed analytical expression. In Table S2, these expressions are given for the local (top) and the global (bottom) dynamics.

\begin{table}[htdp]
\begin{center}
\begin{tabular}{cclcccl}
$^{0}0_{2}^{0}$ & $\qquad x_{t}^{i}$&$=0$ & $\qquad \qquad$ & 
$^{0}2_{2}^{0}$ & $\qquad x_{t}^{i}$&$=x_{0}^{i}$  \\
$^{0}1_{2}^{0}$ & $\qquad x_{t}^{i}$&$=(-1)^{t}x_{0}^{i}+\frac{1}{2}(1-(-1)^{t})$
& $\qquad \qquad$ & 
$^{0}3_{2}^{0}$ & $\qquad x_{t}^{i}$&$=1$ \\
\end{tabular}
\end{center}
\begin{center}
\end{center}
\begin{center}
\begin{tabular}{cclcccl}
$^{0}0_{2}^{0}$ & $\qquad \mathbf{x}_{t}$&$=0$ & $\qquad \qquad$ & 
$^{0}2_{2}^{0}$ & $\qquad \mathbf{x}_{t}$&$=\mathbf{x}_{0}$ \\
$^{0}1_{2}^{0}$ & $\qquad \mathbf{x}_{t}$&$=(-1)^{t}\mathbf{x}_{0}+\frac{1}{2}(1-(-1)^{t})$ & $\qquad \qquad$ & 
$^{0}3_{2}^{0}$ & $\qquad \mathbf{x}_{t}$&$=1$ \\
\end{tabular}
\end{center}
Table S2: Local (top) and global (bottom) dependence in time of the state of the cellular automata for the simplest boolean rules $^{0}R_{2}^{0}$
\end{table}

One can easily prove all entries on Table S2. The proof of the entry for rule $^{0}1_{2}^{0}$ proceeds by induction. Clearly, the expression is valid at $t=0$ and at $t=1$. If one assumes it valid for $t-1$, then, at time $t$ 
\begin{eqnarray}
x_{t}^{i}&=&1-x_{t-1}^{i}=1-\left[(-1)^{t-1}x_{0}^{i}+\frac{1}{2}(1-(-1)^{t-1})\right]=-(-1)^{t-1}x_{0}^{i}+\frac{1}{2}(1+(-1)^{t-1}) \nonumber \\
&=&(-1)^{t}x_{0}^{i}+\frac{1}{2}(1-(-1)^{t}) \qquad \qquad \qquad \qquad \qquad \qquad \qquad \qquad q.e.d. \nonumber \\ \nonumber
\end{eqnarray}

\begin{table}[htdp]
\begin{center}
\begin{tabular}{cclcccl}
Rule & $ \qquad (a_{0},a_{1},a_{2},a_{3})$ & \qquad Map \\
$^{0}0_{2}^{1}$ & \qquad (0,0,0,0) & $\qquad x_{t+1}^{i}=0$ \\
$^{0}1_{2}^{1}$ & \qquad (1,0,0,0) & $\qquad x_{t+1}^{i}=1-x_{t}^{i}-x_{t}^{i-1}+x_{t}^{i}x_{t}^{i-1}$ \\
$^{0}2_{2}^{1}$ & \qquad (0,1,0,0) & $\qquad x_{t+1}^{i}=x_{t}^{i-1}-x_{t}^{i}x_{t}^{i-1}$ \\
$^{0}3_{2}^{1}$ & \qquad (1,1,0,0) & $\qquad x_{t+1}^{i}=1-x_{t}^{i}$ \\
$^{0}4_{2}^{1}$ & \qquad (0,0,1,0) & $\qquad x_{t+1}^{i}=x_{t}^{i}-x_{t}^{i}x_{t}^{i-1}$ \\
$^{0}5_{2}^{1}$ & \qquad (1,0,1,0) & $\qquad x_{t+1}^{i}=1-x_{t}^{i-1}$ \\
$^{0}6_{2}^{1}$ & \qquad (0,1,1,0) & $\qquad x_{t+1}^{i}=x_{t}^{i}+x_{t}^{i-1}-2x_{t}^{i-1}x_{t}^{i}$ \\
$^{0}7_{2}^{1}$ & \qquad (1,1,1,0) & $\qquad x_{t+1}^{i}=1-x_{t}^{i}x_{t}^{i-1}$ \\
$^{0}8_{2}^{1}$ & \qquad (0,0,0,1) & $\qquad x_{t+1}^{i}=x_{t}^{i}x_{t}^{i-1}$ \\
$^{0}9_{2}^{1}$ & \qquad (1,0,0,1) & $\qquad x_{t+1}^{i}=1-x_{t}^{i}-x_{t}^{i-1}+2x_{t}^{i}x_{t}^{i-1}$ \\
$^{0}10_{2}^{1}$ & \qquad (0,1,0,1) & $\qquad x_{t+1}^{i}=x_{t}^{i-1}$ \\
$^{0}11_{2}^{1}$ & \qquad (1,1,0,1) & $\qquad x_{t+1}^{i}=1-x_{t}^{i}+x_{t}^{i-1}x_{t}^{i}$ \\
$^{0}12_{2}^{1}$ & \qquad (0,0,1,1) & $\qquad x_{t+1}^{i}=x_{t}^{i}$ \\
$^{0}13_{2}^{1}$ & \qquad (1,0,1,1) & $\qquad x_{t+1}^{i}=1-x_{t}^{i-1}+x_{t}^{i-1}x_{t}^{i}$ \\
$^{0}14_{2}^{1}$ & \qquad (0,1,1,1) & $\qquad x_{t+1}^{i}=x_{t}^{i}+x_{t}^{i-1}-x_{t}^{i}x_{t}^{i-1}$ \\
$^{0}15_{2}^{1}$ & \qquad (1,1,1,1) & $\qquad x_{t+1}^{i}=1$ \\
\end{tabular}
\end{center}
Table S3: The sixteen maps for the cellular automata implementations of the sixteen logical functions. The maps correspond to rules $^{0}R_{2}^{1}$.
\end{table}
\pagebreak

\begin{table}[htdp]
\begin{center}
\begin{tabular}{cclcccl}
Rule & $ \qquad (a_{0},a_{1},a_{2},a_{3})$ & \qquad Map \\
$^{1}0_{2}^{0}$ & \qquad (0,0,0,0) & $\qquad x_{t+1}^{i}=0$ \\
$^{1}1_{2}^{0}$ & \qquad (1,0,0,0) & $\qquad x_{t+1}^{i}=1-x_{t}^{i}-x_{t}^{i+1}+x_{t}^{i+1}x_{t}^{i}$ \\
$^{1}2_{2}^{0}$ & \qquad (0,1,0,0) & $\qquad x_{t+1}^{i}=x_{t}^{i}-x_{t}^{i}x_{t}^{i+1}$ \\
$^{1}3_{2}^{0}$ & \qquad (1,1,0,0) & $\qquad x_{t+1}^{i}=1-x_{t}^{i+1}$ \\
$^{1}4_{2}^{0}$ & \qquad (0,0,1,0) & $\qquad x_{t+1}^{i}=x_{t}^{i+1}-x_{t}^{i}x_{t}^{i+1}$ \\
$^{1}5_{2}^{0}$ & \qquad (1,0,1,0) & $\qquad x_{t+1}^{i}=1-x_{t}^{i}$ \\
$^{1}6_{2}^{0}$ & \qquad (0,1,1,0) & $\qquad x_{t+1}^{i}=x_{t}^{i}+x_{t}^{i+1}-2x_{t}^{i}x_{t}^{i+1}$ \\
$^{1}7_{2}^{0}$ & \qquad (1,1,1,0) & $\qquad x_{t+1}^{i}=1-x_{t}^{i+1}x_{t}^{i}$ \\
$^{1}8_{2}^{0}$ & \qquad (0,0,0,1) & $\qquad x_{t+1}^{i}=x_{t}^{i+1}x_{t}^{i}$ \\
$^{1}9_{2}^{0}$ & \qquad (1,0,0,1) & $\qquad x_{t+1}^{i}=1-x_{t}^{i}-x_{t}^{i+1}+2x_{t}^{i+1}x_{t}^{i}$ \\
$^{1}10_{2}^{0}$ & \qquad (0,1,0,1) & $\qquad x_{t+1}^{i}=x_{t}^{i}$ \\
$^{1}11_{2}^{0}$ & \qquad (1,1,0,1) & $\qquad x_{t+1}^{i}=1-x_{t}^{i+1}+x_{t}^{i}x_{t}^{i+1}$ \\
$^{1}12_{2}^{0}$ & \qquad (0,0,1,1) & $\qquad x_{t+1}^{i}=x_{t}^{i+1}$ \\
$^{1}13_{2}^{0}$ & \qquad (1,0,1,1) & $\qquad x_{t+1}^{i}=1-x_{t}^{i}+x_{t}^{i}x_{t}^{i+1}$ \\
$^{1}14_{2}^{0}$ & \qquad (0,1,1,1) & $\qquad x_{t+1}^{i}=x_{t}^{i}+x_{t}^{i+1}-x_{t}^{i+1}x_{t}^{i}$ \\
$^{1}15_{2}^{0}$ & \qquad (1,1,1,1) & $\qquad x_{t+1}^{i}=1$ \\
\end{tabular}
\end{center}
Table S3 (cont): The sixteen maps for the cellular automata implementations of the sixteen logical functions. The maps correspond to rules $^{1}R_{2}^{0}$.
\end{table}

\noindent  \textbf{Theorem S9 (Universal CA map for the 16 boolean logical functions):}
\emph{The universal CA map, Eq. (\ref{CA}), reduces to the following expression (boolean logical functions)
\begin{equation}
x_{t}^{i+1}=a_{0}(1-x_{t}^{i})(1-x_{t}^{i-1})+a_{1}x_{t}^{i-1}(1-x_{t}^{i}) + a_{2}x_{t}^{i}(1-x_{t}^{i-1})+a_{3}x_{t}^{i}x_{t}^{i-1} \label{CAlog1}
\end{equation}
for all 16 rules with p=2, l=0, r=1, i.e. the rules $^{0}R_{2}^{1}$, and to
\begin{equation}
x_{t}^{i+1}=a_{0}(1-x_{t}^{i+1})(1-x_{t}^{i})+a_{1}x_{t}^{i}(1-x_{t}^{i+1}) + a_{2}x_{t}^{i+1}(1-x_{t}^{i})+a_{3}x_{t}^{i+1}x_{t}^{i} \label{CAlog2}
\end{equation}
for all 16 rules with p=2, l=1, r=0, i.e. the rules $^{1}R_{2}^{0}$. In both cases $R=\sum_{n=0}^{3}a_{n}2^{n}$.} \\

\noindent \emph{Proof:} By inserting $l=0$, $p=2$, $r=1$, $\Omega=4$ in Eq. (\ref{CA}), 
\begin{equation}
x_{t+1}^{i}=\sum_{n=0}^{3}a_{n}\mathcal{B}\left(2x_{t}^{i}+x_{t}^{i-1}-n,\frac{1}{2}\right) \nonumber
\end{equation}
and by using the definition of $n$ given by Eq. (\ref{confCA}), i.e. $n=2x^{i}+x^{i-1}$, the latter expression can be rewritten as
\begin{equation}
x_{t+1}^{i}=\sum_{x^{i}=0}^{1}\sum_{x^{i-1}=0}^{1}a_{2x^{i}+x^{i-1}}\mathcal{B}\left(2(x_{t}^{i}-x^{i})+(x_{t}^{i-1}-x^{i-1}),\frac{1}{2}\right) \nonumber
\end{equation}
By applying now result $(xii)$ from Appendix A
\begin{equation}
x_{t+1}^{i}=\sum_{x^{i}=0}^{1}\sum_{x^{i-1}=0}^{1}a_{2x^{i}+x^{i-1}}\mathcal{B}\left(x_{t}^{i}-x^{i},\frac{1}{2}\right)\mathcal{B}\left(x_{t}^{i-1}-x^{i-1},\frac{1}{2}\right) \nonumber
\end{equation}
The sum can now be carried explicitly by replacing the corresponding values of $x^{i}$ and $x^{i-1}$ which pick values either 0 or 1 yielding four terms. Then, by applying result $(xv)$ from Appendix A, Eq. (\ref{CAlog1}) is proved.

The proof of Eq. (\ref{CAlog2}) is similar after making the replacements $x_{t}^{i} \to x_{t}^{i+1}$,  $x_{t}^{i-1} \to x_{t}^{i}$, $x^{i} \to x^{i+1}$, $x^{i-1} \to x^{i}$. $\Box$ \\

\begin{figure}
\includegraphics[width=0.8\textwidth]{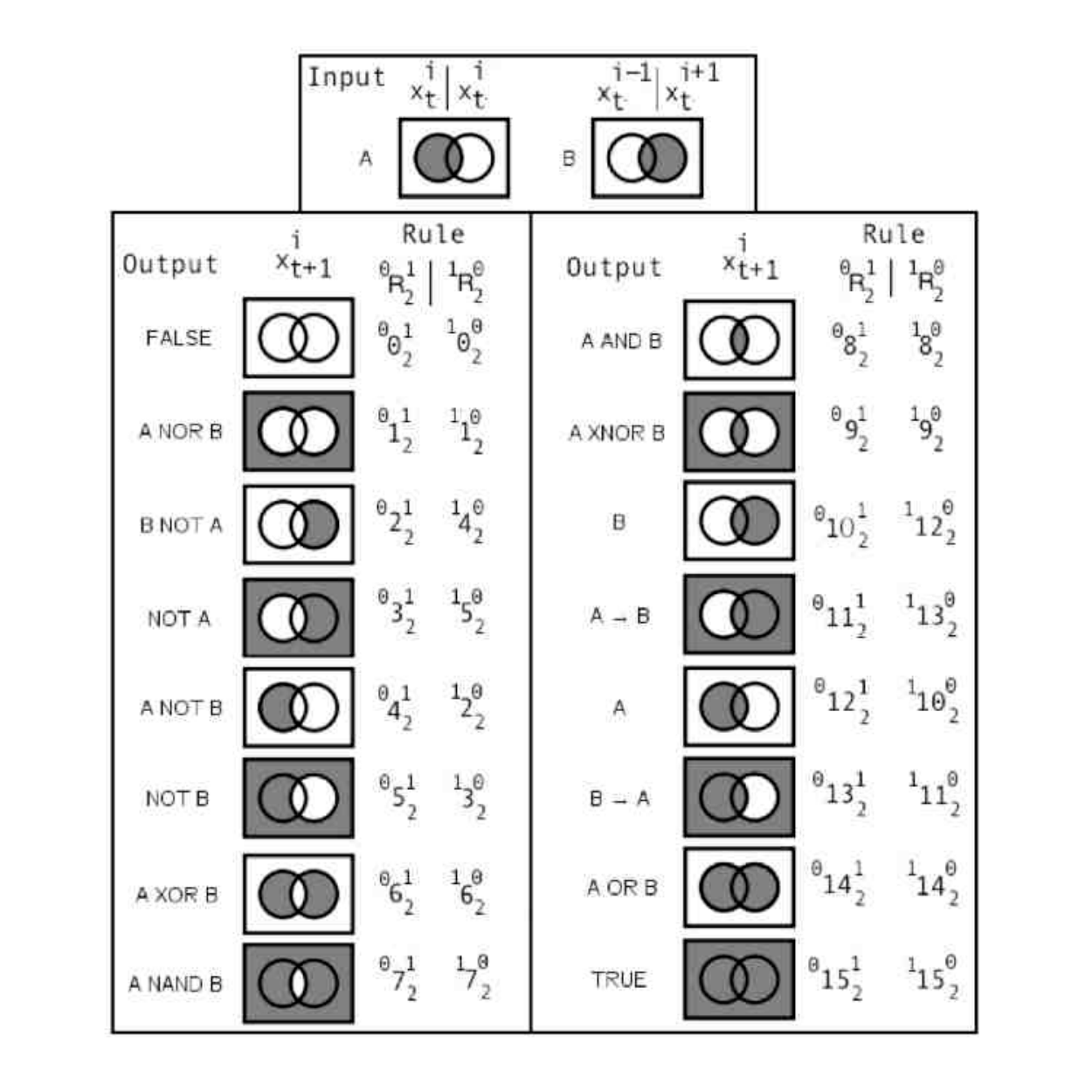}
\begin{center}
Figure S7: The sixteen boolean logical functions implemented through the cellular automata rules $^{0}R_{2}^{1}$ and $^{1}R_{2}^{0}$ indicated in the figure. The associated Venn diagrams are also shown. For both types of rules $^{0}R_{2}^{1}$ and $^{1}R_{2}^{0}$, $A$ denotes the input on site $x_{t}^{i}$ (which is updated at the next time step) while $B$ denotes the input on site $x_{t}^{i-1}$ for rules $^{0}R_{2}^{1}$ or $x_{t}^{i+1}$ for rules $^{1}R_{2}^{0}$.
\end{center} 
\end{figure}

The rules $^{0}R_{2}^{1}$ and $^{1}R_{2}^{0}$ which implement each of the 16 logical boolean functions are shown in Fig. S7 together with their associated Venn diagrams. For both types of rules $^{0}R_{2}^{1}$ and $^{1}R_{2}^{0}$, $A$ denotes the input on site $x_{t}^{i}$ (which is updated at the next time step) while $B$ denotes the input on site $x_{t}^{i-1}$ for rules $^{0}R_{2}^{1}$ or $x_{t}^{i+1}$ for rules $^{1}R_{2}^{0}$. 

As an example the logical function B NOT A, returns 'TRUE' (i.e. 1) if and only if in the previous time step only the site B (and not A) has the value 'TRUE'. This means for rule $^{0}2_{2}^{1}$ that one has the truth table $x_{t+1}^{i}=1$ (i.e. 'TRUE') only if A is 'FALSE' (i.e. if $x_{t}^{i}=0$) and B is 'TRUE' (i.e. $x_{t}^{i-1}=1$). Rule $^{1}4_{2}^{0}$ implements the same logical function, but now B 'TRUE' means $x_{t}^{i+1}=1$.

Because these functions shall prove important in building the theory of complexity presented here, I give in Table S3 the full analytical expressions obtained from Theorem S9.  In Figs. S8 and S9, the spatiotemporal evolutions of these rules are plotted.

\begin{figure}
\begin{center}
\includegraphics[width=0.75\textwidth]{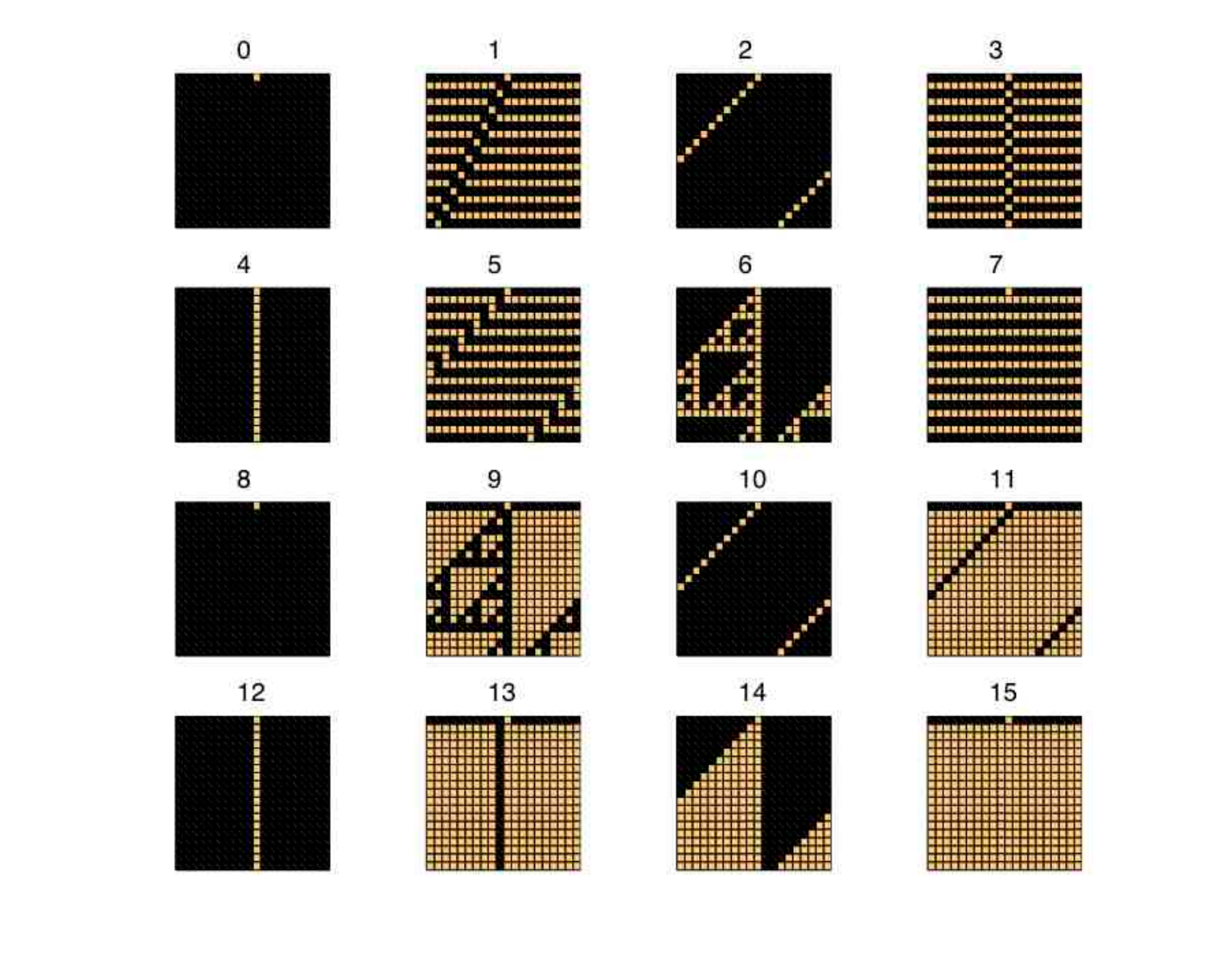}
\includegraphics[width=0.75\textwidth]{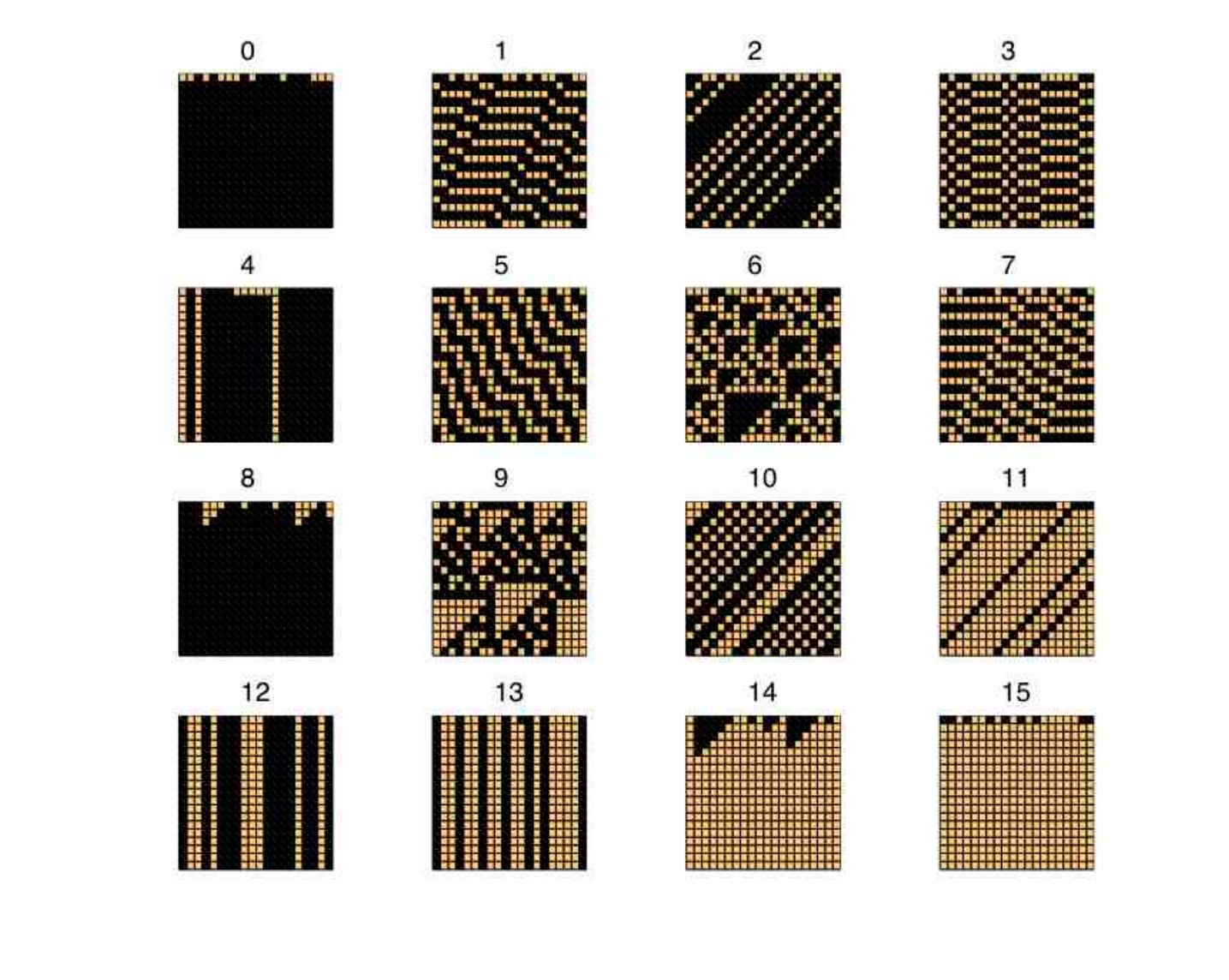}
\end{center}
Figure S8: Spatiotemporal evolution of the sixteen rules $^{0}R_{2}^{1}$ ($R$ is indicated over each panel) for an initial condition consisting on a single seed (top) and a random initial vector (bottom).
\label{FigS5}
\end{figure}

\begin{figure}
\begin{center}
\includegraphics[width=0.75\textwidth]{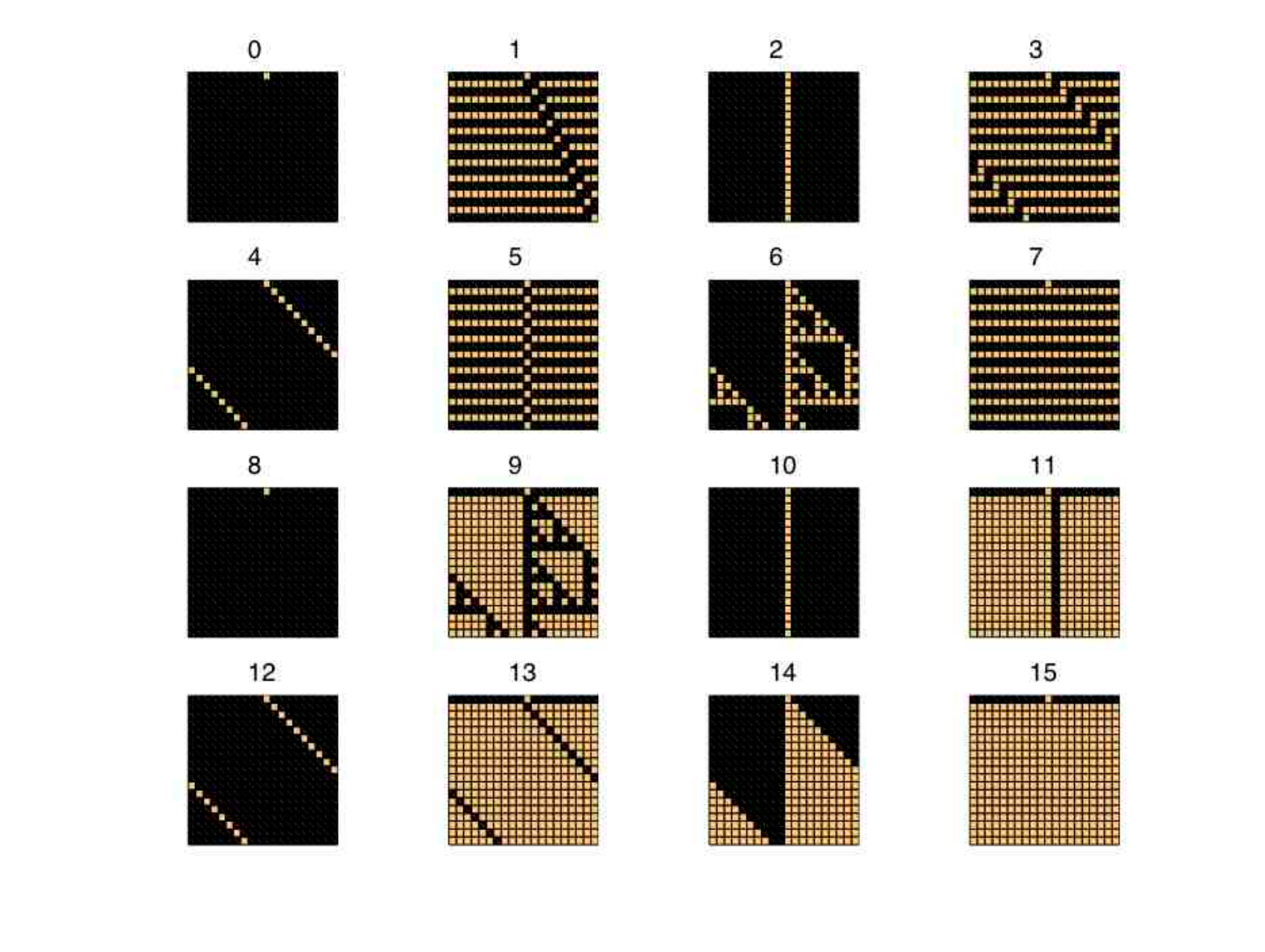}
\includegraphics[width=0.75\textwidth]{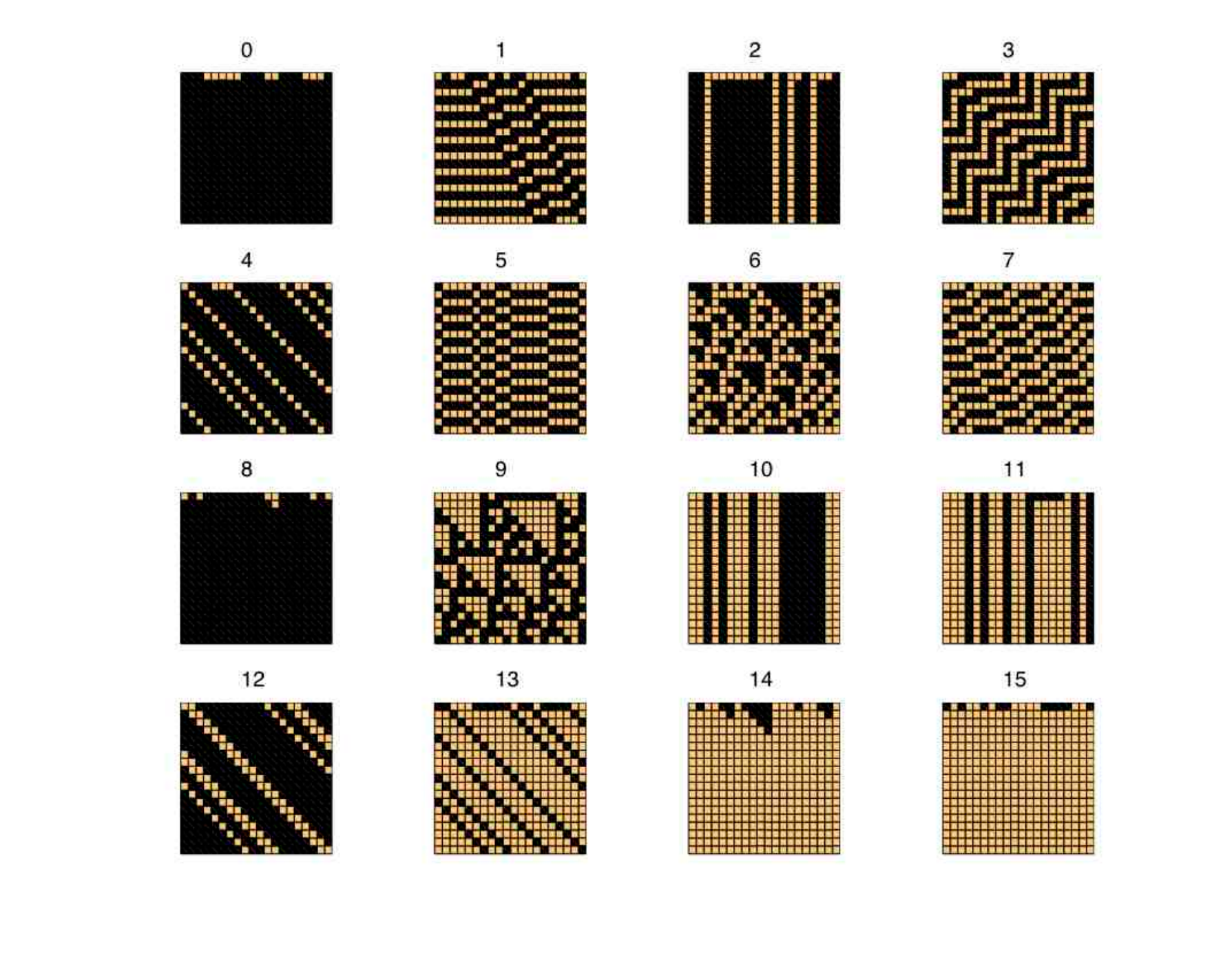}
\end{center}
Figure S9: Spatiotemporal evolution of the sixteen rules $^{1}R_{2}^{0}$ ($R$ is indicated over each panel) for an initial condition consisting on a single seed (top) and a random initial vector (bottom). \label{FigS6}
\end{figure}

Although some of the patterns seem rather complex, their behavior is entirely predictable, as established in the following theorem that I give without proof (all results can be proved by induction -by using the results in Appendix A-  and can be easily checked with a computer). \\

\pagebreak

\noindent  \textbf{Theorem S10 (Predictability of all boolean rules $^{0}R_{2}^{1}$ and $^{1}R_{2}^{0}$):}
\emph{If $\mathbf{x}_{0}=(x_{0}^{1},... x_{0}^{i},...x_{0}^{N_{s}})$ denotes the initial state of a CA at time $t=0$, for time $t \ge 1$ (positive integer), all rules $^{0}R_{2}^{1}$ and $^{1}R_{2}^{0}$ are predictable and the value at the site on a later time $t$, $x_{t}^{i}$, can be known for each rule, as a function of $t$ and the initial site values $\mathbf{x}_{0}$. The orbits (local dependence in time for the state of the cellular automata) for all rules $^{0}R_{2}^{1}$ and $^{1}R_{2}^{0}$ are given in Table S4.} 
\begin{itemize}
\item (i) $\ \ \ \ \ ^{0}0_{2}^{1}$: $\qquad x_{t}^{i}=0$
\item (ii) $\ \ \ \ ^{0}1_{2}^{1}$: $\qquad x_{t}^{i}=\frac{1}{2}(1-(-1)^{t})(1-x_{0}^{i-\frac{t-1}{2}}-x_{0}^{i-\frac{t+1}{2}}+x_{0}^{i-\frac{t-1}{2}}x_{0}^{i-\frac{t+1}{2}})+$ 
\item[]{} $\ \ \ \ \ \ \ \ \ \ \ \ \ \ \ \ \ \ \ \ \ +\frac{1}{2}(1+(-1)^{t})(x_{0}^{i-\frac{t}{2}}+x_{0}^{i-\frac{t-2}{2}}x_{0}^{i-\frac{t+2}{2}}-x_{0}^{i-\frac{t}{2}}x_{0}^{i-\frac{t-2}{2}}x_{0}^{i-\frac{t+2}{2}})$
\item (iii) $\ \ \  ^{0}2_{2}^{1}$: $\qquad x_{t}^{i}=x_{0}^{i-t}-x_{0}^{i-t}x_{0}^{i-t+1}$
\item (iv) $\ \ \ ^{0}3_{2}^{1}$: $\qquad x_{t}^{i}=(-1)^{t}x_{0}^{i}+\frac{1}{2}(1-(-1)^{t})$
\item (v) $\ \ \ \ ^{0}4_{2}^{1}$: $\qquad x_{t}^{i}=x_{0}^{i}-x_{0}^{i}x_{0}^{i-1}$
\item (vi) $\ \ \ ^{0}5_{2}^{1}$: $\qquad x_{t}^{i}=(-1)^{t}x_{0}^{i-t}+\frac{1}{2}(1-(-1)^{t})$
\item (vii) $\ \  ^{0}6_{2}^{1}$: $\qquad x_{t}^{i}=\sum_{m=0}^{2^{t-1}}\mathcal{B}\left(2m+1-\sum_{k=0}^{t}\frac{t!}{k!(t-k)!}x_{0}^{i-t+k},\frac{1}{2}\right)$
\item (viii) $\ ^{0}7_{2}^{1}$: $\qquad x_{t}^{i}=\frac{1}{2}(1-(-1)^{t})(1-x_{0}^{i-\frac{t-1}{2}}x_{0}^{i-\frac{t+1}{2}})+$ 
\item[]{} $\ \ \ \ \ \ \ \ \ \ \ \ \ \ \ \ \ \ \ \ \ +\frac{1}{2}(1+(-1)^{t})(x_{0}^{i-\frac{t}{2}}x_{0}^{i-\frac{t-2}{2}}+x_{0}^{i-\frac{t}{2}}x_{0}^{i-\frac{t+2}{2}}-x_{0}^{i-\frac{t}{2}}x_{0}^{i-\frac{t-2}{2}}x_{0}^{i-\frac{t+2}{2}})$
\item (ix) $\ \ \ ^{0}8_{2}^{1}$: $\qquad x_{t}^{i}=\prod_{k=0}^{t}x_{0}^{i-k}$
\item (x) $\ \ \ \  ^{0}9_{2}^{1}$: $\qquad x_{t}^{i}=\sum_{m=0}^{2^{t-1}}\mathcal{B}\left(2m-\sum_{k=0}^{t}\frac{t!}{k!(t-k)!}(1-x_{0}^{i-t+k}),\frac{1}{2}\right)$
\item (xi) $\ \ ^{0}10_{2}^{1}$: $\qquad x_{t}^{i}=x_{0}^{i-t}$
\item (xii) $\  ^{0}11_{2}^{1}$: $\qquad x_{t}^{i}=1-x_{0}^{i-t+1}+x_{0}^{i-t}x_{0}^{i-t+1}$
\item (xiii) $ ^{0}12_{2}^{1}$: $\qquad x_{t}^{i}=x_{0}^{i}$
\item (xiv) $  ^{0}13_{2}^{1}$: $\qquad x_{t}^{i}=1-x_{0}^{i-1}+x_{0}^{i}x_{0}^{i-1}$
\item (xv) $\ ^{0}14_{2}^{1}$: $\qquad x_{t}^{i}=1-\prod_{k=0}^{t}(1-x_{0}^{i-k})$
\item (xvi) $ ^{0}15_{2}^{1}$: $\qquad x_{t}^{i}=1$
\item (i b) $\ \ \ \ \ ^{1}0_{2}^{0}$: $\qquad x_{t}^{i}=0$
\item (ii b) $\ \ \ \ ^{1}1_{2}^{0}$: $\qquad x_{t}^{i}=\frac{1}{2}(1-(-1)^{t})(1-x_{0}^{i+\frac{t-1}{2}}-x_{0}^{i+\frac{t+1}{2}}+x_{0}^{i+\frac{t-1}{2}}x_{0}^{i+\frac{t+1}{2}})+$ 
\item[]{} $\ \ \ \ \ \ \ \ \ \ \ \ \ \ \ \ \ \ \ \ \ +\frac{1}{2}(1+(-1)^{t})(x_{0}^{i+\frac{t}{2}}+x_{0}^{i+\frac{t-2}{2}}x_{0}^{i+\frac{t+2}{2}}-x_{0}^{i+\frac{t}{2}}x_{0}^{i+\frac{t-2}{2}}x_{0}^{i+\frac{t+2}{2}})$
\item (iii b) $\ \ \  ^{1}2_{2}^{0}$: $\qquad x_{t}^{i}=x_{0}^{i}-x_{0}^{i}x_{0}^{i+1}$
\item (iv b) $\ \ \ ^{1}3_{2}^{0}$: $\qquad x_{t}^{i}=(-1)^{t}x_{0}^{i+t}+\frac{1}{2}(1-(-1)^{t})$
\item (v b) $\ \ \ \ ^{1}4_{2}^{0}$: $\qquad x_{t}^{i}=x_{0}^{i+t}-x_{0}^{i+t}x_{0}^{i+t-1}$
\item (vi b) $\ \ \ ^{1}5_{2}^{0}$: $\qquad x_{t}^{i}=(-1)^{t}x_{0}^{i}+\frac{1}{2}(1-(-1)^{t})$
\item (vii b) $\ \  ^{1}6_{2}^{0}$: $\qquad x_{t}^{i}=\sum_{m=0}^{2^{t-1}}\mathcal{B}\left(2m+1-\sum_{k=0}^{t}\frac{t!}{k!(t-k)!}x_{0}^{i+t-k},\frac{1}{2}\right)$
\item (viii b) $\ ^{1}7_{2}^{0}$: $\qquad x_{t}^{i}=\frac{1}{2}(1-(-1)^{t})(1-x_{0}^{i+\frac{t-1}{2}}x_{0}^{i+\frac{t+1}{2}})+$ 
\item[]{} $\ \ \ \ \ \ \ \ \ \ \ \ \ \ \ \ \ \ \ \ \ +\frac{1}{2}(1+(-1)^{t})(x_{0}^{i+\frac{t}{2}}x_{0}^{i+\frac{t-2}{2}}+x_{0}^{i+\frac{t}{2}}x_{0}^{i+\frac{t+2}{2}}-x_{0}^{i+\frac{t}{2}}x_{0}^{i+\frac{t-2}{2}}x_{0}^{i+\frac{t+2}{2}})$
\item (ix b) $\ \ \ ^{1}8_{2}^{0}$: $\qquad x_{t}^{i}=\prod_{k=0}^{t}x_{0}^{i+k}$
\item (x b) $\ \ \ \  ^{1}9_{2}^{0}$: $\qquad x_{t}^{i}=\sum_{m=0}^{2^{t-1}}\mathcal{B}\left(2m-\sum_{k=0}^{t}\frac{t!}{k!(t-k)!}(1-x_{0}^{i+t-k}),\frac{1}{2}\right)$
\item (xi b) $\ \ ^{1}10_{2}^{0}$: $\qquad x_{t}^{i}=x_{0}^{i}$
\item (xii b) $\  ^{1}11_{2}^{0}$: $\qquad x_{t}^{i}=1-x_{0}^{i+1}+x_{0}^{i+1}x_{0}^{i}$
\item (xiii b) $ ^{1}12_{2}^{0}$: $\qquad x_{t}^{i}=x_{0}^{i+t}$
\item (xiv b) $  ^{1}13_{2}^{0}$: $\qquad x_{t}^{i}=1-x_{0}^{i+t-1}+x_{0}^{i+t}x_{0}^{i+t-1}$
\item (xv b) $\ ^{1}14_{2}^{0}$: $\qquad x_{t}^{i}=1-\prod_{k=0}^{t}(1-x_{0}^{i+k})$
\item (xvi b) $ ^{1}15_{2}^{0}$: $\qquad x_{t}^{i}=1$
\end{itemize}
\begin{center}
Table S4 (cont.): The orbits (local dependence in time for the state of the cellular automata) for all rules $^{0}R_{2}^{1}$ and $^{1}R_{2}^{0}$.
\end{center}

Specially interesting are the rules $^{0}6_{2}^{1}$, $^{0}9_{2}^{1}$, $^{1}6_{2}^{0}$, $^{1}9_{2}^{0}$. Although predictable, the behavior of these rules is far more complex than the one of the other rules. They correspond to the logical XOR function and to the complementary XNOR function. In the following section, it is shown how this is the building block of complexity in physical systems.  These rules represent addition modulo 2 of both site values at a previous instant of time. Therefore, they are also totalistic rules. Rules $^{0}6_{2}^{1}$, $^{1}6_{2}^{0}$, $^{0}9_{2}^{1}$ and $^{1}9_{2}^{0}$ have totalistic codes $^{0}2T_{2}^{1}$, $^{1}2T_{2}^{0}$, $^{0}5T_{2}^{1}$, $^{1}5T_{2}^{0}$ respectively. \\

\noindent  \textbf{Theorem S11 (Universal CA map for the 256 boolean Wolfram Rules):}
\emph{The universal CA map, Eq. (\ref{CA}), reduces to the following expression
\begin{eqnarray}
x_{t}^{i+1}&=&a_{0}(1-x_{t}^{i+1})(1-x_{t}^{i})(1-x_{t}^{i-1})+a_{1}x_{t}^{i-1}(1-x_{t}^{i+1})(1-x_{t}^{i}) +  \nonumber \\
&+& a_{2}x_{t}^{i}(1-x_{t}^{i+1})(1-x_{t}^{i-1}) +a_{3}x_{t}^{i}x_{t}^{i-1}(1-x_{t}^{i+1})+a_{4}x_{t}^{i+1}(1-x_{t}^{i})(1-x_{t}^{i-1})+
 \nonumber \\
&+& a_{5}x_{t}^{i+1}x_{t}^{i-1}(1-x_{t}^{i})+a_{6}x_{t}^{i+1}x_{t}^{i}(1-x_{t}^{i-1})+a_{7}x_{t}^{i+1}x_{t}^{i}x_{t}^{i-1} \label{CAW}
\end{eqnarray}
for all 256 rules with p=2, l=1, r=1 (Wolfram rules), i.e. the rules $^{1}R_{2}^{1}$, where $R=\sum_{n=0}^{7}a_{n}2^{n}$.} \\

\noindent \emph{Proof:} By inserting $l=r=1$, $p=2$, $\Omega=8$ in Eq. (\ref{CA}), 
\begin{equation}
x_{t+1}^{i}=\sum_{n=0}^{7}a_{n}\mathcal{B}\left(4x_{t}^{i+1}+2x_{t}^{i}+x_{t}^{i-1}-n,\frac{1}{2}\right) \label{CAWolfram}
\end{equation}
and by using the definition of $n$ given by Eq. (\ref{confCA}), i.e. $n=4x^{i+1}+2x^{i}+x^{i-1}$, the latter expression can be rewritten as
\begin{equation}
x_{t+1}^{i}=\sum_{x^{i+1}=0}^{1}\sum_{x^{i}=0}^{1}\sum_{x^{i-1}=0}^{1}a_{4x^{i+1}+2x^{i}+x^{i-1}}\mathcal{B}\left(4(x_{t}^{i+1}-x^{i+1})+2(x_{t}^{i}-x^{i})+(x_{t}^{i-1}-x^{i-1}),\frac{1}{2}\right) \nonumber
\end{equation}
By applying now result $(xii)$ from Appendix A
\begin{equation}
x_{t+1}^{i}=\sum_{x^{i+1}=0}^{1}\sum_{x^{i}=0}^{1}\sum_{x^{i-1}=0}^{1}a_{4x^{i+1}+2x^{i}+x^{i-1}}\mathcal{B}\left(x_{t}^{i+1}-x^{i+1},\frac{1}{2}\right)\mathcal{B}\left(x_{t}^{i}-x^{i},\frac{1}{2}\right)\mathcal{B}\left(x_{t}^{i-1}-x^{i-1},\frac{1}{2}\right) \nonumber
\end{equation}
The sum can now be carried explicitly by replacing the corresponding values of $x^{i+1}$, $x^{i}$ and $x^{i-1}$ which pick values either 0 or 1 yielding eight terms in the sum. Then, by applying result $(xv)$ from Appendix A, Eq. (\ref{CAlog1}) is proved.$\Box$ \\

The 256 specific maps obtained from Theorem S11 are provided in Table S6 in Appendix B for each rule  (the values for all $a_{n}$ specifying each rule are there in each case explicitly indicated).    \\

\noindent  \textbf{Theorem S12 (Some equivalence classes of boolean rules):}
\emph{Let $^{l}A^{r}_{2}$ and $^{l}B^{r}_{2}$ be two different boolean rules, with $A=\sum_{n=0}^{\Omega-1}a_{n}2^{n}$ and $B=\sum_{n=0}^{\Omega-1}b_{n}2^{n}$, with $\Omega=2^{l+r+1}$. 
and $n=\sum_{k=-r}^{l}2^{k+r}x^{i+k}$
The following results hold
\begin{itemize}
\item  $(i)$ $\ \ ^{l}B^{r}_{2}$ is related to $^{l}A^{r}_{2}$ through a change of colors transformation (also called global complementation) when $b_{n}=1-a_{\Omega-1-n}$ $\ $ for all $n$.  
\item  $(ii)$ $\ \ ^{r}B^{l}_{2}$ is related to $^{l}A^{r}_{2}$ through reflection when $b_{n}=a_{n-\sum_{k=-r}^{l}(2^{k+r}-2^{l-k})x^{i+k}}$ $\ $ for all $n$ (e.g. for all $x^{i+k}$, where $k \in [-r,l]$).  
\item  $(iii)$ $\ \ ^{r}B^{l}_{2}$ is related to $^{l}A^{r}_{2}$ through reflection following a change of colors transformation when $b_{n}=1-a_{\Omega-1-n+\sum_{k=-r}^{l}(2^{k+r}-2^{l-k})x^{i+k}}$ $\ $ for all $n$ (e.g. for all $x^{i+k}$, where $k \in [-r,l]$). 
\end{itemize}}

\noindent \emph{Proof:} This theorem is indeed a corollary of Theorems S1 and S2, for $p=2$. The only bijective application to change the colors that can be defined is $x \to 1-x$ which coincides with GC.  $\Box$ \\

\noindent \textbf{Example:}  The above theorem can be cursorily checked with the specific case of Wolfram $^{1}110^{1}_{2}$ rule for which $\Omega-1=7$, $(a_{0}, a_{1}, a_{2}, a_{3}, a_{4}, a_{5}, a_{6}, a_{7})=(0,1,1,1,0,1,1,0)$. Therefore, global complementation (from $(i)$ in the above theorem) gives $(b_{0}, b_{1}, b_{2}, b_{3}, b_{4}, b_{5}, b_{6}, b_{7})=(1-a_{7}, 1-a_{6}, 1-a_{5}, 1-a_{4}, 1-a_{3}, 1-a_{2}, 1-a_{1},1-a_{0})=(1,0,0,1,0,0,0,1)$, which corresponds to rule $^{1}137^{1}_{2}$. Under reflection one obtains, from $(ii)$  $(b_{0}, b_{1}, b_{2}, b_{3}, b_{4}, b_{5}, b_{6}, b_{7})=(a_{0}, a_{4}, a_{2}, a_{6}, a_{1}, a_{5}, a_{3}, a_{7})=(0,0,1,1,1,1,1,0)$ which corresponds to rule $^{1}124^{1}_{2}$. Finally, global complementation after reflection gives, from $(iii)$:  $(b_{0}, b_{1}, b_{2}, b_{3}, b_{4}, b_{5}, b_{6}, b_{7})=(1-a_{7}, 1-a_{3}, 1-a_{5}, 1-a_{1}, 1-a_{6}, 1-a_{2}, 1-a_{4}, 1-a_{0})=(1,0,0,0,0,0,1,1)$ which corresponds to rule $^{1}193^{1}_{2}$.  $\Box$ \\

\noindent \textbf{Example:} Consider now $^{0}6^{1}_{2}$, for which. $\Omega-1=3$, $(a_{0}, a_{1}, a_{2}, a_{3})=(0,1,1,0)$. Global complementation (from $(i)$ in the above theorem) gives $(b_{0}, b_{1}, b_{2}, b_{3})=(1-a_{3}, 1-a_{2}, 1-a_{1}, 1-a_{0})=(1,0,0,1)$, which corresponds to rule $^{0}9^{1}_{2}$.  For the rule related through reflection one obtains, from $(ii)$  $(b_{0}, b_{1}, b_{2}, b_{3})=(a_{0}, a_{2}, a_{1}, a_{3})=(0,1,1,0)$ which corresponds to $^{1}6^{0}_{2}$ (note that $l$ and $r$ are exchanged). Finally, global complementation following reflection gives, from $(iii)$:  $(b_{0}, b_{1}, b_{2}, b_{3})=(1-a_{3}, 1-a_{1}, 1-a_{2}, 1-a_{0})=(0,1,1,0)$ which corresponds to rule $^{1}9^{0}_{2}$ (note again that $l$ and $r$ are exchanged compared to the original rule $^{0}6^{1}_{2}$).  $\Box$ \\

Theorem S12 allows the 256 Wolfram rules to be classified in 88 equivalence classes. The shift transformation combined with change of range (see Theorems S4 and S5 and Figure S5) allows to reduce this number even further. Finally, carefully applied and combined with the constructor's theorem S6, Theorem S10 (Table S4) allows even analytical expressions for the spatiotemporal evolution of some of the rules to be easily found, since they are also Wolfram rules after being copied to range $\rho=3$. Analogous expressions as in Theorem S10 (Table S4) might be found for most of the Wolfram rules and this problem will be addressed elsewhere.\\

Rules $^{1}110^{1}_{2}$ and $^{1}54^{1}_{2}$ as well as their class equivalent rules, display a more complex behavior than the rest of the Wolfram rules (of range $\rho=3$). It was observed above that rule $^{0}6^{1}_{2}$ and its class equivalent rules display also a more complex behavior than any other boolean rule of range $\rho=2$.  Rules $^{1}110^{1}_{2}$ is even more complex than rule $^{0}6^{1}_{2}$ because it is $unpredictable$. In the following section a general theory of complexity is introduced, allowing the wide variety of behaviors found in CA to be identified.

\subsection{General theory of complexity}

Based on extensive computer simulations, Stephen Wolfram introduced a classification of CA dynamics in terms of observed behaviors of increasing complexity \footnotemark \footnotetext{See Wolfram, S. \emph{Cellular automata as models of complexity}. Nature, \textbf{311}, 419 (1984) and Wolfram, S. \emph{Universality and Complexity in Cellular Automata}. Physica D, \textbf{10}, 1 (1984).} Although this classification was empirical, based on computer experiments, its robustness is compelling. As shown below, the theory of CA presented here elucidates from  a fundamental (mathematical) point of view this classification, and provides a rigorous definition of complexity for the dynamics of cellular automata. The theory presented here is valid for rules of any range and involving an arbitrary number of symbols. These ideas might be extended to continuum systems: in the last section I show how maps involving real numbers arise from the universal CA map (Eq. \ref{CA}).

Speaking about a generic initial condition, Wolfram classified CA dynamics into four main, broad classes, as a function of the limiting behavior reached by the CA
\begin{itemize}
\item{\textbf{Class 1}: Evolution leads to a spatially homogeneous state} 
\item{\textbf{Class 2}: Evolution leads to a set of separated simple stable or periodic structures} 
\item{\textbf{Class 3}: Evolution leads to a chaotic, aperiodic or nested pattern} 
\item{\textbf{Class 4}: Evolution leads to complex, localized structures, some times long-lived} 
\end{itemize}

This classification, although being entirely descriptive and qualitative, is consistent with a huge number of computer simulations. Some of the rules exhibit a certain coexistence between classes. 

I introduce now some definitions and theorems to proceed further. The proof of Theorem S13 is trivial from the constructor's theorem S6.\\

\noindent  \textbf{Definition S3 (Monotonic rules):}
\emph{A local rule $^{0}R^{0}_{p}$ is monotonic of sign + / - if increasing the value of $x_{t}^{i}$ makes the value of $x_{t+1}^{i}$ to increase/decrease. If it remains constant, the rule is called neutrally monotonic. A general rule $^{l}R^{r}_{p}$ is monotonic, if it is a monotonic function of each of every site values in the neighborhood.} \\

\noindent  \textbf{Theorem S13 (Monotonicity from construction):}
\emph{Let a set of $p$ rules each denoted by $^{l}(A_{m})^{r}_{p}$ with $m\in[0,p-1]$. Then $^{l+1}R^{r}_{p}$, constructed from the left, and $^{l}R^{r+1}_{p}$, constructed from the right, are \emph{monotonic} if all constructing rules $^{l}(A_{m})^{r}_{p}$ are  \textbf{monotonic of the same sign} for every site value (some of the constructing rules can also be neutrally monotonic).} \\

\noindent  \textbf{Definition S4 (Non-monotonic turns in construction layers):}
\emph{A rule $^{l}R^{r}_{p}$ is said to have a non-monotonic turn in the construction layer $k$ if in the construction process starting recursively from the monotonic rules $^{0}R^{0}_{p}$, the rule $^{r'}R'^{\ l'}_{p}$ with $l'+r'=k \le l+r$, constructed after $k$ steps, is non-monotonic. The higher the number of non-monotonic turns, the stronger the non-monotonicity of the rule.} \\

Every rule $^{0}R^{1}_{2}$ and $^{1}R^{0}_{2}$, with the exception of $^{0}6^{1}_{2}$ and its class equivalents, is monotonic. This can be seen from the constructor theorem. Rules $^{0}R^{1}_{2}$ are constructed from rules $^{0}R^{0}_{2}$ adding one site to the right. Because all rules $^{0}R^{0}_{2}$ are monotonic, any construction involving rules with the same monotonicity sign is monotonic as well, as a consequence of Theorem S13.  The only exception is the construction involving simultaneously rules $^{0}1^{0}_{2}$ and $^{0}2^{0}_{2}$: since the former is monotonic of sign - and the latter monotonic of sign +, the constructed rules are not monotonic. The latter rules construct together $^{0}6^{1}_{2}$ and its class equivalents! These rules have, therefore, a non-monotonic turn on the first layer. One can easily see that rule $^{0}6^{1}_{2}$ is $non-monotonic$: let the site on the right have value '0'; then, increasing the site on the left from '0' to '1' makes the output also to $increase$ from '0' to '1'; however, if the site on the right has value '1', increasing the site on the left from '0' to '1' makes the output to $decrease$. 

The analysis of the most simple boolean rules in Theorems S9 and S10 and Figures S6, S8 and S9 showed indeed that rule $^{0}6^{1}_{2}$ and its three class equivalents upon reflection and change of colors is more complex as any other rule $^{0}R^{1}_{2}$. And what the analysis indeed reveals as well is that, since additional degrees of freedom cannot destroy the non-monotonic character achieved by rule $^{0}6^{1}_{2}$, Class 1 and Class 2 cellular automata correspond to monotonic rules, and the most simpler non-monotonic rules begin with the nested structure, which corresponds already to Class 3. This elucidates an observation made by Wolfram 
\footnotemark \footnotetext{See Wolfram, S. \emph{A New Kind of Science}, Wolfram Media (2002), p. 948.} that the number of Class 3 rules increase with increasing range: for an increasing number of construction layers, there are an increased number of possibilities to create non-monotonic turns. And when a non-monotonic turn is created, any rule constructed after that by employing the previous one, is also automatically non-monotonic because non-monotonic turns in previous layers cannot be destroyed.

Another important feature of such a rule as $^{0}6^{1}_{2}$ is that it puts into play every available symbol starting from a single seed. Such rules, yielding nested structures, are our starting point in our aim to understand the emergence of complexity since, being regular and predictable, they lie at the verge of complexity and allow to classify every other rule upon perturbations and symmetry breaking. I show this below.

I first note the following important invariance satisfied by the totalistic CA map, Eq. (\ref{CAtot}) and state it as a theorem. This invariance can be directly checked on Eq. (\ref{CAtot}).\\ 

\noindent  \textbf{Theorem S14 (Invariance upon addition modulo $p$):}
\emph{The totalistic universal CA map, Eq. (\ref{CAtot}) remains invariant after the following set of transformations
\begin{eqnarray}
s & \to & s'=s+mp \label{ad1} \\
\sigma_{s} & \to & \sigma'_{s'}=\sigma_{s} \qquad \qquad \qquad \forall s
\end{eqnarray}}
\emph{where $m$ is an integer number so that $s+mp \in [0,\rho (p-1)]$.}\\

Most rules break this symmetry. The only exceptions are, of course, the ones that satisfy $\sigma_{s}=\sigma_{s+mp}$, $\forall m$ so that $s+mp \in [0,\rho (p-1)]$. And specially interesting within these are those which put every symbol $[0,p-1]$ into play during the time evolution. 
Because of their importance, and for reference, I name these rules invariant upon addition modulo $p$ \emph{Pascal rules}, after the French mathematician Blaise Pascal: these rules, as observed below, reproduce after a bijective application all Pascal simplices modulo $p$. When $\rho=2$, the Pascal simplex coincides with the Pascal triangle. When $\rho >2$ the Pascal simplex is related to the multinomial expansion modulo $p$. Pascal rules are a subset of the so-called additive cellular automata, for which an algebraic theory was formulated. \footnotemark \footnotetext{See Martin, O.; Odlyzko, A. M. and Wolfram S. \emph{Algebraic properties of cellular automata}. Comm. Math. Phys. \textbf{93}, 219-258 (1984).} For reference, I introduce these rules as a definition.\\

\noindent  \textbf{Definition S5 (Pascal rules):}
\emph{A totalistic rule $^{l}RT^{r}_{p}$ is called Pascal rule if it satisfies the following property}
\begin{eqnarray}
\sigma_{s+mp}=\sigma_{s}=s 
\end{eqnarray}
\emph{where $m$ is an integer number so that $s+mp \in [0,\rho (p-1)]$. These rules perform the addition modulo $p$ of all site values contained in the neighborhood.}\\

\noindent  \textbf{Theorem S15 (Pascal rules perform the addition modulo-p):}
\emph{A Pascal rule $^{l}RT^{r}_{p}$ performs the addition modulo $p$ of all site values contained in the neighborhood, i.e. it has the form.}
\begin{eqnarray}
x_{t+1}^{i}=\sum_{m=0}^{\rho} \sum_{s=0}^{p-1} s \mathcal{B}\left(s+mp-\sum_{k=-r}^{l}x_{t}^{i+k},\frac{1}{2} \right)=\mathcal{R}_{p}\left(\sum_{k=-r}^{l}x_{t}^{i+k}\right) \label{wunder}
\end{eqnarray}
\emph{where $\mathcal{R}_{p}(x)$ is the remainder upon division of an integer number $x$ by $p$.}\\

\noindent \emph{Proof:} The first of the equalities in Eq.(\ref{wunder}) is obtained from the totalistic universal map Eq. (\ref{CAtot}) by making the transformation $s \to s+mp$ and using the definition of the Pascal rule above. Then, since $\Theta=\rho(p-1)$
\begin{eqnarray}
x_{t+1}^{i}&=&\sum_{s=0}^{\Theta}\sigma_{s}\mathcal{B}\left(s-\sum_{k=-r}^{l}x_{t}^{i+k},\frac{1}{2}\right) \nonumber \\ 
&=& \sum_{s+mp=0}^{\rho(p-1)}\sigma_{s+mp}\mathcal{B}\left(s+mp-\sum_{k=-r}^{l}x_{t}^{i+k},\frac{1}{2}\right) \nonumber \\
&=& \sum_{m=0}^{\rho}\sum_{s=0}^{p-1}s\mathcal{B}\left(s+mp-\sum_{k=-r}^{l}x_{t}^{i+k},\frac{1}{2}\right)
\end{eqnarray}
The second equality is then proved by using result $(xvi)$ from Appendix A, with $U_{p}=\rho$.$\Box$ \\

The structure of these rules is pretty simple. They have vectors $(\sigma_{0},...,\sigma_{\rho p})$ with the structure $(S,S,S')$, where S is a chain of integers $012...(p-1)$ repeated until the $\rho p$ positions characterizing the rule are filled. $S'$ is the chain $S$ truncated when position $\rho (p-1)$ is reached.\\

\textbf{Examples:} 

\begin{itemize}
\item $^{0}2T^{1}_{2}$ with $(\sigma_{0}, \sigma_{1}, \sigma_{2})=(0,1,0)$, i.e. $x_{t+1}^{i}=\mathcal{R}_{2}\left(\sum_{k=-1}^{0}x_{t}^{i+k}\right)$.
\item $^{1}10T^{1}_{2}$ with $(\sigma_{0}, \sigma_{1}, \sigma_{2}, \sigma_{3})=(0,1,0,1)$, i.e. $x_{t+1}^{i}=\mathcal{R}_{2}\left(\sum_{k=-1}^{1}x_{t}^{i+k}\right)$.
\item $^{1}10T^{2}_{2}$ with $(\sigma_{0}, \sigma_{1}, \sigma_{2},\sigma_{3},\sigma_{4})=(0,1,0,1,0)$, i.e. $x_{t+1}^{i}=\mathcal{R}_{2}\left(\sum_{k=-1}^{2}x_{t}^{i+k}\right)$.
\item $^{0}102T^{1}_{3}$ with $(\sigma_{0}, \sigma_{1}, \sigma_{2}, \sigma_{3}, \sigma_{4})=(0,1,2,0,1)$, i.e. $x_{t+1}^{i}=\mathcal{R}_{3}\left(\sum_{k=-1}^{0}x_{t}^{i+k}\right)$.
\item $^{1}588T^{1}_{3}$ with $(\sigma_{0}, \sigma_{1}, \sigma_{2}, \sigma_{3}, \sigma_{4}, \sigma_{5}, \sigma_{6})=(0,1,2,0,1,2,0)$, \\ i.e. $x_{t+1}^{i}=\mathcal{R}_{3}\left(\sum_{k=-1}^{1}x_{t}^{i+k}\right)$.
\item $^{0}1346680T^{1}_{5}$ with $(\sigma_{0}, \sigma_{1}, \sigma_{2}, \sigma_{3}, \sigma_{4}, \sigma_{5}, \sigma_{6},\sigma_{7},\sigma_{8})=(0,1,2,3,4,0,1,2,3)$, i.e. i.e. $x_{t+1}^{i}=\mathcal{R}_{5}\left(\sum_{k=-1}^{0}x_{t}^{i+k}\right)$.
\item $^{1}546268555T^{1}_{5}$ with $(\sigma_{0}, \sigma_{1}, ..., \sigma_{11}, \sigma_{12})=\ (0,1,2,3,4,0,1,2,3,4,0,1,2)$, i.e. $x_{t+1}^{i}=\mathcal{R}_{5}\left(\sum_{k=-1}^{1}x_{t}^{i+k}\right)$.
\item $^{0}275781750T^{1}_{6}$ with $(\sigma_{0}, \sigma_{1}, ..., \sigma_{9}, \sigma_{10})=\ (0,1,2,3,4,5,0,1,2,3,4)$, i.e. $x_{t+1}^{i}=\mathcal{R}_{6}\left(\sum_{k=-1}^{0}x_{t}^{i+k}\right)$.
\item $^{0}78050441406T^{1}_{7}$ with $(\sigma_{0}, \sigma_{1}, ..., \sigma_{11}, \sigma_{12})=\ (0,1,2,3,4,5,6,0,1,2,3,4,5)$, i.e. $x_{t+1}^{i}=\mathcal{R}_{7}\left(\sum_{k=-1}^{0}x_{t}^{i+k}\right)$.
\end{itemize}

\begin{figure}
\includegraphics[width=1.0\textwidth]{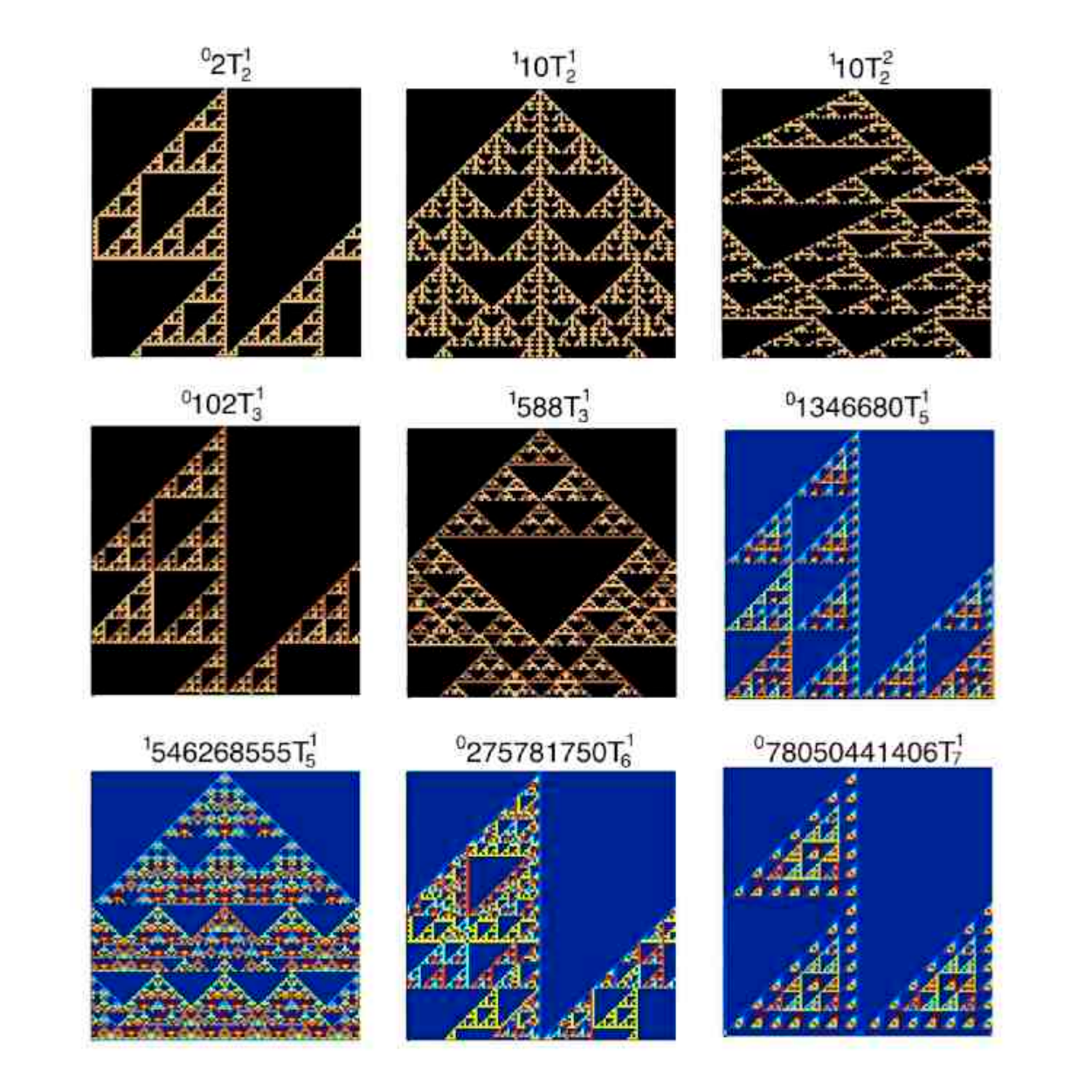}
\begin{center}
Figure S10: Spatiotemporal evolution of some Pascal rules with totalistic codes indicated in the figure. The number of different colors in the figures coincides with $p$. Regular, nested structures arise in every case. 
\end{center} 
\end{figure}

All Pascal rules fall in the Wolfram class 3, since they yield nested structures that can be very complex (see Fig. S10), but that are regular and with predictable features: since they reproduce the Pascal simplices, they are related analytically to the multinomial expansion. The structures that are formed from a single seed are contained in a big triangular structure. The latter is called  \textbf{triangle of expansion} of the Pascal rule. For rules with $\rho=2$, the triangle of expansion coincides with the Pascal triangle modulo $p$. Figure S11 shows a detail of the first time steps of the evolution of rule $^{0}1346680T^{1}_{5}$ showing how, indeed it calculates the Pascal triangle modulo 5.
 
\begin{figure}
\includegraphics[width=0.25\textwidth, angle=270]{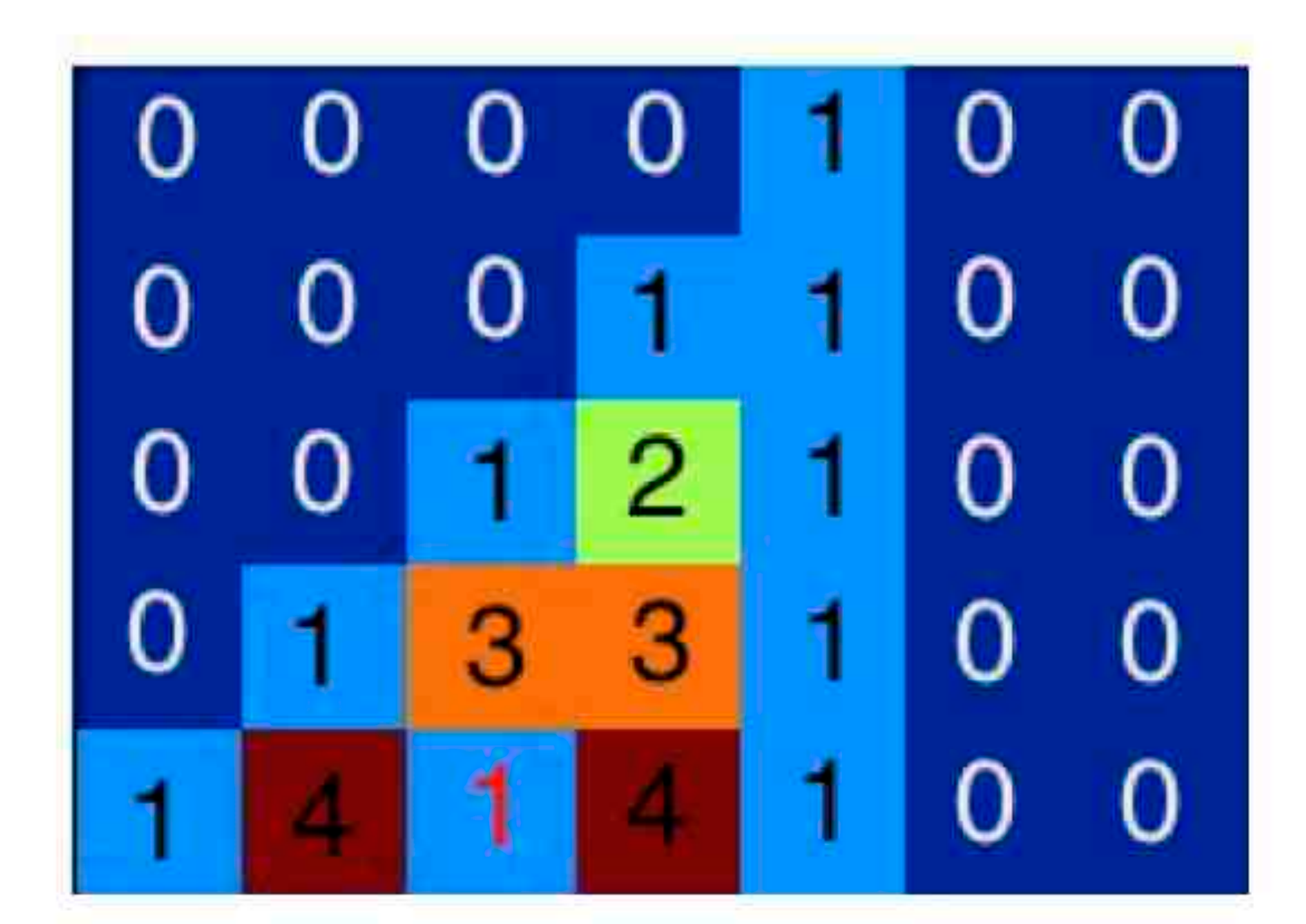}
\begin{center}
Figure S11: Detail of the former stages of the spatiotemporal evolution from a single seed of the Pascal rule with totalistic code $^{0}1346680T^{1}_{5}$. The Pascal structure modulo 5 is clearly recognized. The '1' colored red at the bottom of the figure corresponds to the value 6 in the Pascal triangle, which happens to be equal to 1 modulo 5.
\end{center} 
\end{figure}

Rule $^{0}6^{1}_{2}$ is also a Pascal rule with totalistic code $^{0}2T^{1}_{2}$ (see Fig. S10). Any other rule $^{0}R^{1}_{2}$ (with exception of the class equivalent rule $^{0}9^{1}_{2}$) is simpler than the  Pascal rule: either the triangle of expansion disappears, or it is homogeneously filled, as it is the case with rule $^{0}14^{1}_{2}$. This is the consequence of the absence of the non-monotonic turn, characteristic of rule $^{0}6^{1}_{2}$. 

It is, indeed, possible, to have more complex structures than the Pascal rules. \emph{When a Pascal rule is copied to a higher range and then only a few configurations within the triangle of expansion are tuned so the non-monotonic turn of the copy is destroyed, the symmetry of the rule is broken introducing defects that cause coherent structures to arise and propagate always within the triangle of expansion}. This is, as shown below, a $weak$ symmetry breaking of the addition modulo $p$. Any other way of breaking the symmetry of the Pascal rule is stronger and merely reduces or conserves the complexity of the reference Pascal rule. I state now the central result of this whole research and substantiate it with examples.\\

\begin{center}
\fbox{\fbox{\emph{Class 4 CA arise from a weak symmetry breaking of the addition modulo $p$}}} 
\end{center}
~\\
\noindent \textbf{Example:} Starting with the Pascal rule $^{0}2T^{1}_{2}$, it is first converted to the normal code, which, as known from above is $^{0}6^{1}_{2}$ (in general, Eq. (\ref{totto}), should be used). The rule has vector $(0,1,1,0)$. Its spatiotemporal evolution starting from a simple seed shows that the only configuration of two sites contained within the triangle of expansion yielding its output within the triangle of expansion is '11', i.e. '3' in base 2. Rule $^{0}6^{1}_{2}$ is then copied to the higher range $\rho=3$, obtaining (0,1,1,0,0,1,1,0), i.e. rule $^{1}102^{1}_{2}$ and the output of configuration '011' falling within the triangle of expansion is switched. This amounts to merely make the change $a_{3} \to 1$ in rule $^{1}102^{1}_{2}$. In this way the addition modulo $p$ symmetry of rule $^{0}2T^{1}_{2}$ is weakly broken and, at the same time, the non-monotonic turn of the first substructure of the rule is removed, i.e. (0,1,1,0,...) (non-monotonic) $\to$ (0,1,1,1,...) (monotonic). Then (0,1,1,1,0,1,1,0), is obtained, i.e. rule $^{1}110^{1}_{2}$ which is the celebrated Wolfram rule, known to be a universal Class 4 CA. Figure S12 sketches this process. From the point of view of the construction process, $^{1}110^{1}_{2}$ is non-monotonic because it contains rule $^{0}6^{1}_{2}$ in its second construction layer. But is not so strongly monotonic as  rule $^{1}102^{1}_{2}$ because the latter is entirely constructed by rule $^{0}6^{1}_{2}$ and has, therefore two non-monotonic turns. Rule $^{1}110^{1}_{2}$ contains also the monotonic rule $^{0}14^{1}_{2}$ in its second construction layer. Rule $^{1}110^{1}_{2}$ is \emph{weakly non-monotonic} in contrast with the strongly non-monotonic Pascal rule $^{0}6^{1}_{2}$ (or $^{1}102^{1}_{2}$). \\

\begin{figure}
\includegraphics[width=0.4\textwidth, angle=270]{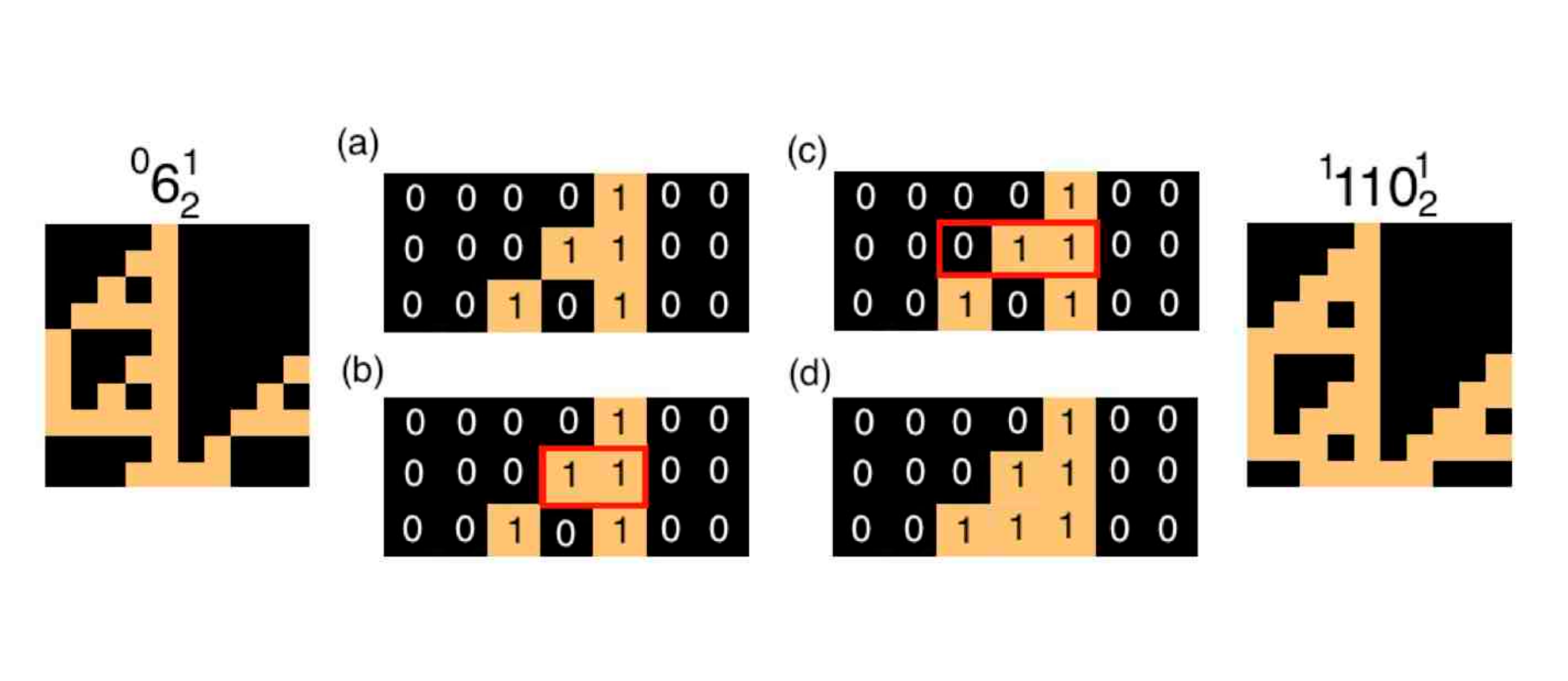}
\begin{center}
Figure S12: Scheme showing how a weak symmetry breaking of the Pascal rule $^{0}2T^{1}_{2}$ ($\equiv \ ^{0}6^{1}_{2}$) leads to Wolfram Class 4 rule $^{1}110^{1}_{2}$. (a) Triangle of expansion for rule $^{0}2T^{1}_{2}$ at the former stages of growth. (b) In range $\rho=2$, only the configuration '11' belongs to the triangle of expansion yielding its output '0', within the triangle of expansion. (c) The rule is copied to range $\rho=3$ yielding the equivalent rule $^{1}102^{1}_{2}$ with an additional degree of freedom that can be used to break the Pascal rule on the configuration '11' (d). The resulting rule $^{1}110^{1}_{2}$ has an unpredictable pattern within the triangle of expansion as a consequence of the weak symmetry breaking of rule $^{0}2T^{1}_{2}$ introduced by switching $a_{3}$ from '0' to '1' in rule $^{1}102^{1}_{2}$.
\end{center} 
\end{figure}

\noindent \textbf{Example:}  Consider the Pascal rule $^{1}10T^{2}_{2}$. First, the rule is converted to the normal code by using Eq. (\ref{totto}) and rule $^{1}27030^{2}_{2}$ with vector $(0,1,1,0,1,0,0,1,1,0,0,1,0,1,1,0)$ is obtained. The triangle of expansion shows that certain configurations yield their output within the triangle and can be tuned breaking non-monotonic turns. Examples are configurations '1010','0011','1111','0010','0001'. The rule is then copied to the higher range $\rho=5$, \\ 

$(0,1,1,0,1,0,0,1,1,0,0,1,0,1,1,0,0,1,1,0,1,0,0,1,1,0,0,1,0,1,1,0)$\\

\noindent which corresponds to rule $^{2}1771465110^{2}_{2}$. Now, if any of the above configurations within the triangle of expansion is considered with the added degree of freedom one has the configurations: '01010','00011','11111','00010','00001', which correspond to numbers '10', '3', '31', '2','1' in the decimal system. Then any of the following positions (or several of them) can be tuned to yield Class 4 behavior: $a_{10}$, $a_{3}$, $a_{31}$, $a_{2}$, and $a_{1}$. Figure 4 in the main text shows some of these possible symmetry breaking tunings and the resulting Class 4 behavior. This result is remarkable: finding these rules by brute force simulations might imply the evaluation of thousands of millions of rules. And with the simple prescription given above, these rules can be \emph{directly} found!\\

\begin{figure}
\includegraphics[width=1.0\textwidth]{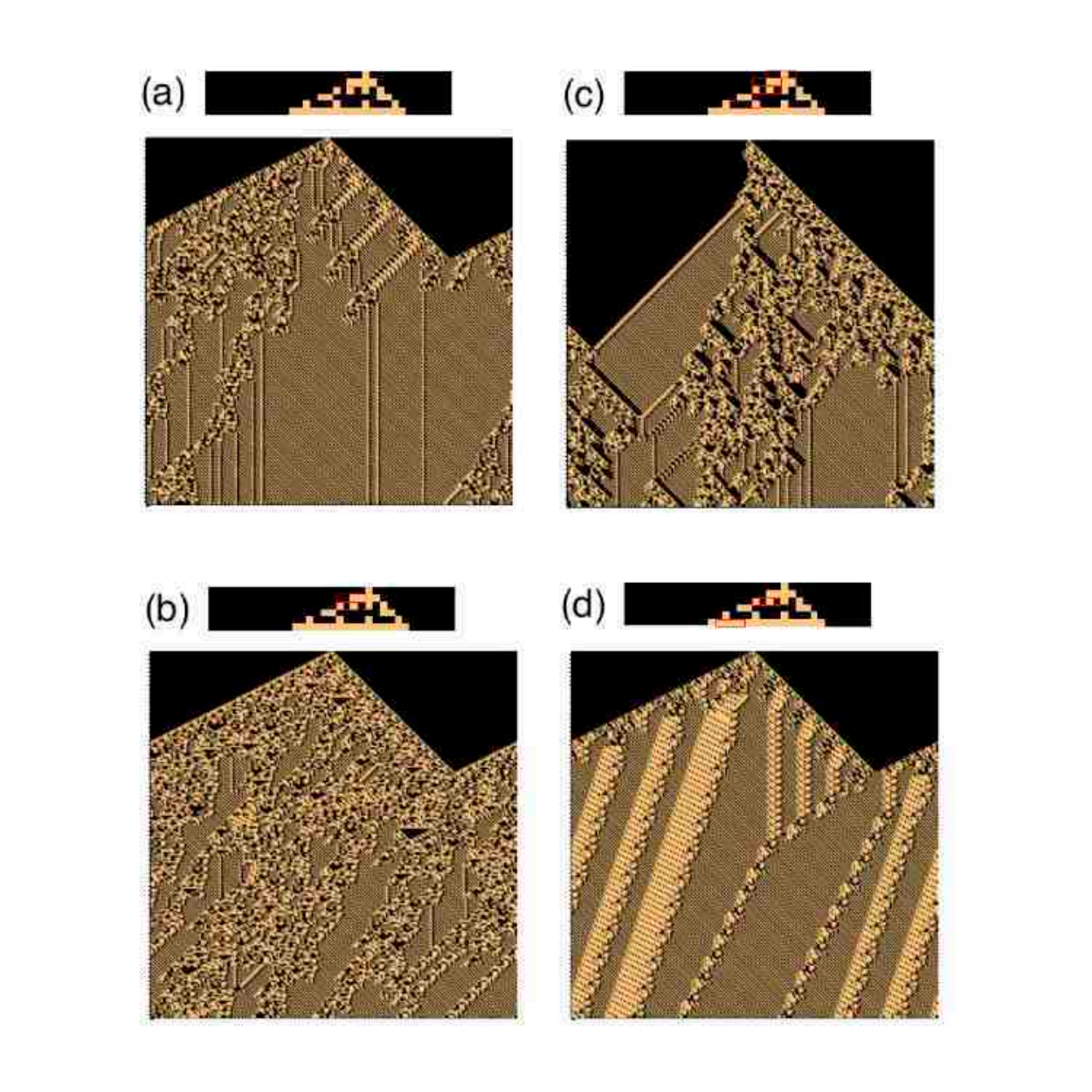}
\begin{center}
Figure S13: Spatiotemporal evolution of Class 4 rules obtained from a weak symmetry breaking of the totalistic Pascal rule $^{1}10T^{2}_{2}$ ($=^{1}27030^{2}_{2}$ with vector $(0,1,1,0,1,0,0,1,1,0,0,1,0,1,1,0)$), whose triangle of expansion is shown above each rule. In red are indicated the configurations that are tuned after the rule is copied to higher range. The rules have following codes: (a) $^{2}1771466134^{2}_{2}$, (b) $^{2}1771466142^{2}_{2}$, (c) $^{2}1771466136^{2}_{2}$ and (d) $^{2}3918949782^{2}_{2}$.  
\end{center} 
\end{figure}

\noindent \textbf{Example:} Also totalistic rules can weakly break the addition modulo $p$. In fact, the following example illustrates more precisely the difference between $weak$ and $strong$ symmetry breaking. While the former eliminates a non-monotonic turn when breaking the modulo $p$ symmetry, the latter does not. As a consequence the Pascal triangle structure is severely distorted and a mixing of all the available symbols ends in chaotic behavior.  This latter behavior is less complex than the one obtained through the, more subtle, weak symmetry breaking. It is nonetheless important, that such complex rules can be engineered using the Pascal simplices modulo $p$ as template. 

\begin{figure}
\includegraphics[width=0.7\textwidth]{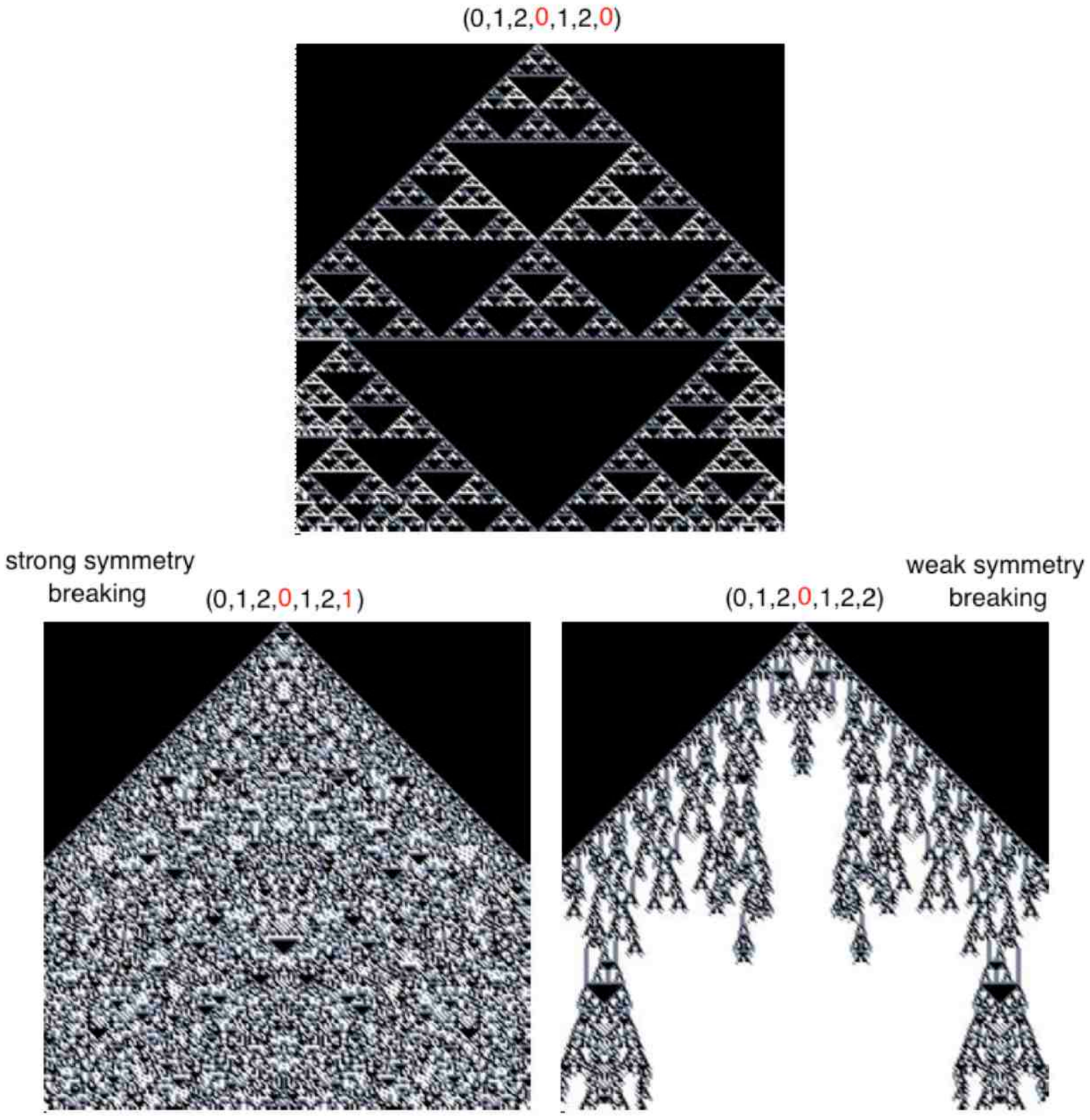}
\begin{center}
Figure S14: Spatiotemporal evolution of Pascal rule with code $^{1}588T^{1}_{3}$ (top). The modulo $p$ symmetry can be broken so that the strong non-monotonicity of the Pascal rule is preserved. In such case rule $^{1}1317T^{1}_{3}$ (bottom left) is obtained and the corresponding dynamical behavior is chaotic (Class 3). If however, the symmetry breaking is weak, so that the modulo $p$ symmetry is destroyed together with the non-monotonic turn, the Class 4 totalistic rule $^{1}2046T^{1}_{3}$ (bottom right) is obtained. The vectors of each totalistic rule are indicated on the figure, and in red are marked the positions that break monotonicity within the Pascal simplex. 
\end{center} 
\end{figure}

Figure S14 illustrates the difference between weak and strong symmetry breaking of a Pascal rule. The Pascal rule has vector $(\sigma_{0},\sigma_{1},\sigma_{2},\sigma_{3},\sigma_{4},\sigma_{5},\sigma_{6})=(0,1,2,0,1,2,0)$ meaning that the output value is $0$ each time a situation in which the sum of the site values is either $3$ or $6$. These positions are responsible for the breaking of the monotonicity of the rule since the natural drive of the sum to increase in the natural numbers is reseted to zero at the sum values which are multiples of 3. The symmetry of the Pascal rule can be now broken by switching the value of one of the inner configurations in the triangle of expansion. If the innermost one, $\sigma_{6}$, is switched it is clear that a structure similar to the Pascal rule propagating unchanged during several time steps is formed. Then a defect occurs and the behavior of this defect decides wether the rule behaves then "chaotically" or with a Class 4 behavior, yielding complex structures. There are two possibilities to destroy the symmetry of the Pascal rule: by tuning $\sigma_{6}$ to '1' or to '2'. The first tuning keeps the non-monotonic turn of the Pascal rule while at the same time destroying the modulo $p$ addition symmetry. The rule is therefore strongly non-monotonic and the resulting rule, with vector $(\sigma_{0},\sigma_{1},\sigma_{2},\sigma_{3},\sigma_{4},\sigma_{5},\sigma_{6})=(0,1,2,0,1,2,1)$  yields seemingly chaotic Class 3 behavior. If now the innermost configuration is tuned as $\sigma_{6} \to 2$,  the modulo $p$ addition is destroyed together with the hard non-monotonic turn, and the gentle monotonicity of the previous steps saturates to a value of two. This illustrates the weak symmetry breaking of the addition modulo $p$. The resulting rule $(\sigma_{0},\sigma_{1},\sigma_{2},\sigma_{3},\sigma_{4},\sigma_{5},\sigma_{6})=(0,1,2,0,1,2,2)$, has Class 4 behavior. In fact, this seems characteristic of all observed totalistic rules of Class 4: the saturation of the sum within the triangle of expansion meaning that, after a certain transient (that can be very long) no complex structures survive anymore.

\begin{table}[htdp]
\begin{center}
\begin{tabular}{cclcccl}
Rule & $ \qquad (a_{0},a_{1},a_{2},a_{3},a_{4},a_{5},a_{6},a_{7})$ & $\qquad GC$ & $\qquad LR$ & $\qquad GCLR$ & $\qquad \kappa $ \\
$^{1}0_{2}^{1}$ & \qquad (0,0,0,0,0,0,0,0) & $\qquad ^{1}255_{2}^{1}$ &\qquad $^{1}0_{2}^{1}$ &\qquad $^{1}255_{2}^{1}$ & $\qquad 0 $ \\
$^{1}1_{2}^{1}$ & \qquad (1,0,0,0,0,0,0,0) & $\qquad ^{1}127_{2}^{1}$ & $\qquad ^{1}1_{2}^{1}$ & $\qquad ^{1}127_{2}^{1}$  & $\qquad 1 $ \\
$^{1}2_{2}^{1}$ & \qquad (0,1,0,0,0,0,0,0) & $\qquad ^{1}191_{2}^{1}$ & $\qquad ^{1}16_{2}^{1}$ & $\qquad ^{1}247_{2}^{1}$ & $\qquad 1 $ \\
$^{1}3_{2}^{1}$ & \qquad (1,1,0,0,0,0,0,0) & $\qquad ^{1}63_{2}^{1}$ & $\qquad ^{1}17_{2}^{1}$ & $\qquad ^{1}119_{2}^{1}$ & $\qquad 1 $  \\
$^{1}4_{2}^{1}$ & \qquad (0,0,1,0,0,0,0,0) & $\qquad ^{1}223_{2}^{1}$ & $\qquad ^{1}4_{2}^{1}$ & $\qquad ^{1}223_{2}^{1}$  & $\qquad 1 $ \\
$^{1}5_{2}^{1}$ & \qquad (1,0,1,0,0,0,0,0) & $\qquad ^{1}95_{2}^{1}$ & $\qquad ^{1}5_{2}^{1}$ & $\qquad ^{1}95_{2}^{1}$  & $\qquad 1 $ \\
$^{1}6_{2}^{1}$ & \qquad (0,1,1,0,0,0,0,0) & $\qquad ^{1}159_{2}^{1}$ & $\qquad ^{1}20_{2}^{1}$ & $\qquad ^{1}215_{2}^{1}$  & $\qquad 2 $ \\
$^{1}7_{2}^{1}$ & \qquad (1,1,1,0,0,0,0,0) & $\qquad ^{1}31_{2}^{1}$ & $\qquad ^{1}21_{2}^{0}$ & $\qquad ^{1}87_{2}^{1}$ & $\qquad 1 $  \\
$^{1}8_{2}^{1}$ & \qquad (0,0,0,1,0,0,0,0) & $\qquad ^{1}239_{2}^{1}$ & $\qquad ^{1}64_{2}^{1}$ & $\qquad ^{1}253_{2}^{1}$ & $\qquad 1 $ \\
$^{1}9_{2}^{1}$ & \qquad (1,0,0,1,0,0,0,0) & $\qquad ^{1}111_{2}^{1}$ & $\qquad ^{1}65_{2}^{1}$ & $\qquad ^{1}125_{2}^{1}$  & $\qquad 2 $ \\
$^{1}10_{2}^{1}$ & \qquad (0,1,0,1,0,0,0,0) & $\qquad ^{1}175_{2}^{1}$ & $\qquad ^{1}80_{2}^{1}$ & $\qquad ^{1}245_{2}^{1}$  & $\qquad 1 $ \\
$^{1}11_{2}^{1}$ & \qquad (1,1,0,1,0,0,0,0) & $\qquad ^{1}47_{2}^{1}$ & $\qquad ^{1}81_{2}^{1}$ & $\qquad ^{1}117_{2}^{1}$  & $\qquad 1 $ \\
$^{1}12_{2}^{1}$ & \qquad (0,0,1,1,0,0,0,0) & $\qquad ^{1}207_{2}^{1}$ & $\qquad ^{1}68_{2}^{1}$ & $\qquad ^{1}221_{2}^{1}$  & $\qquad 1 $ \\
$^{1}13_{2}^{1}$ & \qquad (1,0,1,1,0,0,0,0) & $\qquad ^{1}79_{2}^{1}$ & $\qquad ^{1}69_{2}^{1}$ & $\qquad ^{1}93_{2}^{1}$  & $\qquad 1 $ \\
$^{1}14_{2}^{1}$ & \qquad (0,1,1,1,0,0,0,0) & $\qquad ^{1}143_{2}^{1}$ & $\qquad ^{1}84_{2}^{1}$ & $\qquad ^{1}213_{2}^{1}$  & $\qquad 1 $ \\
$^{1}15_{2}^{1}$ & \qquad (1,1,1,1,0,0,0,0) & $\qquad ^{1}15_{2}^{1}$ & $\qquad ^{1}85_{2}^{1}$ & $\qquad ^{1}85_{2}^{1}$  & $\qquad 1 $ \\
$^{1}18_{2}^{1}$ & \qquad (0,1,0,0,1,0,0,0) & $\qquad ^{1}183_{2}^{1}$ & $\qquad ^{1}18_{2}^{1}$ & $\qquad ^{1}183_{2}^{1}$ & $\qquad 2 $ \\
$^{1}19_{2}^{1}$ & \qquad (1,1,0,0,1,0,0,0) & $\qquad ^{1}55_{2}^{1}$ & $\qquad ^{1}19_{2}^{1}$ & $\qquad ^{1}55_{2}^{1}$  & $\qquad 1 $ \\
$^{1}22_{2}^{1}$ & \qquad (0,1,1,0,1,0,0,0) & $\qquad ^{1}151_{2}^{1}$ & $\qquad ^{1}22_{2}^{1}$ & $\qquad ^{1}151_{2}^{1}$  & $\qquad 2 $ \\
$^{1}23_{2}^{1}$ & \qquad (1,1,1,0,1,0,0,0) & $\qquad ^{1}23_{2}^{1}$ & $\qquad ^{1}23_{2}^{1}$ & $\qquad ^{1}23_{2}^{1}$  & $\qquad 1 $ \\
$^{1}24_{2}^{1}$ & \qquad (0,0,0,1,1,0,0,0) & $\qquad ^{1}231_{2}^{1}$ & $\qquad ^{1}66_{2}^{1}$ & $\qquad ^{1}189_{2}^{1}$  & $\qquad 2 $ \\
\end{tabular}
\end{center}
TABLE S5 : The 88 independent rules under global complementation (GC), left-right transformation (LR) and global complementation of the left right transformation (GCLR) out of the total of 256 rules $^{1}R_{2}^{1}$ and their class-equivalent rules and complexity index.
\end{table}

\begin{table}[htdp]
\begin{center}
\begin{tabular}{cclcccl}
Rule & $ \qquad (a_{0},a_{1},a_{2},a_{3},a_{4},a_{5},a_{6},a_{7})$ & $\qquad GC$ & $\qquad LR$ & $\qquad GCLR$ & $\qquad \kappa $  \\
$^{1}25_{2}^{1}$ & \qquad (1,0,0,1,1,0,0,0) & $\qquad ^{1}103_{2}^{1}$ & $\qquad ^{1}67_{2}^{1}$ & $\qquad ^{1}61_{2}^{1}$ & $\qquad 2 $ \\
$^{1}26_{2}^{1}$ & \qquad (0,1,0,1,1,0,0,0) & $\qquad ^{1}167_{2}^{1}$ & $\qquad ^{1}82_{2}^{1}$ & $\qquad ^{1}181_{2}^{1}$ & $\qquad 2 $ \\
$^{1}27_{2}^{1}$ & \qquad (1,1,0,1,1,0,0,0) & $\qquad ^{1}39_{2}^{1}$ & $\qquad ^{1}83_{2}^{1}$ & $\qquad ^{1}53_{2}^{1}$  & $\qquad 1 $ \\
$^{1}28_{2}^{1}$ & \qquad (0,0,1,1,1,0,0,0) & $\qquad ^{1}199_{2}^{1}$ & $\qquad ^{1}70_{2}^{1}$ & $\qquad ^{1}157_{2}^{1}$  & $\qquad 2 $ \\
$^{1}29_{2}^{1}$ & \qquad (1,0,1,1,1,0,0,0) & $\qquad ^{1}71_{2}^{1}$ & $\qquad ^{1}71_{2}^{1}$ & $\qquad ^{1}29_{2}^{1}$ & $\qquad 1 $ \\
$^{1}30_{2}^{1}$ & \qquad (0,1,1,1,1,0,0,0) & $\qquad ^{1}135_{2}^{1}$ & $\qquad ^{1}86_{2}^{1}$ & $\qquad ^{1}149_{2}^{1}$  & $\qquad 2 $ \\
$^{1}32_{2}^{1}$ & \qquad (0,0,0,0,0,1,0,0) & $\qquad ^{1}251_{2}^{1}$ & $\qquad ^{1}32_{2}^{1}$ & $\qquad ^{1}251_{2}^{1}$  & $\qquad 1 $ \\
$^{1}33_{2}^{1}$ & \qquad (1,0,0,0,0,1,0,0) & $\qquad ^{1}123_{2}^{1}$ & $\qquad ^{1}33_{2}^{1}$ & $\qquad ^{1}123_{2}^{1}$  & $\qquad 2 $ \\
$^{1}34_{2}^{1}$ & \qquad (0,1,0,0,0,1,0,0) & $\qquad ^{1}187_{2}^{1}$ & $\qquad ^{1}48_{2}^{1}$ & $\qquad ^{1}243_{2}^{1}$  & $\qquad 1 $ \\
$^{1}35_{2}^{1}$ & \qquad (1,1,0,0,0,1,0,0) & $\qquad ^{1}59_{2}^{1}$ & $\qquad ^{1}49_{2}^{1}$ & $\qquad ^{1}115_{2}^{1}$  & $\qquad 1 $ \\
$^{1}36_{2}^{1}$ & \qquad (0,0,1,0,0,1,0,0) & $\qquad ^{1}219_{2}^{1}$ & $\qquad ^{1}36_{2}^{1}$ & $\qquad ^{1}219_{2}^{1}$  & $\qquad 2 $ \\
$^{1}37_{2}^{1}$ & \qquad (1,0,1,0,0,1,0,0) & $\qquad ^{1}91_{2}^{1}$ & $\qquad ^{1}37_{2}^{1}$ & $\qquad ^{1}91_{2}^{1}$  & $\qquad 2 $ \\
$^{1}38_{2}^{1}$ & \qquad (0,1,1,0,0,1,0,0) & $\qquad ^{1}155_{2}^{1}$ & $\qquad ^{1}52_{2}^{1}$ & $\qquad ^{1}211_{2}^{1}$  & $\qquad 2 $ \\
$^{1}40_{2}^{1}$ & \qquad (0,0,0,1,0,1,0,0) & $\qquad ^{1}235_{2}^{1}$ & $\qquad ^{1}96_{2}^{1}$ & $\qquad ^{1}249_{2}^{1}$  & $\qquad 2 $ \\
$^{1}41_{2}^{1}$ & \qquad (1,0,0,1,0,1,0,0) & $\qquad ^{1}107_{2}^{1}$ & $\qquad ^{1}97_{2}^{1}$ & $\qquad ^{1}121_{2}^{1}$  & $\qquad 2 $ \\
$^{1}42_{2}^{1}$ & \qquad (0,1,0,1,0,1,0,0) & $\qquad ^{1}171_{2}^{1}$ & $\qquad ^{1}112_{2}^{1}$ & $\qquad ^{1}241_{2}^{1}$  & $\qquad 1 $ \\
$^{1}43_{2}^{1}$ & \qquad (1,1,0,1,0,1,0,0) & $\qquad ^{1}43_{2}^{1}$ & $\qquad ^{1}113_{2}^{1}$ & $\qquad ^{1}113_{2}^{1}$  & $\qquad 1 $ \\
$^{1}44_{2}^{1}$ & \qquad (0,0,1,1,0,1,0,0) & $\qquad ^{1}203_{2}^{1}$ & $\qquad ^{1}100_{2}^{1}$ & $\qquad ^{1}217_{2}^{1}$ & $\qquad 2 $  \\
$^{1}45_{2}^{1}$ & \qquad (1,0,1,1,0,1,0,0) & $\qquad ^{1}75_{2}^{1}$ & $\qquad ^{1}101_{2}^{1}$ & $\qquad ^{1}89_{2}^{1}$ & $\qquad 2 $ \\
$^{1}46_{2}^{1}$ & \qquad (0,1,1,1,0,1,0,0) & $\qquad ^{1}139_{2}^{1}$ & $\qquad ^{1}116_{2}^{1}$ & $\qquad ^{1}209_{2}^{1}$ & $\qquad 1 $ \\
$^{1}50_{2}^{1}$ & \qquad (0,1,0,0,1,1,0,0) & $\qquad ^{1}179_{2}^{1}$ & $\qquad ^{1}50_{2}^{1}$ & $\qquad ^{1}179_{2}^{1}$ & $\qquad 1 $ \\
$^{1}51_{2}^{1}$ & \qquad (1,1,0,0,1,1,0,0) & $\qquad ^{1}51_{2}^{1}$ & $\qquad ^{1}51_{2}^{1}$ & $\qquad ^{1}51_{2}^{1}$  & $\qquad 1 $ \\
$^{1}54_{2}^{1}$ & \qquad (0,1,1,0,1,1,0,0) & $\qquad ^{1}147_{2}^{1}$ & $\qquad ^{1}54_{2}^{1}$ & $\qquad ^{1}147_{2}^{1}$ & $\qquad 2 $ \\
\end{tabular}
\end{center}
TABLE S5 (cont.)
\end{table}

\begin{table}[htdp]
\begin{center}
\begin{tabular}{cclcccl}
Rule & $ \qquad (a_{0},a_{1},a_{2},a_{3},a_{4},a_{5},a_{6},a_{7})$ & $\qquad GC$ & $\qquad LR$ & $\qquad GCLR$ & $\qquad \kappa $  \\
$^{1}56_{2}^{1}$ & \qquad (0,0,0,1,1,1,0,0) & $\qquad ^{1}227_{2}^{1}$ & $\qquad ^{1}98_{2}^{1}$ & $\qquad ^{1}185_{2}^{1}$ & $\qquad 2 $ \\
$^{1}57_{2}^{1}$ & \qquad (1,0,0,1,1,1,0,0) & $\qquad ^{1}99_{2}^{1}$ & $\qquad ^{1}99_{2}^{1}$ & $\qquad ^{1}57_{2}^{1}$  & $\qquad 2 $ \\
$^{1}58_{2}^{1}$ & \qquad (0,1,0,1,1,1,0,0) & $\qquad ^{1}163_{2}^{1}$ & $\qquad ^{1}114_{2}^{1}$ & $\qquad ^{1}177_{2}^{1}$  & $\qquad 1 $ \\
$^{1}60_{2}^{1}$ & \qquad (0,0,1,1,1,1,0,0) & $\qquad ^{1}195_{2}^{1}$ & $\qquad ^{1}102_{2}^{1}$ & $\qquad ^{1}153_{2}^{1}$  & $\qquad 2 $ \\
$^{1}62_{2}^{1}$ & \qquad (0,1,1,1,1,1,0,0) & $\qquad ^{1}131_{2}^{1}$ & $\qquad ^{1}118_{2}^{1}$ & $\qquad ^{1}145_{2}^{1}$ & $\qquad 2 $ \\
$^{1}72_{2}^{1}$ & \qquad (0,0,0,1,0,0,1,0)& $\qquad ^{1}237_{2}^{1}$ & $\qquad ^{1}72_{2}^{1}$ & $\qquad ^{1}237_{2}^{1}$  & $\qquad 2 $ \\
$^{1}73_{2}^{1}$ & \qquad (1,0,0,1,0,0,1,0) & $\qquad ^{1}109_{2}^{1}$ & $\qquad ^{1}73_{2}^{1}$ & $\qquad ^{1}109_{2}^{1}$  & $\qquad 2 $ \\
$^{1}74_{2}^{1}$ & \qquad (0,1,0,1,0,0,1,0) & $\qquad ^{1}173_{2}^{1}$ & $\qquad ^{1}88_{2}^{1}$ & $\qquad ^{1}229_{2}^{1}$  & $\qquad 2 $ \\
$^{1}76_{2}^{1}$ & \qquad (0,0,1,1,0,0,1,0) & $\qquad ^{1}205_{2}^{1}$ & $\qquad ^{1}76_{2}^{1}$ & $\qquad ^{1}205_{2}^{1}$  & $\qquad 1 $\\
$^{1}77_{2}^{1}$ & \qquad (1,0,1,1,0,0,1,0) & $\qquad ^{1}77_{2}^{1}$ & $\qquad ^{1}77_{2}^{1}$ & $\qquad ^{1}77_{2}^{1}$  & $\qquad 1 $ \\
$^{1}78_{2}^{1}$ & \qquad (0,1,1,1,0,0,1,0) & $\qquad ^{1}141_{2}^{1}$ & $\qquad ^{1}92_{2}^{1}$ & $\qquad ^{1}197_{2}^{1}$  & $\qquad 1 $ \\
$^{1}90_{2}^{1}$ & \qquad (0,1,0,1,1,0,1,0) & $\qquad ^{1}165_{2}^{1}$ & $\qquad ^{1}90_{2}^{1}$ & $\qquad ^{1}165_{2}^{1}$  & $\qquad 2 $ \\
$^{1}94_{2}^{1}$ & \qquad (0,1,1,1,1,0,1,0)  & $\qquad ^{1}133_{2}^{1}$ & $\qquad ^{1}94_{2}^{1}$ & $\qquad ^{1}133_{2}^{1}$  & $\qquad 2 $\\
$^{1}104_{2}^{1}$ & \qquad (0,0,0,1,0,1,1,0) & $\qquad ^{1}233_{2}^{1}$ & $\qquad ^{1}104_{2}^{1}$ & $\qquad ^{1}233_{2}^{1}$  & $\qquad 2 $ \\
$^{1}105_{2}^{1}$ & \qquad (1,0,0,1,0,1,1,0) & $\qquad ^{1}105_{2}^{1}$ & $\qquad ^{1}105_{2}^{1}$ & $\qquad ^{1}105_{2}^{1}$  & $\qquad 2 $ \\
$^{1}106_{2}^{1}$ & \qquad (0,1,0,1,0,1,1,0) & $\qquad ^{1}169_{2}^{1}$ & $\qquad ^{1}120_{2}^{1}$ & $\qquad ^{1}225_{2}^{1}$  & $\qquad 2 $ \\
$^{1}108_{2}^{1}$ & \qquad (0,0,1,1,0,1,1,0) & $\qquad ^{1}201_{2}^{1}$ & $\qquad ^{1}108_{2}^{1}$ & $\qquad ^{1}201_{2}^{1}$  & $\qquad 2 $ \\
$^{1}110_{2}^{1}$ & \qquad (0,1,1,1,0,1,1,0) & $\qquad ^{1}137_{2}^{1}$ & $\qquad ^{1}124_{2}^{1}$ & $\qquad ^{1}193_{2}^{1}$  & $\qquad 3 $ \\
$^{1}122_{2}^{1}$ & \qquad (0,1,0,1,1,1,1,0) & $\qquad ^{1}161_{2}^{1}$ & $\qquad ^{1}122_{2}^{1}$ & $\qquad ^{1}161_{2}^{1}$  & $\qquad 2 $ \\
$^{1}126_{2}^{1}$ & \qquad (0,1,1,1,1,1,1,0) & $\qquad ^{1}129_{2}^{1}$ & $\qquad ^{1}126_{2}^{1}$ & $\qquad ^{1}129_{2}^{1}$  & $\qquad 2 $ \\
$^{1}128_{2}^{1}$ & \qquad (0,0,0,0,0,0,0,1) & $\qquad ^{1}254_{2}^{1}$ & $\qquad ^{1}128_{2}^{1}$ & $\qquad ^{1}254_{2}^{1}$  & $\qquad 1 $ \\
$^{1}130_{2}^{1}$ & \qquad (0,1,0,0,0,0,0,1) & $\qquad ^{1}190_{2}^{1}$ & $\qquad ^{1}144_{2}^{1}$ & $\qquad ^{1}246_{2}^{1}$  & $\qquad 2 $ \\
$^{1}132_{2}^{1}$ & \qquad (0,0,1,0,0,0,0,1) & $\qquad ^{1}222_{2}^{1}$ & $\qquad ^{1}132_{2}^{1}$ & $\qquad ^{1}222_{2}^{1}$  & $\qquad 2 $ \\
\end{tabular}
\end{center}
TABLE S5 (cont.)
\end{table}

\begin{table}[htdp]
\begin{center}
\begin{tabular}{cclcccl}
Rule & $ \qquad (a_{0},a_{1},a_{2},a_{3},a_{4},a_{5},a_{6},a_{7})$ & $\qquad GC$ & $\qquad LR$ & $\qquad GCLR$ & $\qquad \kappa $  \\
$^{1}134_{2}^{1}$ & \qquad (0,1,1,0,0,0,0,1) & $\qquad ^{1}158_{2}^{1}$ & $\qquad ^{1}148_{2}^{1}$ & $\qquad ^{1}214_{2}^{1}$  & $\qquad 2 $ \\
$^{1}136_{2}^{1}$ & \qquad (0,0,0,1,0,0,0,1) & $\qquad ^{1}238_{2}^{1}$ & $\qquad ^{1}192_{2}^{1}$ & $\qquad ^{1}252_{2}^{1}$  & $\qquad 1 $ \\
$^{1}138_{2}^{1}$ & \qquad (0,1,0,1,0,0,0,1) & $\qquad ^{1}174_{2}^{1}$ & $\qquad ^{1}208_{2}^{1}$ & $\qquad ^{1}244_{2}^{1}$  & $\qquad 1 $ \\
$^{1}140_{2}^{1}$ & \qquad (0,0,1,1,0,0,0,1) & $\qquad ^{1}206_{2}^{1}$ & $\qquad ^{1}196_{2}^{1}$ & $\qquad ^{1}220_{2}^{1}$  & $\qquad 1 $ \\
$^{1}142_{2}^{1}$ & \qquad (0,1,1,1,0,0,0,1) & $\qquad ^{1}142_{2}^{1}$ & $\qquad ^{1}212_{2}^{1}$ & $\qquad ^{1}212_{2}^{1}$  & $\qquad 1 $ \\
$^{1}146_{2}^{1}$ & \qquad (0,1,0,0,1,0,0,1) & $\qquad ^{1}182_{2}^{1}$ & $\qquad ^{1}146_{2}^{1}$ & $\qquad ^{1}182_{2}^{1}$  & $\qquad 2 $ \\
$^{1}150_{2}^{1}$ & \qquad (0,1,1,0,1,0,0,1) & $\qquad ^{1}150_{2}^{1}$ & $\qquad ^{1}150_{2}^{1}$ & $\qquad ^{1}150_{2}^{1}$  & $\qquad 2 $ \\
$^{1}152_{2}^{1}$ & \qquad (0,0,0,1,1,0,0,1) & $\qquad ^{1}230_{2}^{1}$ & $\qquad ^{1}194_{2}^{1}$ & $\qquad ^{1}188_{2}^{1}$  & $\qquad 2 $ \\
$^{1}154_{2}^{1}$ & \qquad (0,1,0,1,1,0,0,1) & $\qquad ^{1}166_{2}^{1}$ & $\qquad ^{1}210_{2}^{1}$ & $\qquad ^{1}180_{2}^{1}$  & $\qquad 2 $ \\
$^{1}156_{2}^{1}$ & \qquad (0,0,1,1,1,0,0,1) & $\qquad ^{1}198_{2}^{1}$ & $\qquad ^{1}198_{2}^{1}$ & $\qquad ^{1}156_{2}^{1}$  & $\qquad 2 $ \\
$^{1}160_{2}^{1}$ & \qquad (0,0,0,0,0,1,0,1) & $\qquad ^{1}250_{2}^{1}$ & $\qquad ^{1}160_{2}^{1}$ & $\qquad ^{1}250_{2}^{1}$  & $\qquad 1 $ \\
$^{1}162_{2}^{1}$ & \qquad (0,1,0,0,0,1,0,1) & $\qquad ^{1}186_{2}^{1}$ & $\qquad ^{1}176_{2}^{1}$ & $\qquad ^{1}242_{2}^{1}$  & $\qquad 1 $ \\
$^{1}164_{2}^{1}$ & \qquad (0,0,1,0,0,1,0,1) & $\qquad ^{1}218_{2}^{1}$ & $\qquad ^{1}164_{2}^{1}$ & $\qquad ^{1}218_{2}^{1}$  & $\qquad 2 $ \\
$^{1}168_{2}^{1}$ & \qquad (0,0,0,1,0,1,0,1) & $\qquad ^{1}234_{2}^{1}$ & $\qquad ^{1}224_{2}^{1}$ & $\qquad ^{1}248_{2}^{1}$  & $\qquad 1 $ \\
$^{1}170_{2}^{1}$ & \qquad (0,1,0,1,0,1,0,1) & $\qquad ^{1}170_{2}^{1}$ & $\qquad ^{1}240_{2}^{1}$ & $\qquad ^{1}240_{2}^{1}$  & $\qquad 2 $ \\
$^{1}172_{2}^{1}$ & \qquad (0,0,1,1,0,1,0,1) & $\qquad ^{1}202_{2}^{1}$ & $\qquad ^{1}228_{2}^{1}$ & $\qquad ^{1}216_{2}^{1}$  & $\qquad 1 $ \\
$^{1}178_{2}^{1}$ & \qquad (0,1,0,0,1,1,0,1) & $\qquad ^{1}178_{2}^{1}$ & $\qquad ^{1}178_{2}^{1}$ & $\qquad ^{1}178_{2}^{1}$  & $\qquad 1 $ \\
$^{1}184_{2}^{1}$ & \qquad (0,0,0,1,1,1,0,1) & $\qquad ^{1}226_{2}^{1}$ & $\qquad ^{1}226_{2}^{1}$ & $\qquad ^{1}184_{2}^{1}$  & $\qquad 1 $ \\
$^{1}200_{2}^{1}$ & \qquad (0,0,0,1,0,0,1,1) & $\qquad ^{1}236_{2}^{1}$ & $\qquad ^{1}200_{2}^{1}$ & $\qquad ^{1}236_{2}^{1}$  & $\qquad 1 $ \\
$^{1}204_{2}^{1}$ & \qquad (0,0,1,1,0,0,1,1) & $\qquad ^{1}204_{2}^{1}$ & $\qquad ^{1}204_{2}^{1}$ & $\qquad ^{1}204_{2}^{1}$  & $\qquad 1 $ \\
$^{1}232_{2}^{1}$ & \qquad (0,0,0,1,0,1,1,1) & $\qquad ^{1}232_{2}^{1}$ & $\qquad ^{1}232_{2}^{1}$ & $\qquad ^{1}232_{2}^{1}$  & $\qquad 1 $ \\
\end{tabular}
\end{center}
TABLE S5 (cont.)
~\\~\\~\\~\\~\\~\\~\\~\\
\end{table}

It is to be noted that the whole discussion about complexity in this section applies to all members in a given equivalence class under change of colors, reflection, shift, etc.

I introduce a complexity index $\kappa$ with values 1, 2 or 3, according to the following prescription

\begin{itemize}
\item{$\kappa=1$: Neutrally monotonic and other monotonic rules (Class 1 and Class 2 behaviors)} 
\item{$\kappa=2$: Strongly non-monotonic rules, Pascal rules and non-monotonic rules that strongly break the addition modulo $p$ (Class 3 behavior)} 
\item{$\kappa=3$: Weakly non-monotonic rules that weakly break the addition modulo $p$ (Class 4 behavior)}
\end{itemize}

This index allows to classify every CA rule. As an example, in Table S5, the complexity index $\kappa$ is listed for the 256 Wolfram classes such as they fall in the 88 equivalence classes under change of colors and reflection. (As noted above, shift invariance also implies that not all of these 88 rules are independent since rules 12 and 34 are also equivalent). The index is calculated easily from the construction theorem. \\

\noindent \textbf{Example}: Wolfram Rule $^{1}30^{1}_{2}$ has vector $(a_{0}, a_{1}, a_{2}, a_{3}, a_{4}, a_{5}, a_{6}, a_{7})=(0,1,1,1,1,0,0,0)$. To obtain the construction from the left simply separate $(0,1,1,1,1,0,0,0)$ into $p=2$ consecutive parts with same size. Rules $(0,1,1,1)$ and $(1,0,0,0)$ are obtained, which correspond, respectively, to rules $^{0}14^{1}_{2}$ and $^{0}1^{1}_{2}$. The latter rules are constructed from the right by rules $(0,1)$ and $(1,1)$ (Rule $^{0}14^{1}_{2}$) and $(1,0)$ and $(0,0)$ (Rule $^{0}1^{1}_{2}$). The rules in the first construction layer are both monotonic, but they have different monotonicity sign: rule $^{0}14^{1}_{2}$ has monotonicity sign + while rule $^{0}1^{1}_{2}$ has monotonicity sign -. As a result, rule $^{1}30^{1}_{2}$ has a non-monotonic turn in its second layer and belongs to either Class 3 or Class 4. Compared to the closest Pascal rule $^{0}6^{1}_{2}$ copied to higher range, i.e. $^{0}102^{1}_{2}$, one sees that $a_{3}$, $a_{4}$, $a_{5}$ and $a_{6}$ are switched.  The positions controlled by $a_{4}$, $a_{5}$ and $a_{6}$ affect the central spine of the triangle of expansion of the associated Pascal rule, and break $strongly$ the addition modulo $p$. Rule $^{1}30^{1}_{2}$ belongs, therefore, to Class 3 and has complexity index $\kappa=2$. $\Box$ \\



\subsection{CA in higher dimensions}

The above results can be easily generalized to an arbitrary number of dimensions. In 2D, for example, by using the site that is updated after each time step as a pivoting site and defining 1D codes for each spatial direction counterclockwise, universal maps for all possible deterministic CA in 2D (depending on the topology of the interactions in the lattice) can be derived. The most popular neighborhoods in 2D are shown in Fig. S15. The von Neumann neighborhood can be specified giving two binary codes $n_{1}$ and $n_{2}$ as indicated in the figure. The hexagonal neighborhood requires the specification of three codes to describe each configuration. Finally, the Moore neighborhood requires four 1D codes. In each neighborhood, the site on the center is updated on the next time step.

\begin{figure}
\begin{center}
\includegraphics[width=0.7\textwidth, angle=270]{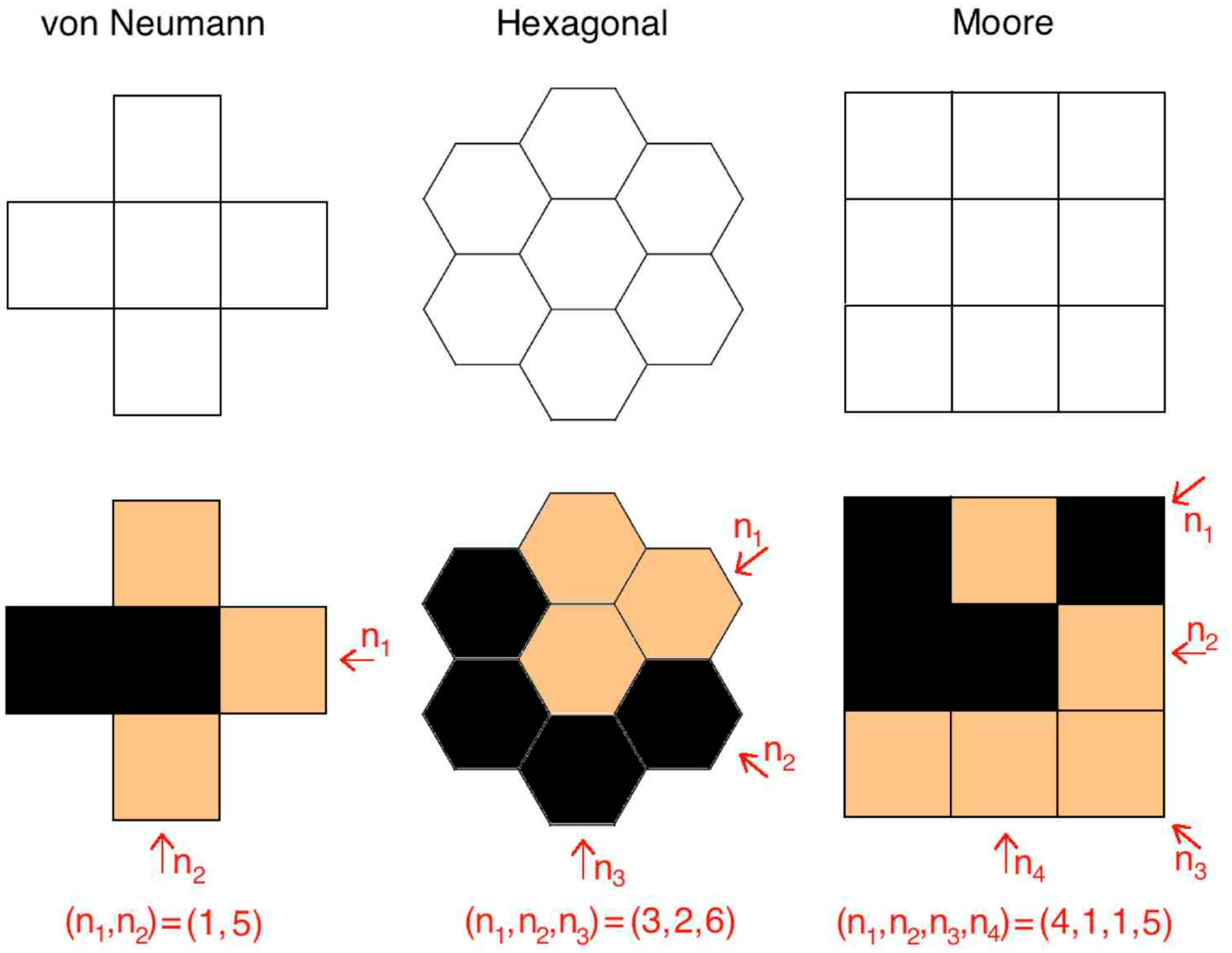}
\end{center}
Figure S15: Symmetrical Von Neumann, hexagonal and Moore neighborhoods in two dimensions. A configuration can be specified by giving two, three, or four 1D codes, respectively.   
\end{figure}

Since a configuration is now composed of several 1D configurations that must hold simultaneously, a product of boxcar functions is needed to fully account for the dynamical state. For a von Neumann neighborhood with horizontal and vertical ranges given by the pairs $\rho_{1}=l_{1}+r_{1}+1$ and $\rho_{2}=l_{2}+r_{2}+1$ ($\Omega_{1}=p^{\rho_{1}}$ and $\Omega_{2}=p^{\rho_{2}}$) respectively, it is obtained, instead of Eq. (\ref{CA})
\begin{equation}
x_{t+1}^{i,j}=\sum_{n_{1}=0}^{\Omega_{1}-1}\sum_{n_{2}=0}^{\Omega_{2}-1}a_{n_{1},n_{2}}\mathcal{B}\left(n_{1}-\sum_{k=-r_{1}}^{l_{1}}p^{k+r_{1}}x_{t}^{i+k,j},\frac{1}{2}\right)\mathcal{B}\left(n_{2}-\sum_{k=-r_{2}}^{l_{2}}p^{k+r_{2}}x_{t}^{i,j+k},\frac{1}{2}\right)
\end{equation}
(note that in the figure only the entirely symmetrical neighborhood with $\rho_{1}=\rho_{2}=3$ is shown). For a totalistic rule over a von Neumann neighborhood 
\begin{equation}
x_{t+1}^{i,j}=\sum_{s=0}^{(\rho_{1}+\rho_{2}-1)(p-1)}\sigma_{s}\mathcal{B}\left(s+1-\sum_{k=-r_{1}}^{l_{1}}x_{t}^{i+k,j}-\sum_{k=-r_{2}}^{l_{2}}x_{t}^{i,j+k},\frac{1}{2}\right)
\end{equation}
For an hexagonal neighborhood
\begin{eqnarray}
x_{t+1}^{i,j}&=&\sum_{n_{1}=0}^{\Omega_{1}-1}\sum_{n_{2}=0}^{\Omega_{2}-1}\sum_{n_{3}=0}^{\Omega_{3}-1}
a_{n_{1},n_{2},n_{3}}\mathcal{B}\left(n_{1}-\sum_{k=-r_{1}}^{l_{1}}p^{k+r_{1}}x_{t}^{i+k,j-k},\frac{1}{2}\right) \times \nonumber \\
&& \times\mathcal{B}\left(n_{2}-\sum_{k=-r_{2}}^{l_{2}}p^{k+r_{2}}x_{t}^{i+k,j+k},\frac{1}{2}\right)
\mathcal{B}\left(n_{3}-\sum_{k=-r_{3}}^{l_{3}}p^{k+r_{3}}x_{t}^{i,j+k},\frac{1}{2}\right)
\end{eqnarray}
and for the totalistic case
\begin{equation}
x_{t+1}^{i,j}=\sum_{s=0}^{(\rho_{1}+\rho_{2}+\rho_{3}-2)(p-1)}\sigma_{s}\mathcal{B}\left(s+2-\sum_{k=-r_{1}}^{l_{1}}x_{t}^{i+k,j-k}-\sum_{k=-r_{2}}^{l_{2}}x_{t}^{i+k,j+k}-\sum_{k=-r_{3}}^{l_{3}}x_{t}^{i,j+k},\frac{1}{2}\right)
\end{equation}
Finally, for the Moore neighborhood one has, for the general case,
\begin{eqnarray}
x_{t+1}^{i,j}&=&\sum_{n_{1}=0}^{\Omega_{1}-1}\sum_{n_{2}=0}^{\Omega_{2}-1}\sum_{n_{3}=0}^{\Omega_{3}-1}\sum_{n_{4}=0}^{\Omega_{4}-1}
a_{n_{1},n_{2},n_{3},n_{4}}\mathcal{B}\left(n_{1}-\sum_{k=-r_{1}}^{l_{1}}p^{k+r_{1}}x_{t}^{i+k,j-k},\frac{1}{2}\right) \times \nonumber \\
&& \times\mathcal{B}\left(n_{2}-\sum_{k=-r_{2}}^{l_{2}}p^{k+r_{2}}x_{t}^{i+k,j},\frac{1}{2}\right)
\mathcal{B}\left(n_{3}-\sum_{k=-r_{3}}^{l_{3}}p^{k+r_{3}}x_{t}^{i+k,j+k},\frac{1}{2}\right) \nonumber \\
&& \times\mathcal{B}\left(n_{4}-\sum_{k=-r_{4}}^{l_{4}}p^{k+r_{4}}x_{t}^{i,j+k},\frac{1}{2}\right)
\end{eqnarray}
and for the totalistic one
\begin{eqnarray}
x_{t+1}^{i,j}=\sum_{s=0}^{(\rho_{1}+\rho_{2}+\rho_{3}+\rho_{4}-3)(p-1)}\sigma_{s}\mathcal{B}&&\left(s+3-\sum_{k=-r_{1}}^{l_{1}}x_{t}^{i+k,j-k}-\sum_{k=-r_{2}}^{l_{2}}x_{t}^{i+k,j}-\sum_{k=-r_{3}}^{l_{3}}x_{t}^{i+k,j+k}-\right. \nonumber \\
&&-\left.\sum_{k=-r_{4}}^{l_{4}}x_{t}^{i,j+k},\frac{1}{2}\right)
\end{eqnarray}
Above I considered the most general situation, where each spatial direction can be non-symmetrical respect to the site $i,j$. In many interesting cases, however, symmetrical neighborhoods as the ones in Fig. S15 are considered, when particularized to them, these expressions reduce to very simple forms. Taking now, for example, a Moore neighborhood with $l_{1}=l_{2}=l_{3}=l_{4}=r_{1}=r_{2}=r_{3}=r_{4}=1$, as in Fig. S2b, the latter expression simplifies to
\begin{equation}
x_{t+1}^{i,j}=\sum_{s=0}^{9}\sigma_{s}\mathcal{B}\left(s-\sum_{k,m=-1}^{1}x_{t}^{i+k,j+m},\frac{1}{2}\right)
\end{equation}
A famous totalistic CA in such a Moore neighborhood due to Gerard Vichniac is the so-called ``vote'' and is given by a vector $(\sigma_{0},\sigma_{1},...,\sigma_{9})=(0,0,0,0,1,1,1,1,1)$. For such CA, since the sum over the cells in the neighborhood can only be at most '9' and the CA only returns '1' when $s \ge 5$ and zero otherwise, 
\begin{equation}
x_{t+1}^{i,j}=\sum_{s=5}^{9}\mathcal{B}\left(s-\sum_{k,m=-1}^{1}x_{t}^{i+k,j+m},\frac{1}{2}\right)=\sum_{s=5}^{\infty}\mathcal{B}\left(s-\sum_{k,m=-1}^{1}x_{t}^{i+k,j+m},\frac{1}{2}\right)
\end{equation}
where the sum has been extended to infinity (a trick that is possible to employ here for this specific CA because of its structure). By using result $(x)$ from Appendix A
\begin{equation}
x_{t+1}^{i,j}=H\left(\sum_{k,m=-1}^{1}x_{t}^{i+k,j+m}-\frac{9}{2}\right)
\end{equation}
The r.h.s. of the latter expression is only one when the sum over the cells is higher or equal than $5$.

Another famous semitotalistic 2D boolean CA called ``Game of life'' and invented by J. H. Conway is also defined on such a Moore neighborhood in terms of the following rules, governing the behavior of the site in the center of the neighborhood (this site is called a ``live cell'' if it has value '1' and a ``dead cell'' if it has value '0')
\begin{itemize}
\item   1. Any live cell with fewer than two live neighbors dies, as if caused by under-population.
\item   2. Any live cell with two or three live neighbors lives on to the next generation.
\item   3. Any live cell with more than three live neighbors dies, as if by overcrowding.
\item   4. Any dead cell with exactly three live neighbors becomes a live cell, as if by reproduction.
\end{itemize}
This set of rules can indeed be reduced to three
\begin{itemize}
\item   1. If on the entire neighborhood the sum of all site values is '3' the cell in the center lives (or becomes a live cell) in the next generation.
\item   2. If on the entire neighborhood the sum of all site values is '4' the cell in the center lives in the next generation only if it is already alive.
\item   3. The cell in the center is dead on the next generation if neither 1. nor 2. holds
\end{itemize}
By employing Eq. (\ref{FormGen}) an analytical map for this CA can be written $directly$, translating the rules into the mathematical language introduced here
\begin{equation}
x_{t+1}^{i,j}=\mathcal{B}\left(3-\sum_{k,m=-1}^{1}x_{t}^{i+k,j+m},\frac{1}{2}\right)+x_{t}^{i,j}\mathcal{B}\left(4-\sum_{k,m=-1}^{1}x_{t}^{i+k,j+m},\frac{1}{2}\right) \label{Life}
\end{equation}
This is a local map for each site $i,j$ in the lattice that implements the Game of Life.

\subsection{Continuum limit}

Although, as defined, Eqs. (\ref{CAtot}) and Eq.(\ref{CA}) apply to a finite set of integers $\in [0,p-1]$ which are mapped to the integers in the same interval, if $p \to \infty$ the above expression may reproduce with arbitrary precision any continuous map. I  assume $l=r=0$ and a number of symbols $p$. Then $\Omega=p^{\rho}=p$ (since $\rho=l+r+1=1$). $p$ can now play the role of the precision in the sense that a real number is given in terms of multiples of $1/p$ and then a mapping of the rational numbers contained in the interval $[0,1]$ to itself is performed. This can be seen from Eq. (\ref{CA}) since
\begin{eqnarray}
x_{t+1}^{i}&=&\sum_{n=0}^{\Omega-1}a_{n}\mathcal{B}\left(n-x_{t}^{i},\frac{1}{2}\right) \nonumber  \\
&=&\sum_{n=0}^{p-1}a_{n}\mathcal{B}\left(\frac{n-x_{t}^{i}}{p-1},\frac{1}{2(p-1)}\right) \label{parti}
\end{eqnarray}\\
where result $(iii)$ from Appendix A has been used. By defining the real quantities $n'=n/(p-1)$, $y_{t}^{i}=x_{t}^{i}/(p-1)$ and $y_{t+1}^{i}=x_{t+1}^{i}/(p-1)$, by taking the limit $p \to \infty$, and by using $(v)$ in Appendix A  
\begin{eqnarray}
y_{t+1}^{i}&=&\lim_{p \to \infty}\int_{0}^{1}dn'\frac{a_{n' p}}{(p-1)^{2}}\delta\left(n'-y_{t}^{i}\right)
 \label{CAreal}
\end{eqnarray}\\
The latter expression, involving real quantities constitutes the limiting behavior of a CA map involving an infinitely large number of integers. The logistic map, for example, can be reproduced if 
\begin{equation}
a_{n'(p-1)}=r(p-1)^{2}n'(1-n')
\end{equation}
since in such case Eq. (\ref{CAreal}) becomes
\begin{equation}
y_{t+1}^{i}=ry_{t}^{i}(1-y_{t}^{i})
\end{equation}
In terms of the original map involving only integers a CA rule $^{0}R_{\infty}^{0}$ described by Eq. (\ref{CA}) with
\begin{equation}
a_{n}=r(p-1)n\left(1-\frac{n}{p-1}\right)
\end{equation}
and where the code $R$, defined by Eq.(\ref{code}) is given by
\begin{equation}
R \equiv \sum_{n=0}^{\Omega-1}a_{n}p^{n}=\lim_{p \to \infty}\sum_{n=0}^{p-1}rn(p-1)(1-n/(p-1))p^{n}
\end{equation}
coincides with the logistic map in its dynamical behavior. Simulations show that if the limit is not taken and $p=1000$, Eq.  (\ref{parti}) (obtained from Eq. (\ref{CA}) in this case) already provides an excellent approximation to the logistic map.


\section*{Appendix A: Some results involving the function $\mathcal{B}(x,\epsilon)$ }

In this Appendix several results involving the function $\mathcal{B}$ are outlined. Proofs are skipped for most of the statements, although they can be given easily by induction and exhaustion. 

The following definitions of $\mathcal{B}(x,\epsilon)$ are equivalent to Definition S1
\begin{eqnarray}
\mathcal{B}(x,\epsilon)&=&\frac{1}{2}\left[sign(x+\epsilon)-sign(x-\epsilon)\right] \label{d2} \\
&=& H(x+\epsilon)-H(x-\epsilon) \label{d3} \\
&=& sign(\epsilon) \ H\left(1-\frac{x^{2}}{\epsilon^2}\right) \label{d4} \\
&=& sign(\epsilon) \ \chi_{(-|\epsilon|, |\epsilon|)}(x) \label{d5}
\end{eqnarray}
where $sign(x)=\left\{
\begin{array}{cc}
1  & \quad x > 0    \\
-1  & \quad x < 0     
\end{array}
\right. $
is the sign function, 
$H(x)=\left\{
\begin{array}{cc}
1  & \quad x > 0    \\
0  & \quad x < 0     
\end{array}
\right. $ is the Heaviside function, and $\chi_{(-|\epsilon|, |\epsilon|)}(x)=\left\{
\begin{array}{cc}
1  & \quad x \in (-|\epsilon|, |\epsilon|)    \\
0  & \quad otherwise    
\end{array}
\right. $ is the indicator function for the interval $(-|\epsilon|, |\epsilon|)$. \\

\noindent  \textbf{Results:}
\emph{The function $\mathcal{B}(x,\epsilon)$ satisfies
\begin{itemize}
\item (i) ~ $\mathcal{B}(-x,\epsilon)=\mathcal{B}(x,\epsilon)$
\item (ii) ~ $\mathcal{B}(x,-\epsilon)=-\mathcal{B}(x,\epsilon)$
\item (iii) ~ $\mathcal{B}(ax,\epsilon)=\mathcal{B}(x,\frac{\epsilon}{a})$, for real $a$.
\item (iv) ~ $\mathcal{B}^{n}(x,\epsilon)=\left(\frac{\epsilon}{|\epsilon|}\right)^{n}\mathcal{B}(x,|\epsilon|)$, for $n$ non-negative integer.
\item (v) ~ $\lim_{\epsilon \to 0} \frac{\mathcal{B}(x,\epsilon)}{2\epsilon}=\delta(x)$, where $\delta(x)$ is the Dirac delta function.
\item (vi) ~ $\sum_{k=0}^{N-1}\mathcal{B}(x-2k\epsilon,\epsilon)=\mathcal{B}(x-(N-1)\epsilon,N\epsilon)$, ~ for $0<\epsilon \le 1/2$.
\item (vii) ~ $\sum_{k=0}^{N-1}\mathcal{B}(x-2k\epsilon,\epsilon)=1$, ~ for integer $x \in [-\epsilon,(2N-1)\epsilon]$ and $0<\epsilon \le 1/2$.
\item (viii) ~ $\sum_{k=0}^{N-1}\mathcal{B}\left(x-k,\frac{1}{2}\right)=1$ and $\sum_{k=0}^{N-1}k\mathcal{B}\left(x-k,\frac{1}{2}\right)=x$, ~ for integer $x \in [-1/2,(2N-1)/2]$.
\item (ix) ~ $\sum_{k=0}^{N-1}f(k)\mathcal{B}\left(x-k,\frac{1}{2}\right)=f(x)$, ~ for integer $x \in [-1/2,(2N-1)/2]$.
\item (x) ~ $\sum_{k=0}^{\infty}\mathcal{B}(x-2k\epsilon,\epsilon)=H(x+\epsilon)$,   for $0<\epsilon \le 1/2$.
\item (xi) ~ $\sum_{k=-\infty}^{\infty}\mathcal{B}(x-2k\epsilon,\epsilon)=1$,   for $0<\epsilon \le 1/2$.
\item (xii) ~ $\mathcal{B}(an+bm,\epsilon)=\mathcal{B}(n,\epsilon)\mathcal{B}(m,\epsilon)$, ~ for  any integers $a, b, n, m$ such that $|an| \ne |bm|$ when either $n$ or $m$ is non-zero and $0<\epsilon \le 1/2$. $(xiib)$ If $b=0$ one then also has $\mathcal{B}(an,\epsilon)=\mathcal{B}(n,\epsilon)$ for any integer $a \ne 0$.
\item (xiii) ~ $\mathcal{B}(n,\epsilon)+\mathcal{B}(m,\epsilon)-\mathcal{B}(n,\epsilon)\mathcal{B}(m,\epsilon)=\mathcal{B}(nm,\epsilon)$, ~ for  any integers $n, m$ and $0<\epsilon \le 1/2$. 
\item (xiv) ~ $\mathcal{B}(n-a,\epsilon)+\mathcal{B}(n-b,\epsilon)=\mathcal{B}\left((n-a)(n-b),\epsilon \right)$, ~ for  any integers $n, a, b$, $(a \ne b)$ and $0<\epsilon \le 1/2$. 
\item (xv) $\mathcal{B}(x,\frac{1}{2})=1-x$ and $\mathcal{B}(x-1,\frac{1}{2})=x$, when $x$ can be only equal to zero or one.
\item (xvi) $\mathcal{R}_{p}(a) = \sum_{m=0}^{U_{p}} \sum_{k=0}^{p-1} k \mathcal{B}\left(a-k-mp,\frac{1}{2} \right)$ where  $\mathcal{R}_{p}(a)$ is the remainder of the division of a non-negative integer $a$ by $p$ and $U_{p}$ is a suitable upper-bound (positive integer).
\end{itemize}
} 

\noindent \emph{Proof:} Results $(i)$ to $(iv)$ follow directly from Definition S1. Result $(v)$ is a direct consequence of Eq. (\ref{d3}) and from the standard definition of a derivative of a function of one real variable $x$ (with $\epsilon \to 0$ playing the role of an infinitesimal increment). Result $(vi)$ can be proved by induction: For $N=1$ the result holds. If the result is considered valid for $N-1$, one has, for $N$
\begin{eqnarray}
\sum_{k=0}^{N-1}\mathcal{B}(x-2k\epsilon,\epsilon)&=&\mathcal{B}(x-2(N-1)\epsilon,\epsilon)+
\sum_{k=0}^{N-2}\mathcal{B}(x-2k\epsilon,\epsilon) \nonumber \\
&& \nonumber\\
&=&\mathcal{B}(x-2(N-1)\epsilon,\epsilon)+ \mathcal{B}(x-(N-2)\epsilon,(N-1)\epsilon) \nonumber \\
&& \nonumber\\
&=&\frac{1}{2}\left(\frac{x-(2N-3)\epsilon}{|x-(2N-3)\epsilon|}-\frac{x-(2N-1)\epsilon}{|x-(2N-1)\epsilon|}+\frac{x+\epsilon}{|x+\epsilon|}-\frac{x-(2N-3)\epsilon}{|x-(2N-3)\epsilon|}\right)
 \nonumber \\
 && \nonumber\\
&=& \frac{1}{2}\left(\frac{x+\epsilon}{|x+\epsilon|}-\frac{x-(2N-1)\epsilon}{|x-(2N-1)\epsilon|}\right)=  \mathcal{B}(x-(N-1)\epsilon,N\epsilon) \nonumber
\end{eqnarray}
which proves $(vi)$. Results $(vii)$ to $(ix)$ come then from $(vi)$. Result $(x)$ can also be proved using $(vi)$ since 
\begin{eqnarray}
\sum_{k=0}^{\infty}\mathcal{B}(x-2k\epsilon,\epsilon)&=&\lim_{N\to \infty}
\sum_{k=0}^{N-1}\mathcal{B}(x-2k\epsilon,\epsilon)=\lim_{N\to \infty}\mathcal{B}(x-(N-1)\epsilon,N\epsilon)\nonumber \\
&& \nonumber\\
&=&\lim_{N\to \infty}\frac{1}{2}\left(\frac{x+\epsilon}{|x+\epsilon|}-\frac{x-(2N-1)\epsilon}{|x-(2N-1)\epsilon|}\right) = \frac{1}{2}\left(\frac{x+\epsilon}{|x+\epsilon|}+1\right)\nonumber \\
&& \nonumber\\
&=&H(x+\epsilon) \nonumber
\end{eqnarray}
The proof of $(xi)$ proceeds in a similar way. $(xii)$ can be proved by taking into account that both the l.h.s and the r.h.s. are one only when both $n$ and $m$ vanish and zero otherwise (since in the latter case $|an|\ne |bm|$). The proof of $(xiii)$ proceeds by exhaustion: considering the possibilities a) both n and m zero or b) either n or m or both nonzero, the equality is always valid and these are the only possibilities. Result $(xiv)$ is a trivial consequence of $(xiii)$. Result $(xv)$ is directly obtained from $(viii)$ and can be easily checked from the Definition S1 by applying it to both possible values of the variable $x$. 

For a positive integer $p$, two integers $a \ge k \ge 0$ are said to be congruent modulo $p$, if their difference $a - k$ is an integer multiple of $p$, i.e. if there exist another integer $m$ such that $a-k-mp=0$. If such is the case, the $r.h.s.$ of the expression in result $(xvi)$ is equal to one only for one integer $m$ in the interval $[0,a]$ called $quotient$ and for one integer $k$, called $remainder$. Since the two sums are carried over all possible values for these integers, it is clear, by applying $(viii)$  that the r.h.s of $(xvi)$ provides the remainder.  Note that $U_{p}=a$ would be a rather conservative upper bound for the integers $m$ over which the sum is carried and usually a value $U_{p} < a$ can be found, depending of the specific context on which the result is applied. A non-negative integer number $a$ is $divisible$ by $p$ when $\mathcal{R}_{p}(a)=0$. $\Box$ \\

\pagebreak

\section*{Appendix B: Maps for the 256 Wolfram rules.}

\pagebreak

\begin{table}[htdp]
\begin{center}
\begin{tabular}{cclcccl}
Rule & $ \qquad (a_{0},a_{1},a_{2},a_{3},a_{4},a_{5},a_{6},a_{7})$ & \qquad Map \\
$^{1}0_{2}^{1}$ & \qquad (0,0,0,0,0,0,0,0) & $\qquad x_{t+1}^{i}=0$ \\
$^{1}1_{2}^{1}$ & \qquad (1,0,0,0,0,0,0,0) & $\qquad x_{t+1}^{i}=1-x_{t}^{i}-x_{t}^{i-1}-x_{t}^{i+1}
+x_{t}^{i}x_{t}^{i-1}+x_{t}^{i}x_{t}^{i+1}+$\\
&& $\quad \qquad \qquad +x_{t}^{i+1}x_{t}^{i-1}-x_{t}^{i-1}x_{t}^{i}x_{t}^{i+1}$ \\
$^{1}2_{2}^{1}$ & \qquad (0,1,0,0,0,0,0,0) & $\qquad x_{t+1}^{i}=x_{t}^{i-1} - x_{t}^{i+1} x_{t}^{i-1} - x_{t}^{i} x_{t}^{i-1} + x_{t}^{i+1} x_{t}^{i} x_{t}^{i-1}$ \\
$^{1}3_{2}^{1}$ & \qquad (1,1,0,0,0,0,0,0) & $\qquad x_{t+1}^{i}=1 - x_{t}^{i+1} - x_{t}^{i} + x_{t}^{i+1} x_{t}^{i}$ \\
$^{1}4_{2}^{1}$ & \qquad (0,0,1,0,0,0,0,0) & $\qquad x_{t+1}^{i}=x_{t}^{i} - x_{t}^{i+1} x_{t}^{i} - x_{t}^{i} x_{t}^{i-1} + x_{t}^{i+1} x_{t}^{i} x_{t}^{i-1}$ \\
$^{1}5_{2}^{1}$ & \qquad (1,0,1,0,0,0,0,0) & $\qquad x_{t+1}^{i}=1 - x_{t}^{i+1} - x_{t}^{i-1} + x_{t}^{i+1} x_{t}^{i-1}$ \\
$^{1}6_{2}^{1}$ & \qquad (0,1,1,0,0,0,0,0) & $\qquad x_{t+1}^{i}=x_{t}^{i} - x_{t}^{i+1} x_{t}^{i} + x_{t}^{i-1} - x_{t}^{i+1} x_{t}^{i-1} - 2 x_{t}^{i} x_{t}^{i-1} + 2 x_{t}^{i+1} x_{t}^{i} x_{t}^{i-1}$ \\
$^{1}7_{2}^{1}$ & \qquad (1,1,1,0,0,0,0,0) & $\qquad x_{t+1}^{i}=1 - x_{t}^{i+1} - x_{t}^{i} x_{t}^{i-1} + x_{t}^{i+1} x_{t}^{i} x_{t}^{i-1}$ \\
$^{1}8_{2}^{1}$ & \qquad (0,0,0,1,0,0,0,0) & $\qquad x_{t+1}^{i}=x_{t}^{i} x_{t}^{i-1} - x_{t}^{i+1} x_{t}^{i} x_{t}^{i-1}$ \\
$^{1}9_{2}^{1}$ & \qquad (1,0,0,1,0,0,0,0) & $\qquad x_{t+1}^{i}=1 - x_{t}^{i+1} - x_{t}^{i} + x_{t}^{i+1} x_{t}^{i} - x_{t}^{i-1} + x_{t}^{i+1} x_{t}^{i-1} + $\\ && $\quad \qquad \qquad +2 x_{t}^{i} x_{t}^{i-1} - 2 x_{t}^{i+1} x_{t}^{i} x_{t}^{i-1}$ \\
$^{1}10_{2}^{1}$ & \qquad (0,1,0,1,0,0,0,0) & $\qquad x_{t+1}^{i}=x_{t}^{i-1} - x_{t}^{i+1} x_{t}^{i-1}$ \\
$^{1}11_{2}^{1}$ & \qquad (1,1,0,1,0,0,0,0) & $\qquad x_{t+1}^{i}=1 - x_{t}^{i+1}- x_{t}^{i} + x_{t}^{i+1}x_{t}^{i} + x_{t}^{i} x_{t}^{i-1} - x_{t}^{i+1}x_{t}^{i} x_{t}^{i-1}$ \\
$^{1}12_{2}^{1}$ & \qquad (0,0,1,1,0,0,0,0) & $\qquad x_{t+1}^{i}=x_{t}^{i} - x_{t}^{i+1}x_{t}^{i}$\\
$^{1}13_{2}^{1}$ & \qquad (1,0,1,1,0,0,0,0) & $\qquad x_{t+1}^{i}=1 - x_{t}^{i+1}- x_{t}^{i-1} + x_{t}^{i+1}x_{t}^{i-1} + x_{t}^{i} x_{t}^{i-1} - x_{t}^{i+1}x_{t}^{i} x_{t}^{i-1}$ \\
$^{1}14_{2}^{1}$ & \qquad (0,1,1,1,0,0,0,0) & $\qquad x_{t+1}^{i}=x_{t}^{i} - x_{t}^{i+1}x_{t}^{i} + x_{t}^{i-1} - x_{t}^{i+1}x_{t}^{i-1} - x_{t}^{i} x_{t}^{i-1} + x_{t}^{i+1}x_{t}^{i} x_{t}^{i-1}$ \\
$^{1}15_{2}^{1}$ & \qquad (1,1,1,1,0,0,0,0) & $\qquad x_{t+1}^{i}=1 - x_{t}^{i+1}$ \\
$^{1}16_{2}^{1}$ & \qquad (0,0,0,0,1,0,0,0) & $\qquad x_{t+1}^{i}=x_{t}^{i+1}- x_{t}^{i+1}x_{t}^{i} - x_{t}^{i+1}x_{t}^{i-1} + x_{t}^{i+1}x_{t}^{i} x_{t}^{i-1}$ \\
$^{1}17_{2}^{1}$ & \qquad (1,0,0,0,1,0,0,0) & $\qquad x_{t+1}^{i}=1 - x_{t}^{i} - x_{t}^{i-1} + x_{t}^{i} x_{t}^{i-1}$ \\
$^{1}18_{2}^{1}$ & \qquad (0,1,0,0,1,0,0,0) & $\qquad x_{t+1}^{i}=x_{t}^{i+1}- x_{t}^{i+1}x_{t}^{i} + x_{t}^{i-1} - 2 x_{t}^{i+1}x_{t}^{i-1} - x_{t}^{i} x_{t}^{i-1} + 2 x_{t}^{i+1}x_{t}^{i} x_{t}^{i-1}$ \\
$^{1}19_{2}^{1}$ & \qquad (1,1,0,0,1,0,0,0) & $\qquad x_{t+1}^{i}=1 - x_{t}^{i} - x_{t}^{i+1}x_{t}^{i-1} + x_{t}^{i+1}x_{t}^{i} x_{t}^{i-1}$ \\
$^{1}20_{2}^{1}$ & \qquad (0,0,1,0,1,0,0,0) & $\qquad x_{t+1}^{i}=x_{t}^{i+1}+ x_{t}^{i} - 2 x_{t}^{i+1}x_{t}^{i} - x_{t}^{i+1}x_{t}^{i-1} - x_{t}^{i} x_{t}^{i-1} + 2 x_{t}^{i+1}x_{t}^{i} x_{t}^{i-1}$ \\
$^{1}21_{2}^{1}$ & \qquad (1,0,1,0,1,0,0,0) & $\qquad x_{t+1}^{i}=1 - x_{t}^{i+1}x_{t}^{i} - x_{t}^{i-1} + x_{t}^{i+1}x_{t}^{i} x_{t}^{i-1}$ \\
\end{tabular}
\end{center}
Table S6: The cellular automata maps for the 256 Wolfram boolean rules $^{1}R_{2}^{1}$ as obtained from Eq. (\ref{CA}) 
\end{table}

\begin{table}[htdp]
\begin{center}
\begin{tabular}{cclcccl}
Rule & $ \qquad (a_{0},a_{1},a_{2},a_{3},a_{4},a_{5},a_{6},a_{7})$ & \qquad Map \\
$^{1}22_{2}^{1}$ & \qquad (0,1,1,0,1,0,0,0) & $\qquad x_{t+1}^{i}=x_{t}^{i+1}+ x_{t}^{i} - 2 x_{t}^{i+1}x_{t}^{i} + x_{t}^{i-1} - 2 x_{t}^{i+1}x_{t}^{i-1} - 2 x_{t}^{i} x_{t}^{i-1}+$ \\  
&& $\quad \qquad \qquad + 3 x_{t}^{i+1}x_{t}^{i} x_{t}^{i-1}$ \\
$^{1}23_{2}^{1}$ & \qquad (1,1,1,0,1,0,0,0) & $\qquad x_{t+1}^{i}=1 - x_{t}^{i+1}x_{t}^{i} - x_{t}^{i+1}x_{t}^{i-1} - x_{t}^{i} x_{t}^{i-1} + 2 x_{t}^{i+1}x_{t}^{i} x_{t}^{i-1}$ \\
$^{1}24_{2}^{1}$ & \qquad (0,0,0,1,1,0,0,0) & $\qquad x_{t+1}^{i}=x_{t}^{i+1}- x_{t}^{i+1}x_{t}^{i} - x_{t}^{i+1}x_{t}^{i-1} + x_{t}^{i} x_{t}^{i-1}$ \\
$^{1}25_{2}^{1}$ & \qquad (1,0,0,1,1,0,0,0) & $\qquad x_{t+1}^{i}=1 - x_{t}^{i} - x_{t}^{i-1} + 2 x_{t}^{i} x_{t}^{i-1} - x_{t}^{i+1}x_{t}^{i} x_{t}^{i-1}$ \\
$^{1}26_{2}^{1}$ & \qquad (0,1,0,1,1,0,0,0) & $\qquad x_{t+1}^{i}=x_{t}^{i+1}- x_{t}^{i+1}x_{t}^{i} + x_{t}^{i-1} - 2 x_{t}^{i+1}x_{t}^{i-1} + x_{t}^{i+1}x_{t}^{i} x_{t}^{i-1}$ \\
$^{1}27_{2}^{1}$ & \qquad (1,1,0,1,1,0,0,0) & $\qquad x_{t+1}^{i}=1 - x_{t}^{i} - x_{t}^{i+1}x_{t}^{i-1} + x_{t}^{i} x_{t}^{i-1}$ \\
$^{1}28_{2}^{1}$ & \qquad (0,0,1,1,1,0,0,0) & $\qquad x_{t+1}^{i}=x_{t}^{i+1}+ x_{t}^{i} - 2 x_{t}^{i+1}x_{t}^{i} - x_{t}^{i+1}x_{t}^{i-1} + x_{t}^{i+1}x_{t}^{i} x_{t}^{i-1}
$ \\
$^{1}29_{2}^{1}$ & \qquad (1,0,1,1,1,0,0,0) & $\qquad x_{t+1}^{i}=1 - x_{t}^{i+1}x_{t}^{i} - x_{t}^{i-1} + x_{t}^{i} x_{t}^{i-1}$ \\
$^{1}30_{2}^{1}$ & \qquad (0,1,1,1,1,0,0,0) & $\qquad x_{t+1}^{i}=x_{t}^{i+1}+ x_{t}^{i} - 2 x_{t}^{i+1}x_{t}^{i} + x_{t}^{i-1} - 2 x_{t}^{i+1}x_{t}^{i-1} - x_{t}^{i} x_{t}^{i-1}+$ \\
&& $\quad \qquad \qquad  + 2 x_{t}^{i+1}x_{t}^{i} x_{t}^{i-1}$ \\
$^{1}31_{2}^{1}$ & \qquad (1,1,1,1,1,0,0,0) & $\qquad x_{t+1}^{i}=1 - x_{t}^{i+1}x_{t}^{i} - x_{t}^{i+1}x_{t}^{i-1} + x_{t}^{i+1}x_{t}^{i} x_{t}^{i-1}$ \\
$^{1}32_{2}^{1}$ & \qquad (0,0,0,0,0,1,0,0) & $\qquad x_{t+1}^{i}=x_{t}^{i+1}x_{t}^{i-1} - x_{t}^{i+1}x_{t}^{i} x_{t}^{i-1}$\\
$^{1}33_{2}^{1}$ & \qquad (1,0,0,0,0,1,0,0) & $\qquad x_{t+1}^{i}=1 - x_{t}^{i+1}- x_{t}^{i} + x_{t}^{i+1}x_{t}^{i} - x_{t}^{i-1} + 2 x_{t}^{i+1}x_{t}^{i-1} + x_{t}^{i} x_{t}^{i-1}-$\\
&& $\quad \qquad \qquad  - 2 x_{t}^{i+1}x_{t}^{i} x_{t}^{i-1}$ \\
$^{1}34_{2}^{1}$ & \qquad (0,1,0,0,0,1,0,0) & $\qquad x_{t+1}^{i}=x_{t}^{i-1} - x_{t}^{i} x_{t}^{i-1}$ \\
$^{1}35_{2}^{1}$ & \qquad (1,1,0,0,0,1,0,0) & $\qquad x_{t+1}^{i}=1 - x_{t}^{i+1}- x_{t}^{i} + x_{t}^{i+1}x_{t}^{i} + x_{t}^{i+1}x_{t}^{i-1} - x_{t}^{i+1}x_{t}^{i} x_{t}^{i-1}$ \\
$^{1}36_{2}^{1}$ & \qquad (0,0,1,0,0,1,0,0) & $\qquad x_{t+1}^{i}=x_{t}^{i} - x_{t}^{i+1}x_{t}^{i} + x_{t}^{i+1}x_{t}^{i-1} - x_{t}^{i} x_{t}^{i-1}$ \\
$^{1}37_{2}^{1}$ & \qquad (1,0,1,0,0,1,0,0) & $\qquad x_{t+1}^{i}=1 - x_{t}^{i+1}- x_{t}^{i-1} + 2 x_{t}^{i+1}x_{t}^{i-1} - x_{t}^{i+1}x_{t}^{i} x_{t}^{i-1}$ \\
$^{1}38_{2}^{1}$ & \qquad (0,1,1,0,0,1,0,0) & $\qquad x_{t+1}^{i}=x_{t}^{i} - x_{t}^{i+1}x_{t}^{i} + x_{t}^{i-1} - 2 x_{t}^{i} x_{t}^{i-1} + x_{t}^{i+1}x_{t}^{i} x_{t}^{i-1}$ \\
$^{1}39_{2}^{1}$ & \qquad (1,1,1,0,0,1,0,0) & $\qquad x_{t+1}^{i}=1 - x_{t}^{i+1}+ x_{t}^{i+1}x_{t}^{i-1} - x_{t}^{i} x_{t}^{i-1}$ \\
$^{1}40_{2}^{1}$ & \qquad (0,0,0,1,0,1,0,0) & $\qquad x_{t+1}^{i}=x_{t}^{i+1}x_{t}^{i-1} + x_{t}^{i} x_{t}^{i-1} - 2 x_{t}^{i+1}x_{t}^{i} x_{t}^{i-1}$ \\
$^{1}41_{2}^{1}$ & \qquad (1,0,0,1,0,1,0,0)  & $\qquad x_{t+1}^{i}=1 - x_{t}^{i+1} - x_{t}^{i} + x_{t}^{i+1} x_{t}^{i} - x_{t}^{i-1} + 2 x_{t}^{i+1} x_{t}^{i-1} + 2 x_{t}^{i} x_{t}^{i-1} - $\\
&& $\quad \qquad \qquad -3 x_{t}^{i+1} x_{t}^{i} x_{t}^{i-1}$ \\
$^{1}42_{2}^{1}$ & \qquad (0,1,0,1,0,1,0,0)  & $\qquad x_{t+1}^{i}=x_{t}^{i-1} - x_{t}^{i+1} x_{t}^{i} x_{t}^{i-1}$ \\
\end{tabular}
\end{center}
Table S6 (cont.)
\end{table}
\begin{table}[htdp]
\begin{center}
\begin{tabular}{cclcccl}
Rule & $ \qquad (a_{0},a_{1},a_{2},a_{3},a_{4},a_{5},a_{6},a_{7})$ & \qquad Map \\
$^{1}43_{2}^{1}$ & \qquad (1,1,0,1,0,1,0,0)  & $\qquad x_{t+1}^{i}=1 - x_{t}^{i+1} - x_{t}^{i} + x_{t}^{i+1} x_{t}^{i} + x_{t}^{i+1} x_{t}^{i-1} + x_{t}^{i} x_{t}^{i-1} -$\\
&& $\quad \qquad \qquad - 2 x_{t}^{i+1} x_{t}^{i} x_{t}^{i-1}$ \\
$^{1}44_{2}^{1}$ & \qquad (0,0,1,1,0,1,0,0)  & $\qquad x_{t+1}^{i}=x_{t}^{i} - x_{t}^{i+1} x_{t}^{i} + x_{t}^{i+1} x_{t}^{i-1} - x_{t}^{i+1} x_{t}^{i} x_{t}^{i-1}$ \\
$^{1}45_{2}^{1}$ & \qquad (1,0,1,1,0,1,0,0)  & $\qquad x_{t+1}^{i}=1 - x_{t}^{i+1} - x_{t}^{i-1} + 2 x_{t}^{i+1} x_{t}^{i-1} + x_{t}^{i} x_{t}^{i-1} - 2 x_{t}^{i+1} x_{t}^{i} x_{t}^{i-1}$ \\
$^{1}46_{2}^{1}$ & \qquad (0,1,1,1,0,1,0,0)  & $\qquad x_{t+1}^{i}=x_{t}^{i} - x_{t}^{i+1} x_{t}^{i} + x_{t}^{i-1} - x_{t}^{i} x_{t}^{i-1}$ \\
$^{1}47_{2}^{1}$ & \qquad (1,1,1,1,0,1,0,0)  & $\qquad x_{t+1}^{i}=1 - x_{t}^{i+1} + x_{t}^{i+1} x_{t}^{i-1} - x_{t}^{i+1} x_{t}^{i} x_{t}^{i-1}$ \\
$^{1}48_{2}^{1}$ & \qquad (0,0,0,0,1,1,0,0)  & $\qquad x_{t+1}^{i}=x_{t}^{i+1} - x_{t}^{i+1} x_{t}^{i}$ \\
$^{1}49_{2}^{1}$ & \qquad (1,0,0,0,1,1,0,0) & $\qquad x_{t+1}^{i}=1 - x_{t}^{i} - x_{t}^{i-1} + x_{t}^{i+1} x_{t}^{i-1} + x_{t}^{i} x_{t}^{i-1} - x_{t}^{i+1} x_{t}^{i} x_{t}^{i-1}$ \\
$^{1}50_{2}^{1}$ & \qquad (0,1,0,0,1,1,0,0) & $\qquad x_{t+1}^{i}=x_{t}^{i+1} - x_{t}^{i+1} x_{t}^{i} + x_{t}^{i-1} - x_{t}^{i+1} x_{t}^{i-1} - x_{t}^{i} x_{t}^{i-1} + x_{t}^{i+1} x_{t}^{i} x_{t}^{i-1}$ \\
$^{1}51_{2}^{1}$ & \qquad (1,1,0,0,1,1,0,0) & $\qquad x_{t+1}^{i}=1 - x_{t}^{i}$ \\
$^{1}52_{2}^{1}$ & \qquad (0,0,1,0,1,1,0,0) & $\qquad x_{t+1}^{i}=x_{t}^{i+1} + x_{t}^{i} - 2 x_{t}^{i+1} x_{t}^{i} - x_{t}^{i} x_{t}^{i-1} + x_{t}^{i+1} x_{t}^{i} x_{t}^{i-1}$\\
$^{1}53_{2}^{1}$ & \qquad (1,0,1,0,1,1,0,0) & $\qquad x_{t+1}^{i}=1 - x_{t}^{i+1} x_{t}^{i} - x_{t}^{i-1} + x_{t}^{i+1} x_{t}^{i-1}$ \\
$^{1}54_{2}^{1}$ & \qquad (0,1,1,0,1,1,0,0) & $\qquad x_{t+1}^{i}=x_{t}^{i+1} + x_{t}^{i} - 2 x_{t}^{i+1} x_{t}^{i} + x_{t}^{i-1} - x_{t}^{i+1} x_{t}^{i-1} - 2 x_{t}^{i} x_{t}^{i-1} +$\\
&& $\quad \qquad \qquad + 2 x_{t}^{i+1} x_{t}^{i} x_{t}^{i-1}$ \\
$^{1}55_{2}^{1}$ & \qquad (1,1,1,0,1,1,0,0) & $\qquad x_{t+1}^{i}=1 - x_{t}^{i+1} x_{t}^{i} - x_{t}^{i} x_{t}^{i-1} + x_{t}^{i+1} x_{t}^{i} x_{t}^{i-1}$ \\
$^{1}56_{2}^{1}$ & \qquad (0,0,0,1,1,1,0,0) & $\qquad x_{t+1}^{i}=x_{t}^{i+1} - x_{t}^{i+1} x_{t}^{i} + x_{t}^{i} x_{t}^{i-1} - x_{t}^{i+1} x_{t}^{i} x_{t}^{i-1}$ \\
$^{1}57_{2}^{1}$ & \qquad (1,0,0,1,1,1,0,0) & $\qquad x_{t+1}^{i}=1 - x_{t}^{i} - x_{t}^{i-1} + x_{t}^{i+1} x_{t}^{i-1} + 2 x_{t}^{i} x_{t}^{i-1} - 2 x_{t}^{i+1} x_{t}^{i} x_{t}^{i-1}$ \\
$^{1}58_{2}^{1}$ & \qquad (0,1,0,1,1,1,0,0) & $\qquad x_{t+1}^{i}=x_{t}^{i+1} - x_{t}^{i+1} x_{t}^{i} + x_{t}^{i-1} - x_{t}^{i+1} x_{t}^{i-1}$ \\
$^{1}59_{2}^{1}$ & \qquad (1,1,0,1,1,1,0,0) & $\qquad x_{t+1}^{i}=1 - x_{t}^{i} + x_{t}^{i} x_{t}^{i-1} - x_{t}^{i+1} x_{t}^{i} x_{t}^{i-1}$ \\
$^{1}60_{2}^{1}$ & \qquad (0,0,1,1,1,1,0,0) & $\qquad x_{t+1}^{i}=x_{t}^{i+1} + x_{t}^{i} - 2 x_{t}^{i+1} x_{t}^{i}$ \\
$^{1}61_{2}^{1}$ & \qquad (1,0,1,1,1,1,0,0) & $\qquad x_{t+1}^{i}=1 - x_{t}^{i+1} x_{t}^{i} - x_{t}^{i-1} + x_{t}^{i+1} x_{t}^{i-1} + x_{t}^{i} x_{t}^{i-1} - x_{t}^{i+1} x_{t}^{i} x_{t}^{i-1}$ \\
$^{1}62_{2}^{1}$ & \qquad (0,1,1,1,1,1,0,0) & $\qquad x_{t+1}^{i}=x_{t}^{i+1} + x_{t}^{i} - 2 x_{t}^{i+1} x_{t}^{i} + x_{t}^{i-1} - x_{t}^{i+1} x_{t}^{i-1} - x_{t}^{i} x_{t}^{i-1} +$\\
&& $\quad \qquad \qquad + x_{t}^{i+1} x_{t}^{i} x_{t}^{i-1}$ \\
$^{1}63_{2}^{1}$ & \qquad (1,1,1,1,1,1,0,0) & $\qquad x_{t+1}^{i}=1 - x_{t}^{i+1} x_{t}^{i}$ \\
$^{1}64_{2}^{1}$ & \qquad (0,0,0,0,0,0,1,0) & $\qquad x_{t+1}^{i}=x_{t}^{i+1} x_{t}^{i} - x_{t}^{i+1} x_{t}^{i} x_{t}^{i-1}$ \\
\end{tabular}
\end{center}
Table S6 (cont.)
\end{table}
\begin{table}[htdp]
\begin{center}
\begin{tabular}{cclcccl}
Rule & $ \qquad (a_{0},a_{1},a_{2},a_{3},a_{4},a_{5},a_{6},a_{7})$ & \qquad Map \\
$^{1}65_{2}^{1}$ & \qquad (1,0,0,0,0,0,1,0) & $\qquad x_{t+1}^{i}=1 - x_{t}^{i+1} - x_{t}^{i} + 2 x_{t}^{i+1} x_{t}^{i} - x_{t}^{i-1} + x_{t}^{i+1} x_{t}^{i-1} + x_{t}^{i} x_{t}^{i-1} - $\\
&& $\quad \qquad \qquad -2 x_{t}^{i+1} x_{t}^{i} x_{t}^{i-1}$ \\
$^{1}66_{2}^{1}$ & \qquad (0,1,0,0,0,0,1,0) & $\qquad x_{t+1}^{i}=x_{t}^{i+1} x_{t}^{i} + x_{t}^{i-1} - x_{t}^{i+1} x_{t}^{i-1} - x_{t}^{i} x_{t}^{i-1}$ \\
$^{1}67_{2}^{1}$ & \qquad (1,1,0,0,0,0,1,0) & $\qquad x_{t+1}^{i}=1 - x_{t}^{i+1} - x_{t}^{i} + 2 x_{t}^{i+1} x_{t}^{i} - x_{t}^{i+1} x_{t}^{i} x_{t}^{i-1}$ \\
$^{1}68_{2}^{1}$ & \qquad (0,0,1,0,0,0,1,0) & $\qquad x_{t+1}^{i}=x_{t}^{i} - x_{t}^{i} x_{t}^{i-1}$ \\
$^{1}69_{2}^{1}$ & \qquad (1,0,1,0,0,0,1,0) & $\qquad x_{t+1}^{i}=1 - x_{t}^{i+1} + x_{t}^{i+1} x_{t}^{i} - x_{t}^{i-1} + x_{t}^{i+1} x_{t}^{i-1} - x_{t}^{i+1} x_{t}^{i} x_{t}^{i-1}$ \\
$^{1}70_{2}^{1}$ & \qquad (0,1,1,0,0,0,1,0) & $\qquad x_{t+1}^{i}=x_{t}^{i} + x_{t}^{i-1} - x_{t}^{i+1} x_{t}^{i-1} - 2 x_{t}^{i} x_{t}^{i-1} + x_{t}^{i+1} x_{t}^{i} x_{t}^{i-1}$ \\
$^{1}71_{2}^{1}$ & \qquad (1,1,1,0,0,0,1,0) & $\qquad x_{t+1}^{i}=1 - x_{t}^{i+1} + x_{t}^{i+1} x_{t}^{i} - x_{t}^{i} x_{t}^{i-1}$ \\
$^{1}72_{2}^{1}$ & \qquad (0,0,0,1,0,0,1,0) & $\qquad x_{t+1}^{i}=x_{t}^{i+1} x_{t}^{i} + x_{t}^{i} x_{t}^{i-1} - 2 x_{t}^{i+1} x_{t}^{i} x_{t}^{i-1}$ \\
$^{1}73_{2}^{1}$ & \qquad (1,0,0,1,0,0,1,0) & $\qquad x_{t+1}^{i}=1 - x_{t}^{i+1} - x_{t}^{i} + 2 x_{t}^{i+1} x_{t}^{i} - x_{t}^{i-1} + x_{t}^{i+1} x_{t}^{i-1} + 2 x_{t}^{i} x_{t}^{i-1} - $\\
&& $\quad \qquad \qquad - 3 x_{t}^{i+1} x_{t}^{i} x_{t}^{i-1}$ \\
$^{1}74_{2}^{1}$ & \qquad (0,1,0,1,0,0,1,0) & $\qquad x_{t+1}^{i}=x_{t}^{i+1} x_{t}^{i} + x_{t}^{i-1} - x_{t}^{i+1} x_{t}^{i-1} - x_{t}^{i+1} x_{t}^{i} x_{t}^{i-1}$ \\
$^{1}75_{2}^{1}$ & \qquad (1,1,0,1,0,0,1,0) & $\qquad x_{t+1}^{i}=1 - x_{t}^{i+1} - x_{t}^{i} + 2 x_{t}^{i+1} x_{t}^{i} + x_{t}^{i} x_{t}^{i-1} - $\\
&& $\quad \qquad \qquad -2 x_{t}^{i+1} x_{t}^{i} x_{t}^{i-1}$ \\
$^{1}76_{2}^{1}$ & \qquad (0,0,1,1,0,0,1,0) & $\qquad x_{t+1}^{i}=x_{t}^{i} - x_{t}^{i+1} x_{t}^{i} x_{t}^{i-1}$\\
$^{1}77_{2}^{1}$ & \qquad (1,0,1,1,0,0,1,0) & $\qquad x_{t+1}^{i}=1 - x_{t}^{i+1} + x_{t}^{i+1} x_{t}^{i} - x_{t}^{i-1} + x_{t}^{i+1} x_{t}^{i-1} + x_{t}^{i} x_{t}^{i-1} -$\\
&& $\quad \qquad \qquad - 2 x_{t}^{i+1} x_{t}^{i} x_{t}^{i-1}$ \\
$^{1}78_{2}^{1}$ & \qquad (0,1,1,1,0,0,1,0) & $\qquad x_{t+1}^{i}=x_{t}^{i} + x_{t}^{i-1} - x_{t}^{i+1} x_{t}^{i-1} - x_{t}^{i} x_{t}^{i-1}$ \\
$^{1}79_{2}^{1}$ & \qquad (1,1,1,1,0,0,1,0) & $\qquad x_{t+1}^{i}=1 - x_{t}^{i+1} + x_{t}^{i+1} x_{t}^{i} - x_{t}^{i+1} x_{t}^{i} x_{t}^{i-1}$ \\
$^{1}80_{2}^{1}$ & \qquad (0,0,0,0,1,0,1,0) & $\qquad x_{t+1}^{i}=x_{t}^{i+1} - x_{t}^{i+1} x_{t}^{i-1}$ \\
$^{1}81_{2}^{1}$ & \qquad (1,0,0,0,1,0,1,0) & $\qquad x_{t+1}^{i}=1 - x_{t}^{i} + x_{t}^{i+1} x_{t}^{i} - x_{t}^{i-1} + x_{t}^{i} x_{t}^{i-1} - x_{t}^{i+1} x_{t}^{i} x_{t}^{i-1}$ \\
$^{1}82_{2}^{1}$ & \qquad (0,1,0,0,1,0,1,0) & $\qquad x_{t+1}^{i}=x_{t}^{i+1} + x_{t}^{i-1} - 2 x_{t}^{i+1} x_{t}^{i-1} - x_{t}^{i} x_{t}^{i-1} + x_{t}^{i+1} x_{t}^{i} x_{t}^{i-1}$ \\
$^{1}83_{2}^{1}$ & \qquad (1,1,0,0,1,0,1,0) & $\qquad x_{t+1}^{i}=1 - x_{t}^{i} + x_{t}^{i+1} x_{t}^{i} - x_{t}^{i+1} x_{t}^{i-1}$ \\
$^{1}84_{2}^{1}$ & \qquad (0,0,1,0,1,0,1,0) & $\qquad x_{t+1}^{i}=x_{t}^{i+1} + x_{t}^{i} - x_{t}^{i+1} x_{t}^{i} - x_{t}^{i+1} x_{t}^{i-1} - x_{t}^{i} x_{t}^{i-1} + x_{t}^{i+1} x_{t}^{i} x_{t}^{i-1}$ \\
$^{1}85_{2}^{1}$ & \qquad (1,0,1,0,1,0,1,0) & $\qquad x_{t+1}^{i}=1 - x_{t}^{i-1}$ \\
\end{tabular}
\end{center}
Table S6 (cont.)
\end{table}

\begin{table}[htdp]
\begin{center}
\begin{tabular}{cclcccl}
Rule & $ \qquad (a_{0},a_{1},a_{2},a_{3},a_{4},a_{5},a_{6},a_{7})$ & \qquad Map \\
$^{1}86_{2}^{1}$ & \qquad (0,1,1,0,1,0,1,0) & $\qquad x_{t+1}^{i}=x_{t}^{i+1} + x_{t}^{i} - x_{t}^{i+1} x_{t}^{i} + x_{t}^{i-1} - 2 x_{t}^{i+1} x_{t}^{i-1} - 2 x_{t}^{i} x_{t}^{i-1} +$\\
&& $\quad \qquad \qquad + 2 x_{t}^{i+1} x_{t}^{i} x_{t}^{i-1}$ \\
$^{1}87_{2}^{1}$ & \qquad (1,1,1,0,1,0,1,0) & $\qquad x_{t+1}^{i}=1 - x_{t}^{i+1} x_{t}^{i-1} - x_{t}^{i} x_{t}^{i-1} + x_{t}^{i+1} x_{t}^{i} x_{t}^{i-1}$ \\
$^{1}88_{2}^{1}$ & \qquad (0,0,0,1,1,0,1,0) & $\qquad x_{t+1}^{i}=x_{t}^{i+1} - x_{t}^{i+1} x_{t}^{i-1} + x_{t}^{i} x_{t}^{i-1} - x_{t}^{i+1} x_{t}^{i} x_{t}^{i-1}$ \\
$^{1}89_{2}^{1}$ & \qquad (1,0,0,1,1,0,1,0) & $\qquad x_{t+1}^{i}=1 - x_{t}^{i} + x_{t}^{i+1} x_{t}^{i} - x_{t}^{i-1} + 2 x_{t}^{i} x_{t}^{i-1} - 2 x_{t}^{i+1} x_{t}^{i} x_{t}^{i-1}$ \\
$^{1}90_{2}^{1}$ & \qquad (0,1,0,1,1,0,1,0) & $\qquad x_{t+1}^{i}=x_{t}^{i+1} + x_{t}^{i-1} - 2 x_{t}^{i+1} x_{t}^{i-1}$ \\
$^{1}91_{2}^{1}$ & \qquad (1,1,0,1,1,0,1,0) & $\qquad x_{t+1}^{i}=1 - x_{t}^{i} + x_{t}^{i+1} x_{t}^{i} - x_{t}^{i+1} x_{t}^{i-1} + x_{t}^{i} x_{t}^{i-1} - x_{t}^{i+1} x_{t}^{i} x_{t}^{i-1}$ \\
$^{1}92_{2}^{1}$ & \qquad (0,0,1,1,1,0,1,0) & $\qquad x_{t+1}^{i}=x_{t}^{i+1} + x_{t}^{i} - x_{t}^{i+1} x_{t}^{i} - x_{t}^{i+1} x_{t}^{i-1}
$ \\
$^{1}93_{2}^{1}$ & \qquad (1,0,1,1,1,0,1,0) & $\qquad x_{t+1}^{i}=1 - x_{t}^{i-1} + x_{t}^{i} x_{t}^{i-1} - x_{t}^{i+1} x_{t}^{i} x_{t}^{i-1}$ \\
$^{1}94_{2}^{1}$ & \qquad (0,1,1,1,1,0,1,0) & $\qquad x_{t+1}^{i}=x_{t}^{i+1} + x_{t}^{i} - x_{t}^{i+1} x_{t}^{i} + x_{t}^{i-1} - 2 x_{t}^{i+1} x_{t}^{i-1} - x_{t}^{i} x_{t}^{i-1} + $\\
&& $\quad \qquad \qquad +x_{t}^{i+1} x_{t}^{i} x_{t}^{i-1}$ \\
$^{1}95_{2}^{1}$ & \qquad (1,1,1,1,1,0,1,0) & $\qquad x_{t+1}^{i}=1 - x_{t}^{i+1} x_{t}^{i-1}$ \\
$^{1}96_{2}^{1}$ & \qquad (0,0,0,0,0,1,1,0) & $\qquad x_{t+1}^{i}=x_{t}^{i+1} x_{t}^{i} + x_{t}^{i+1} x_{t}^{i-1} - 2 x_{t}^{i+1} x_{t}^{i} x_{t}^{i-1}$\\
$^{1}97_{2}^{1}$ & \qquad (1,0,0,0,0,1,1,0) & $\qquad x_{t+1}^{i}=1 - x_{t}^{i+1} - x_{t}^{i} + 2 x_{t}^{i+1} x_{t}^{i} - x_{t}^{i-1} + 2 x_{t}^{i+1} x_{t}^{i-1} + x_{t}^{i} x_{t}^{i-1} -$\\
&& $\quad \qquad \qquad - 3 x_{t}^{i+1} x_{t}^{i} x_{t}^{i-1}$ \\
$^{1}98_{2}^{1}$ & \qquad (0,1,0,0,0,1,1,0) & $\qquad x_{t+1}^{i}=x_{t}^{i+1} x_{t}^{i} + x_{t}^{i-1} - x_{t}^{i} x_{t}^{i-1} - x_{t}^{i+1} x_{t}^{i} x_{t}^{i-1}$ \\
$^{1}99_{2}^{1}$ & \qquad (1,1,0,0,0,1,1,0) & $\qquad x_{t+1}^{i}=1 - x_{t}^{i+1} - x_{t}^{i} + 2 x_{t}^{i+1} x_{t}^{i} + x_{t}^{i+1} x_{t}^{i-1} - 2 x_{t}^{i+1} x_{t}^{i} x_{t}^{i-1}$ \\
$^{1}100_{2}^{1}$ & \qquad (0,0,1,0,0,1,1,0) & $\qquad x_{t+1}^{i}=x_{t}^{i} + x_{t}^{i+1} x_{t}^{i-1} - x_{t}^{i} x_{t}^{i-1} - x_{t}^{i+1} x_{t}^{i} x_{t}^{i-1}$ \\
$^{1}101_{2}^{1}$ & \qquad (1,0,1,0,0,1,1,0) & $\qquad x_{t+1}^{i}=1 - x_{t}^{i+1} + x_{t}^{i+1} x_{t}^{i} - x_{t}^{i-1} + 2 x_{t}^{i+1} x_{t}^{i-1} - 2 x_{t}^{i+1} x_{t}^{i} x_{t}^{i-1}$ \\
$^{1}102_{2}^{1}$ & \qquad (0,1,1,0,0,1,1,0) & $\qquad x_{t+1}^{i}=x_{t}^{i} + x_{t}^{i-1} - 2 x_{t}^{i} x_{t}^{i-1}$ \\
$^{1}103_{2}^{1}$ & \qquad (1,1,1,0,0,1,1,0) & $\qquad x_{t+1}^{i}=1 - x_{t}^{i+1} + x_{t}^{i+1} x_{t}^{i} + x_{t}^{i+1} x_{t}^{i-1} - x_{t}^{i} x_{t}^{i-1} - x_{t}^{i+1} x_{t}^{i} x_{t}^{i-1}$ \\
$^{1}104_{2}^{1}$ & \qquad (0,0,0,1,0,1,1,0) & $\qquad x_{t+1}^{i}=x_{t}^{i+1} x_{t}^{i} + x_{t}^{i+1} x_{t}^{i-1} + x_{t}^{i} x_{t}^{i-1} - 3 x_{t}^{i+1} x_{t}^{i} x_{t}^{i-1}$ \\
$^{1}105_{2}^{1}$ & \qquad (1,0,0,1,0,1,1,0) & $\qquad x_{t+1}^{i}=1 - x_{t}^{i+1} - x_{t}^{i} + 2 x_{t}^{i+1} x_{t}^{i} - x_{t}^{i-1} + 2 x_{t}^{i+1} x_{t}^{i-1} +$\\
&& $\quad \qquad \qquad + 2 x_{t}^{i} x_{t}^{i-1} - 4 x_{t}^{i+1} x_{t}^{i} x_{t}^{i-1}$ \\
$^{1}106_{2}^{1}$ & \qquad (0,1,0,1,0,1,1,0)  & $\qquad x_{t+1}^{i}=x_{t}^{i+1} x_{t}^{i} + x_{t}^{i-1} - 2 x_{t}^{i+1} x_{t}^{i} x_{t}^{i-1}$ \\
\end{tabular}
\end{center}
Table S6 (cont.)
\end{table}

\begin{table}[htdp]
\begin{center}
\begin{tabular}{cclcccl}
Rule & $ \qquad (a_{0},a_{1},a_{2},a_{3},a_{4},a_{5},a_{6},a_{7})$ & \qquad Map \\
$^{1}107_{2}^{1}$ & \qquad (1,1,0,1,0,1,1,0)  & $\qquad x_{t+1}^{i}=1 - x_{t}^{i+1} - x_{t}^{i} + 2 x_{t}^{i+1} x_{t}^{i} + x_{t}^{i+1} x_{t}^{i-1} + x_{t}^{i} x_{t}^{i-1} -$\\
&& $\quad \qquad \qquad - 3 x_{t}^{i+1} x_{t}^{i} x_{t}^{i-1}$ \\
$^{1}108_{2}^{1}$ & \qquad (0,0,1,1,0,1,1,0)  & $\qquad x_{t+1}^{i}=x_{t}^{i} + x_{t}^{i+1} x_{t}^{i-1} - 2 x_{t}^{i+1} x_{t}^{i} x_{t}^{i-1}$ \\
$^{1}109_{2}^{1}$ & \qquad (1,0,1,1,0,1,1,0)  & $\qquad x_{t+1}^{i}=1 - x_{t}^{i+1} + x_{t}^{i+1} x_{t}^{i} - x_{t}^{i-1} + 2 x_{t}^{i+1} x_{t}^{i-1} + x_{t}^{i} x_{t}^{i-1} -$\\
&& $\quad \qquad \qquad - 3 x_{t}^{i+1} x_{t}^{i} x_{t}^{i-1}$ \\
$^{1}110_{2}^{1}$ & \qquad (0,1,1,1,0,1,1,0)  & $\qquad x_{t+1}^{i}=x_{t}^{i} + x_{t}^{i-1} - x_{t}^{i} x_{t}^{i-1} - x_{t}^{i+1} x_{t}^{i} x_{t}^{i-1}$ \\
$^{1}111_{2}^{1}$ & \qquad (1,1,1,1,0,1,1,0)  & $\qquad x_{t+1}^{i}=1 - x_{t}^{i+1} + x_{t}^{i+1} x_{t}^{i} + x_{t}^{i+1} x_{t}^{i-1} - 2 x_{t}^{i+1} x_{t}^{i} x_{t}^{i-1}$ \\
$^{1}112_{2}^{1}$ & \qquad (0,0,0,0,1,1,1,0)  & $\qquad x_{t+1}^{i}=x_{t}^{i+1} - x_{t}^{i+1} x_{t}^{i} x_{t}^{i-1}$ \\
$^{1}113_{2}^{1}$ & \qquad (1,0,0,0,1,1,1,0) & $\qquad x_{t+1}^{i}=1 - x_{t}^{i} + x_{t}^{i+1} x_{t}^{i} - x_{t}^{i-1} + x_{t}^{i+1} x_{t}^{i-1} + x_{t}^{i} x_{t}^{i-1} -$\\
&& $\quad \qquad \qquad - 2 x_{t}^{i+1} x_{t}^{i} x_{t}^{i-1}$ \\
$^{1}114_{2}^{1}$ & \qquad (0,1,0,0,1,1,1,0) & $\qquad x_{t+1}^{i}=x_{t}^{i+1} + x_{t}^{i-1} - x_{t}^{i+1} x_{t}^{i-1} - x_{t}^{i} x_{t}^{i-1}$ \\
$^{1}115_{2}^{1}$ & \qquad (1,1,0,0,1,1,1,0) & $\qquad x_{t+1}^{i}=1 - x_{t}^{i} + x_{t}^{i+1} x_{t}^{i} - x_{t}^{i+1} x_{t}^{i} x_{t}^{i-1}$ \\
$^{1}116_{2}^{1}$ & \qquad (0,0,1,0,1,1,1,0) & $\qquad x_{t+1}^{i}=x_{t}^{i+1} + x_{t}^{i} - x_{t}^{i+1} x_{t}^{i} - x_{t}^{i} x_{t}^{i-1}$\\
$^{1}117_{2}^{1}$ & \qquad (1,0,1,0,1,1,1,0) & $\qquad x_{t+1}^{i}=1 - x_{t}^{i-1} + x_{t}^{i+1} x_{t}^{i-1} - x_{t}^{i+1} x_{t}^{i} x_{t}^{i-1}$ \\
$^{1}118_{2}^{1}$ & \qquad (0,1,1,0,1,1,1,0) & $\qquad x_{t+1}^{i}=x_{t}^{i+1} + x_{t}^{i} - x_{t}^{i+1} x_{t}^{i} + x_{t}^{i-1} - x_{t}^{i+1} x_{t}^{i-1} - 2 x_{t}^{i} x_{t}^{i-1} +$\\
&& $\quad \qquad \qquad + x_{t}^{i+1} x_{t}^{i} x_{t}^{i-1}$ \\
$^{1}119_{2}^{1}$ & \qquad (1,1,1,0,1,1,1,0) & $\qquad x_{t+1}^{i}=1 - x_{t}^{i} x_{t}^{i-1}$ \\
$^{1}120_{2}^{1}$ & \qquad (0,0,0,1,1,1,1,0) & $\qquad x_{t+1}^{i}=x_{t}^{i+1} + x_{t}^{i} x_{t}^{i-1} - 2 x_{t}^{i+1} x_{t}^{i} x_{t}^{i-1}$ \\
$^{1}121_{2}^{1}$ & \qquad (1,0,0,1,1,1,1,0) & $\qquad x_{t+1}^{i}=1 - x_{t}^{i} + x_{t}^{i+1} x_{t}^{i} - x_{t}^{i-1} + x_{t}^{i+1} x_{t}^{i-1} + 2 x_{t}^{i} x_{t}^{i-1} -$\\
&& $\quad \qquad \qquad - 3 x_{t}^{i+1} x_{t}^{i} x_{t}^{i-1}$ \\
$^{1}122_{2}^{1}$ & \qquad (0,1,0,1,1,1,1,0) & $\qquad x_{t+1}^{i}=x_{t}^{i+1} + x_{t}^{i-1} - x_{t}^{i+1} x_{t}^{i-1} - x_{t}^{i+1} x_{t}^{i} x_{t}^{i-1}$ \\
$^{1}123_{2}^{1}$ & \qquad (1,1,0,1,1,1,1,0) & $\qquad x_{t+1}^{i}=1 - x_{t}^{i} + x_{t}^{i+1} x_{t}^{i} + x_{t}^{i} x_{t}^{i-1} - 2 x_{t}^{i+1} x_{t}^{i} x_{t}^{i-1}$ \\
$^{1}124_{2}^{1}$ & \qquad (0,0,1,1,1,1,1,0) & $\qquad x_{t+1}^{i}=x_{t}^{i+1} + x_{t}^{i} - x_{t}^{i+1} x_{t}^{i} - x_{t}^{i+1} x_{t}^{i} x_{t}^{i-1}$ \\
$^{1}125_{2}^{1}$ & \qquad (1,0,1,1,1,1,1,0) & $\qquad x_{t+1}^{i}=1 - x_{t}^{i-1} + x_{t}^{i+1} x_{t}^{i-1} + x_{t}^{i} x_{t}^{i-1} - 2 x_{t}^{i+1} x_{t}^{i} x_{t}^{i-1}$ \\
$^{1}126_{2}^{1}$ & \qquad (0,1,1,1,1,1,1,0) & $\qquad x_{t+1}^{i}=x_{t}^{i+1} + x_{t}^{i} - x_{t}^{i+1} x_{t}^{i} + x_{t}^{i-1} - x_{t}^{i+1} x_{t}^{i-1} - x_{t}^{i} x_{t}^{i-1}$ \\
\end{tabular}
\end{center}
Table S6 (cont.)
\end{table}

\begin{table}[htdp]
\begin{center}
\begin{tabular}{cclcccl}
Rule & $ \qquad (a_{0},a_{1},a_{2},a_{3},a_{4},a_{5},a_{6},a_{7})$ & \qquad Map \\
$^{1}127_{2}^{1}$ & \qquad (1,1,1,1,1,1,1,0) & $\qquad x_{t+1}^{i}=1 - x_{t}^{i+1} x_{t}^{i} x_{t}^{i-1}$ \\
$^{1}128_{2}^{1}$ & \qquad (0,0,0,0,0,0,0,1) & $\qquad x_{t+1}^{i}=x_{t}^{i+1} x_{t}^{i} x_{t}^{i-1}$ \\
$^{1}129_{2}^{1}$ & \qquad (1,0,0,0,0,0,0,1) & $\qquad x_{t+1}^{i}=1 - x_{t}^{i+1} - x_{t}^{i} + x_{t}^{i+1} x_{t}^{i} - x_{t}^{i-1} + x_{t}^{i+1} x_{t}^{i-1} + x_{t}^{i} x_{t}^{i-1}$ \\
$^{1}130_{2}^{1}$ & \qquad (0,1,0,0,0,0,0,1) & $\qquad x_{t+1}^{i}=x_{t}^{i-1} - x_{t}^{i+1} x_{t}^{i-1} - x_{t}^{i} x_{t}^{i-1} + 2 x_{t}^{i+1} x_{t}^{i} x_{t}^{i-1}$ \\
$^{1}131_{2}^{1}$ & \qquad (1,1,0,0,0,0,0,1) & $\qquad x_{t+1}^{i}=1 - x_{t}^{i+1} - x_{t}^{i} + x_{t}^{i+1} x_{t}^{i} + x_{t}^{i+1} x_{t}^{i} x_{t}^{i-1}$ \\
$^{1}132_{2}^{1}$ & \qquad (0,0,1,0,0,0,0,1) & $\qquad x_{t+1}^{i}=x_{t}^{i} - x_{t}^{i+1} x_{t}^{i} - x_{t}^{i} x_{t}^{i-1} + 2 x_{t}^{i+1} x_{t}^{i} x_{t}^{i-1}$ \\
$^{1}133_{2}^{1}$ & \qquad (1,0,1,0,0,0,0,1) & $\qquad x_{t+1}^{i}=1 - x_{t}^{i+1} - x_{t}^{i-1} + x_{t}^{i+1} x_{t}^{i-1} + x_{t}^{i+1} x_{t}^{i} x_{t}^{i-1}$ \\
$^{1}134_{2}^{1}$ & \qquad (0,1,1,0,0,0,0,1) & $\qquad x_{t+1}^{i}=x_{t}^{i} - x_{t}^{i+1} x_{t}^{i} + x_{t}^{i-1} - x_{t}^{i+1} x_{t}^{i-1} - 2 x_{t}^{i} x_{t}^{i-1} +$\\
&& $\quad \qquad \qquad + 3 x_{t}^{i+1} x_{t}^{i} x_{t}^{i-1}$ \\
$^{1}135_{2}^{1}$ & \qquad (1,1,1,0,0,0,0,1) & $\qquad x_{t+1}^{i}=1 - x_{t}^{i+1} - x_{t}^{i} x_{t}^{i-1} + 2 x_{t}^{i+1} x_{t}^{i} x_{t}^{i-1}$ \\
$^{1}136_{2}^{1}$ & \qquad (0,0,0,1,0,0,0,1) & $\qquad x_{t+1}^{i}=x_{t}^{i} x_{t}^{i-1}$ \\
$^{1}137_{2}^{1}$ & \qquad (1,0,0,1,0,0,0,1) & $\qquad x_{t+1}^{i}=1 - x_{t}^{i+1} - x_{t}^{i} + x_{t}^{i+1} x_{t}^{i} - x_{t}^{i-1} + x_{t}^{i+1} x_{t}^{i-1} +$\\
&& $\quad \qquad \qquad + 2 x_{t}^{i} x_{t}^{i-1} - x_{t}^{i+1} x_{t}^{i} x_{t}^{i-1}$ \\
$^{1}138_{2}^{1}$ & \qquad (0,1,0,1,0,0,0,1) & $\qquad x_{t+1}^{i}=x_{t}^{i-1} - x_{t}^{i+1} x_{t}^{i-1} + x_{t}^{i+1} x_{t}^{i} x_{t}^{i-1}$ \\
$^{1}139_{2}^{1}$ & \qquad (1,1,0,1,0,0,0,1) & $\qquad x_{t+1}^{i}=1 - x_{t}^{i+1} - x_{t}^{i} + x_{t}^{i+1} x_{t}^{i} + x_{t}^{i} x_{t}^{i-1}$ \\
$^{1}140_{2}^{1}$ & \qquad (0,0,1,1,0,0,0,1) & $\qquad x_{t+1}^{i}=x_{t}^{i} - x_{t}^{i+1} x_{t}^{i} + x_{t}^{i+1} x_{t}^{i} x_{t}^{i-1}$\\
$^{1}141_{2}^{1}$ & \qquad (1,0,1,1,0,0,0,1) & $\qquad x_{t+1}^{i}=1 - x_{t}^{i+1} - x_{t}^{i-1} + x_{t}^{i+1} x_{t}^{i-1} + x_{t}^{i} x_{t}^{i-1}$ \\
$^{1}142_{2}^{1}$ & \qquad (0,1,1,1,0,0,0,1) & $\qquad x_{t+1}^{i}=x_{t}^{i} - x_{t}^{i+1} x_{t}^{i} + x_{t}^{i-1} - x_{t}^{i+1} x_{t}^{i-1} - x_{t}^{i} x_{t}^{i-1} + 2 x_{t}^{i+1} x_{t}^{i} x_{t}^{i-1}$ \\
$^{1}143_{2}^{1}$ & \qquad (1,1,1,1,0,0,0,1) & $\qquad x_{t+1}^{i}=1 - x_{t}^{i+1} + x_{t}^{i+1} x_{t}^{i} x_{t}^{i-1}$ \\
$^{1}144_{2}^{1}$ & \qquad (0,0,0,0,1,0,0,1) & $\qquad x_{t+1}^{i}=x_{t}^{i+1} - x_{t}^{i+1} x_{t}^{i} - x_{t}^{i+1} x_{t}^{i-1} + 2 x_{t}^{i+1} x_{t}^{i} x_{t}^{i-1}$ \\
$^{1}145_{2}^{1}$ & \qquad (1,0,0,0,1,0,0,1) & $\qquad x_{t+1}^{i}=1 - x_{t}^{i} - x_{t}^{i-1} + x_{t}^{i} x_{t}^{i-1} + x_{t}^{i+1} x_{t}^{i} x_{t}^{i-1}$ \\
$^{1}146_{2}^{1}$ & \qquad (0,1,0,0,1,0,0,1) & $\qquad x_{t+1}^{i}=x_{t}^{i+1} - x_{t}^{i+1} x_{t}^{i} + x_{t}^{i-1} - 2 x_{t}^{i+1} x_{t}^{i-1} - x_{t}^{i} x_{t}^{i-1} +$\\
&& $\quad \qquad \qquad + 3 x_{t}^{i+1} x_{t}^{i} x_{t}^{i-1}$ \\
$^{1}147_{2}^{1}$ & \qquad (1,1,0,0,1,0,0,1) & $\qquad x_{t+1}^{i}=1 - x_{t}^{i} - x_{t}^{i+1} x_{t}^{i-1} + 2 x_{t}^{i+1} x_{t}^{i} x_{t}^{i-1}$ \\
$^{1}148_{2}^{1}$ & \qquad (0,0,1,0,1,0,0,1) & $\qquad x_{t+1}^{i}=x_{t}^{i+1} + x_{t}^{i} - 2 x_{t}^{i+1} x_{t}^{i} - x_{t}^{i+1} x_{t}^{i-1} - x_{t}^{i} x_{t}^{i-1} + 3 x_{t}^{i+1} x_{t}^{i} x_{t}^{i-1}$ \\
\end{tabular}
\end{center}
Table S6 (cont.)
\end{table}

\begin{table}[htdp]
\begin{center}
\begin{tabular}{cclcccl}
Rule & $ \qquad (a_{0},a_{1},a_{2},a_{3},a_{4},a_{5},a_{6},a_{7})$ & \qquad Map \\
$^{1}149_{2}^{1}$ & \qquad (1,0,1,0,1,0,0,1) & $\qquad x_{t+1}^{i}=1 - x_{t}^{i+1} x_{t}^{i} - x_{t}^{i-1} + 2 x_{t}^{i+1} x_{t}^{i} x_{t}^{i-1}$ \\
$^{1}150_{2}^{1}$ & \qquad (0,1,1,0,1,0,0,1) & $\qquad x_{t+1}^{i}=x_{t}^{i+1} + x_{t}^{i} + x_{t}^{i-1}- 2 x_{t}^{i+1} x_{t}^{i} - 2 x_{t}^{i+1} x_{t}^{i-1} $\\
&& $\quad \qquad \qquad - 2 x_{t}^{i} x_{t}^{i-1} + 4 x_{t}^{i+1} x_{t}^{i} x_{t}^{i-1}$ \\
$^{1}151_{2}^{1}$ & \qquad (1,1,1,0,1,0,0,1) & $\qquad x_{t+1}^{i}=1 - x_{t}^{i+1} x_{t}^{i} - x_{t}^{i+1} x_{t}^{i-1} - x_{t}^{i} x_{t}^{i-1} + 3 x_{t}^{i+1} x_{t}^{i} x_{t}^{i-1}$ \\
$^{1}152_{2}^{1}$ & \qquad (0,0,0,1,1,0,0,1) & $\qquad x_{t+1}^{i}=x_{t}^{i+1} - x_{t}^{i+1} x_{t}^{i} - x_{t}^{i+1} x_{t}^{i-1} + x_{t}^{i} x_{t}^{i-1} + x_{t}^{i+1} x_{t}^{i} x_{t}^{i-1}$ \\
$^{1}153_{2}^{1}$ & \qquad (1,0,0,1,1,0,0,1) & $\qquad x_{t+1}^{i}=1 - x_{t}^{i} - x_{t}^{i-1} + 2 x_{t}^{i} x_{t}^{i-1}$ \\
&& $\quad \qquad \qquad - 2 x_{t}^{i} x_{t}^{i-1} + 4 x_{t}^{i+1} x_{t}^{i} x_{t}^{i-1}$ \\
$^{1}154_{2}^{1}$ & \qquad (0,1,0,1,1,0,0,1) & $\qquad x_{t+1}^{i}=x_{t}^{i+1} - x_{t}^{i+1} x_{t}^{i} + x_{t}^{i-1} - 2 x_{t}^{i+1} x_{t}^{i-1} + 2 x_{t}^{i+1} x_{t}^{i} x_{t}^{i-1}$ \\
$^{1}155_{2}^{1}$ & \qquad (1,1,0,1,1,0,0,1) & $\qquad x_{t+1}^{i}=1 - x_{t}^{i} - x_{t}^{i+1} x_{t}^{i-1} + x_{t}^{i} x_{t}^{i-1} + x_{t}^{i+1} x_{t}^{i} x_{t}^{i-1}$ \\
$^{1}156_{2}^{1}$ & \qquad (0,0,1,1,1,0,0,1) & $\qquad x_{t+1}^{i}=x_{t}^{i+1} + x_{t}^{i} - 2 x_{t}^{i+1} x_{t}^{i} - x_{t}^{i+1} x_{t}^{i-1} + 2 x_{t}^{i+1} x_{t}^{i} x_{t}^{i-1}
$ \\
$^{1}157_{2}^{1}$ & \qquad (1,0,1,1,1,0,0,1) & $\qquad x_{t+1}^{i}=1 - x_{t}^{i+1} x_{t}^{i} - x_{t}^{i-1} + x_{t}^{i} x_{t}^{i-1} + x_{t}^{i+1} x_{t}^{i} x_{t}^{i-1}$ \\
$^{1}158_{2}^{1}$ & \qquad (0,1,1,1,1,0,0,1) & $\qquad x_{t+1}^{i}=x_{t}^{i+1} + x_{t}^{i} - 2 x_{t}^{i+1} x_{t}^{i} + x_{t}^{i-1} - 2 x_{t}^{i+1} x_{t}^{i-1} - $\\
&& $\quad \qquad \qquad - x_{t}^{i} x_{t}^{i-1} + 3 x_{t}^{i+1} x_{t}^{i} x_{t}^{i-1}$ \\
$^{1}159_{2}^{1}$ & \qquad (1,1,1,1,1,0,0,1) & $\qquad x_{t+1}^{i}=1 - x_{t}^{i+1} x_{t}^{i} - x_{t}^{i+1} x_{t}^{i-1} + 2 x_{t}^{i+1} x_{t}^{i} x_{t}^{i-1}$ \\
$^{1}160_{2}^{1}$ & \qquad (0,0,0,0,0,1,0,1) & $\qquad x_{t+1}^{i}=x_{t}^{i+1} x_{t}^{i-1}$\\
$^{1}161_{2}^{1}$ & \qquad (1,0,0,0,0,1,0,1) & $\qquad x_{t+1}^{i}=1 - x_{t}^{i+1} - x_{t}^{i} + x_{t}^{i+1} x_{t}^{i} - x_{t}^{i-1} + 2 x_{t}^{i+1} x_{t}^{i-1} +$\\
&& $\quad \qquad \qquad + x_{t}^{i} x_{t}^{i-1} - x_{t}^{i+1} x_{t}^{i} x_{t}^{i-1}$ \\
$^{1}162_{2}^{1}$ & \qquad (0,1,0,0,0,1,0,1) & $\qquad x_{t+1}^{i}=x_{t}^{i-1} - x_{t}^{i} x_{t}^{i-1} + x_{t}^{i+1} x_{t}^{i} x_{t}^{i-1}$ \\
$^{1}163_{2}^{1}$ & \qquad (1,1,0,0,0,1,0,1) & $\qquad x_{t+1}^{i}=1 - x_{t}^{i+1} - x_{t}^{i} + x_{t}^{i+1} x_{t}^{i} + x_{t}^{i+1} x_{t}^{i-1}$ \\
$^{1}164_{2}^{1}$ & \qquad (0,0,1,0,0,1,0,1) & $\qquad x_{t+1}^{i}=x_{t}^{i} - x_{t}^{i+1} x_{t}^{i} + x_{t}^{i+1} x_{t}^{i-1} - x_{t}^{i} x_{t}^{i-1} + x_{t}^{i+1} x_{t}^{i} x_{t}^{i-1}$ \\
$^{1}165_{2}^{1}$ & \qquad (1,0,1,0,0,1,0,1) & $\qquad x_{t+1}^{i}=1 - x_{t}^{i+1} - x_{t}^{i-1} + 2 x_{t}^{i+1} x_{t}^{i-1}$ \\
$^{1}166_{2}^{1}$ & \qquad (0,1,1,0,0,1,0,1) & $\qquad x_{t+1}^{i}=x_{t}^{i} - x_{t}^{i+1} x_{t}^{i} + x_{t}^{i-1} - 2 x_{t}^{i} x_{t}^{i-1} + 2 x_{t}^{i+1} x_{t}^{i} x_{t}^{i-1}$ \\
$^{1}167_{2}^{1}$ & \qquad (1,1,1,0,0,1,0,1) & $\qquad x_{t+1}^{i}=1 - x_{t}^{i+1} + x_{t}^{i+1} x_{t}^{i-1} - x_{t}^{i} x_{t}^{i-1} + x_{t}^{i+1} x_{t}^{i} x_{t}^{i-1}$ \\
$^{1}168_{2}^{1}$ & \qquad (0,0,0,1,0,1,0,1) & $\qquad x_{t+1}^{i}=x_{t}^{i+1} x_{t}^{i-1} + x_{t}^{i} x_{t}^{i-1} - x_{t}^{i+1} x_{t}^{i} x_{t}^{i-1}$ \\
$^{1}169_{2}^{1}$ & \qquad (1,0,0,1,0,1,0,1)  & $\qquad x_{t+1}^{i}=1 - x_{t}^{i+1} - x_{t}^{i} + x_{t}^{i+1} x_{t}^{i} - x_{t}^{i-1} + 2 x_{t}^{i+1} x_{t}^{i-1} +$\\
&& $\quad \qquad \qquad + 2 x_{t}^{i} x_{t}^{i-1} - 2 x_{t}^{i+1} x_{t}^{i} x_{t}^{i-1}$ \\
\end{tabular}
\end{center}
Table S6 (cont.)
\end{table}

\begin{table}[htdp]
\begin{center}
\begin{tabular}{cclcccl}
Rule & $ \qquad (a_{0},a_{1},a_{2},a_{3},a_{4},a_{5},a_{6},a_{7})$ & \qquad Map \\
$^{1}170_{2}^{1}$ & \qquad (0,1,0,1,0,1,0,1)  & $\qquad x_{t+1}^{i}=x_{t}^{i-1}$ \\
$^{1}171_{2}^{1}$ & \qquad (1,1,0,1,0,1,0,1)  & $\qquad x_{t+1}^{i}=1 - x_{t}^{i+1} - x_{t}^{i} + x_{t}^{i+1} x_{t}^{i} + x_{t}^{i+1} x_{t}^{i-1} + x_{t}^{i} x_{t}^{i-1} -$\\
&& $\quad \qquad \qquad - x_{t}^{i+1} x_{t}^{i} x_{t}^{i-1}$ \\
$^{1}172_{2}^{1}$ & \qquad (0,0,1,1,0,1,0,1)  & $\qquad x_{t+1}^{i}=x_{t}^{i} - x_{t}^{i+1} x_{t}^{i} + x_{t}^{i+1} x_{t}^{i-1}$ \\
$^{1}173_{2}^{1}$ & \qquad (1,0,1,1,0,1,0,1)  & $\qquad x_{t+1}^{i}=1 - x_{t}^{i+1} - x_{t}^{i-1} + 2 x_{t}^{i+1} x_{t}^{i-1} + x_{t}^{i} x_{t}^{i-1} - x_{t}^{i+1} x_{t}^{i} x_{t}^{i-1}$ \\
$^{1}174_{2}^{1}$ & \qquad (0,1,1,1,0,1,0,1)  & $\qquad x_{t+1}^{i}=x_{t}^{i} - x_{t}^{i+1} x_{t}^{i} + x_{t}^{i-1} - x_{t}^{i} x_{t}^{i-1} + x_{t}^{i+1} x_{t}^{i} x_{t}^{i-1}$ \\
$^{1}175_{2}^{1}$ & \qquad (1,1,1,1,0,1,0,1)  & $\qquad x_{t+1}^{i}=1 - x_{t}^{i+1} + x_{t}^{i+1} x_{t}^{i-1}$ \\
$^{1}176_{2}^{1}$ & \qquad (0,0,0,0,1,1,0,1)  & $\qquad x_{t+1}^{i}=x_{t}^{i+1} - x_{t}^{i+1} x_{t}^{i} + x_{t}^{i+1} x_{t}^{i} x_{t}^{i-1}$ \\
$^{1}177_{2}^{1}$ & \qquad (1,0,0,0,1,1,0,1) & $\qquad x_{t+1}^{i}=1 - x_{t}^{i} - x_{t}^{i-1} + x_{t}^{i+1} x_{t}^{i-1} + x_{t}^{i} x_{t}^{i-1}$ \\
$^{1}178_{2}^{1}$ & \qquad (0,1,0,0,1,1,0,1) & $\qquad x_{t+1}^{i}=x_{t}^{i+1} - x_{t}^{i+1} x_{t}^{i} + x_{t}^{i-1} - x_{t}^{i+1} x_{t}^{i-1} - x_{t}^{i} x_{t}^{i-1} +$\\
&& $\quad \qquad \qquad + 2 x_{t}^{i+1} x_{t}^{i} x_{t}^{i-1}$ \\
$^{1}179_{2}^{1}$ & \qquad (1,1,0,0,1,1,0,1) & $\qquad x_{t+1}^{i}=1 - x_{t}^{i} + x_{t}^{i+1} x_{t}^{i} x_{t}^{i-1}$ \\
$^{1}180_{2}^{1}$ & \qquad (0,0,1,0,1,1,0,1) & $\qquad x_{t+1}^{i}=x_{t}^{i+1} + x_{t}^{i} - 2 x_{t}^{i+1} x_{t}^{i} - x_{t}^{i} x_{t}^{i-1} + 2 x_{t}^{i+1} x_{t}^{i} x_{t}^{i-1}$\\
$^{1}181_{2}^{1}$ & \qquad (1,0,1,0,1,1,0,1) & $\qquad x_{t+1}^{i}=1 - x_{t}^{i+1} x_{t}^{i} - x_{t}^{i-1} + x_{t}^{i+1} x_{t}^{i-1} + x_{t}^{i+1} x_{t}^{i} x_{t}^{i-1}$ \\
$^{1}182_{2}^{1}$ & \qquad (0,1,1,0,1,1,0,1) & $\qquad x_{t+1}^{i}=x_{t}^{i+1} + x_{t}^{i} - 2 x_{t}^{i+1} x_{t}^{i} + x_{t}^{i-1} - x_{t}^{i+1} x_{t}^{i-1} -$\\
&& $\quad \qquad \qquad - 2 x_{t}^{i} x_{t}^{i-1} + 3 x_{t}^{i+1} x_{t}^{i} x_{t}^{i-1}$ \\
$^{1}183_{2}^{1}$ & \qquad (1,1,1,0,1,1,0,1) & $\qquad x_{t+1}^{i}=1 - x_{t}^{i+1} x_{t}^{i} - x_{t}^{i} x_{t}^{i-1} + 2 x_{t}^{i+1} x_{t}^{i} x_{t}^{i-1}$ \\
$^{1}184_{2}^{1}$ & \qquad (0,0,0,1,1,1,0,1) & $\qquad x_{t+1}^{i}=x_{t}^{i+1} - x_{t}^{i+1} x_{t}^{i} + x_{t}^{i} x_{t}^{i-1}$ \\
$^{1}185_{2}^{1}$ & \qquad (1,0,0,1,1,1,0,1) & $\qquad x_{t+1}^{i}=1 - x_{t}^{i} - x_{t}^{i-1} + x_{t}^{i+1} x_{t}^{i-1} + 2 x_{t}^{i} x_{t}^{i-1} - x_{t}^{i+1} x_{t}^{i} x_{t}^{i-1}$ \\
$^{1}186_{2}^{1}$ & \qquad (0,1,0,1,1,1,0,1) & $\qquad x_{t+1}^{i}=x_{t}^{i+1} - x_{t}^{i+1} x_{t}^{i} + x_{t}^{i-1} - x_{t}^{i+1} x_{t}^{i-1} + x_{t}^{i+1} x_{t}^{i} x_{t}^{i-1}$ \\
$^{1}187_{2}^{1}$ & \qquad (1,1,0,1,1,1,0,1) & $\qquad x_{t+1}^{i}=1 - x_{t}^{i} + x_{t}^{i} x_{t}^{i-1}$ \\
$^{1}188_{2}^{1}$ & \qquad (0,0,1,1,1,1,0,1) & $\qquad x_{t+1}^{i}=x_{t}^{i+1} + x_{t}^{i} - 2 x_{t}^{i+1} x_{t}^{i} + x_{t}^{i+1} x_{t}^{i} x_{t}^{i-1}$ \\
$^{1}189_{2}^{1}$ & \qquad (1,0,1,1,1,1,0,1) & $\qquad x_{t+1}^{i}=1 - x_{t}^{i+1} x_{t}^{i} - x_{t}^{i-1} + x_{t}^{i+1} x_{t}^{i-1} + x_{t}^{i} x_{t}^{i-1}$ \\
$^{1}190_{2}^{1}$ & \qquad (0,1,1,1,1,1,0,1) & $\qquad x_{t+1}^{i}=x_{t}^{i+1} + x_{t}^{i} - 2 x_{t}^{i+1} x_{t}^{i} + x_{t}^{i-1} - x_{t}^{i+1} x_{t}^{i-1} - x_{t}^{i} x_{t}^{i-1} +$\\
&& $\quad \qquad \qquad + 2 x_{t}^{i+1} x_{t}^{i} x_{t}^{i-1}$ \\
$^{1}191_{2}^{1}$ & \qquad (1,1,1,1,1,1,0,1) & $\qquad x_{t+1}^{i}=1 - x_{t}^{i+1} x_{t}^{i} + x_{t}^{i+1} x_{t}^{i} x_{t}^{i-1}$ \\
\end{tabular}
\end{center}
Table S6 (cont.)
\end{table}

\begin{table}[htdp]
\begin{center}
\begin{tabular}{cclcccl}
Rule & $ \qquad (a_{0},a_{1},a_{2},a_{3},a_{4},a_{5},a_{6},a_{7})$ & \qquad Map \\
$^{1}192_{2}^{1}$ & \qquad (0,0,0,0,0,0,1,1) & $\qquad x_{t+1}^{i}=x_{t}^{i+1} x_{t}^{i}$ \\
$^{1}193_{2}^{1}$ & \qquad (1,0,0,0,0,0,1,1) & $\qquad x_{t+1}^{i}=1 - x_{t}^{i+1} - x_{t}^{i} + 2 x_{t}^{i+1} x_{t}^{i} - x_{t}^{i-1} + x_{t}^{i+1} x_{t}^{i-1} +$\\
&& $\quad \qquad \qquad + x_{t}^{i} x_{t}^{i-1} - x_{t}^{i+1} x_{t}^{i} x_{t}^{i-1}$ \\
$^{1}194_{2}^{1}$ & \qquad (0,1,0,0,0,0,1,1) & $\qquad x_{t+1}^{i}=x_{t}^{i+1} x_{t}^{i} + x_{t}^{i-1} - x_{t}^{i+1} x_{t}^{i-1} - x_{t}^{i} x_{t}^{i-1} + x_{t}^{i+1} x_{t}^{i} x_{t}^{i-1}$ \\
$^{1}195_{2}^{1}$ & \qquad (1,1,0,0,0,0,1,1) & $\qquad x_{t+1}^{i}=1 - x_{t}^{i+1} - x_{t}^{i} + 2 x_{t}^{i+1} x_{t}^{i}$ \\
$^{1}196_{2}^{1}$ & \qquad (0,0,1,0,0,0,1,1) & $\qquad x_{t+1}^{i}=x_{t}^{i} - x_{t}^{i} x_{t}^{i-1} + x_{t}^{i+1} x_{t}^{i} x_{t}^{i-1}$ \\
$^{1}197_{2}^{1}$ & \qquad (1,0,1,0,0,0,1,1) & $\qquad x_{t+1}^{i}=1 - x_{t}^{i+1} + x_{t}^{i+1} x_{t}^{i} - x_{t}^{i-1} + x_{t}^{i+1} x_{t}^{i-1}$ \\
$^{1}198_{2}^{1}$ & \qquad (0,1,1,0,0,0,1,1) & $\qquad x_{t+1}^{i}=x_{t}^{i} + x_{t}^{i-1} - x_{t}^{i+1} x_{t}^{i-1} - 2 x_{t}^{i} x_{t}^{i-1} + 2 x_{t}^{i+1} x_{t}^{i} x_{t}^{i-1}$ \\
$^{1}199_{2}^{1}$ & \qquad (1,1,1,0,0,0,1,1) & $\qquad x_{t+1}^{i}=1 - x_{t}^{i+1} + x_{t}^{i+1} x_{t}^{i} - x_{t}^{i} x_{t}^{i-1} + x_{t}^{i+1} x_{t}^{i} x_{t}^{i-1}$ \\
$^{1}200_{2}^{1}$ & \qquad (0,0,0,1,0,0,1,1) & $\qquad x_{t+1}^{i}=x_{t}^{i+1} x_{t}^{i} + x_{t}^{i} x_{t}^{i-1} - x_{t}^{i+1} x_{t}^{i} x_{t}^{i-1}$ \\
$^{1}201_{2}^{1}$ & \qquad (1,0,0,1,0,0,1,1) & $\qquad x_{t+1}^{i}=1 - x_{t}^{i+1} - x_{t}^{i} + 2 x_{t}^{i+1} x_{t}^{i} - x_{t}^{i-1} + x_{t}^{i+1} x_{t}^{i-1} + 2 x_{t}^{i} x_{t}^{i-1}- $\\
&& $\quad \qquad \qquad - 2 x_{t}^{i+1} x_{t}^{i} x_{t}^{i-1}$ \\
$^{1}202_{2}^{1}$ & \qquad (0,1,0,1,0,0,1,1) & $\qquad x_{t+1}^{i}=x_{t}^{i+1} x_{t}^{i} + x_{t}^{i-1} - x_{t}^{i+1} x_{t}^{i-1}$ \\
$^{1}203_{2}^{1}$ & \qquad (1,1,0,1,0,0,1,1) & $\qquad x_{t+1}^{i}=1 - x_{t}^{i+1} - x_{t}^{i} + 2 x_{t}^{i+1} x_{t}^{i} + x_{t}^{i} x_{t}^{i-1} - x_{t}^{i+1} x_{t}^{i} x_{t}^{i-1}$ \\
$^{1}204_{2}^{1}$ & \qquad (0,0,1,1,0,0,1,1) & $\qquad x_{t+1}^{i}=x_{t}^{i}$\\
$^{1}205_{2}^{1}$ & \qquad (1,0,1,1,0,0,1,1) & $\qquad x_{t+1}^{i}=1 - x_{t}^{i+1} + x_{t}^{i+1} x_{t}^{i} - x_{t}^{i-1} + x_{t}^{i+1} x_{t}^{i-1} + x_{t}^{i} x_{t}^{i-1} -$\\
&& $\quad \qquad \qquad - x_{t}^{i+1} x_{t}^{i} x_{t}^{i-1}$ \\
$^{1}206_{2}^{1}$ & \qquad (0,1,1,1,0,0,1,1) & $\qquad x_{t+1}^{i}=x_{t}^{i} + x_{t}^{i-1} - x_{t}^{i+1} x_{t}^{i-1} - x_{t}^{i} x_{t}^{i-1} + x_{t}^{i+1} x_{t}^{i} x_{t}^{i-1}$ \\
$^{1}207_{2}^{1}$ & \qquad (1,1,1,1,0,0,1,1) & $\qquad x_{t+1}^{i}=1 - x_{t}^{i+1} + x_{t}^{i+1} x_{t}^{i}$ \\
$^{1}208_{2}^{1}$ & \qquad (0,0,0,0,1,0,1,1) & $\qquad x_{t+1}^{i}=x_{t}^{i+1} - x_{t}^{i+1} x_{t}^{i-1} + x_{t}^{i+1} x_{t}^{i} x_{t}^{i-1}$ \\
$^{1}209_{2}^{1}$ & \qquad (1,0,0,0,1,0,1,1) & $\qquad x_{t+1}^{i}=1 - x_{t}^{i} + x_{t}^{i+1} x_{t}^{i} - x_{t}^{i-1} + x_{t}^{i} x_{t}^{i-1}$ \\
$^{1}210_{2}^{1}$ & \qquad (0,1,0,0,1,0,1,1) & $\qquad x_{t+1}^{i}=x_{t}^{i+1} + x_{t}^{i-1} - 2 x_{t}^{i+1} x_{t}^{i-1} - x_{t}^{i} x_{t}^{i-1} + 2 x_{t}^{i+1} x_{t}^{i} x_{t}^{i-1}$ \\
$^{1}211_{2}^{1}$ & \qquad (1,1,0,0,1,0,1,1) & $\qquad x_{t+1}^{i}=1 - x_{t}^{i} + x_{t}^{i+1} x_{t}^{i} - x_{t}^{i+1} x_{t}^{i-1} + x_{t}^{i+1} x_{t}^{i} x_{t}^{i-1}$ \\
$^{1}212_{2}^{1}$ & \qquad (0,0,1,0,1,0,1,1) & $\qquad x_{t+1}^{i}=x_{t}^{i+1} + x_{t}^{i} - x_{t}^{i+1} x_{t}^{i} - x_{t}^{i+1} x_{t}^{i-1} - x_{t}^{i} x_{t}^{i-1} + 2 x_{t}^{i+1} x_{t}^{i} x_{t}^{i-1}$ \\
$^{1}213_{2}^{1}$ & \qquad (1,0,1,0,1,0,1,1) & $\qquad x_{t+1}^{i}=1 - x_{t}^{i-1} + x_{t}^{i+1} x_{t}^{i} x_{t}^{i-1}$ \\
\end{tabular}
\end{center}
Table S6 (cont.)
\end{table}

\begin{table}[htdp]
\begin{center}
\begin{tabular}{cclcccl}
Rule & $ \qquad (a_{0},a_{1},a_{2},a_{3},a_{4},a_{5},a_{6},a_{7})$ & \qquad Map \\
$^{1}214_{2}^{1}$ & \qquad (0,1,1,0,1,0,1,1) & $\qquad x_{t+1}^{i}=x_{t}^{i+1} + x_{t}^{i} - x_{t}^{i+1} x_{t}^{i} + x_{t}^{i-1} - 2 x_{t}^{i+1} x_{t}^{i-1} - 2 x_{t}^{i} x_{t}^{i-1} +$\\
&& $\quad \qquad \qquad + 3 x_{t}^{i+1} x_{t}^{i} x_{t}^{i-1}$ \\
$^{1}215_{2}^{1}$ & \qquad (1,1,1,0,1,0,1,1) & $\qquad x_{t+1}^{i}=1 - x_{t}^{i+1} x_{t}^{i-1} - x_{t}^{i} x_{t}^{i-1} + 2 x_{t}^{i+1} x_{t}^{i} x_{t}^{i-1}$ \\
$^{1}216_{2}^{1}$ & \qquad (0,0,0,1,1,0,1,1) & $\qquad x_{t+1}^{i}=x_{t}^{i+1} - x_{t}^{i+1} x_{t}^{i-1} + x_{t}^{i} x_{t}^{i-1}$ \\
$^{1}217_{2}^{1}$ & \qquad (1,0,0,1,1,0,1,1) & $\qquad x_{t+1}^{i}=1 - x_{t}^{i} + x_{t}^{i+1} x_{t}^{i} - x_{t}^{i-1} + 2 x_{t}^{i} x_{t}^{i-1} - x_{t}^{i+1} x_{t}^{i} x_{t}^{i-1}$ \\
$^{1}218_{2}^{1}$ & \qquad (0,1,0,1,1,0,1,1) & $\qquad x_{t+1}^{i}=x_{t}^{i+1} + x_{t}^{i-1} - 2 x_{t}^{i+1} x_{t}^{i-1} + x_{t}^{i+1} x_{t}^{i} x_{t}^{i-1}$ \\
$^{1}219_{2}^{1}$ & \qquad (1,1,0,1,1,0,1,1) & $\qquad x_{t+1}^{i}=1 - x_{t}^{i} + x_{t}^{i+1} x_{t}^{i} - x_{t}^{i+1} x_{t}^{i-1} + x_{t}^{i} x_{t}^{i-1}$ \\
$^{1}220_{2}^{1}$ & \qquad (0,0,1,1,1,0,1,1) & $\qquad x_{t+1}^{i}=x_{t}^{i+1} + x_{t}^{i} - x_{t}^{i+1} x_{t}^{i} - x_{t}^{i+1} x_{t}^{i-1} + x_{t}^{i+1} x_{t}^{i} x_{t}^{i-1}
$ \\
$^{1}221_{2}^{1}$ & \qquad (1,0,1,1,1,0,1,1) & $\qquad x_{t+1}^{i}=1 - x_{t}^{i-1} + x_{t}^{i} x_{t}^{i-1}$ \\
$^{1}222_{2}^{1}$ & \qquad (0,1,1,1,1,0,1,1) & $\qquad x_{t+1}^{i}=x_{t}^{i+1} + x_{t}^{i} - x_{t}^{i+1} x_{t}^{i} + x_{t}^{i-1} - 2 x_{t}^{i+1} x_{t}^{i-1} - x_{t}^{i} x_{t}^{i-1} + $\\
&& $\quad \qquad \qquad + 2 x_{t}^{i+1} x_{t}^{i} x_{t}^{i-1}$ \\
$^{1}223_{2}^{1}$ & \qquad (1,1,1,1,1,0,1,1) & $\qquad x_{t+1}^{i}=1 - x_{t}^{i+1} x_{t}^{i-1} + x_{t}^{i+1} x_{t}^{i} x_{t}^{i-1}$ \\
$^{1}224_{2}^{1}$ & \qquad (0,0,0,0,0,1,1,1) & $\qquad x_{t+1}^{i}=x_{t}^{i+1} x_{t}^{i} + x_{t}^{i+1} x_{t}^{i-1} - x_{t}^{i+1} x_{t}^{i} x_{t}^{i-1}$\\
$^{1}225_{2}^{1}$ & \qquad (1,0,0,0,0,1,1,1) & $\qquad x_{t+1}^{i}=1 - x_{t}^{i+1} - x_{t}^{i} + 2 x_{t}^{i+1} x_{t}^{i} - x_{t}^{i-1} + 2 x_{t}^{i+1} x_{t}^{i-1} + x_{t}^{i} x_{t}^{i-1} -$\\
&& $\quad \qquad \qquad - 2 x_{t}^{i+1} x_{t}^{i} x_{t}^{i-1}$ \\
$^{1}226_{2}^{1}$ & \qquad (0,1,0,0,0,1,1,1) & $\qquad x_{t+1}^{i}=x_{t}^{i+1} x_{t}^{i} + x_{t}^{i-1} - x_{t}^{i} x_{t}^{i-1}$ \\
$^{1}227_{2}^{1}$ & \qquad (1,1,0,0,0,1,1,1) & $\qquad x_{t+1}^{i}=1 - x_{t}^{i+1} - x_{t}^{i} + 2 x_{t}^{i+1} x_{t}^{i} + x_{t}^{i+1} x_{t}^{i-1} - x_{t}^{i+1} x_{t}^{i} x_{t}^{i-1}$ \\
$^{1}228_{2}^{1}$ & \qquad (0,0,1,0,0,1,1,1) & $\qquad x_{t+1}^{i}=x_{t}^{i} + x_{t}^{i+1} x_{t}^{i-1} - x_{t}^{i} x_{t}^{i-1}$ \\
$^{1}229_{2}^{1}$ & \qquad (1,0,1,0,0,1,1,1) & $\qquad x_{t+1}^{i}=1 - x_{t}^{i+1} + x_{t}^{i+1} x_{t}^{i} - x_{t}^{i-1} + 2 x_{t}^{i+1} x_{t}^{i-1} - x_{t}^{i+1} x_{t}^{i} x_{t}^{i-1}$ \\
$^{1}230_{2}^{1}$ & \qquad (0,1,1,0,0,1,1,1) & $\qquad x_{t+1}^{i}=x_{t}^{i} + x_{t}^{i-1} - 2 x_{t}^{i} x_{t}^{i-1} + x_{t}^{i+1} x_{t}^{i} x_{t}^{i-1}$ \\
$^{1}231_{2}^{1}$ & \qquad (1,1,1,0,0,1,1,1) & $\qquad x_{t+1}^{i}=1 - x_{t}^{i+1} + x_{t}^{i+1} x_{t}^{i} + x_{t}^{i+1} x_{t}^{i-1} - x_{t}^{i} x_{t}^{i-1}$ \\
$^{1}232_{2}^{1}$ & \qquad (0,0,0,1,0,1,1,1) & $\qquad x_{t+1}^{i}=x_{t}^{i+1} x_{t}^{i} + x_{t}^{i+1} x_{t}^{i-1} + x_{t}^{i} x_{t}^{i-1} - 2 x_{t}^{i+1} x_{t}^{i} x_{t}^{i-1}$ \\
$^{1}233_{2}^{1}$ & \qquad (1,0,0,1,0,1,1,1) & $\qquad x_{t+1}^{i}=1 - x_{t}^{i+1} - x_{t}^{i} + 2 x_{t}^{i+1} x_{t}^{i} - x_{t}^{i-1} + 2 x_{t}^{i+1} x_{t}^{i-1} +$\\
&& $\quad \qquad \qquad + 2 x_{t}^{i} x_{t}^{i-1} - 3 x_{t}^{i+1} x_{t}^{i} x_{t}^{i-1}$ \\
$^{1}234_{2}^{1}$ & \qquad (0,1,0,1,0,1,1,1)  & $\qquad x_{t+1}^{i}=x_{t}^{i+1} x_{t}^{i} + x_{t}^{i-1} - x_{t}^{i+1} x_{t}^{i} x_{t}^{i-1}$ \\
\end{tabular}
\end{center}
Table S6 (cont.)
\end{table}

\begin{table}[htdp]
\begin{center}
\begin{tabular}{cclcccl}
Rule & $ \qquad (a_{0},a_{1},a_{2},a_{3},a_{4},a_{5},a_{6},a_{7})$ & \qquad Map \\
$^{1}235_{2}^{1}$ & \qquad (1,1,0,1,0,1,1,1)  & $\qquad x_{t+1}^{i}=1 - x_{t}^{i+1} - x_{t}^{i} + 2 x_{t}^{i+1} x_{t}^{i} + x_{t}^{i+1} x_{t}^{i-1} + x_{t}^{i} x_{t}^{i-1} -$\\
&& $\quad \qquad \qquad - 2 x_{t}^{i+1} x_{t}^{i} x_{t}^{i-1}$ \\
$^{1}236_{2}^{1}$ & \qquad (0,0,1,1,0,1,1,1)  & $\qquad x_{t+1}^{i}=x_{t}^{i} + x_{t}^{i+1} x_{t}^{i-1} - x_{t}^{i+1} x_{t}^{i} x_{t}^{i-1}$ \\
$^{1}237_{2}^{1}$ & \qquad (1,0,1,1,0,1,1,1)  & $\qquad x_{t+1}^{i}=1 - x_{t}^{i+1} + x_{t}^{i+1} x_{t}^{i} - x_{t}^{i-1} + 2 x_{t}^{i+1} x_{t}^{i-1} + x_{t}^{i} x_{t}^{i-1} -$\\
&& $\quad \qquad \qquad - 2 x_{t}^{i+1} x_{t}^{i} x_{t}^{i-1}$ \\
$^{1}238_{2}^{1}$ & \qquad (0,1,1,1,0,1,1,1)  & $\qquad x_{t+1}^{i}=x_{t}^{i} + x_{t}^{i-1} - x_{t}^{i} x_{t}^{i-1}$ \\
$^{1}239_{2}^{1}$ & \qquad (1,1,1,1,0,1,1,1)  & $\qquad x_{t+1}^{i}=1 - x_{t}^{i+1} + x_{t}^{i+1} x_{t}^{i} + x_{t}^{i+1} x_{t}^{i-1} - x_{t}^{i+1} x_{t}^{i} x_{t}^{i-1}$ \\

$^{1}240_{2}^{1}$ & \qquad (0,0,0,0,1,1,1,1)  & $\qquad x_{t+1}^{i}=x_{t}^{i+1}$ \\
$^{1}241_{2}^{1}$ & \qquad (1,0,0,0,1,1,1,1) & $\qquad x_{t+1}^{i}=1 - x_{t}^{i} + x_{t}^{i+1} x_{t}^{i} - x_{t}^{i-1} + x_{t}^{i+1} x_{t}^{i-1} + x_{t}^{i} x_{t}^{i-1} -$\\
&& $\quad \qquad \qquad - x_{t}^{i+1} x_{t}^{i} x_{t}^{i-1}$ \\
$^{1}242_{2}^{1}$ & \qquad (0,1,0,0,1,1,1,1) & $\qquad x_{t+1}^{i}=x_{t}^{i+1} + x_{t}^{i-1} - x_{t}^{i+1} x_{t}^{i-1} - x_{t}^{i} x_{t}^{i-1} + x_{t}^{i+1} x_{t}^{i} x_{t}^{i-1}$ \\
$^{1}243_{2}^{1}$ & \qquad (1,1,0,0,1,1,1,1) & $\qquad x_{t+1}^{i}=1 - x_{t}^{i} + x_{t}^{i+1} x_{t}^{i}$ \\
$^{1}244_{2}^{1}$ & \qquad (0,0,1,0,1,1,1,1) & $\qquad x_{t+1}^{i}=x_{t}^{i+1} + x_{t}^{i} - x_{t}^{i+1} x_{t}^{i} - x_{t}^{i} x_{t}^{i-1} + x_{t}^{i+1} x_{t}^{i} x_{t}^{i-1}$\\
$^{1}245_{2}^{1}$ & \qquad (1,0,1,0,1,1,1,1) & $\qquad x_{t+1}^{i}=1 - x_{t}^{i-1} + x_{t}^{i+1} x_{t}^{i-1}$ \\
$^{1}246_{2}^{1}$ & \qquad (0,1,1,0,1,1,1,1) & $\qquad x_{t+1}^{i}=x_{t}^{i+1} + x_{t}^{i} - x_{t}^{i+1} x_{t}^{i} + x_{t}^{i-1} - x_{t}^{i+1} x_{t}^{i-1} -$\\
&& $\quad \qquad \qquad - 2 x_{t}^{i} x_{t}^{i-1} + 2 x_{t}^{i+1} x_{t}^{i} x_{t}^{i-1}$ \\
$^{1}247_{2}^{1}$ & \qquad (1,1,1,0,1,1,1,1) & $\qquad x_{t+1}^{i}=1 - x_{t}^{i} x_{t}^{i-1} + x_{t}^{i+1} x_{t}^{i} x_{t}^{i-1}$ \\
$^{1}248_{2}^{1}$ & \qquad (0,0,0,1,1,1,1,1) & $\qquad x_{t+1}^{i}=x_{t}^{i+1} + x_{t}^{i} x_{t}^{i-1} - x_{t}^{i+1} x_{t}^{i} x_{t}^{i-1}$ \\
$^{1}249_{2}^{1}$ & \qquad (1,0,0,1,1,1,1,1) & $\qquad x_{t+1}^{i}=1 - x_{t}^{i} + x_{t}^{i+1} x_{t}^{i} - x_{t}^{i-1} + x_{t}^{i+1} x_{t}^{i-1}+ $\\
&& $\quad \qquad \qquad + 2 x_{t}^{i} x_{t}^{i-1} - 2 x_{t}^{i+1} x_{t}^{i} x_{t}^{i-1}$ \\
$^{1}250_{2}^{1}$ & \qquad (0,1,0,1,1,1,1,1) & $\qquad x_{t+1}^{i}=x_{t}^{i+1} + x_{t}^{i-1} - x_{t}^{i+1} x_{t}^{i-1}$ \\
$^{1}251_{2}^{1}$ & \qquad (1,1,0,1,1,1,1,1) & $\qquad x_{t+1}^{i}=1 - x_{t}^{i} + x_{t}^{i+1} x_{t}^{i} + x_{t}^{i} x_{t}^{i-1} - x_{t}^{i+1} x_{t}^{i} x_{t}^{i-1}$ \\
$^{1}252_{2}^{1}$ & \qquad (0,0,1,1,1,1,1,1) & $\qquad x_{t+1}^{i}=x_{t}^{i+1} + x_{t}^{i} - x_{t}^{i+1} x_{t}^{i}$ \\
$^{1}253_{2}^{1}$ & \qquad (1,0,1,1,1,1,1,1) & $\qquad x_{t+1}^{i}=1 - x_{t}^{i-1} + x_{t}^{i+1} x_{t}^{i-1} + x_{t}^{i} x_{t}^{i-1} - x_{t}^{i+1} x_{t}^{i} x_{t}^{i-1}$ \\
$^{1}254_{2}^{1}$ & \qquad (0,1,1,1,1,1,1,1) & $\qquad x_{t+1}^{i}=x_{t}^{i+1} + x_{t}^{i} - x_{t}^{i+1} x_{t}^{i} + x_{t}^{i-1} - x_{t}^{i+1} x_{t}^{i-1} - x_{t}^{i} x_{t}^{i-1} +$\\
&& $\quad \qquad \qquad + x_{t}^{i+1} x_{t}^{i} x_{t}^{i-1}$ \\
$^{1}255_{2}^{1}$ & \qquad (1,1,1,1,1,1,1,1) & $\qquad x_{t+1}^{i}=1$ \\
\end{tabular}
\end{center}
\end{table}

\begin{figure}
\begin{center}
\includegraphics[width=0.68 \textwidth]{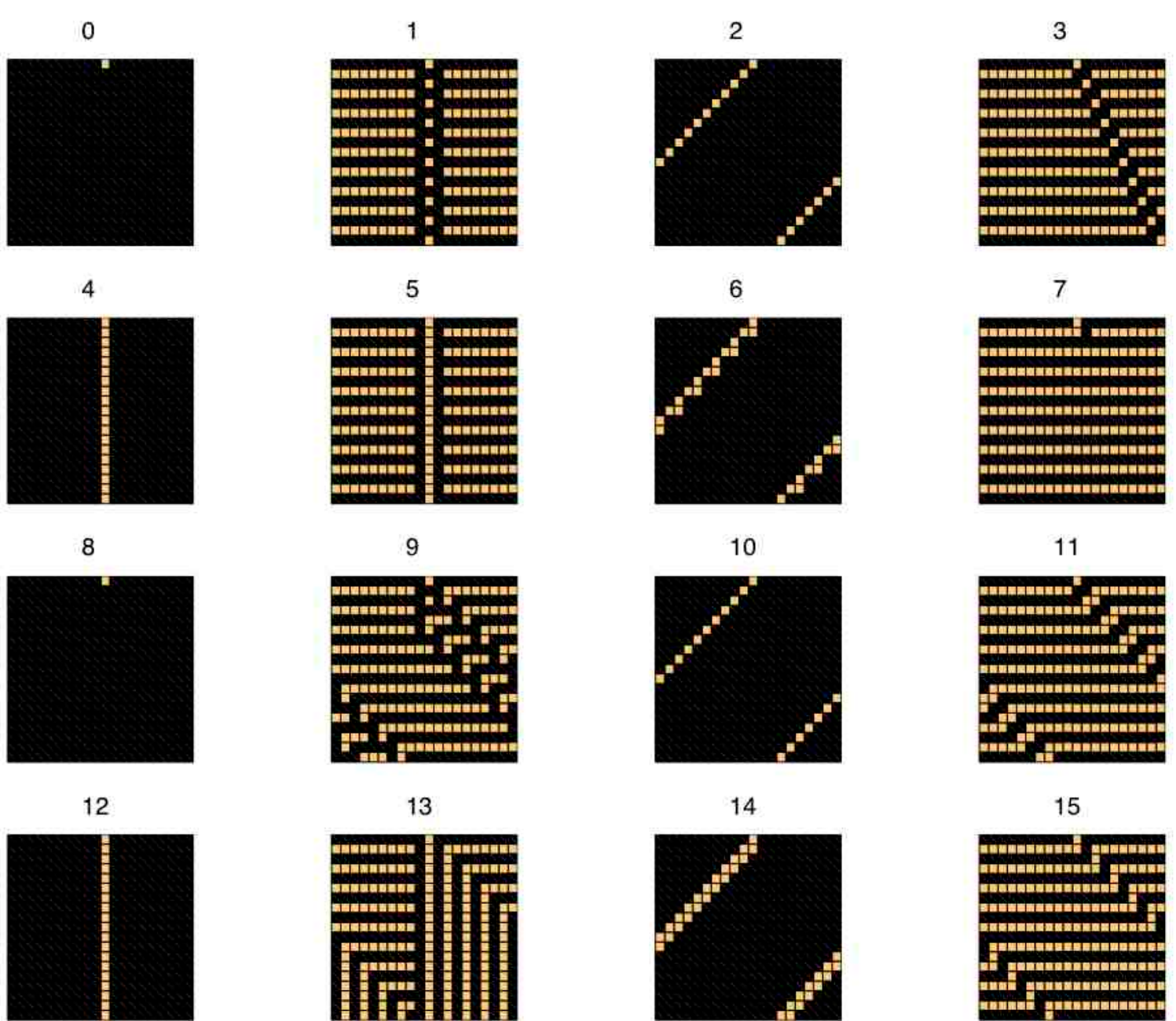}
~\\ ~\\
\includegraphics[width=0.68 \textwidth]{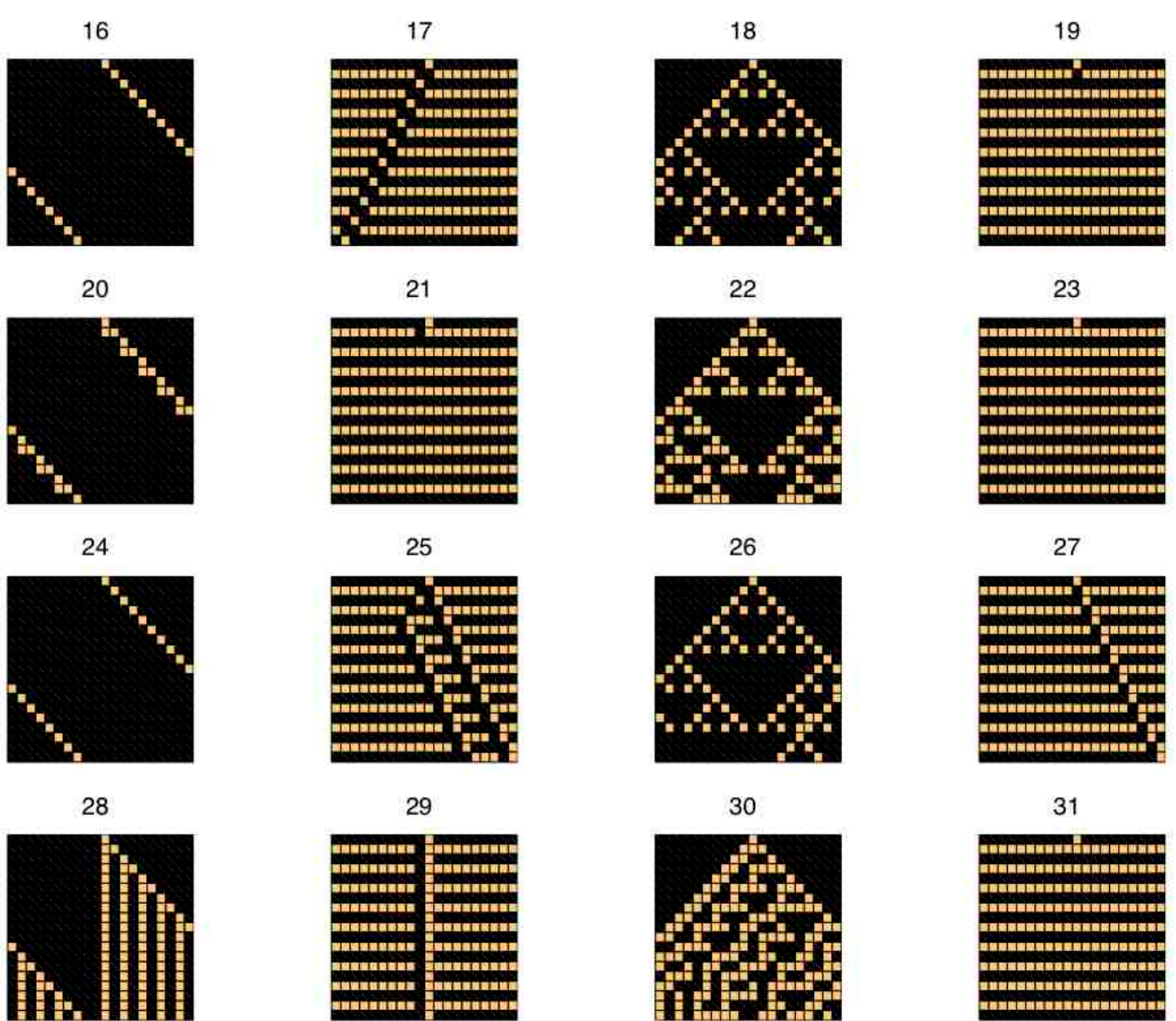}
\end{center}
Figure S16: Spatiotemporal evolution of the 256 Wolfram rules $^{1}R_{2}^{1}$ ($R$ indicated over each panel) 
\end{figure}

\begin{figure}
\begin{center}
\includegraphics[width=0.68 \textwidth]{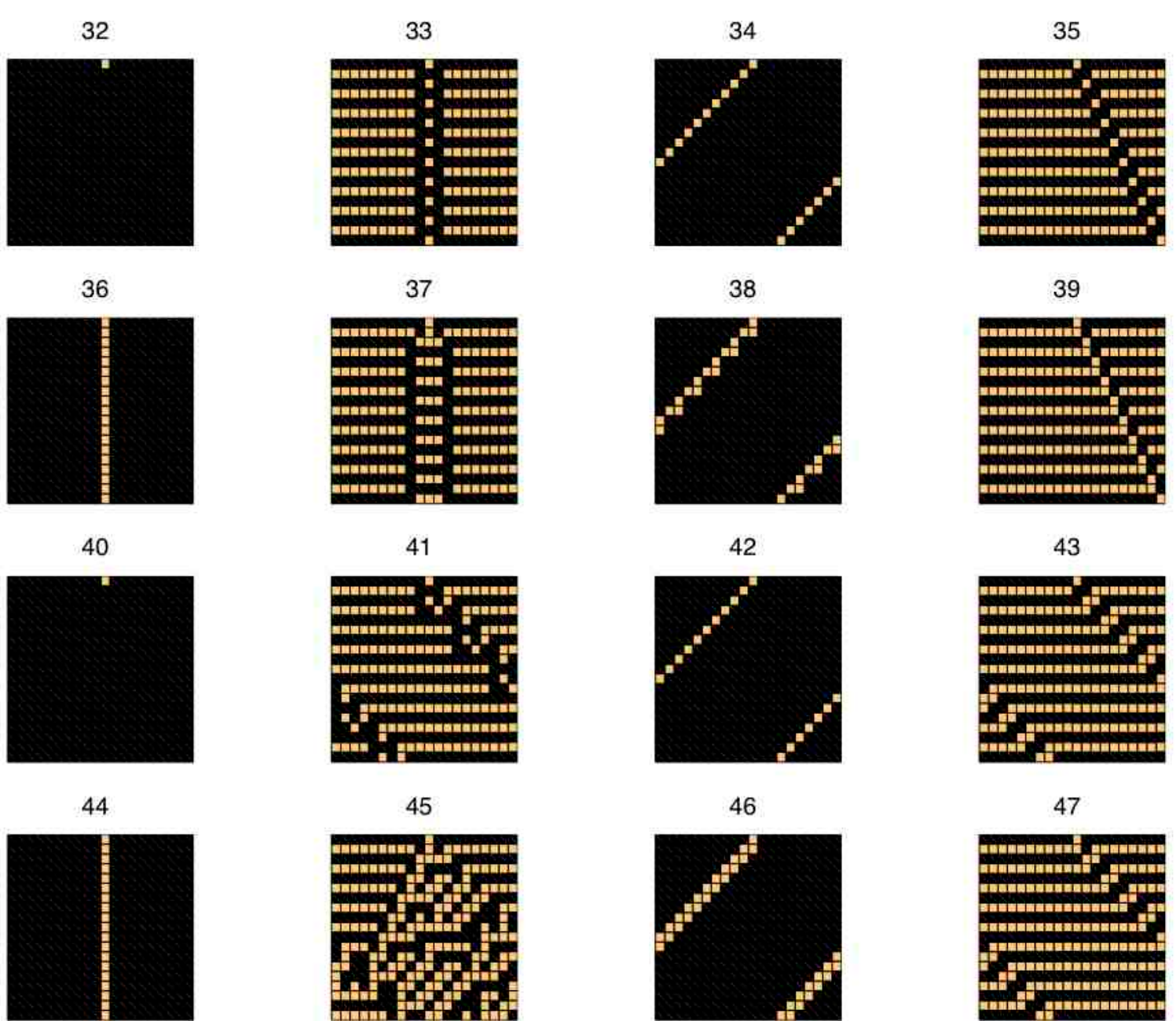}
~\\ ~\\
\includegraphics[width=0.68 \textwidth]{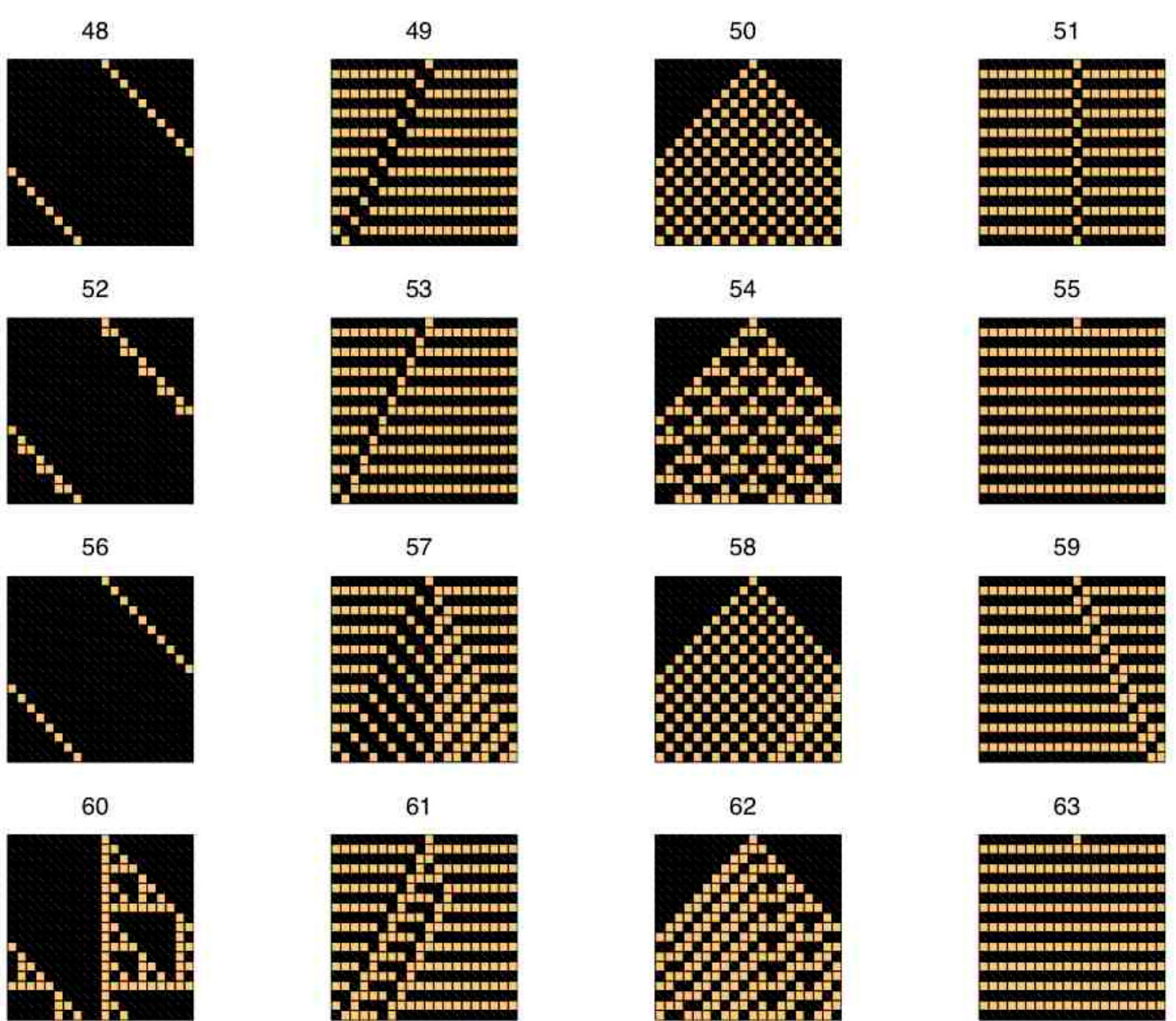}
\end{center}
Figure S16 (cont.): Spatiotemporal evolution of the 256 Wolfram rules $^{1}R_{2}^{1}$ 
\end{figure}

\begin{figure}
\begin{center}
\includegraphics[width=0.68 \textwidth]{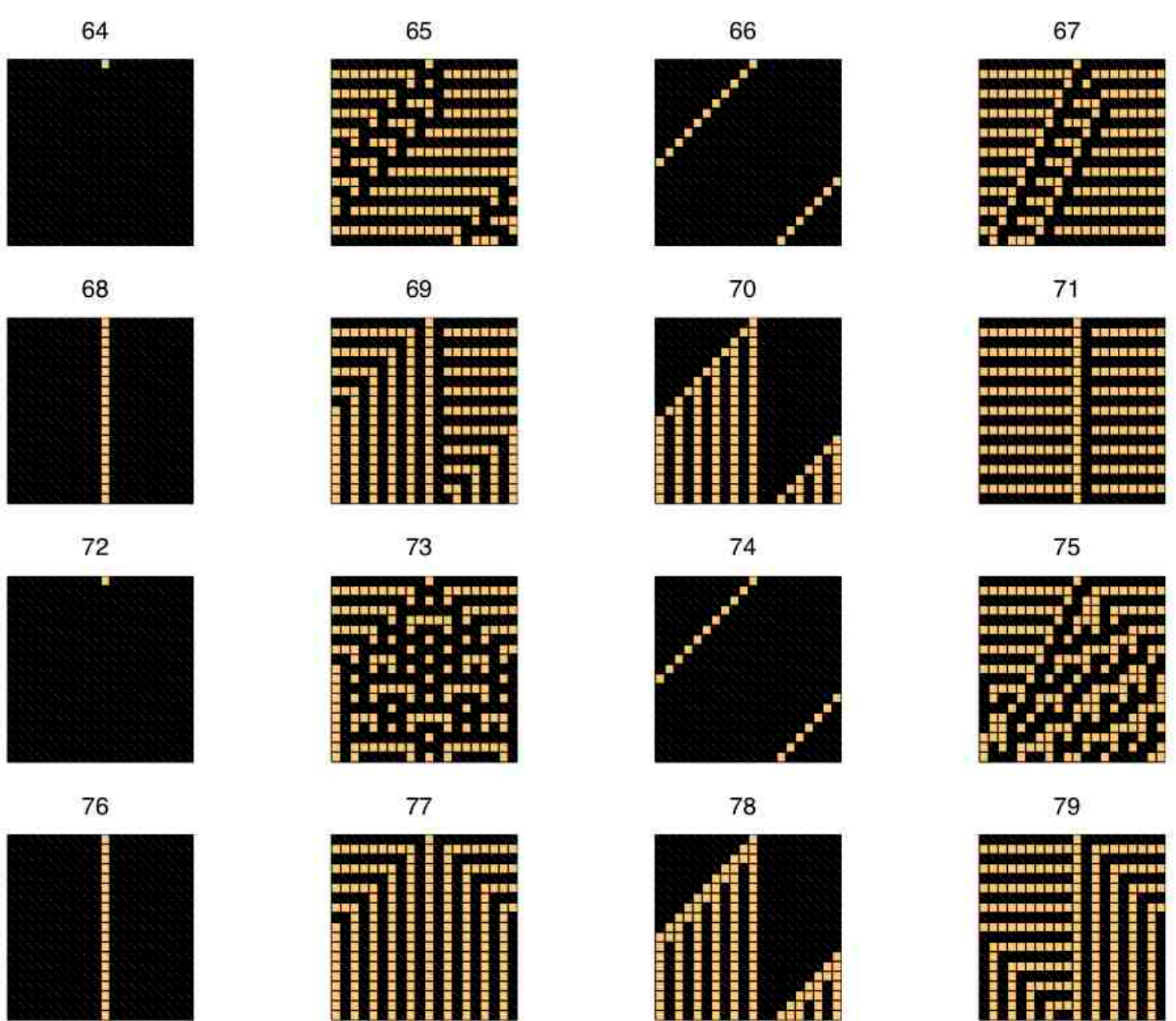}
~\\ ~\\
\includegraphics[width=0.68 \textwidth]{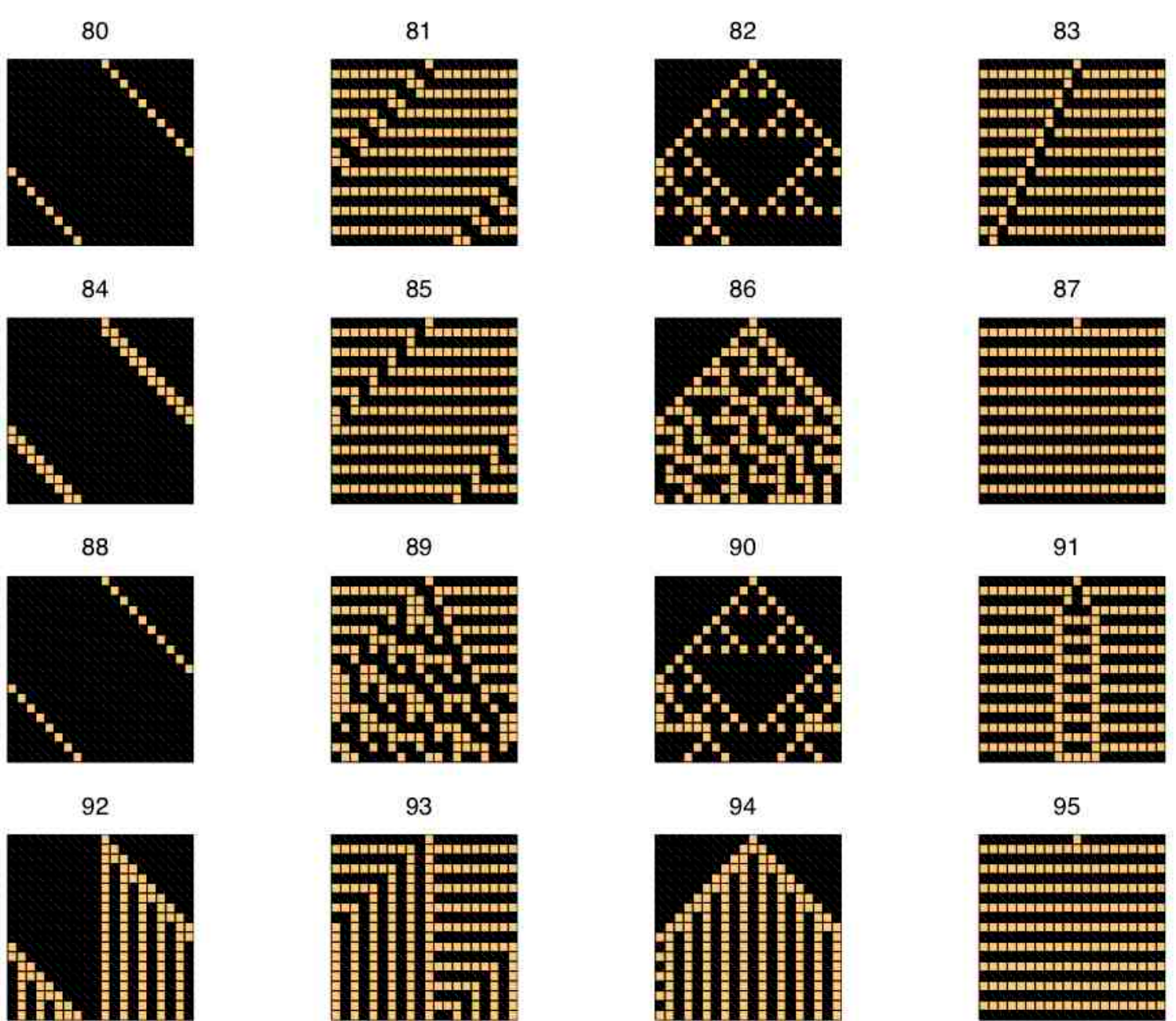}
\end{center}
Figure S16 (cont.): Spatiotemporal evolution of the 256 Wolfram rules $^{1}R_{2}^{1}$ 
\end{figure}

\begin{figure}
\begin{center}
\includegraphics[width=0.68 \textwidth]{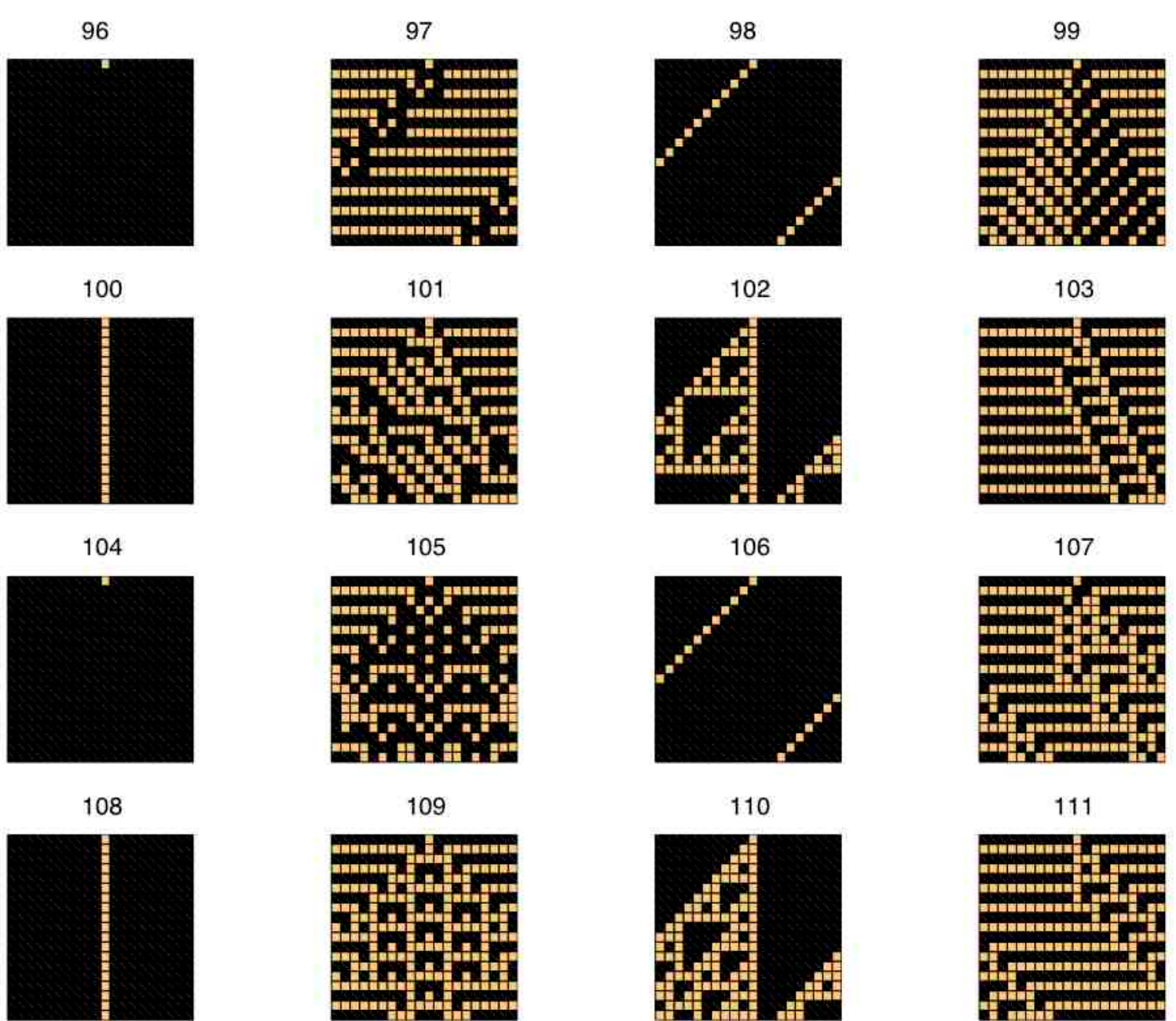}
~\\ ~\\
\includegraphics[width=0.68 \textwidth]{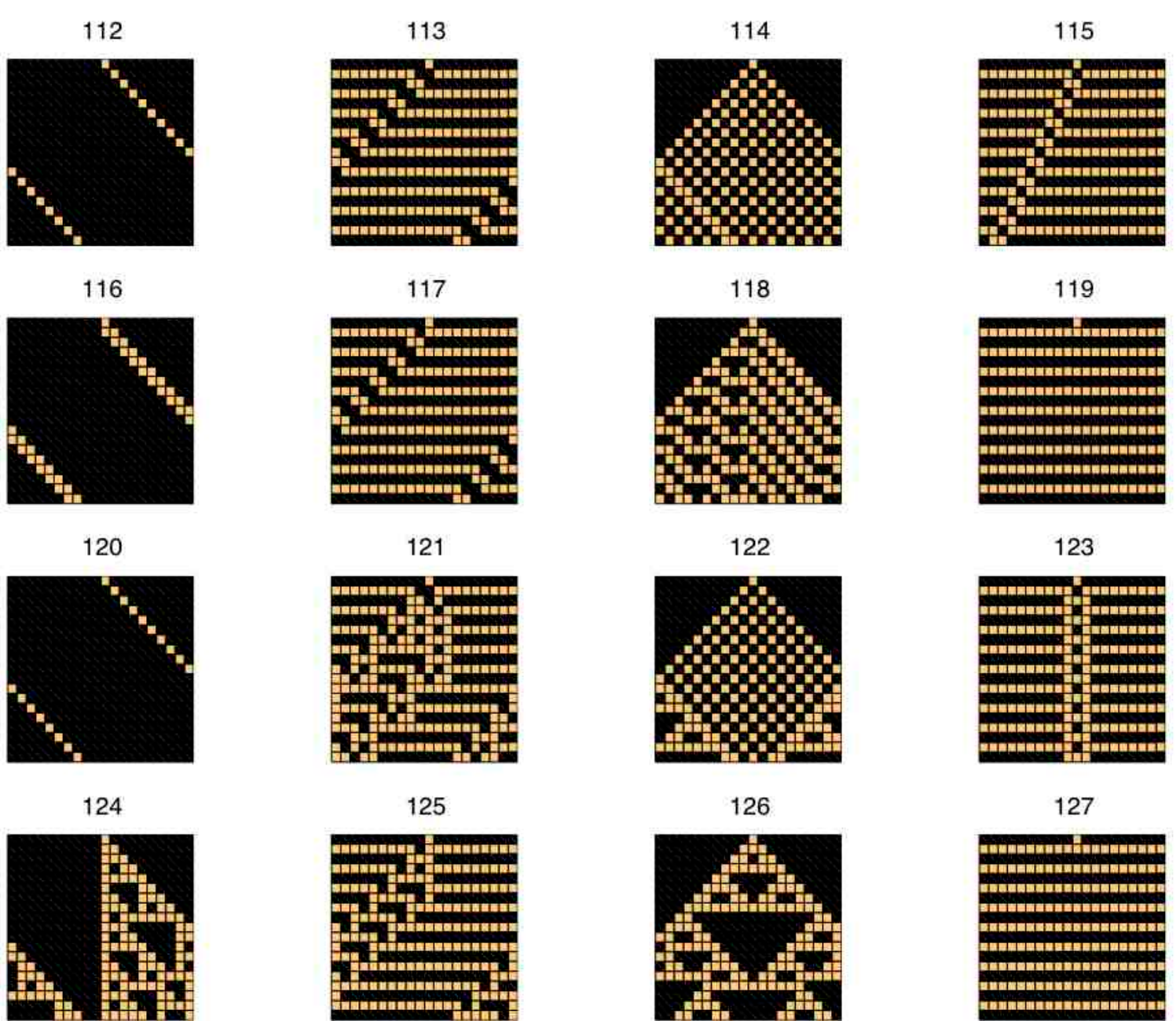}
\end{center}
Figure S16 (cont.): Spatiotemporal evolution of the 256 Wolfram rules $^{1}R_{2}^{1}$ 
\end{figure}

\begin{figure}
\begin{center}
\includegraphics[width=0.68 \textwidth]{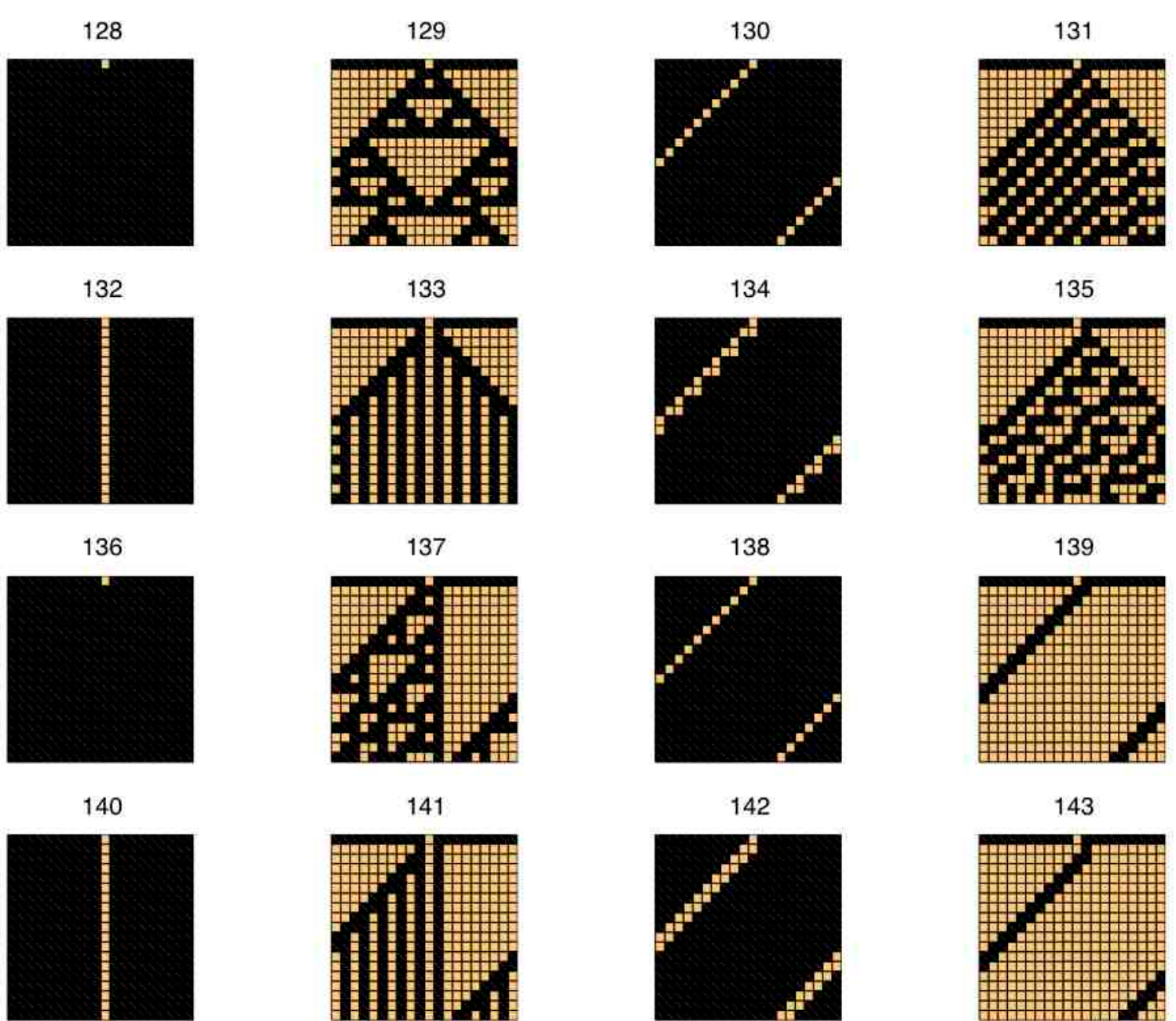}
~\\ ~\\
\includegraphics[width=0.68 \textwidth]{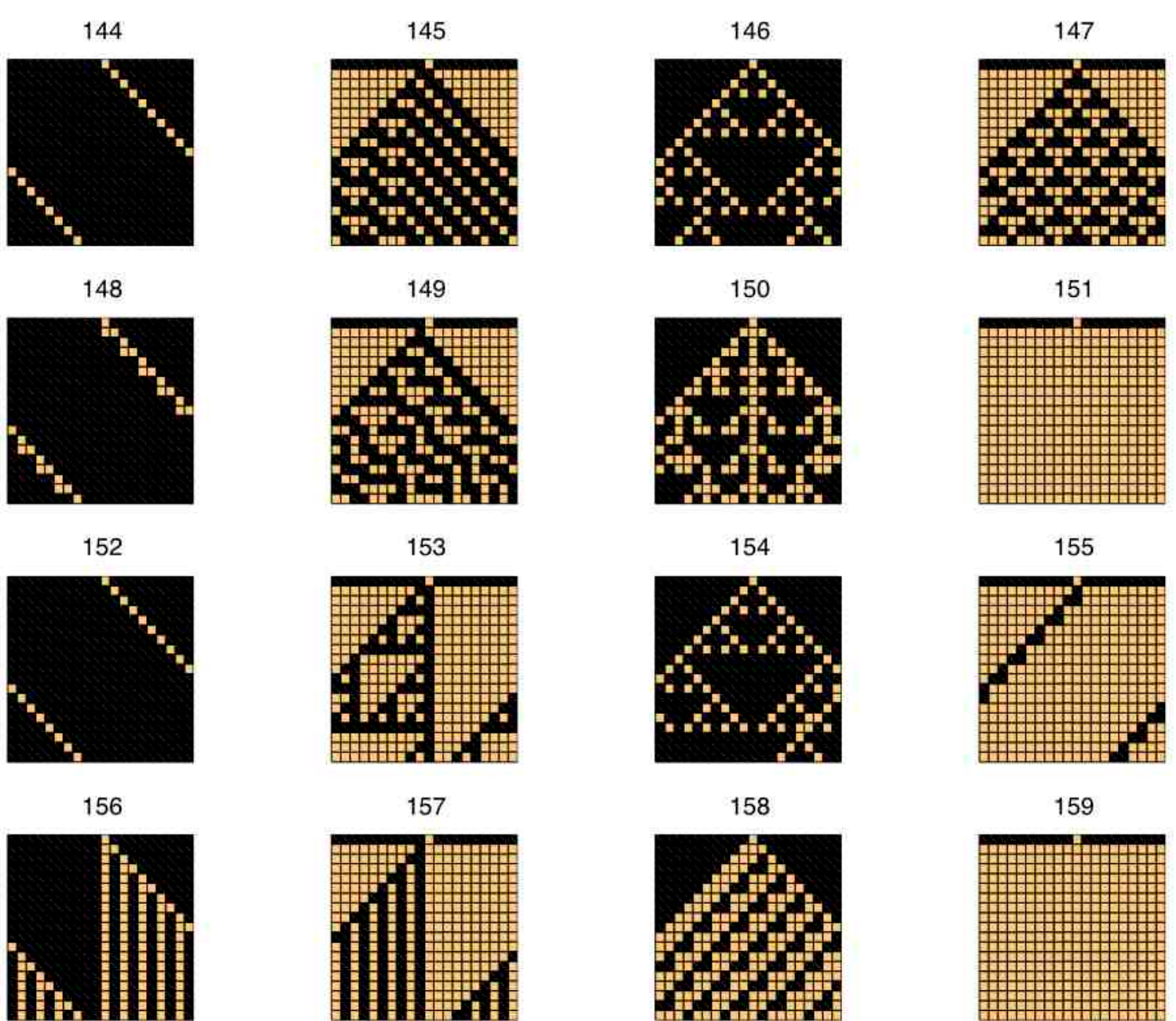}
\end{center}
Figure S16 (cont.): Spatiotemporal evolution of the 256 Wolfram rules $^{1}R_{2}^{1}$ 
\end{figure}

\begin{figure}
\begin{center}
\includegraphics[width=0.68 \textwidth]{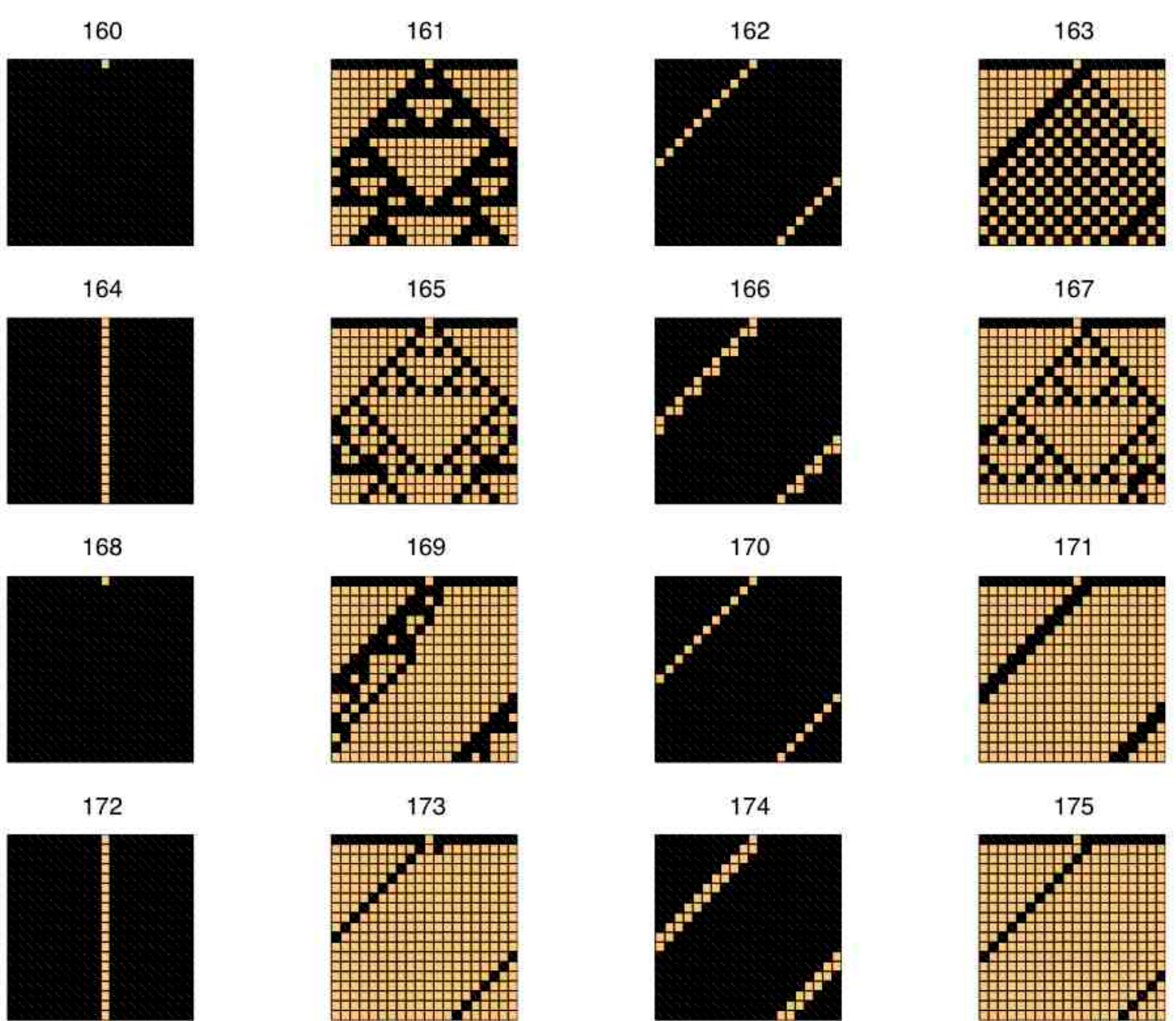}
~\\ ~\\
\includegraphics[width=0.68 \textwidth]{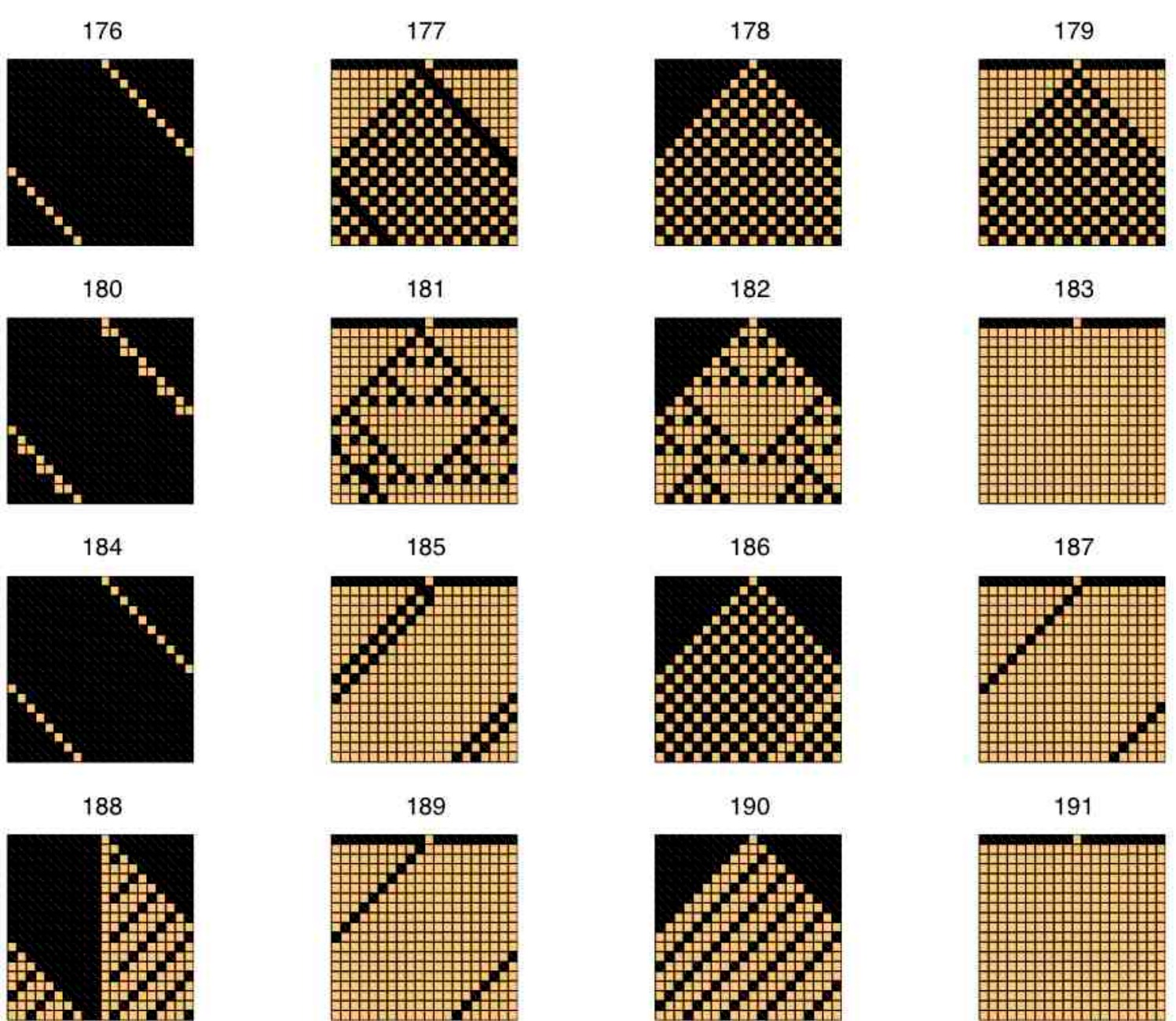}
\end{center}
Figure S16 (cont.): Spatiotemporal evolution of the 256 Wolfram rules $^{1}R_{2}^{1}$ 
\end{figure}

\begin{figure}
\begin{center}
\includegraphics[width=0.68 \textwidth]{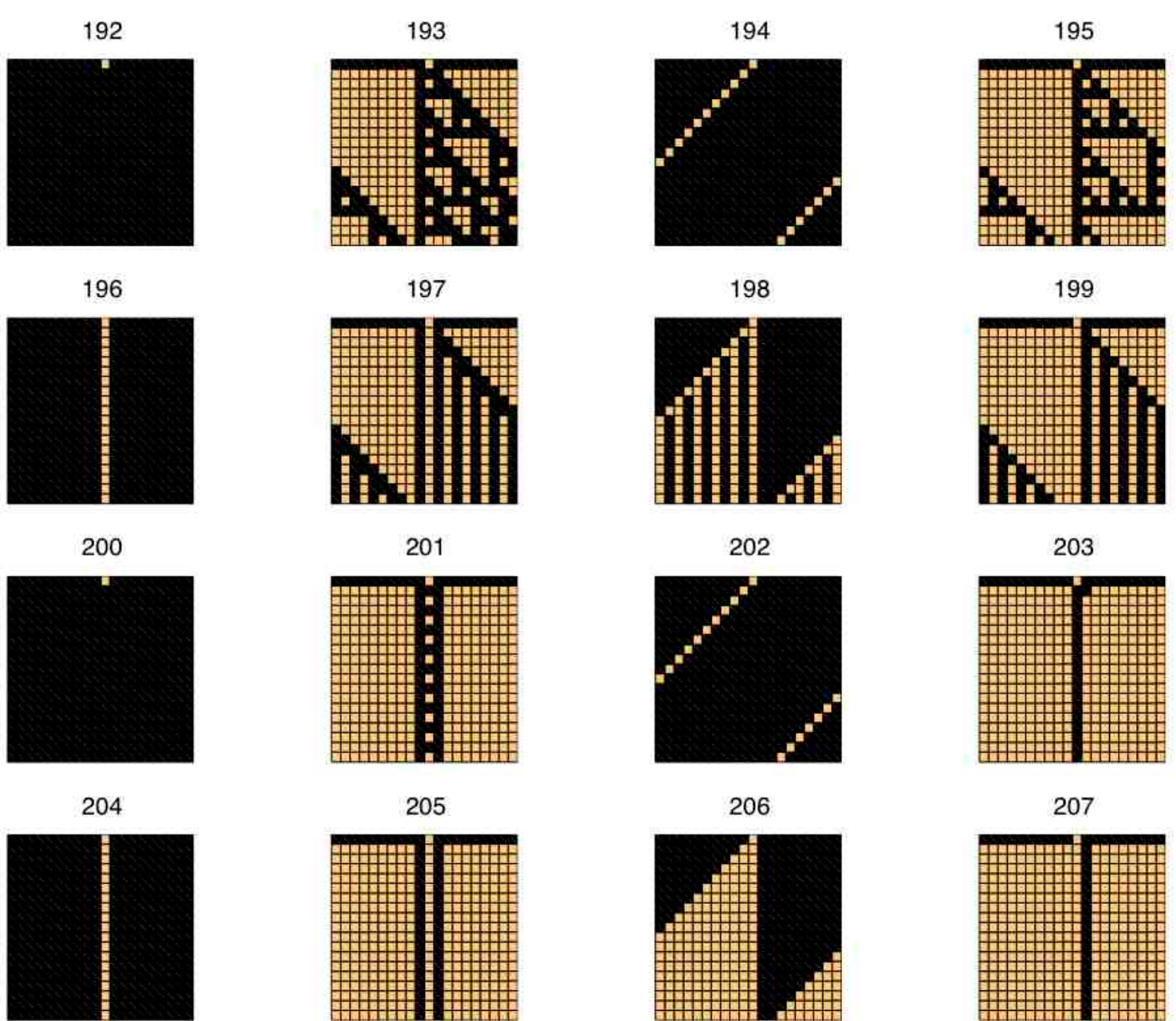}
~\\ ~\\
\includegraphics[width=0.68 \textwidth]{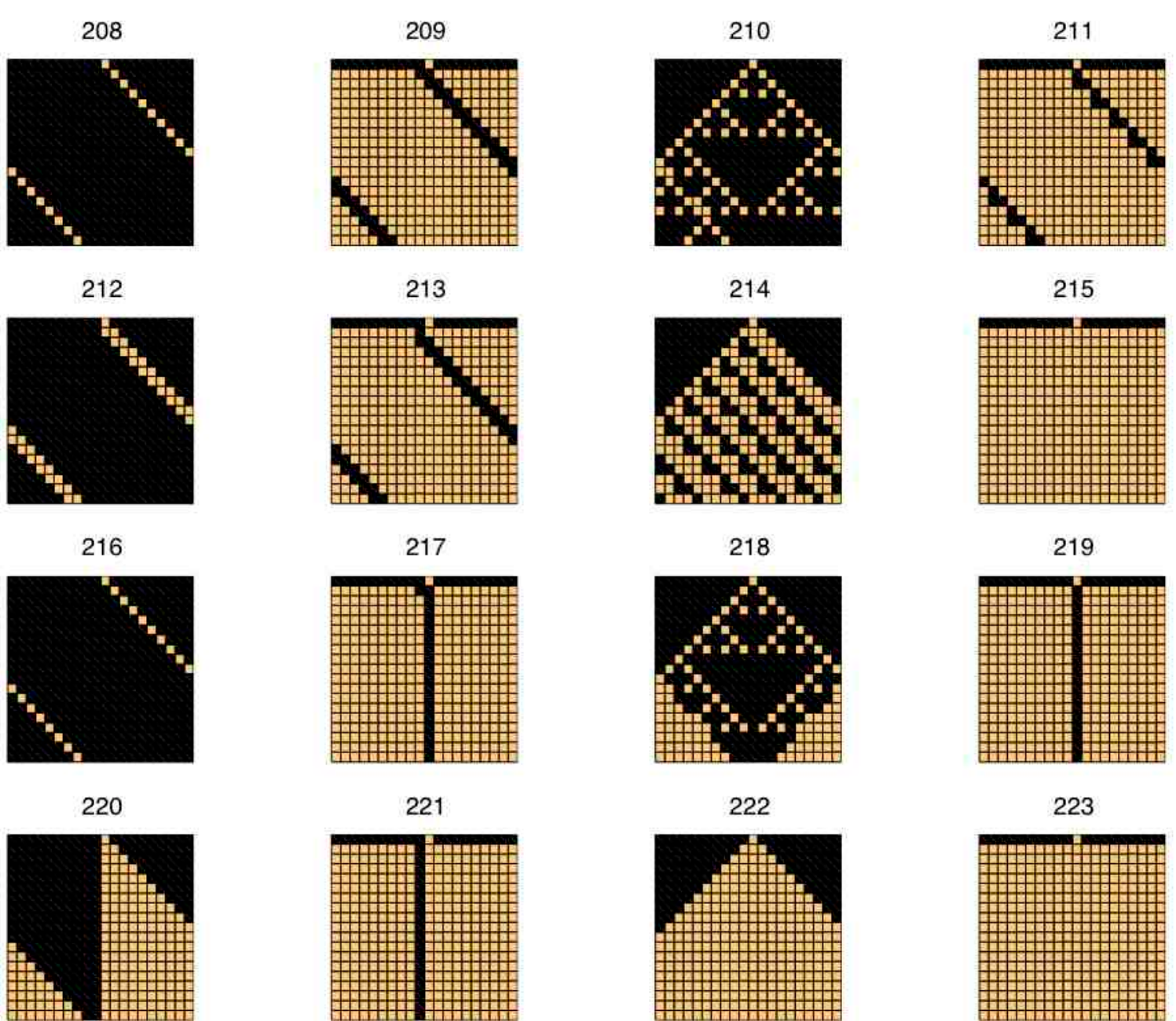}
\end{center}
Figure S16 (cont.): Spatiotemporal evolution of the 256 Wolfram rules $^{1}R_{2}^{1}$ 
\end{figure}
\begin{figure}
\begin{center}
\includegraphics[width=0.68 \textwidth]{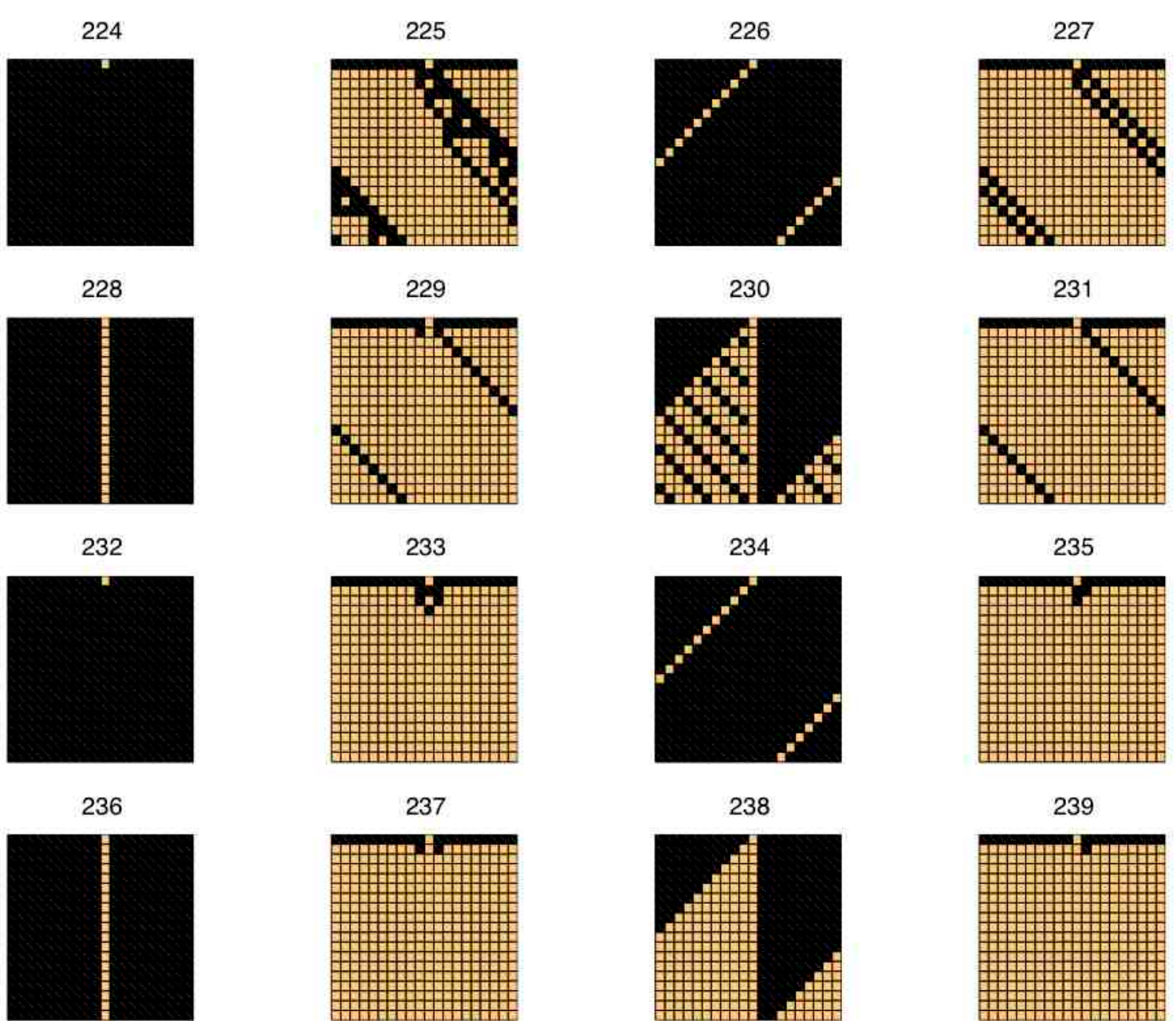}
~\\ ~\\
\includegraphics[width=0.68 \textwidth]{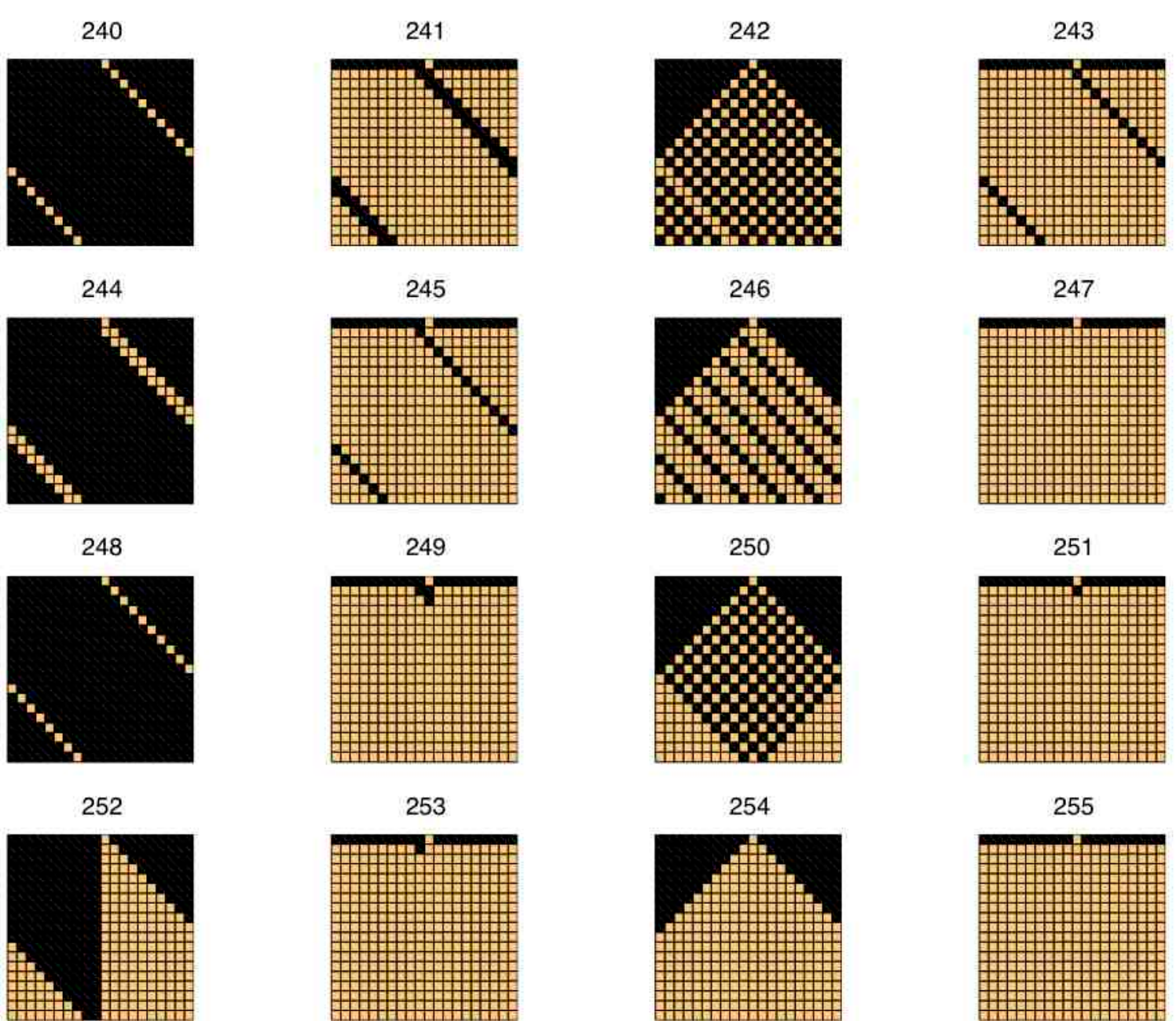}
\end{center}
Figure S16 (cont.): Spatiotemporal evolution of the 256 Wolfram rules $^{1}R_{2}^{1}$ 
\end{figure}


\begin{thebibliography}{}
\bibitem{Wolfram1} S. Wolfram, {\it A New Kind of Science} (Wolfram Media Inc., Champaign, IL, 2002).
\bibitem{WolframB} S. Wolfram, {\it Cellular Automata and Complexity: Collected Papers} (Addison-Wesley, Reading MA, 1994).
\bibitem{PhysicaD} Physica (Amsterdam) D issues No. 10 and No. 45 are devoted to CA.
\bibitem{Wolfram2}
S. Wolfram, Nature {\bf 311}, 419 (1984).
\bibitem{Wolfram3}
 S. Wolfram, Rev. Mod. Phys. {\bf 55}, 601 (1983).
\bibitem{Wolfram4} S. Wolfram, Physica D {\bf 10}, 1 (1984).
\bibitem{Gerhardt} M. Gerhardt, H. Schuster, J. J. Tyson, Science {\bf 247}, 1563 (1990).
\bibitem{Neumann}
 J. von Neumann, in: A. W. Burks (Ed.) {\it Theory of Self-Reproducing Automata} (University of Illinois Press, Champaign, IL, 1966).
\bibitem{Codd}
E. F. Codd, {\it Cellular Automata} (Academic Press, New York, 1968).
\bibitem{Wolfram5} S. Wolfram, Phys. Rev. Lett. {\bf 54}, 735 (1985).
\bibitem{Wolfram6} S. Wolfram, Phys. Rev. Lett. {\bf 55}, 449 (1985).
\bibitem{Wuensche}
A. Wuensche and M. Lesser, {\it The Global Dynamics of Cellular Automata} (Addison-Wesley, Reading, MA, 1992).
\bibitem{Adamatzky}
A. Adamatzky, {\it Identification of Cellular Automata} (Taylor and Francis, London, 1994).
\bibitem{Chua} 
L. O. Chua, {\it A Nonlinear Dynamics Perspective of Wolfram's New Kind of Science, vol. I-IV} (World Scientific, Singapore, 2006-2011).
\bibitem{Langton} 
C. G. Langton, Physica D {\bf 12}, 42 (1990).
\bibitem{Li}
W. Li, N. H. Packard, C. G. Langton, Physica D {\bf 45}, 77 (1990).
\bibitem{Toffoli} 
T. Toffoli, Physica D {\bf 10}, 117 (1984).
\bibitem{Ilachinski} 
A. Ilachinski, {\it Cellular Automata: A Discrete Universe} (World Scientific, Singapore, 2001).
\bibitem{Griffeath} 
D. Griffeath, C. Moore, eds. {\it New Constructions in Cellular Automata} (Oxford University Press, New York, 2003).
\bibitem{group}  T. Ceccherini-Silberstein, M. Coornaert, {\it Cellular Automata and Groups} (Springer Verlag, Heidelberg, 2010).
\bibitem{Mcintosh}
 H. V. McIntosh, {\it One Dimensional Cellular Automata} (Luniver Press, Frome, UK, 2009).
\bibitem{Fredkin} 
E. Fredkin, Physica D {\bf 45},
 254 (1990).
 \bibitem{supp}
 The supporting information follows in the forthcoming pages.
 \bibitem{Martin} 
 O. Martin, A. M. Odlyzko and S. Wolfram, Comm. Math. Phys. {\bf 93},
 219 (1984).
 \bibitem{Cook} 
 M. Cook, Complex Systems {\bf 15}, 1 (2004).
 \bibitem{Gardner1} 
 M. Gardner, Sci. Am. {\bf 223}, 120 (1970).
\bibitem{Gardner2} 
M. Gardner, Sci. Am. {\bf 224}, 112 (1971).
\bibitem{Goldenfeld}
N. Israeli and N. Goldenfeld, Phys. Rev. Lett. {\bf 92}, 74105 (2004).
\bibitem{rusos} 
G. M. Zaslavsky, R. Z. Sagdeev, D. A. Usikov, A. A. Chernikov, {\it Weak Chaos and Quasi-Regular Patterns} (Cambridge University Press, Cambridge, UK, 1991).
\end{thebibliography}
\end{document}